\documentclass[hyper,a4paper]{JHEP3} 

\usepackage{epsfig}
\usepackage{latexsym}
\usepackage{graphicx}
\usepackage{amsmath}
\usepackage{amsfonts}   
\usepackage{amssymb}    

\def\prl#1{Phys.~Rev.~Lett.~{\bf #1}}

\def\mt{{\ifmmode M^{eff}_T\else $M^{eff}_T$\fi}}

\def\e{\epsilon}

\def\l{\left}
\def\r{\right}

\def\ra{\rangle}

\def\ln#1{\mbox{ln}\l(#1\r)}

\def\e3{$\epsilon_3$}

\def\ch2{$\chi^2$}

\def\co#1{{\ifmmode{\cal O}_{#1}\else${\cal O}_{#1}$\fi}}

\newdimen\unit
\def\point#1 #2 #3{\vbox to0pt{\kern-#2\unit
  \hbox{\kern#1\unit#3}\vss}
 \nointerlineskip}

\newcommand{\be}{\begin{equation}}
\newcommand{\ee}{\end{equation}}
\newcommand{\bea}{\begin{eqnarray}}
\newcommand{\eea}{\end{eqnarray}}

\newcommand{\mev}{\mbox{ MeV }}
\newcommand{\gev}{\mbox{ GeV}}
\newcommand{\tev}{\mbox{ TeV}}

\newcommand{\cl}{\text{CL}}

\newcommand{\dasusy}{\delta a_{\mu}^{\text{SUSY}}}
\newcommand{\alphaemmz}{\alpha_{\text{em}}(M_Z)^{\overline{MS}}}
\newcommand{\alphas}{\alpha_s(M_Z)^{\overline{MS}}}

\newcommand{\data}{d}

\newcount\hour
\newcount\minute
\newtoks\amorpm
\hour=\time\divide\hour by60 \minute=\time{\multiply\hour by60
\global\advance\minute by- \hour}
\edef\standardtime{{\ifnum\hour<12 \global\amorpm={am}%
    \else\global\amorpm={pm}\advance\hour by-12 \fi
    \ifnum\hour=0 \hour=12 \fi
    \number\hour:\ifnum\minute<100\fi\number\minute\the\amorpm}}
\edef\militarytime{\number\hour:\ifnum\minute<100\fi\number\minute}
\def\bold#1{\setbox0=\hbox{$#1$}%
     \kern-.025em\copy0\kern-\wd0
     \kern.05em\copy0\kern-\wd0
     \kern-.025em\raise.0433em\box0 }

\newcommand{\newc}{\newcommand}
\newc\eg{{\rm {e.g.}}}  \newc\etal{{\rm {et al.}}} \newc\ie{{\rm i.e.}}
\newc\etc{{\rm {etc}}}
\newcommand\lsim{\mathrel{\rlap{\lower4pt\hbox{\hskip1pt$\sim$}}
    \raise1pt\hbox{$<$}}}
\newcommand\gsim{\mathrel{\rlap{\lower4pt\hbox{\hskip1pt$\sim$}}
    \raise1pt\hbox{$>$}}}
\newc{\mhalf}{m_{1/2}}      \newc{\mzero}{m_0}
\newc{\tanb}{\tan\beta}
\newc{\azero}{A_0}
\newc{\at}{A_t} \newc{\ab}{A_b} \newc{\atau}{A_\tau}
\newc{\bmu}{B\mu}           \newc{\sgn}{{\rm sgn}}
\newc{\mone}{M_1}           \newc{\mtwo}{M_2}

\newc{\charone}{\chi_1^\pm} \newc{\mcharone}{m_{\chi_1^\pm}}

\newc{\hl}{h}               \newc{\mhl}{m_{\hl}}   \newc{\gammahl}{\Gamma_{\hl}}
\newc{\hh}{H}               \newc{\mhh}{m_{\hh}}   \newc{\gammahh}{\Gamma_{\hh}}
\newc{\ha}{A}               \newc{\mha}{m_{\ha}}   \newc{\gammaha}{\Gamma_{\ha}}
\newc{\hpm}{H^{\pm}}        \newc{\mhpm}{m_{\hpm}} \newc{\gammahpm}{\Gamma_{\hpm}}
\newc{\hp}{H^{+}} \newc{\mhp}{m_{\hp}} \newc{\hm}{H^{-}}
\newc{\mhm}{m_{\hm}}         
\newc{\xt}{X_{t}}           \newc{\xb}{X_{b}}

\newc{\qzero}{Q_0}          \newc{\qstop}{Q_{\widetilde t}}
\newc{\amu}{a_{\mu}}        \newc{\amususy}{a_{\mu}^{\text{SUSY}}}
\newc{\amuexpt}{a_{\mu}^{\text{expt}}}        \newc{\amusm}{a_{\mu}^{\text{SM}}}
\newc{\deltaamususy}{\delta a_{\mu}^{\text{SUSY}}}
\newc\gmtwo{(g-2)_{\mu}} \newc\deltaamu{\Delta a_{\mu}}
\newc{\msbar}{\overline{MS}} \newc{\drbar}{\overline{DR}}
\newc{\yt}{h_t} \newc{\yb}{h_b} \newc{\ytau}{h_{\tau}}

\newc{\mtop}{m_t}               \newc{\mtpole}{M_t}
\newc{\mtaupole}{m_{\tau}^{\text{pole}}}
\newc{\mtmtsmmsbar}{m_t(m_t)^{\msbar}_{{\text{SM}}}}
\newc{\mtmtsmdrbar}{m_t(m_t)^{\drbar}_{{\text{SM}}}}
\newc{\mtmtmssmdrbar}{m_t(m_t)^{\drbar}_{{\text{SUSY}}}}

\newc{\mbmbmsbar}{m_b(m_b)^{\msbar} }

\newc{\mbmbsmmsbar}{m_b(m_b)^{\msbar}_{{\text{SM}}}}
\newc{\mbmzsmmsbar}{m_b(\mz)^{\msbar}_{{\text{SM}}}}
\newc{\mbmzsmdrbar}{m_b(\mz)^{\drbar}_{{\text{SM}}}}
\newc{\mbmzmssmdrbar}{m_b(\mz)^{\drbar}_{{\text{SUSY}}}}

\newc{\mtaumzsmmsbar}{m_{\tau}(\mz)^{\msbar}_{{\text{SM}}}}
\newc{\mtaumzsmdrbar}{m_{\tau}(\mz)^{\drbar}_{{\text{SM}}}}
\newc{\mtaumzmssmdrbar}{m_{\tau}(\mz)^{\drbar}_{{\text{SUSY}}}}

\newc{\mgut}{M_{\rm GUT}}
\newc{\mplanck}{M_{\rm P}}      \newc{\mpl}{M_{\text{Pl}}}
\newc{\msusy}{M_{\rm SUSY}}      \newc{\ms}{M_{\text{S}}}
\newc{\jxf}{J({\xf})}
\newc{\jxfexact}{J_{\rm exact}({\xf})}  \newc{\jxfexp}{J_{\rm exp}({\xf})}
\newc{\VEV}[1]{\langle #1 \rangle}
\newc{\xf}{x_f}
\newc\vrel{v_{\rm rel}}
              
\newc\sell{{\widetilde e}_L}      \newc\msell{m_{\sell}}
\newc\selr{{\widetilde e}_R}      \newc\mselr{m_{\selr}}
\newc\snue{{\widetilde \nu}_e}      \newc\msnue{m_{\snue}}
\newc\snutau{{\widetilde \nu}_\tau}      \newc\msnutau{m_{\snutau}}
\newc\supl{{\widetilde u}_L}      \newc\msupl{m_{\supl}}
\newc\supr{{\widetilde u}_R}      \newc\msupr{m_{\supr}}
\newc\sdl{{\widetilde d}_L}      \newc\msdl{m_{\sdl}}
\newc\sdr{{\widetilde d}_R}      \newc\msdr{m_{\sdr}}
\newcommand\stopq{{\widetilde t}}   
\newcommand\mstopave{ {\overline m}_{\stopq}}
\newcommand\stopone{{\widetilde t}_1}   \newcommand\mstopone{m_{\stopone}}
\newcommand\stoptwo{{\widetilde t}_2}   \newcommand\mstoptwo{m_{\stoptwo}}

\newcommand\sbotq{{\widetilde b}}   
\newcommand\msbotave{ {\overline m}_{\sbotq}}

\newc\sfermion{\tilde f}  \newc\msfermion{m_{\sfermion}}
\newc\kmeter{{\rm km}}
\newc\second{{\rm sec}}
\newc{\gstar}{g_\ast}           \newc{\gsstar}{g_{s\ast}}
\newc{\geff}{g_{\rm eff}}
\newcommand\mz{m_{Z}}

\newc{\sthw}{\sin\theta_W}              \newc{\cthw}{\cos\theta_W}
\newc{\bino}{\widetilde B}              \newc{\wino}{\widetilde W_30}
\newc{\higgsinob}{{\widetilde H}^0_b}   \newc{\higgsinot}{{\widetilde H}^0_t}
\newc{\abund}{\Omega h^2}
\newc{\abundchi}{\Omega_\chi h^2}
\newc{\abundcdm}{\Omega_{\text{CDM}} h^2}
\newc{\omegam}{\Omega_{M}}       \newc{\abundm}{\Omega_{M} h^2}
\newc{\omegab}{\Omega_{b}}       \newc{\abundb}{\Omega_{b} h^2}
\newc{\omegacdm}{\Omega_{CDM}}
\newc{\omegatot}{\Omega_{TOT}}
\newc{\rhocrit}{\rho_{crit}}
\newc{\rhochi}{\rho_{\chi}}
 \newcommand\fb{\,\mbox{fb}}

\newc\BR{BR}
\newc\bsgamma{b\rightarrow s \gamma }
\newc\bxsgamma{\overline{B}\rightarrow X_{s}\gamma}
\newc\brbsgamma{\BR(\overline{B}\rightarrow X_s\gamma)}

\newcommand\brbsmumu{\BR(\overline{B}_s\to\mu^+\mu^-)}

\newcommand\bbbarmix{\overline{B}_s\mbox{-}B_s}      
\newcommand\delmbs{\Delta M_{B_s}}

\newc{\beq}{\begin{equation}}
\newc{\eeq}{\end{equation}}

\newc\stoponetwo{{\widetilde t}_{1,2}}
\newc\sbotonetwo{{\widetilde b}_{1,2}}
\newc\stauonetwo{{\widetilde \tau}_{1,2}}

\newc{\sigsip}{\sigma^{SI}_{p}} \newc{\sigsin}{\sigma^{SI}_{n}}
\newc{\sigsiN}{\sigma^{SI}_{N}}
\newc{\sigsdp}{\sigma^{SD}_{p}} \newc{\sigsdn}{\sigma^{SD}_{n}}
\newc{\sigsiA}{\sigma^{SI}_{A}}

\newc\xilim{\xi_{\rm lim}} 
\newc\tlim{t_{\rm lim}} 
\newc\zetalim{\zeta_{\rm lim}} 

\newc\zetah{\zeta_h}
\newc{\relprobone}[1]{p({#1} \vert d)}
\newc{\relprobtwo}[2]{p({#1},{#2} \vert d)}

\long\def\begincomment#1\endcomment{%
        \begingroup\sf\baselineskip12pt#1\endgroup}

\newc\AP[3]
                {{\em Ann. Phys.} {\bf #1} (#2) #3}
\newc\APJ[3]
                {{\em Astropart. J.} {\bf #1} (#2) #3}
\newc\APP[3]
                {{\em Astropart. Phys.} {\bf #1} (#2) #3}
\newc\APS[3]
                {{\em Astrophys. J. Suppl.} {\bf #1} (#2) #3}
\newc\ARNPS[3]
                {{\em Ann. Rev. Nucl. Part. Sci.} {\bf C#1} (#2) #3}
\newc\CPC[3]
                {{\em Comp. Phys. Comm.} {\bf C#1} (#2) #3}
\newc\EPJ[3]
                {{\em Eur. Phys. J.} {\bf C#1} (#2) #3}
\newc\JCAP[3]
                {{\em JCAP} {\bf #1} (#2) #3}
\newc\JHEP[3]
                {{\em JHEP} {\bf #1} (#2) #3}
\newc\IJMP[3]
                {{\em Int. J. Mod. Phys.} {\bf A#1} (#2) #3}
\newc\MNRAS[3]
                 {{\em Mon. Not. Roy. Astron. Soc.} {\bf #1} (#2) #3}
\newc\MPL[3]
                {{\em Mod. Phys. Lett.} {\bf A#1} (#2) #3}
\newc\NCA[3]
                {{\em Nuovo Cimento} {\bf #1} (#2) #3}
\newc\NIM[3]
                {{\em Nucl. Instrum. Methods} {\bf #1} (#2) #3}
\newc\NIMA[3]
                {{\em Nucl. Instrum. Methods} {\bf A#1} (#2) #3}
\newc\NAT[3]
                {{\em Nature} {\bf #1} (#2) #3}
\newc\NPB[3]
                {{\em Nucl. Phys.} {\bf B#1} (#2) #3}
\newc\PL[3]
                {{\em Phys. Lett.} {\bf B#1} (#2) #3}
\newc\PLB[3]
                {{\em Phys. Lett.} {\bf B#1} (#2) #3}
\newc\PR[3]
                {{\em Phys. Rep.} {\bf #1} (#2) #3}
\newc\PRL[3]
                {{\em Phys. Rev. Lett.} {\bf B#1} (#2) #3}
\newc\PRD[3]
                {{\em Phys. Rev.} {\bf D#1} (#2) #3}
\newc\PTP[3]
                {{\em Prog. Theor. Phys.} {\bf #1} (#2) #3}
\newc\PPNP[3] 
                {{\em Prog. Part. Nucl. Phys.} {\bf #1} (#2) #3}
\newc\RMP[3]
                {{\em Rev. Mod. Phys.} {\bf #1} (#2) #3 }
\newc\RPP[3]
                {{\em Rept. Prog. Phys.} {\bf #1} (#2) #3 }
\newc\SC[3]
                {{\em Science} {\bf #1} (#2) #3 }
\newc\ZPC[3]
                {{\em Z. Phys.} {\bf C#1} (#2) #3}
\newc\Err[3]
           {{\em Erratum-ibid.} {\bf #1} (#2) #3 }

\newcommand{\squishlist}{
   \begin{list}{$\bullet$}
    { \setlength{\itemsep}{0pt}      \setlength{\parsep}{3pt}
      \setlength{\topsep}{3pt}       \setlength{\partopsep}{0pt}
      \setlength{\leftmargin}{1.em} \setlength{\labelwidth}{1em}
      \setlength{\labelsep}{0.5em} } }
\newcommand{\squishend}{
    \end{list}  }
        \newcommand\mW{m_{W}}

\newcommand{\tauptaum}{\tau^+\tau^-}

\newcommand{\qqbar}{q\bar{q}} \newcommand{\ppbar}{p\bar{p}}
\newcommand{\bbbar}{b\bar{b}} 
\newcommand{\ffbar}{f\bar{f}} 

\newcommand{\alphaeff}{\alpha_{\text{eff}}}

\newcommand{\nuis}{\psi}
\newcommand{\params}{\theta}
\newcommand{\basis}{m}
\newcommand{\derived}{\xi}

\newcommand{\sineff}{\sin^2 \theta_{\text{eff}}}

\title{On the detectability of the CMSSM light Higgs boson at the Tevatron}
\author{Leszek Roszkowski\\
        Department of Physics and Astronomy, University of Sheffield,\\
        Sheffield S3 7RH, England, and\\
Theory Division, CERN, CH-1211 Geneva 23, Switzerland\\
        E-mail: \email{L.Roszkowski@sheffield.ac.uk}}
\author{Roberto Ruiz de Austri\\
        Departamento de F\'{\i}sica Te\'{o}rica C-XI
        and Instituto de F\'{\i}sica Te\'{o}rica C-XVI,\\
        Universidad Aut\'{o}noma de Madrid, Cantoblanco,
        28049 Madrid, Spain\\
        E-mail: \email{rruiz@delta.ft.uam.es}}
\author{Roberto Trotta\\
        Astrophysics Department, Oxford University \\
        Denys Wilkinson Building,  Keble Road, Oxford OX1 3RH, United Kingdom\\
         E-mail: \email{rxt@astro.ox.ac.uk}}

\abstract{\small We examine the prospects of detecting the light Higgs
  boson $\hl^0$ of the Constrained MSSM at the Tevatron. To this end
  we explore the CMSSM parameter space with $\mu>0$, using a Markov
  Chain Monte Carlo technique, and apply all relevant collider and
  cosmological constraints including their uncertainties, as well as
  those of the Standard Model parameters.  Taking $50\gev <
  \mhalf,\mzero < 4 \tev$, $|\azero| < 7\tev$ and $2 < \tanb < 62$ as
  flat priors and using the formalism of Bayesian statistics we find
  that the 68\% posterior probability region for the $\hl^0$ mass lies
  between $115.4\gev$ and $120.4\gev$. Otherwise, $\hl^0$ is very
  similar to the Standard Model Higgs boson. Nevertheless, we point
  out some enhancements in its couplings to bottom and tau pairs,
  ranging from a few per cent in most of the CMSSM parameter space, up
  to several per cent in the favored region of $\tanb\sim50$ and
  the pseudoscalar Higgs mass of $\mha\lsim1\tev$. We also find that
  the other Higgs bosons are typically heavier, although not
  necessarily much heavier. For values of 
  $\hl^0$ mass
  within the 95\% probability range as determined by our analysis, a
  95\%~\cl\ exclusion limit can be set with about $2\fb^{-1}$ of
  integrated luminosity per experiment, or else with $4\fb^{-1}$
  ($12\fb^{-1}$) a $3\sigma$ evidence ($5\sigma$ discovery) will be
  guaranteed.  We also emphasize that the alternative statistical
  measure of the mean quality--of--fit favors a somewhat lower Higgs
  mass range; this implies even more optimistic prospects for the
  CMSSM light Higgs search than the more conservative Bayesian
  approach. In conclusion, for the above CMSSM parameter ranges,
  especially $\mzero$,
  either some evidence will be found at the Tevatron for the light
  Higgs boson or, at a high confidence level, the CMSSM will be ruled
  out.}

\keywords{Supersymmetric Effective Theories, Cosmology of Theories
beyond the SM, Dark Matter}
\preprint{CERN-PH-TH/2006-224}
\begin{document}


\section{Introduction}\label{sec:intro}

One of the main goals of the Tevatron and the LHC experimental
programmes is to detect a Higgs boson. In contrast to the Standard
Model (SM), in models with softly broken low energy supersymmetry
(SUSY), the mass of one Higgs boson is restricted to be fairly low,
$\mhl \lsim 150 \gev$,\footnote{For recent extensive reviews and
further references see, e.g.,~\cite{ch02,djouadi05}.} which allows for
a more focused search. On the other hand, even in the Minimal
Supersymmetric Standard Model (MSSM) there are as many as five Higgs
mass states: two scalars, $\hl^0$ and $\hh^0$, a pseudoscalar, $\ha^0$
and a pair of charged bosons, $\hpm$, which makes the experimental
search more involved. While Higgs boson tree--level mass parameters
obey some well--known relations, top and stop loop--dominated
radiative corrections introduce large modifications to Higgs masses
and couplings in terms of several unknown SUSY parameters. In the
general MSSM most soft SUSY--breaking parameters remain fairly
unrestricted, which makes it difficult to conduct a thorough
exploration of the parameter space.  Instead, many studies in the
general MSSM, including the Higgs discovery potential at the Tevatron,
often adopt rather arbitrary choices for MSSM parameter values.

It is therefore interesting and worthwhile to assess Higgs
observability in more constrained and well--motivated low energy
supersymmetric models. One particularly popular framework is the
Constrained MSSM (CMSSM), introduced in
ref.~\cite{kkrw94},\footnote{One well--known implementation of the
CMSSM is the minimal supergravity model~\cite{sugra-reviews}.} which
is defined in terms of the usual four free parameters: the ratio of
Higgs vacuum expectation values $\tanb$, as well as the common soft
SUSY--breaking parameters for gauginos $\mhalf$, scalars $\mzero$ and
tri--linear couplings $\azero$. The parameters $\mhalf$, $\mzero$ and
$\azero$ are specified at the GUT scale $\mgut\simeq 2\times
10^{16}\gev$ and serve as boundary conditions for evolving the MSSM
Renormalization Group Equations (RGEs) down to a low energy scale
$\msusy\equiv \sqrt{\mstopone\mstoptwo}$ (where
$m_{\stopone,\stoptwo}$ denote the masses of the scalar partners of
the top quark), chosen so as to minimize higher order loop
corrections. At $\msusy$ the conditions of electroweak symmetry
breaking (EWSB) are imposed and the SUSY spectrum is computed at $\mz$. The
sign of the Higgs/higgsino mass parameter $\mu$, however, remains
undetermined.

Prospects for Higgs collider searches in the CMSSM and other unified
models have been explored in several recent
analyses~\cite{Higgs-CMSSM-colliders,ehow04-Higgs-CMSSM}.  A usual
approach is to perform a fixed grid (``frequentist'') scan in some of
the CMSSM parameters (typically $\mhalf$ and $\mzero$) while keeping
the remaining ones (typically $\tanb$ and $\azero$) and also SM parameters
fixed.  The resulting ``allowed regions'' of parameter space then
often largely underestimate the true extent of the uncertainties,
mainly because of the existence of degeneracies in parameter space
that this procedure does not account for. On the other hand, a full
scan over a parameter space of even moderate dimensionality using grid
techniques is highly inefficient. Not only the size of (and time spent
on) the scan grows as a power--law with each new parameter added to the
scan, but there are other limitations.  It is difficult to incorporate
residual error--bars of relevant SM parameters, which are often simply
fixed at their central values. Experimental limits on SUSY are applied
at some arbitrary confidence level, e.g., at 1 or $2\sigma$. As a
result, it is very difficult, if not impossible, to derive a global
picture of the most probable ranges of SUSY parameters.

Recently more efficient exploration methods based on the Markov Chain
Monte Carlo (MCMC) technique~\cite{mcmc} have been successfully
applied to studying SUSY phenomenology and are becoming increasingly
popular~\cite{bg04,al05,allanach06,rtr1,alw06}. The MCMC technique
allows one to make a thorough scan of a model's full
multi--dimensional parameter space. Additionally, by combining the
MCMC algorithm with the formalism of Bayesian statistics, maps of
probability distributions can be drawn not only for the model's
parameters but also for all the observables (and their combinations)
included in the analysis.

In our first paper~\cite{rtr1} we applied this approach to performing
a full analysis of the CMSSM. As in a similar (and concurrent, but
independent) work of Allanach and Lester~\cite{al05}, we applied all
relevant constraints on Higgs and superpartner masses from collider
searches, from the rare processes $\brbsgamma$ and $\brbsmumu$, from the
anomalous magnetic moment of the muon $\gmtwo$, and also from
cosmology, on the relic abundance of the lightest neutralino
$\abundchi$ assumed, in the presence of $R$--parity, to be the cold
dark matter in the Universe. We also took into account residual error
bars in the pole top mass, the bottom mass and $\alphas$. Going beyond
the work of ref.~\cite{al05}, we further included in our analysis the
experimental error in the fine structure constant measured at $\mz$
(which had a sizable impact on $\abundchi$), explored wider ranges of
$\mzero$ up to $4\tev$ (which allowed us to explore the focus point
region) and computed the spin--independent cross section for dark
matter neutralino scattering off nuclei $\sigsip$ (but did not use it
as a constraint on the model because of astrophysical
uncertainties). We further included constraints from contributions to
$\mW$ and $\sineff$ from full one--loop SM and MSSM corrections and
from two--loop SM corrections involving the top Yukawa. Last but not
least, we also emphasized the difference between high posterior
probability regions of the parameters in the Bayesian language and
those of the high mean quality--of--fit, i.e., possibly limited
ranges of parameters that give the best fit to the data. We have
found that in the CMSSM the two can be rather different, which is a
consequence of the complex dependence of the model's parameters on the
applied constraints.
The main results of both analyses~\cite{al05} and~\cite{rtr1} came
out remarkably consistent with each other, in spite of the above
differences and, additionally, of some nuances in computing the likelihood
function.  In particular, high probability regions showed preference
for $\mhalf,\mzero\lsim1\tev$, but not  for values nearly as low as
claimed in ref.~\cite{ehow06} based on a $\chi^2$ analysis.  

In our present work we apply an important recent shift in the SM value
of $\brbsgamma$. In~\cite{rtr1} we used the previous SM prediction
$(3.70\pm 0.30)\times 10^{-4}$~\cite{bsgsm}, which included a full NLO
calculation and partial charm mass contribution. Recently, partial
NNLO contributions, most importantly  an
approximate charm mass one, have been obtained in~\cite{ms06-bsg} which
led to a rather dramatic shift down to $(3.15 \pm 0.26) \times
10^{-4}$, with a further slight decrease to
$(2.98 \pm 0.26) \times 10^{-4}$ after
including some additional subtle effects due to a treatment of a photon energy cut
$E_\gamma>1.6\gev$~\cite{bn06}. At the same time, the experimental world average
has somewhat increased from $(3.39^{+0.30}_{-0.27})\times 10^{-4}$ to
the current range of $(3.55 \pm 0.26) \times
10^{-4}$~\cite{bsgexp}. This leads to some discrepancy, at the
level of $1.4\sigma$, between the SM and the experimental
average. More importantly, because the SM value for the branching
ratio has moved down below the measured one, any potential overall SUSY
contributions should now preferably be constructive, in contrast to
the situation before. This will lead to significant shift in the
preferred regions of the CMSSM parameter space.

In the present work we further include full two--loop and available
higher order SM corrections, as well as dominant two--loop MSSM gluon
corrections, to $\mW$ and $\sineff$. Assuming for comparison the
previous values of $\brbsgamma$, including these observables does not,
however, lead to any appreciable differences with respect
to~\cite{al05,rtr1}, as was also shown in very recent updated analysis
of Allanach \etal,~\cite{alw06} who included them at comparable level.
We also update several experimental constraints, as discussed
below. We additionally compute $B_s$ mixing, $\delmbs$, which has
recently been precisely measured at the Tevatron by the
CDF~Collaboration~\cite{cdf-deltambs}.

We devote this paper to a study of the light Higgs boson in the CMSSM
and to the prospects for its detection at the Tevatron. 
Results of our new analysis regarding the CMSSM parameters will be
presented elsewhere~\cite{rrt3}.

Our main results can be summarized as follows. We find that, imposing
flat priors on wide ranges of the CMSSM parameters: $50\gev <
\mhalf,\mzero < 4 \tev$, $|\azero| < 7\tev$ and $2 < \tanb < 62$, the
68\% probability region for the mass of the lightest Higgs is given by
$115.4\gev < \mhl < 120.4\gev$, the 95\% probability range being
$112.5\gev < \mhl < 121.9\gev$. Its couplings generally closely match
those of the SM Higgs boson with the same mass, although we find some
differences at the level of a few to several per cent. Ensuing
prospects for experimental Higgs search at the Tevatron look
excellent.  So far, with about $1\fb^{-1}$ of data analyzed, both CDF
and D0 Collaborations have been able to put interesting limits on the
Higgs cross sections~\cite{fnal-Higgs-discovery} for some specific
choices of the general MSSM
parameters~\cite{tev-higgs-bench-chww,chww05-fnal-Higgs-discovery},
which are, however, not representative of unified models. As more data
are coming in, both collaborations will soon be in a position to start
probing unification--based models, including the CMSSM.  In the whole
Higgs mass range given above, a 95\%~\cl\ exclusion limit on the
SM--like Higgs boson can be set with about $2\fb^{-1}$ of integrated
luminosity per experiment. In the CMSSM a $3\sigma$ ($5\sigma$) signal
should be seen in this mass range with about $4\fb^{-1}$
($12\fb^{-1}$) of data per experiment. On the other hand, with about
$8\fb^{-1}$ of integrated luminosity eventually expected per
experiment, a $5\sigma$ discovery will be possible, should the light
Higgs mass be around $115\gev$.  In conclusion, if the CMSSM (or
another supersymmetric model with similar light Higgs boson
properties) has been chosen by nature, then the Higgs boson with
SM--like properties will be discovered at the Tevatron.

The paper is organized as follows. In section~\ref{sec:outline} we
briefly summarize the main features of our Bayesian analysis and then
provide our updated list of experimental constraints. We then proceed
to present, in section~\ref{sec:higgsprperties}, our results for Higgs
mass distribution and other properties. Section~\ref{sec:higgsattev}
is devoted to a discussion of light Higgs production and decay at the
Tevatron. In section~\ref{sec:summary} we present our summary and
conclusions.


\section{An Outline of the Phenomenological Analysis}\label{sec:outline}

\subsection{Posterior probabilities}\label{sec:pdf}
Our procedure based on MCMC scans and Bayesian analysis has been presented in detail
in~\cite{rtr1}. Here, for completeness, we summarize its main features.

We are interested in delineating high probability regions of the CMSSM
parameters.  We fix $\text{sign}(\mu)= +1$ throughout and denote
the remaining four free CMSSM parameters by the set
\be \label{indeppars:eq}
\params =  (\mzero, \mhalf, \azero, \tanb ).
\ee
As demonstrated in~\cite{al05,rtr1}, the values of the relevant SM
parameters can strongly influence some of the CMSSM predictions, and,
in contrast to common practice, should not be just kept fixed at their
central values. We thus introduce a set $\nuis$ of so--called {\em
``nuisance parameters''}. Those most relevant to our analysis are
\be \label{nuipars:eq} \nuis = ( \mtpole,
\mbmbmsbar, \alphaemmz, \alphas ), \ee
where $\mtpole$ is the pole top quark mass. The other three
parameters:  $\mbmbmsbar$ --- the bottom
quark mass at $m_b$, $\alphaemmz$ and $\alphas$ --- respectively the
electromagnetic and the strong coupling constants at the $Z$ pole mass
$M_Z$ --  are all computed in the $\msbar$ scheme.

The set of parameters $\params$ and $\nuis$ form an 8--dimensional set
$\basis$ of our {\em ``basis parameters''} $\basis = (\params,
\nuis)$.\footnote{In~\cite{rtr1} we denoted our basis parameters with a
symbol $\eta$. } 
In terms of the basis parameters we compute a number of
collider and cosmological observables, which we call {\em ``derived
variables''} and which we collectively denote by the set $\derived=
(\derived_1, \derived_2, \ldots)$.  The observables, which are listed
below, will be used to compare CMSSM predictions with a set of
experimental data $\data$, which is currently available either in the
form of positive measurements or as limits.

In order to map out high probability regions of the CMSSM, we compute
the {\em posterior probability density functions} (pdf's) $p(\basis |\data)$
for the basis parameters $\basis$ and for several observables. 
The posterior pdf represents our state of knowledge
about the parameters $\basis$ after we have
taken the data into consideration (hence the name). Using
Bayes' theorem, the posterior pdf is given by
\be \label{eq:bayes}
 p(\basis | \data) = \frac{p(\data |
\derived) \pi(\basis)}{p(\data)}. \ee
On the r.h.s. of eq.~\eqref{eq:bayes}, the quantity $p(\data |
\derived)$, taken as a function of $\data$ for a given $\basis$, and
hence a given $\derived(\basis)$, is called  a ``sampling
distribution''. It represents the probability of reproducing the data
$\data$ for a fixed value of $\derived(\basis)$. Considered instead as
a function of $\derived$ for {\em fixed data} $\data$,
$p(\data|\derived)$ is called the {\em likelihood} (where the
dependence of $\derived$ on $\basis$ is understood).  The likelihood
supplies the information provided by the data and, for the purpose of
our analysis, it is constructed in Sec.~3.1 of ref.~\cite{rtr1}.  The
quantity $\pi(\basis)$ denotes a {\em prior 
probability density function} (hereafter called simply {\em a prior})
which encodes our state of knowledge about the values of the
parameters in $\basis$ before we see the data. The state of knowledge
is then updated to the posterior via the likelihood.  Finally, the
quantity in the denominator is called {\em evidence} or {\em model
likelihood}. In the context of this analysis it is only a
normalization constant, independent of $\basis$, and therefore will be
dropped in the following.  

As in ref.~\cite{rtr1}, our posterior pdf's
presented below will be normalized to their maximum values, and {\em not} in
such a way as to give a total probability of 1. Accordingly we
will use the name of a ``relative posterior pdf'', or simply of ``relative
probability density''.

\subsection{Likelihood function and constraints}\label{sec:constraints}

\begin{table}
\centering    . 

\begin{tabular}{|l | l l | l|}
\hline
SM (nuisance) parameter  &   Mean value  & \multicolumn{1}{c|}{Uncertainty} & Ref. \\
 &   $\mu$      & ${\sigma}$ (exper.)  &  \\ \hline
$\mtpole$           &  171.4 GeV    & 2.1 GeV&  \cite{cdf+dzero-mtop-06} \\
$m_b (m_b)^{\overline{MS}}$ &4.20 GeV  & 0.07 GeV &  \cite{pdg06} \\
$\alphas$       &   0.1176   & 0.002 &  \cite{pdg06}\\
$1/\alphaemmz$  & 127.955 & 0.018 &  \cite{pdg06} \\ \hline
\end{tabular}
\caption{Experimental mean $\mu$ and standard deviation $\sigma$
 adopted for the likelihood function for SM (nuisance) parameters,
 assumed to be described by a Gaussian distribution.
\label{tab:meas}} 
\end{table}

\begin{table}
\centering
\begin{tabular}{|l | l l l | l|}
\hline
Observable &   Mean value & \multicolumn{2}{c|}{Uncertainties} & Ref. \\
 &   $\mu$      & ${\sigma}$ (exper.)  & $\tau$ (theor.) & \\\hline
 $M_W$     &  $80.392\gev$   & $29\mev$ & $15\mev$ & \cite{lepwwg} \\
$\sineff{}$    &  $0.23153$      & $16\times10^{-5}$
                & $15\times10^{-5}$ &  \cite{lepwwg}  \\
$\dasusy \times 10^{10}$       &  28 & 8.1 &  1 & \cite{pdg06} \\
 $\brbsgamma \times 10^{4}$ &
 3.55 & 0.26 & 0.21 & \cite{bsgexp} \\
$\delmbs$     &  17.33  & 0.12  & 4.8
& \cite{cdf-deltambs} \\
$\abundchi$ &  0.104 & 0.009 & $0.1\,\abundchi$& \cite{wmap3yr} \\\hline
   &  Limit (95\%~\cl)  & \multicolumn{2}{r|}{$\tau$ (theor.)} & Ref. \\ \hline
$\brbsmumu$ &  $ <1.0\times 10^{-7}$
& \multicolumn{2}{r|}{14\%}  & \cite{cdf-bsmumu}\\ 
$\mhl$  & $>114.4\gev$\ ($91.0\gev$)  & \multicolumn{2}{r|}{$3 \gev$}
& \cite{lhwg} \\ 
$\zetah^2$  & $f(m_h)$ & \multicolumn{2}{r|}{negligible}  & \cite{lhwg} \\ 
sparticle masses  &  \multicolumn{3}{c|}{See table~4 in ref.~\cite{rtr1}.}  & \\ \hline
\end{tabular}
\caption{Summary of the observables used in the analysis. Upper part:
Observables for which a positive measurement has been
made. $\deltaamususy$ denotes the discrepancy between the SM
prediction and the experimental value of the anomalous magnetic moment
of the muon $\gmtwo$. For central values of the SM input parameters
used here, the SM value of $\brbsgamma$ is $3.11\times10^{-4}$, while
the theoretical error of $0.21\times10^{-4}$ includes uncertainties
other than the parametric dependence on the SM nuisance parameters,
especially $\mtpole$ and $\alphas$.  As explained in the text, for
each quantity we use a likelihood function with mean $\mu$ and
standard deviation $s = \sqrt{\sigma^2+ \tau^2}$, where $\sigma$ is
the experimental uncertainty and $\tau$ represents our estimate of the
theoretical uncertainty. Lower part: Observables for which only limits
currently exist.  The likelihood function has been constructed as in
ref.~\cite{rtr1}, including in particular a smearing out of
experimental errors and limits to include an appropriate theoretical
uncertainty in the observables.
\label{tab:measderived}}
\end{table}

We scan over very wide ranges of CMSSM parameters; compare table~1 of
ref.~\cite{rtr1}. In particular we take flat priors on the ranges
$50\gev < \mhalf,\mzero < 4 \tev$, $|\azero| < 7\tev$ and $2 < \tanb <
62$. (In ref.~\cite{rtr1} we called this the ``4\tev\ range''.) For
the SM (nuisance) parameters, we adopt a Gaussian likelihood with mean
and standard deviation as given in table~\ref{tab:meas}, and we assume
flat priors over wide ranges of their values~\cite{rtr1}. Note that,
with respect to ref.~\cite{rtr1}, we have updated the values of all
the parameters, including the recent shift in $\mtpole$ based on
Tevatron's Run--II $1\fb^{-1}$ of data.

The experimental values of the collider and cosmological observables
(our derived variables) are listed in table~\ref{tab:measderived} and
in table~4 of ref.~\cite{rtr1}, with updates where applicable.  In
particular, in addition to (most importantly) $\brbsgamma$ summarized above,
we update an experimental constraint from the anomalous magnetic
moment of the muon $\gmtwo$ (denoted here by $\deltaamususy$) for
which a discrepancy between measurement and SM predictions (based on
$e^+e^-$ data) persists at the level of 2--3$\sigma$~\cite{pdg06}. We
note here, however, that the impact of this (somewhat controversial)
constraint on our findings will be rather limited.  We also apply the
new values for the measured branching ratio for
$\bsgamma$~\cite{bsgexp}, and an improved 95\%~\cl\ limit $\brbsmumu
<1.0\times 10^{-7}$~\cite{cdf-bsmumu}. In constraining the relic
abundance $\abundchi$ of the lightest neutralino we use the 3--year
data from WMAP~\cite{wmap3yr}.  As a new constraint, we add a recent
value of $\bbbarmix$ mixing, $\delmbs$, which has recently been
precisely measured at the Tevatron by the
CDF~Collaboration~\cite{cdf-deltambs}. In both cases we use
expressions from ref.~\cite{for2+3} which include dominant large
$\tanb$--enhanced beyond--LO SUSY contributions from Higgs penguin
diagrams.  Unfortunately, theoretical uncertainties, especially in
lattice evaluations of $f_{B_s}$ are still very large (as reflected in
table~\ref{tab:measderived} in the estimated theoretical error for
$\delmbs$), which makes the impact of this precise measurement on
constraining SUSY parameter space somewhat limited.\footnote{On the
other hand, in the MSSM with general flavor mixing the bound from
$\delmbs$ is in many cases much more constraining than from other rare
processes~\cite{for4}.}

For all the quantities for which positive measurements have been made
(as listed in the upper part of table~\ref{tab:measderived}), we
assume a Gaussian likelihood function with a variance given by the sum
of the theoretical and experimental variances, as motivated by
eq.~(3.3) in ref.~\cite{rtr1}. For the observables for which only
lower or upper limits are available (as listed in the bottom part of
table~\ref{tab:measderived}) we use a smoothed--out version of the
likelihood function that accounts for the theoretical error in the
computation of the observable, see eq.~(3.5) and fig.~1 in~\cite{rtr1}.

The likelihood function for the CMSSM light Higgs boson requires a
more refined treatment. The final LEP--II lower bound of $114.4\gev$
(95\%~\cl)~\cite{lhwg} is applicable for the case of the SM Higgs
boson. It also applies to the lightest Higgs boson $h$ of the MSSM
when its coupling to the $Z$ boson is SM--like, i.e., when $\zetah^2
\equiv g^2(\hl ZZ)_{\text{MSSM}}/g^2(\hl ZZ)_{\text{SM}} \simeq 1$
which holds in a so--called decoupling regime of $\mha \gg \mz$. For
arbitrary values of $\mha$, the LEP--II Collaboration has set
95\%~\cl\ bounds on $\mhl$ and $\mha$ as a function of
$\zetah^2$~\cite{lhwg}, with the lower bound of $\mhl > 91 \gev$ for
$\mhl \sim \mha$ and $\zetah^2\ll 1$~\cite{lhwg}.  In this case we use
a cubic spline to interpolate between selected points in $\mhl$ and
translate the above bound into the corresponding 95\%~\cl\ bound in
the $(\mhl, \zetah^2)$ plane. We then add a theoretical uncertainty
$\tau(\mhl) = 3\gev$, following eq.~(3.5) in ref.~\cite{rtr1}.  (Notice
that the parametric uncertainties coming from the errors in top quark
mass and the strong coupling constant have already been fully
accounted for by including them as nuisance parameters.)  We then
simultaneously constrain the values of $(\mhl, \zetah^2)$ obtained in
the CMSSM by comparing them with the 2--dim likelihood function for
these two variables from the LEP results. Since we find $\zeta_h^2
\simeq 1$ with very high accuracy basically everywhere in the parameter space
(as we will see later), introducing an extra theoretical uncertainty
in $\zetah^2$ (which could be implemented by extending eq.~(3.5) in ref.~\cite{rtr1}
to a 2--dim case) would not affect our results in any appreciable
way.\footnote{We note that, in contrast to
refs.~\cite{ehow04-Higgs-CMSSM,alw06,ehow06}, in translating LEP--II
bounds in the low $\mha$ regime into our likelihood function we do not
assume {\em a priori} that the CMSSM light Higgs scalar is SM--like.}
This procedure results in a conservative likelihood function for
$\mhl$, which does not simply cut away points below the 95\%~\cl\ limit
of LEP--II, but instead assigns to them a lower probability that
gradually goes to zero for lower masses.

As mentioned in the Introduction, here we make some additional
improvements in our treatment of the radiative corrections to the
electroweak observables $M_W$ and $\sineff$. We now include full
two--loop and known higher order SM corrections as computed in
ref.~\cite{awramik-acfw04}, as well as the gluonic two--loop MSSM
ones~\cite{dghhjw97}. (In the CMSSM two--loop gluino corrections are
typically subdominant because the colored superpartners tend to be
rather heavy.)  These updates and improvements lead, however, to
fairly minor changes in the overall distribution of most probable
CMSSM parameter regions relative to refs.~\cite{al05,rtr1}, in
agreement with ref.~\cite{alw06}. On the other hand, as we have
mentioned, the recent downwards shift in the SM value of $\brbsgamma$ has
caused a corresponding big change in the pdf distribution in the CMSSM
parameters~\cite{rrt3}.

Finally, points that do not fulfil the conditions of radiative EWSB
and/or give non--physical (tachyonic) solutions are discarded.  We
adopt the same convergence and mixing criteria as described in
appendix~A2 of ref.~\cite{rtr1}, while our sampling procedure is
described in appendix~A1 of ref.~\cite{rtr1}. We have the total of
$N=10$ MC chains, with a merged number of
samples $3\times10^5$, and an acceptance rate of about 2\%. More
details of our numerical MCMC scan can be found in~\cite{rtr1}.

\begin{figure}[!t]
\begin{center}
\begin{tabular}{c c}
 	\includegraphics[width=0.4\textwidth]{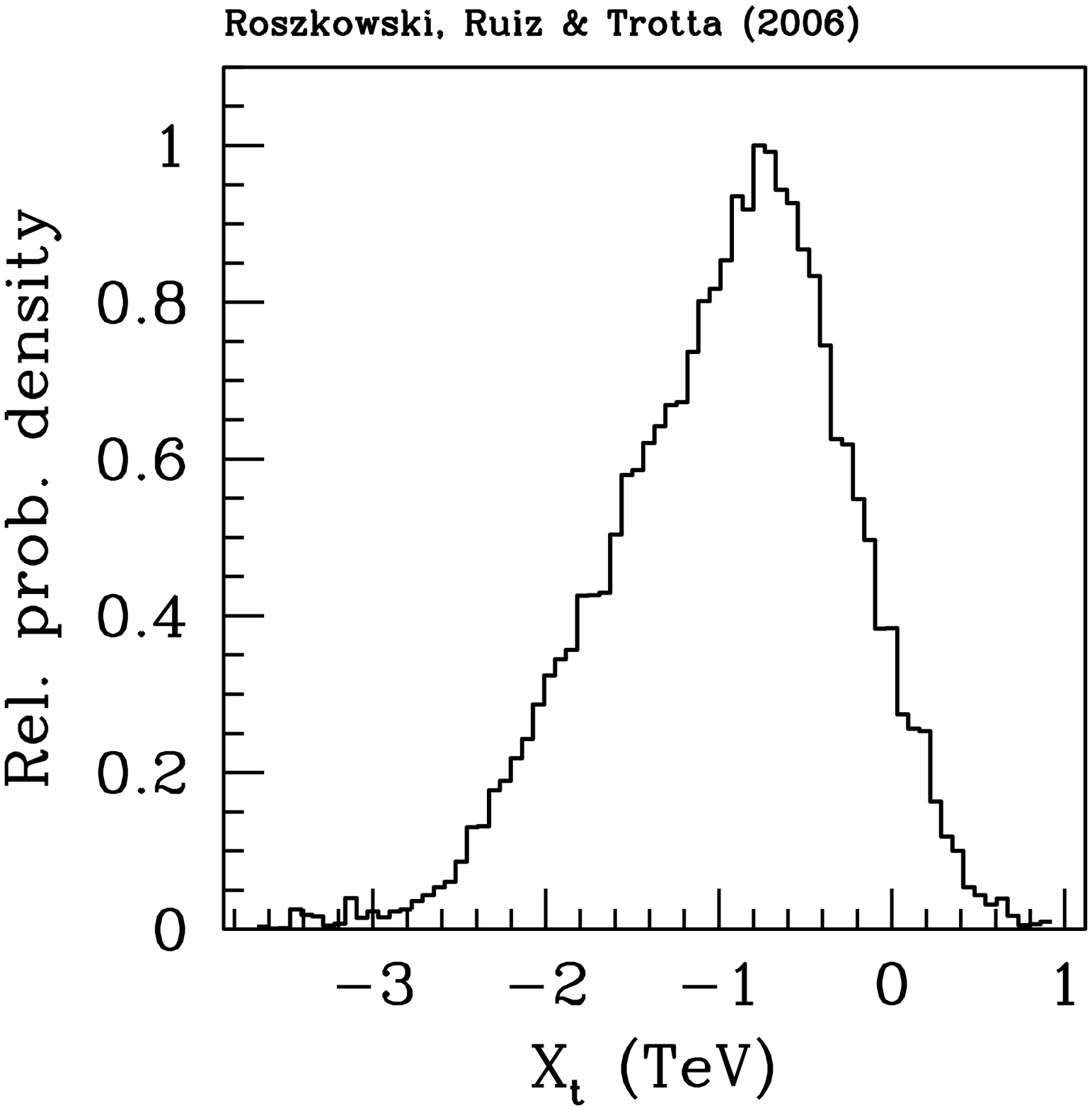}
 &	\includegraphics[width=0.4\textwidth]{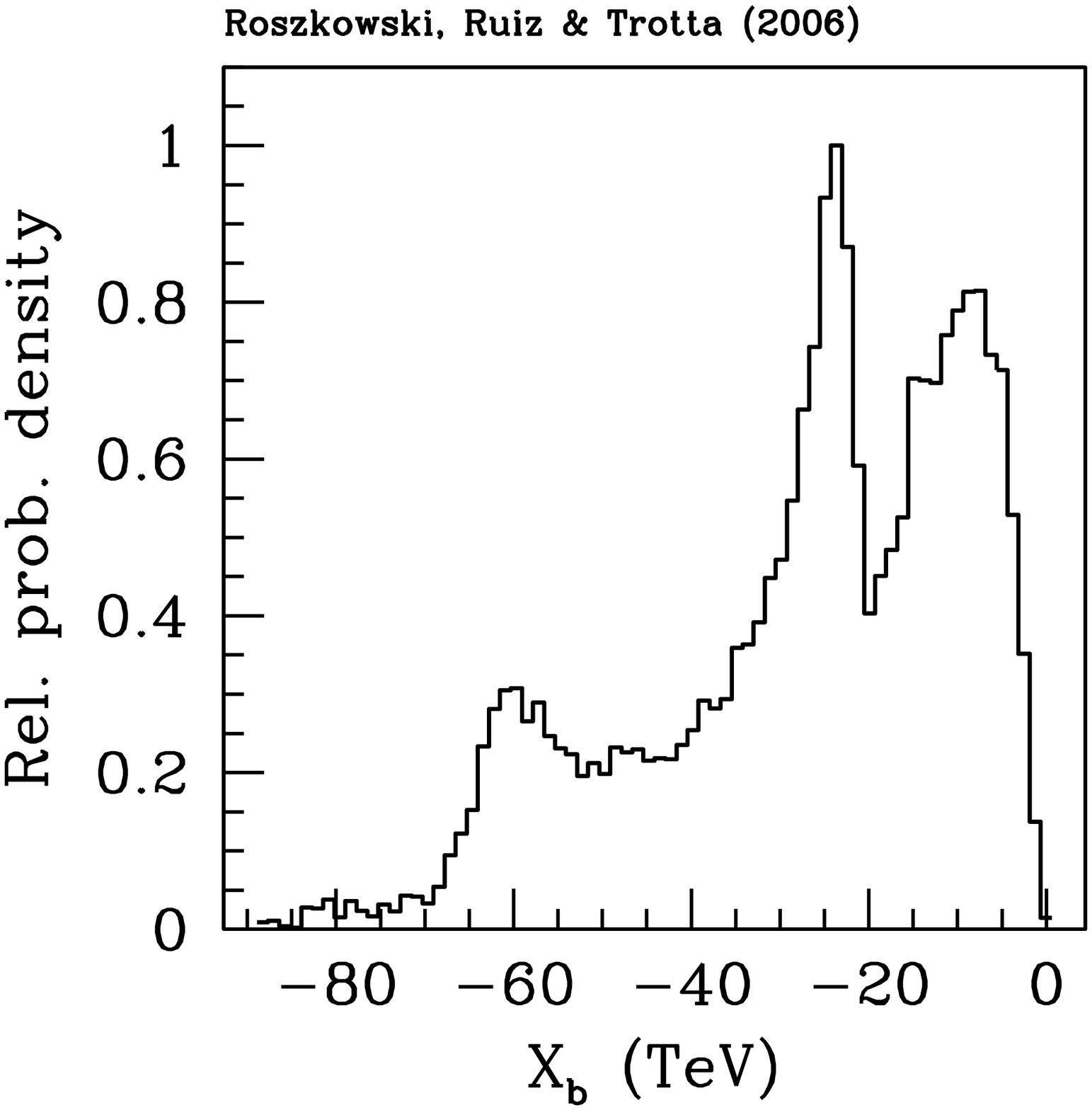}
\end{tabular}
\end{center}
\caption{The 1--dim relative probability densities for $\xt$ (left panel)
  and $\xb$ (right panel).  Here and in all subsequent figures all parameters
  which are not shown have been marginalized (i.e., integrated) over. 
\label{fig:xtandxb} }
\end{figure}
\begin{figure}[!tbh]
\begin{center}
\begin{tabular}{r}
\includegraphics[width=0.5\textwidth]{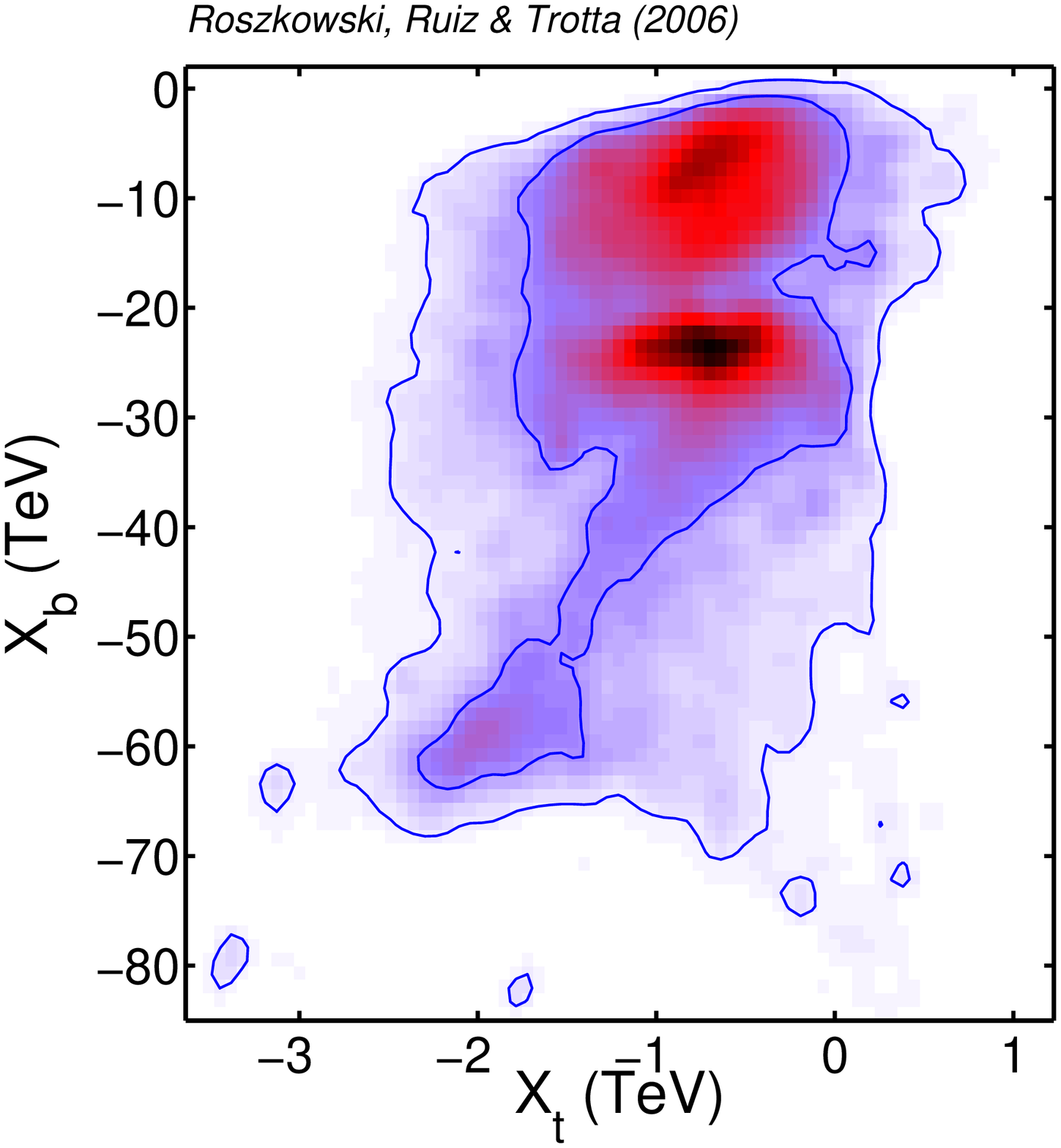}\\
\includegraphics[width=0.425\textwidth]{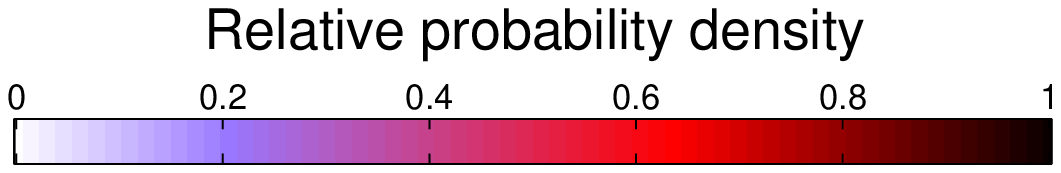} 
\end{tabular}
\end{center} 
\caption{The 2--dim relative probability density in the plane of $\xt$
  and $\xb$. The probability color code shown here applies also to all
  subsequent figures showing 2--dim relative probability density. The inner
  (outer) solid contours delimit the regions of 68\% and 95\%
  total probability, respectively.
  \label{fig:rrt2-Xt-vs-Xb} }
\end{figure}

\section{Properties of the lightest Higgs boson in the CMSSM}\label{sec:higgsprperties}

In this work we are particularly interested in the properties of
the lightest neutral Higgs boson. At the tree level, the Higgs sector
of the MSSM is determined by $\tanb$ and $\mha$. The (by far dominant)
one--loop radiative corrections are generated by diagrams involving
the top quark and its scalar partners, and, at large $\tanb$, also the
bottom quark and its scalar partners. Both are proportional to their
respective Yukawa couplings.  The radiative corrections have been
computed using several different methods.  Full one--loop expressions
are known~\cite{mh-1loop-early,mh-full1loop, bmpz}. Leading two--loop
corrections have been computed using renormalization group~\cite{rg}
and two--loop effective potential methods~\cite{eff,eff-sft}, and in
the Feynman--diagrammatic approach~\cite{hhw}.  Furthermore the
tadpole corrections, needed to minimise the effective scalar potential,
have been calculated at one loop~\cite{bmpz, veff-1loop} and the
leading ones at two loops~\cite{ds03, martin}. The remaining
theoretical uncertainty in the light Higgs mass $\mhl$ has
conservatively been estimated at $\lsim 3\gev$~\cite{adkps04,hhrw04}.

In computing the Higgs (and SUSY) mass spectrum we employ the code
SOFTSUSY~v2.08~\cite{softsusy}, which implements radiative corrections
in the modified Dimensional Reduction  scheme, $\drbar$,
based on the results of~\cite{bmpz, eff-sft, ds03}. For comparison, in
FeynHiggs~\cite{feynhiggs} the On--Shell scheme approach is adopted.
Both are in good agreement~\cite{adkps04}, i.e., within the above
theoretical uncertainty.

In the literature one often considers the cases of ``maximal mixing''
and ``no mixing'' (or ``minimal mixing'') to describe the impact on
the Higgs sector of the off--diagonal terms $m_q X_q$ ($q=t,b$)
in the stop and sbottom mass matrices
relative to their diagonal entries. In terms of
\beq 
\xt=A_t -\mu\cot\beta,~~~~~~~~~~~~~~~~\xb=A_b -\mu\tanb, 
\label{eq:xtxbdef} 
\eeq
where $A_{t,b}$ are the stop/sbottom trilinear soft parameters, the
``no mixing'' case, in particular, corresponds to $\xt=0$ and $\xb=0$.
In fig.~\ref{fig:xtandxb} we show in the left panel the 1--dim
relative probability densities for $\xt$ and in the right panel those for
$\xb$ in the CMSSM. Their 2--dim relative probability density
(marginalized over all other parameters) is shown in
fig.~\ref{fig:rrt2-Xt-vs-Xb}.  One can see that both variables are
typically negative, with $|\xb|\gg |\xt|$. Very large negative values
of $\xb$ are predominantly caused by the fact that the relative
probability density of $\tanb$ is strongly peaked at large values of
$\simeq 52$~\cite{al05,rtr1,rrt3}; compare fig.~2 of~\cite{rtr1}. Such
large values of both $\xt$ and $\xb$ do not, however, necessarily
imply large mixings in the stop and sbottom sectors since in the
corresponding mass matrices they are multiplied by their respective
quark masses.  In fact, in the CMSSM in the sbottom sector we find a
nearly strict no--mixing limit while in the stop sector we find a
spread of values (mild mixing) but again with the peak in the
probability distribution close to no mixing. Note that the ranges of
$\xt$ and $\xb$ in figs.~\ref{fig:xtandxb} and~\ref{fig:rrt2-Xt-vs-Xb}
are very different from the values proposed for general MSSM Higgs
searches at the
Tevatron~\cite{tev-higgs-bench-chww,chww05-fnal-Higgs-discovery}.

\begin{figure}[!tbh]
\begin{center}
 	\includegraphics[width=0.6\textwidth]{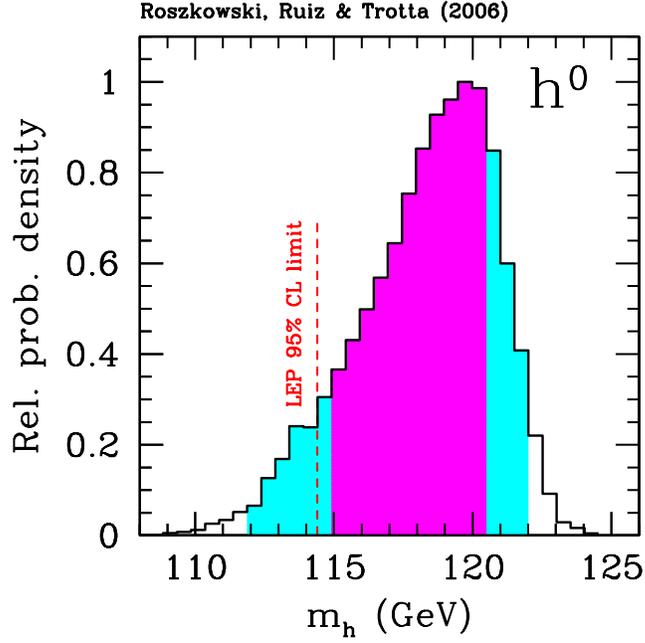}
\end{center}
\caption{The 1--dim relative probability density for the light Higgs
boson mass $\mhl$.  Magenta (dark shaded) and cyan (shaded) bands
delimit 68\% and 95\% posterior probability (2 tails) ranges of
$\mhl$, respectively.
\label{fig:hlmass} }
\end{figure}

The values of $\xt$ and $\xb$ determine to some
extent the upper bound on $\mhl$~\cite{ch02,djouadi05}:
\be
\mhl^2\lsim \mz^2 +\frac{3g_2^2\mtop^4}{8\pi^2\mW^2}
\left[ \ln{\frac{\mstopave^2}{\mtop^2}} +  \frac{\xt^2}{\mstopave^2}
\left(1-\frac{\xt^2}{12\mstopave^2}
\right) \right] + \left( t\to b, \stopq\to\sbotq\right),
\label{mhllimit}
\ee
where $g_2$ is the $SU(2)_L$ gauge coupling and
$\mstopave^2=(\mstopone^2+\mstoptwo^2)/2$ is an average stop
mass--squared, and analogously for the sbottoms. From
fig.~\ref{fig:rrt2-Xt-vs-Xb} one can easily see that the
bottom--sbottom contribution to~\eqref{mhllimit} can be comparable to
the top--stop one.

\begin{figure}[!t]
\begin{center}
\begin{tabular}{c c}
 	\includegraphics[width=0.4\textwidth]{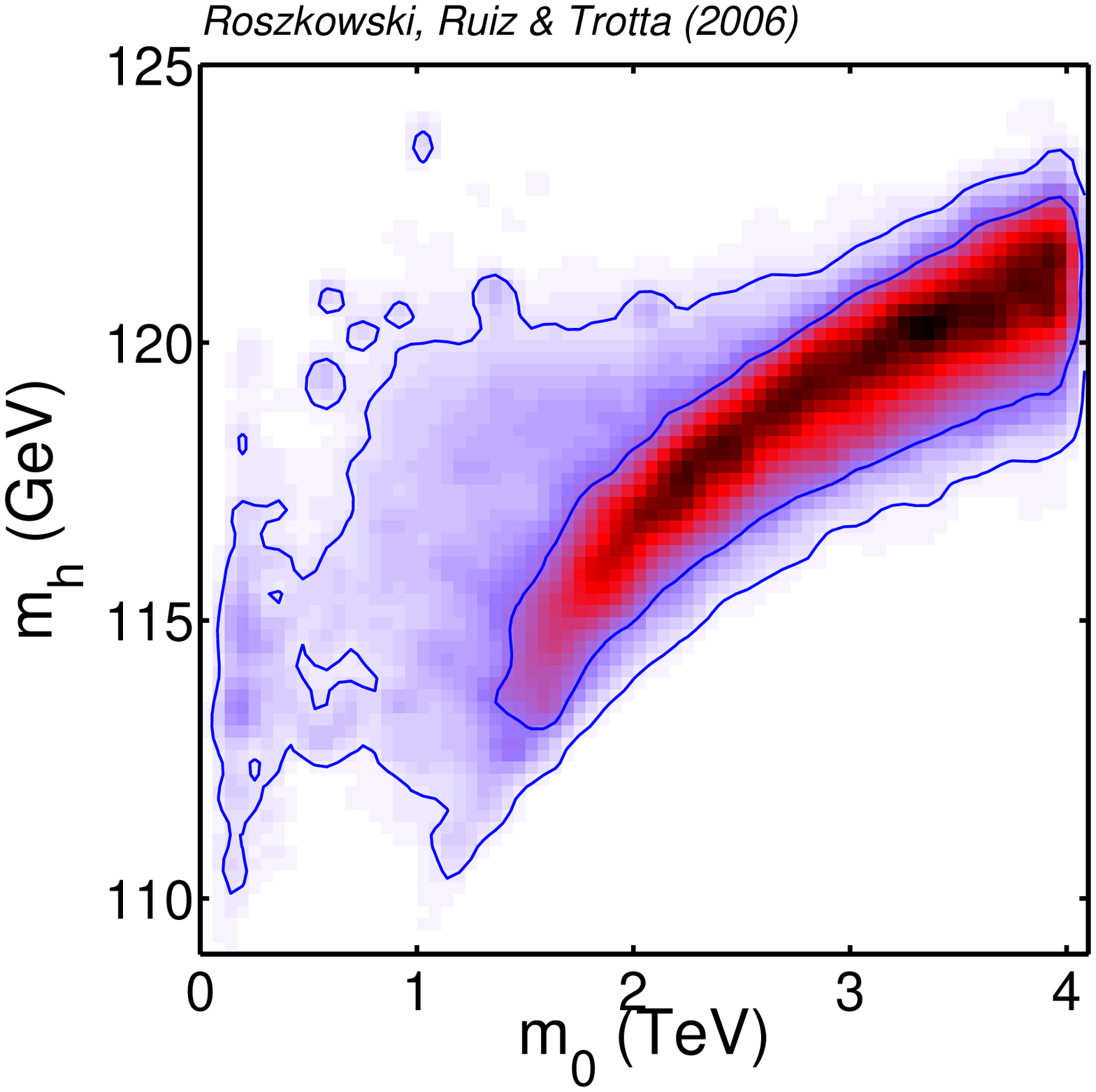}
 &	\includegraphics[width=0.4\textwidth]{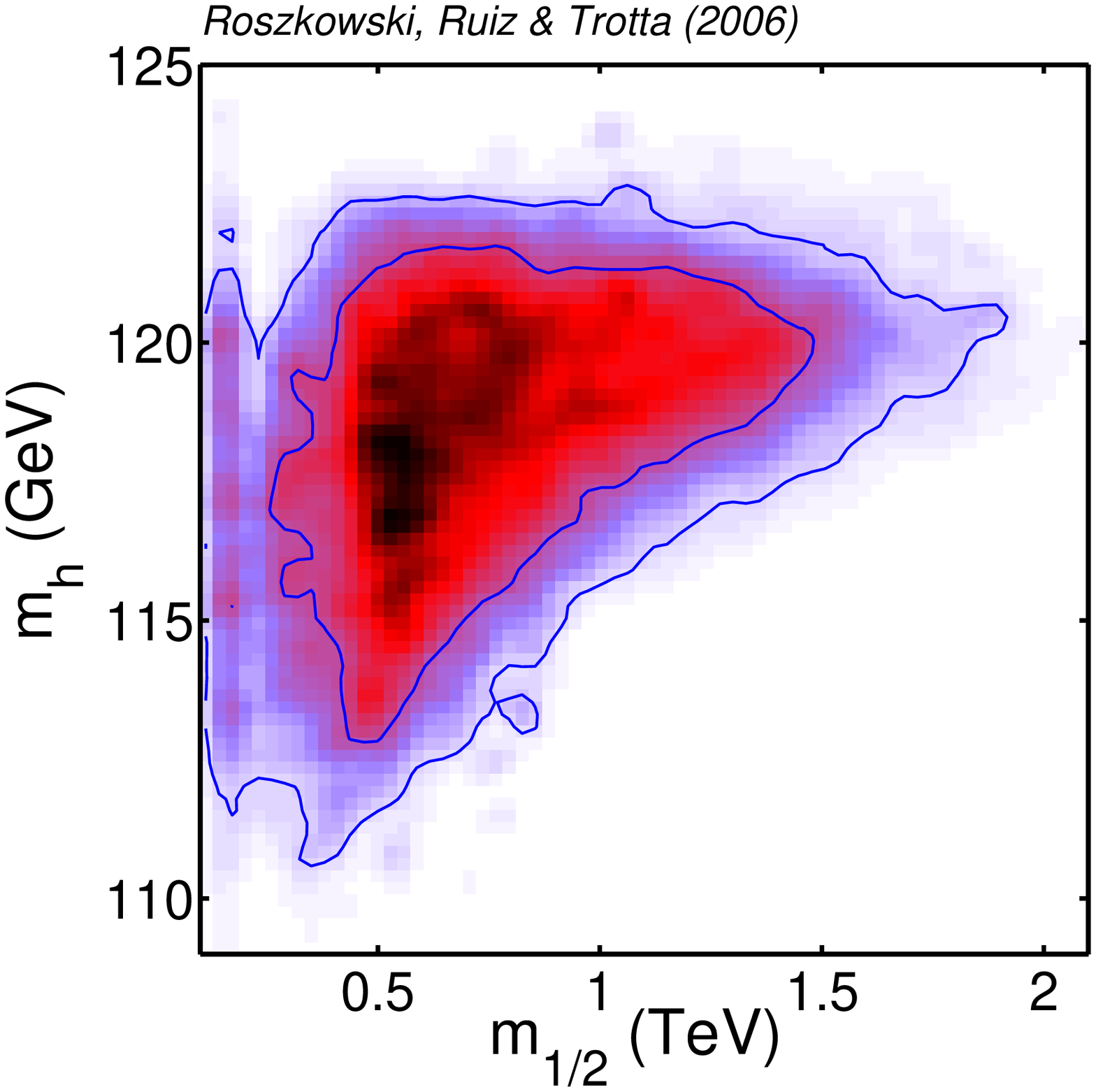}
\end{tabular}
\end{center}
\caption{The 2--dim relative probability densities in the plane of $(\mzero,\mhl)$
  (left panel)  and $(\mhalf,\mhl)$ (right panel). The inner
  (outer) solid contours delimit the regions of 68\% and 95\%
  total probability, respectively.
\label{fig:mhvsm0m12} }
\end{figure}
\begin{figure}[!tbh]
\begin{center}
\begin{tabular}{c c c}
	\includegraphics[width=0.3\textwidth]{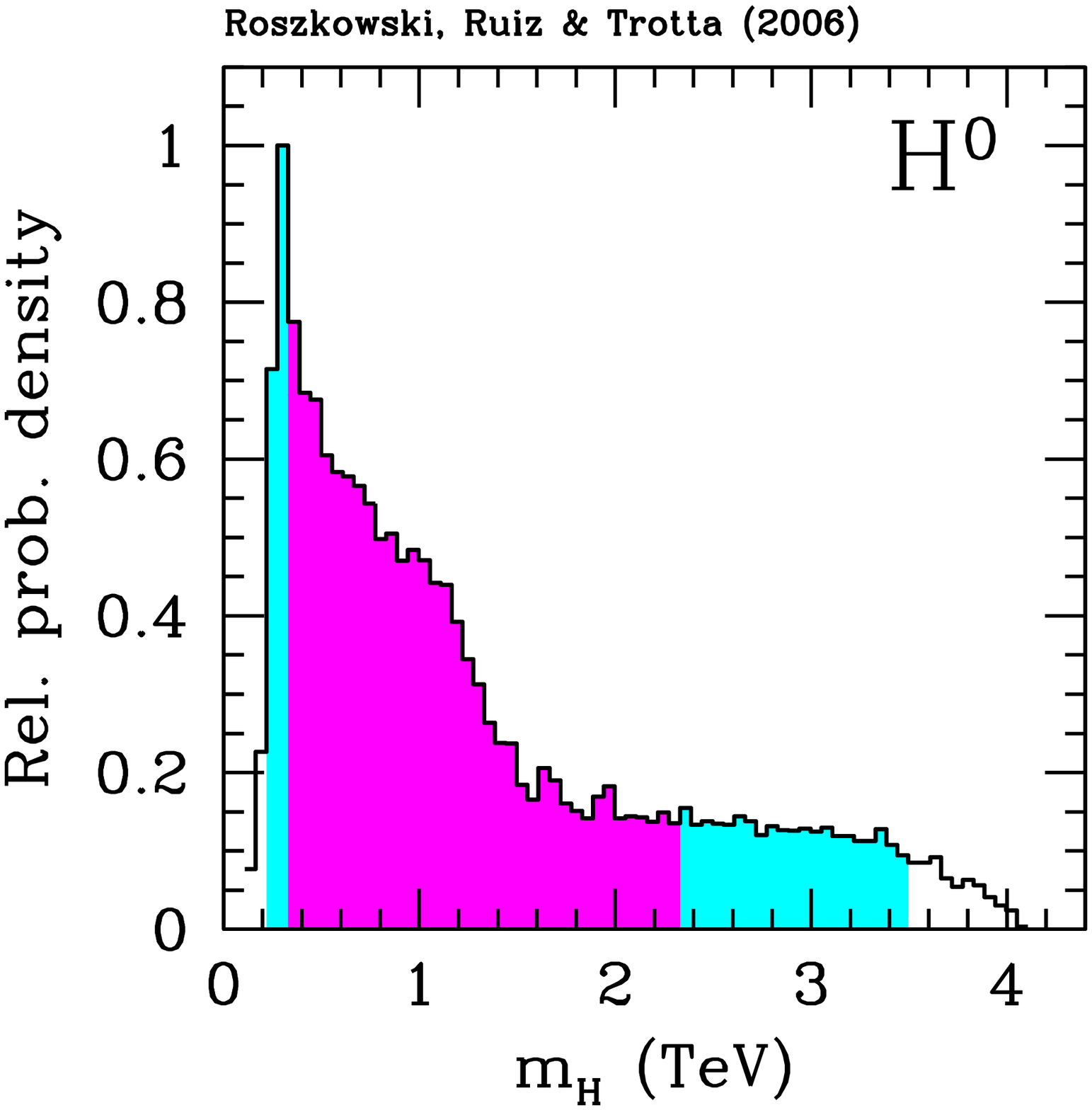}
&	\includegraphics[width=0.3\textwidth]{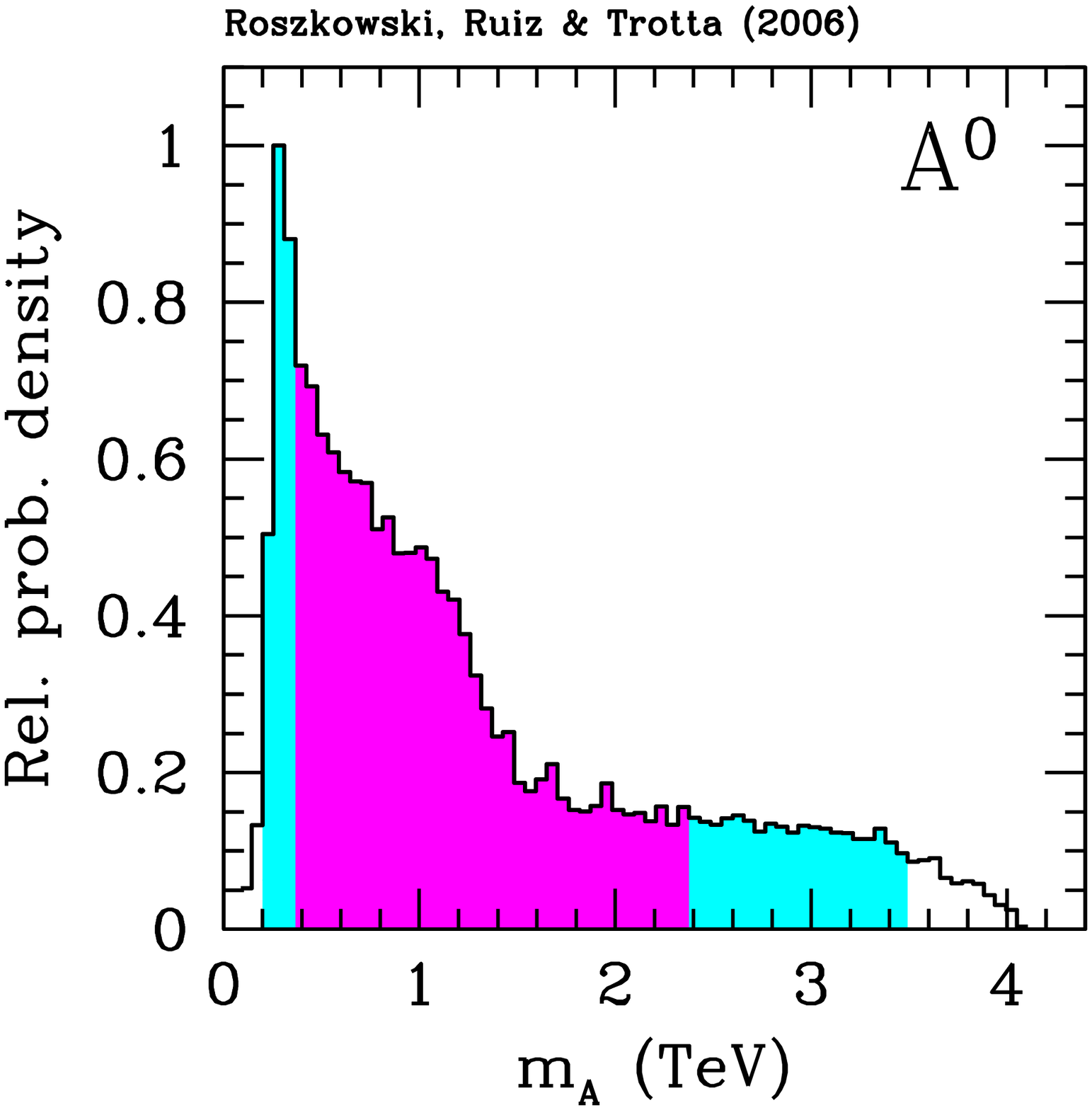}
&	\includegraphics[width=0.3\textwidth]{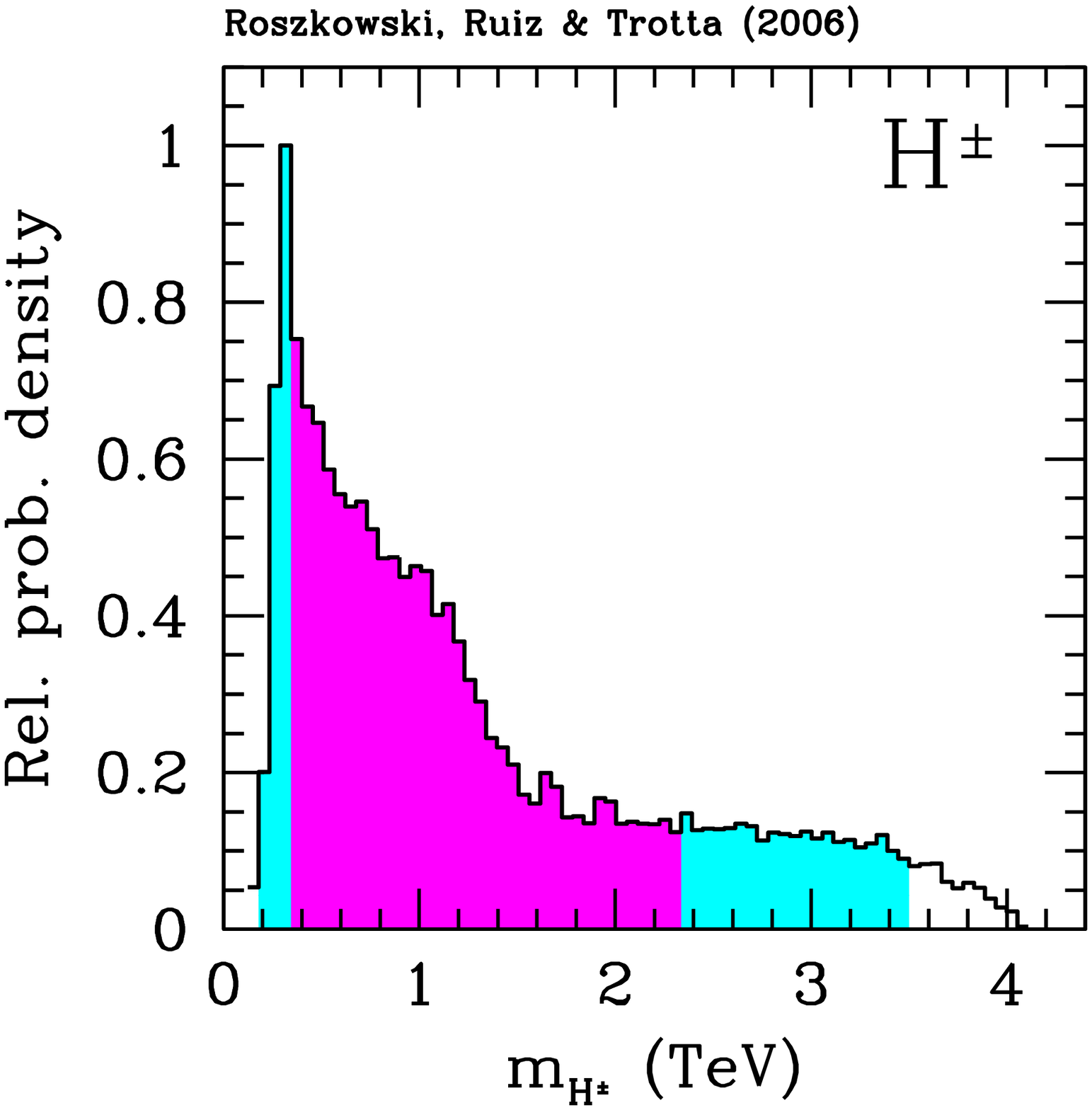}
\end{tabular}
\end{center}
\caption{The 1--dim relative probability densities for the
masses of $\hh$, $\ha$ and $\hpm$, respectively. Magenta (dark shaded)
and cyan (shaded) bands delimit mass ranges of 68\% and 95\% of
posterior probability (2 tails), respectively.
\label{fig:heavyhiggsmass} }
\end{figure}

In fig.~\ref{fig:hlmass} we display the 1--dim relative probability
density of the light Higgs scalar mass. It is clearly well confined,
with the ranges of posterior probability given by
\be
\begin{aligned}
115.4\gev < \mhl < 120.4\gev\quad & (68\% \text{ region}), \\
112.5\gev < \mhl < 121.9\gev\quad & (95\% \text{ region}).
\label{eq:lighthiggsmassrange}
\end{aligned}
\ee

The finite tail on the l.h.s. of the relative probability density in
fig.~\ref{fig:hlmass}, below the final LEP--II lower bound of
$114.4\gev$ (95\%~\cl) is a consequence of the fact that our
likelihood function does not simply cut off points with $\mhl$ below
some arbitrary~\cl, but instead it assigns to them a lower
probability, as described above.  On the other hand, the sharp
drop--off on the r.h.s. of the relative probability density is mostly
caused by the assumed upper bound on $\mzero<4\tev$. This is shown in
the left panel of fig.~\ref{fig:mhvsm0m12}.  The upper bound on $\mhl$
increases with $\mstopave$ and $\msbotave$~(\ref{mhllimit}), whose
largest values in turn depend on the maximum allowed value of
$\mzero$, as we could already see in fig.~5 of~\cite{rtr1}. With the
new SM value for $\brbsgamma$ in the current analysis, the dependence
on the prior range of $\mzero$ has, unfortunately, become even
stronger. 
For instance, adopting a much more generous upper limit $\mzero<8\tev$
would lead to changing the ranges~\ref{eq:lighthiggsmassrange} to roughly
$120.4\gev \lsim \mhl \lsim 124.4\gev$ (68\%~\cl) and $115.4\gev \lsim
\mhl \lsim 125.6 \gev$ (95\%~\cl).
We will come back to this issue when we discuss light Higgs
detection prospects at the Tevatron.

On the other hand, the upper bound on $\mhl$ does not depend
on the upper limit imposed in the prior for $\mhalf$, as can be seen in the right panel of
fig.~\ref{fig:mhvsm0m12} (compare also fig.~2 in
ref.~\cite{rtr1}). Basically, for very large $\tanb\gsim60$ the bottom
quark Yukawa running coupling becomes non--perturbative below the
unification scale and it is no longer possible to find consistent mass
spectra using the RGEs. This upper bound on $\tanb$ limits from above
the values of $\mhalf$ that can still be consistent with
$\abundchi$~\cite{rtr1}.

Fig.~\ref{fig:hlmass} confirms the well--known fact that in the CMSSM
the largest values of $\mhl$ are typically much lower than in the general
MSSM where stop and sbottom masses can to a large extent be treated as
free parameters.  The shape of the relative probability density also
agrees rather well with ref.~\cite{alw06}.
\begin{figure}[!bht]
\begin{center}
\begin{tabular}{c c c}
	\includegraphics[height=0.3\textwidth]{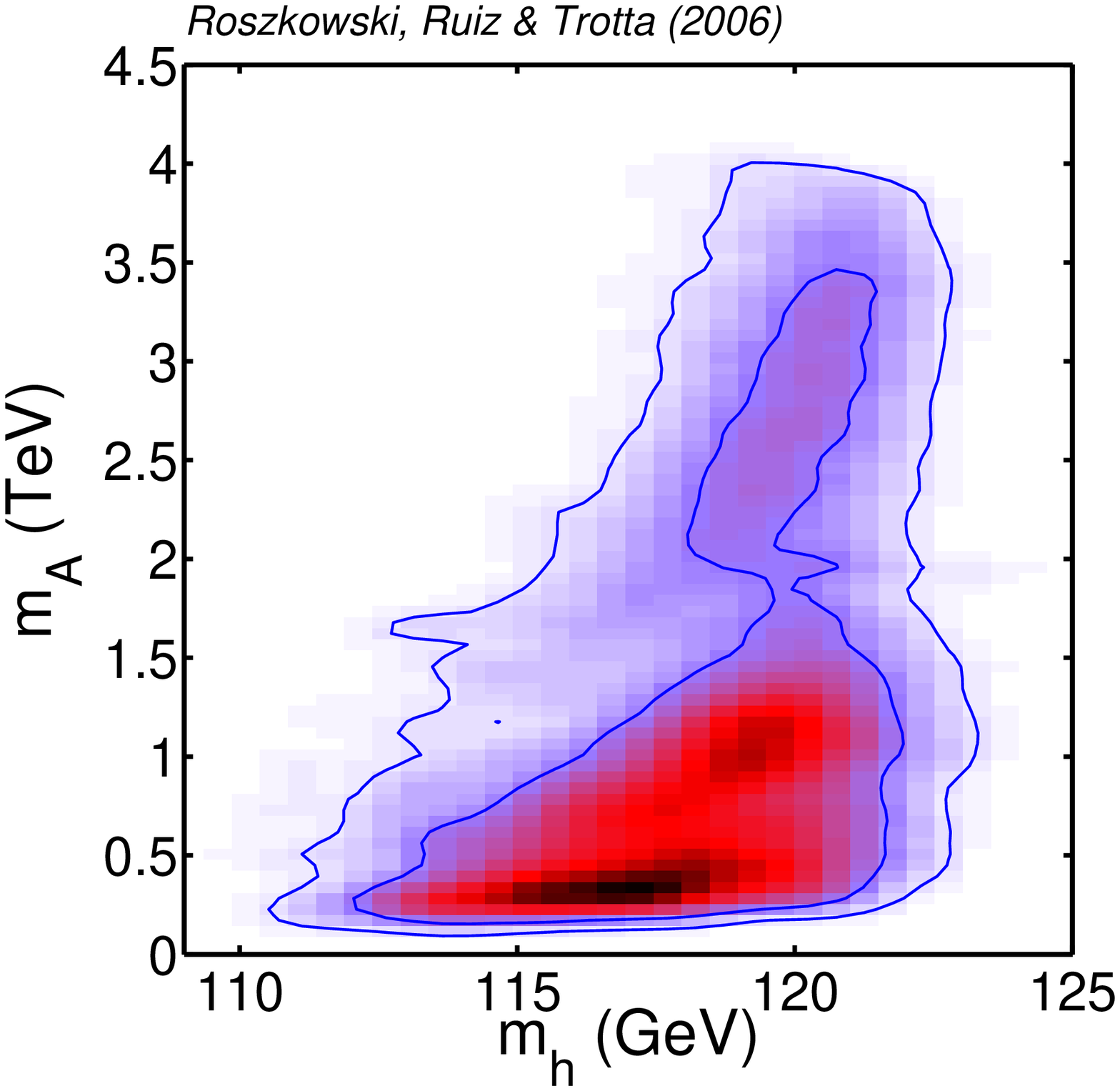}
&	\includegraphics[height=0.3\textwidth]{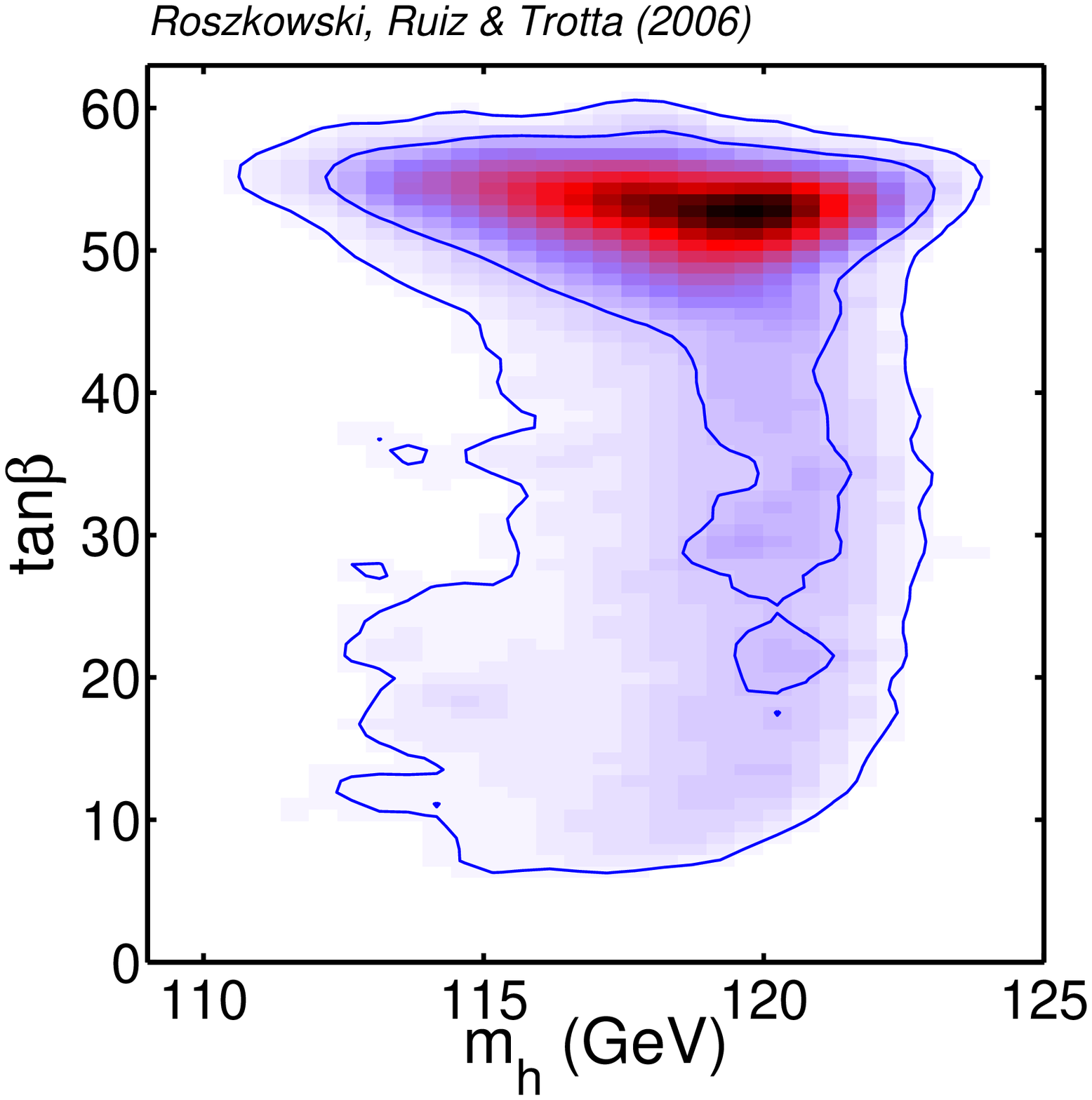}
&	\includegraphics[height=0.3\textwidth]{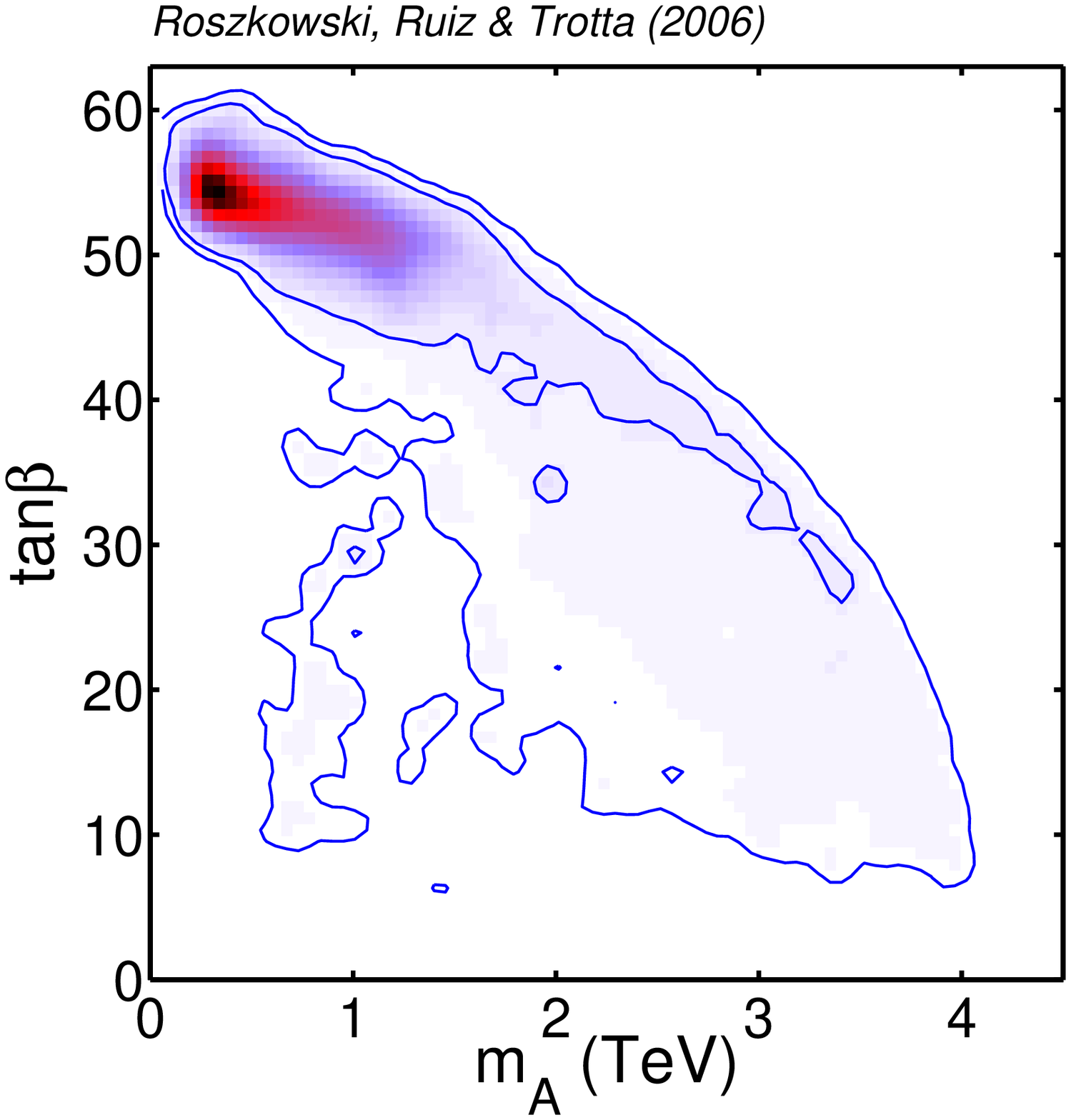}
\end{tabular}
\end{center}
\caption{The 2--dim relative probability density in the plane of
  $(\mhl,\mha)$ (left panel), $(\mhl,\tanb)$ (middle panel) and $(\mha,\tanb)$
  (right panel). 
\label{fig:rrt2-mh-mA-tanb} 
}
\end{figure}

The other Higgs bosons are typically somewhat, but not necessarily
much, heavier. This can be seen in fig.~\ref{fig:heavyhiggsmass},
where we show the 1--dim relative probability densities for the masses
of $\hh$, $\ha$ and $\hpm$, respectively. Note that the shapes of
their relative probability densities are almost identical since the
masses of the three Higgs bosons are nearly degenerate.  Their
posterior probability regions are given by
\be
\begin{aligned}
0.4\tev < m_{\hh,\ha,\hpm} < 2.5\tev\quad & (68\% \text{ region}), \\
0.2\tev < m_{\hh,\ha,\hpm} < 3.6\tev\quad & (95\% \text{ region}).
\end{aligned}
\label{eq:heavyhiggsmassrange}
\ee

\begin{figure}[!t]
\begin{center}
\begin{tabular}{c c}
 	\includegraphics[width=0.3\textwidth]{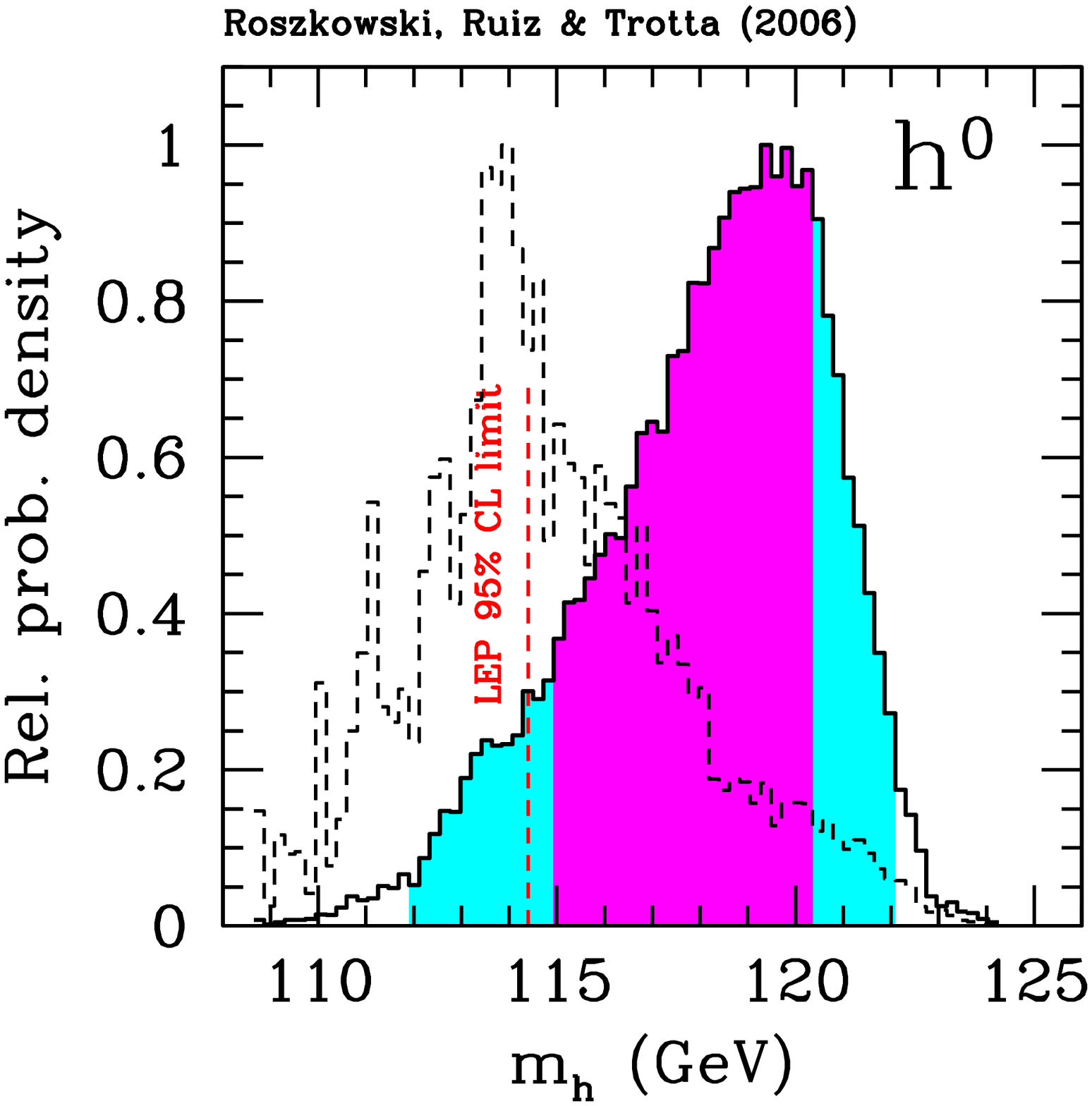}
 &	\includegraphics[width=0.3\textwidth]{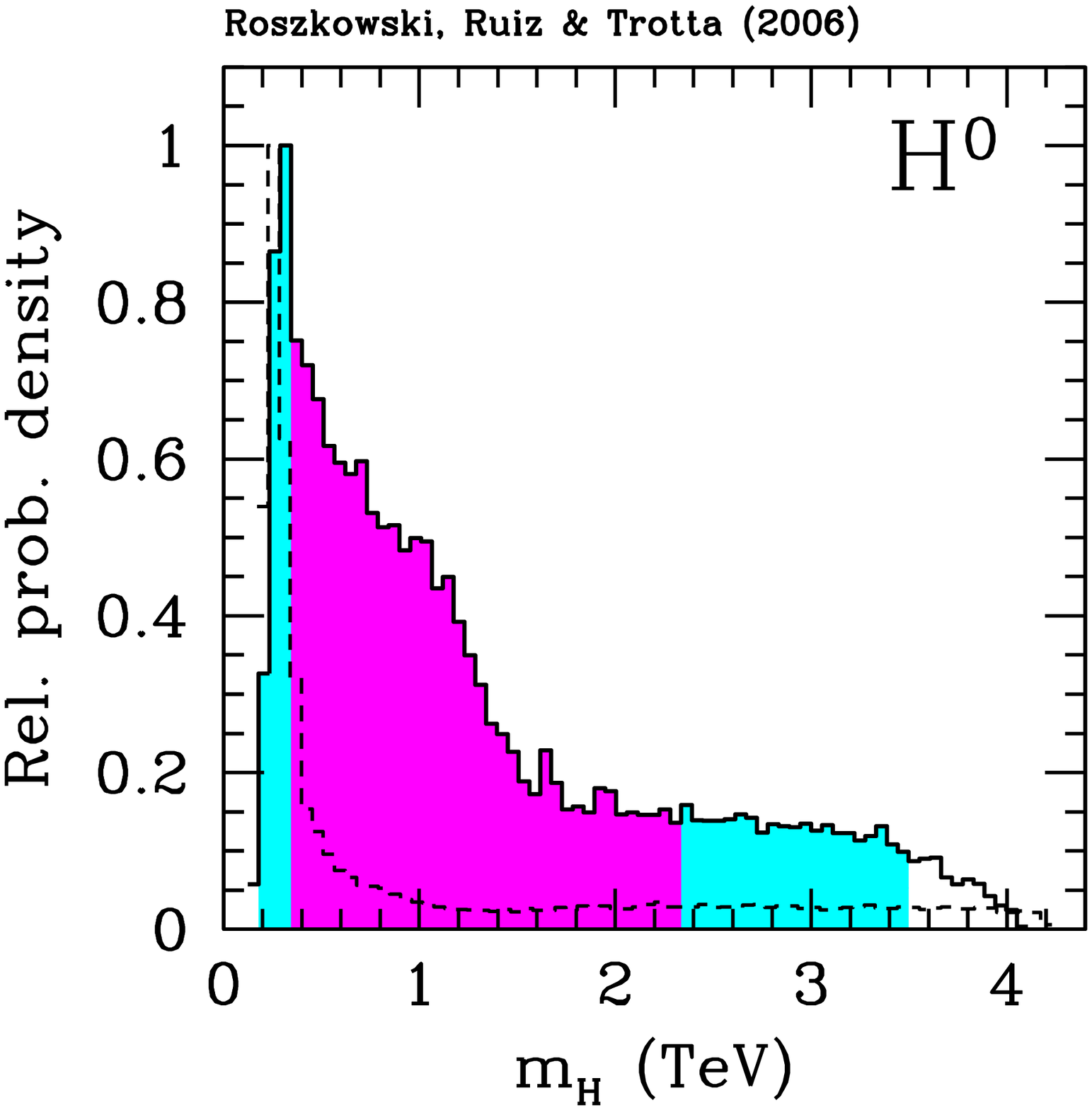}\\
 	\includegraphics[width=0.3\textwidth]{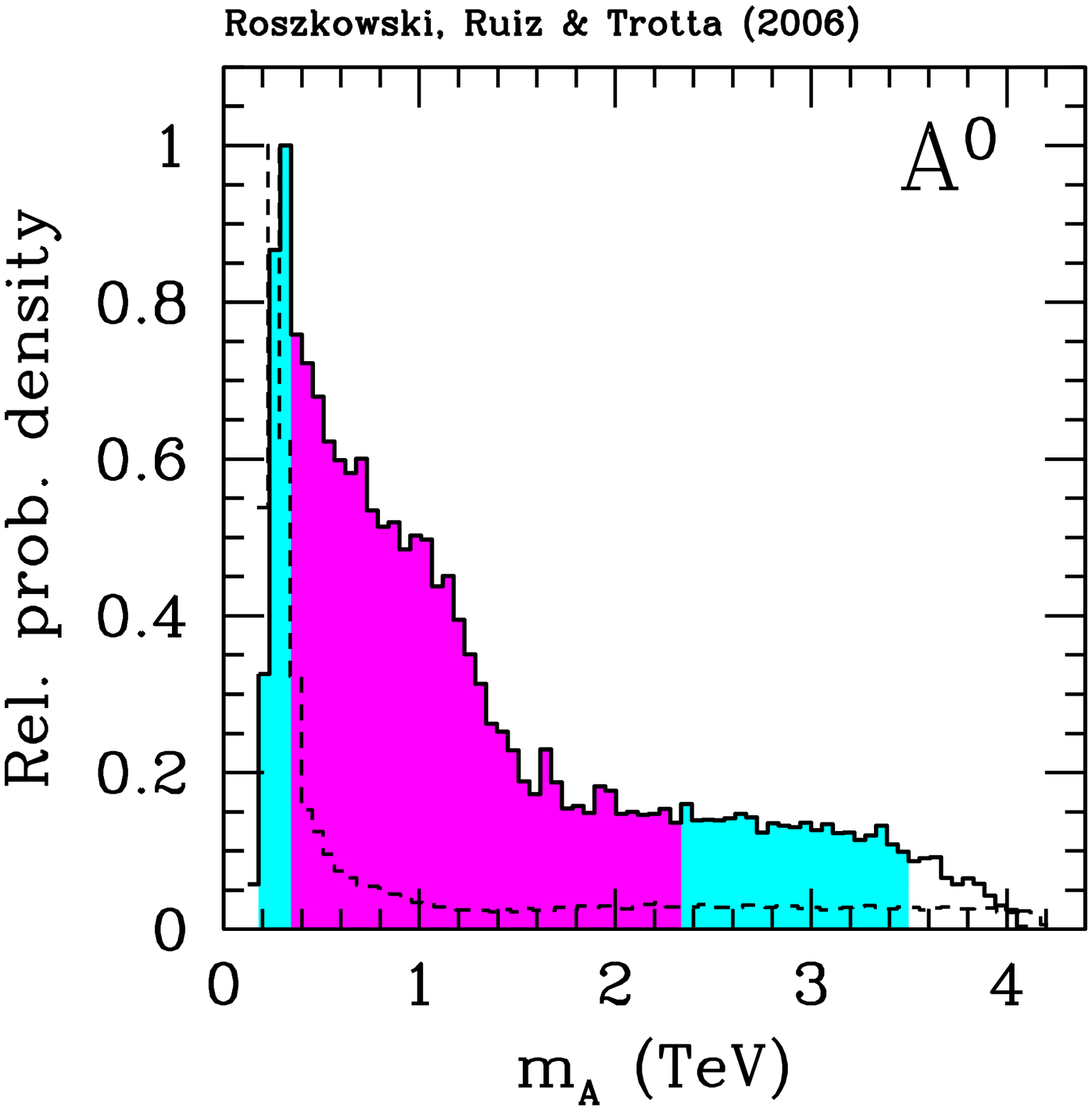}
 &	\includegraphics[width=0.3\textwidth]{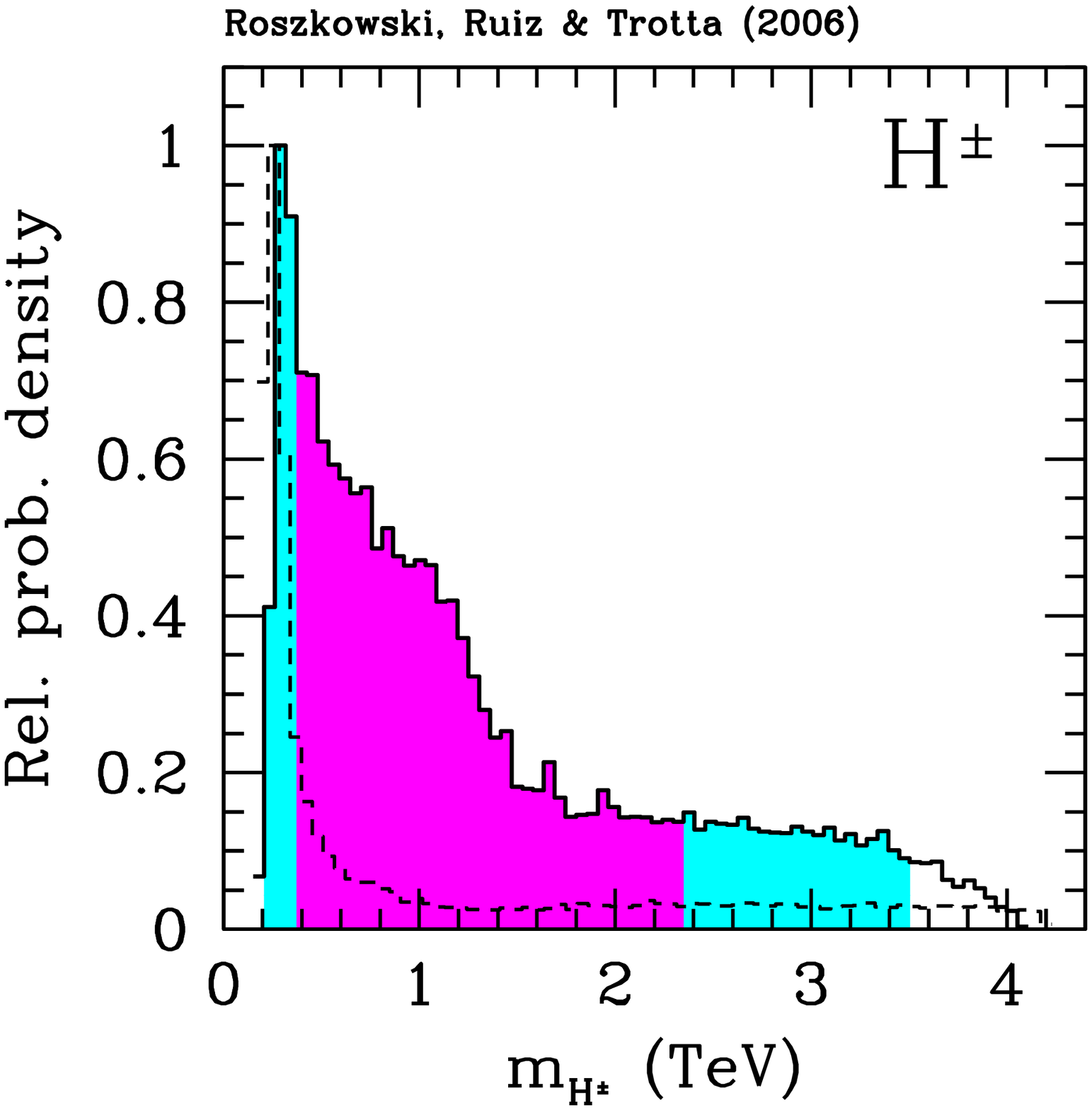}\\
\end{tabular}
\end{center}
\caption{The same as in
  figs.~\protect\ref{fig:hlmass} and~\protect\ref{fig:heavyhiggsmass} but with the mean
  quality--of--fit (dashed line) added for comparison.
\label{fig:hmass-like} }
\end{figure}
\begin{figure}[!tb]
\begin{center}
\begin{tabular}{c c c}
 	\includegraphics[height=0.3\textwidth]{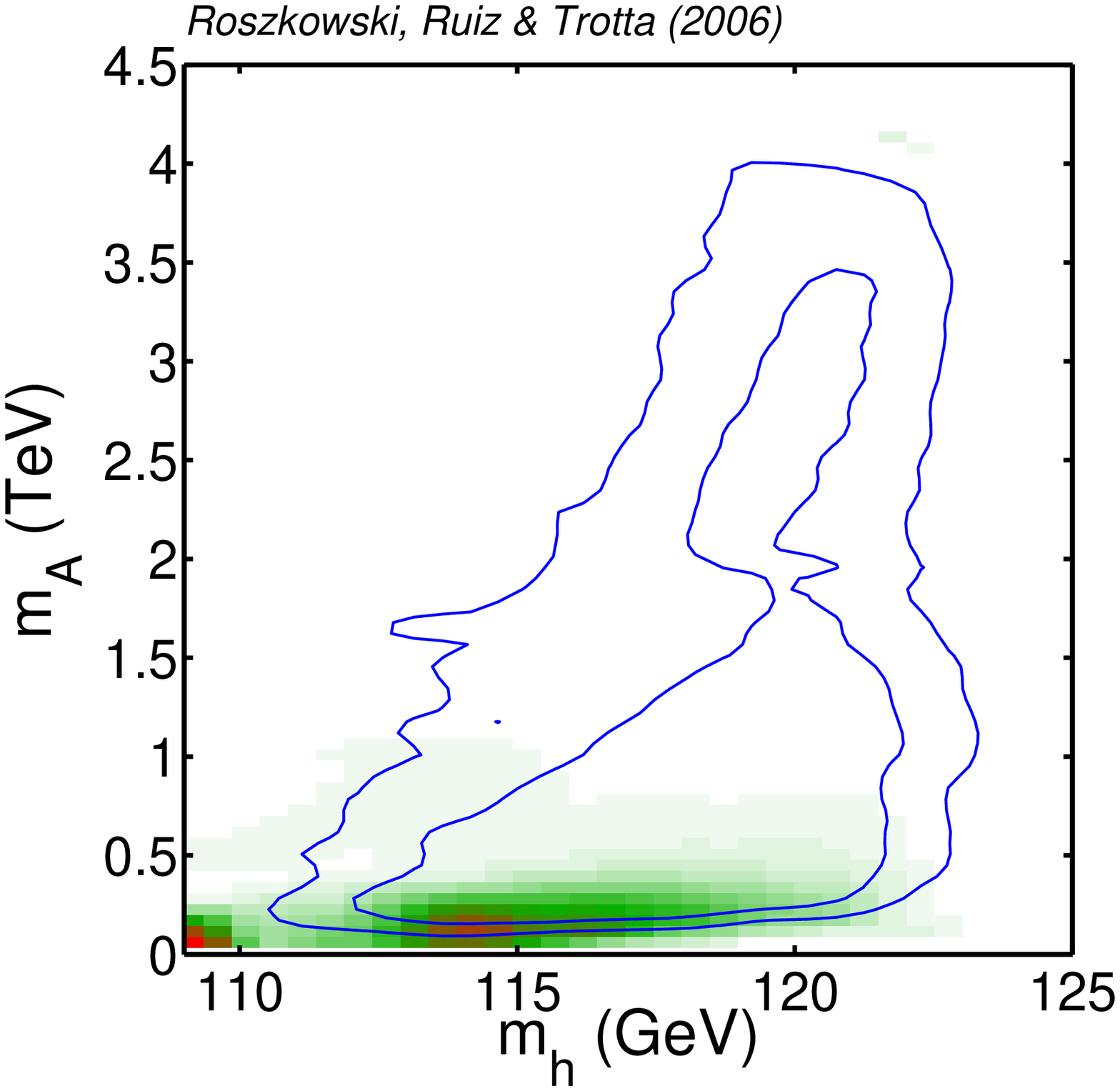}
& 	\includegraphics[height=0.3\textwidth]{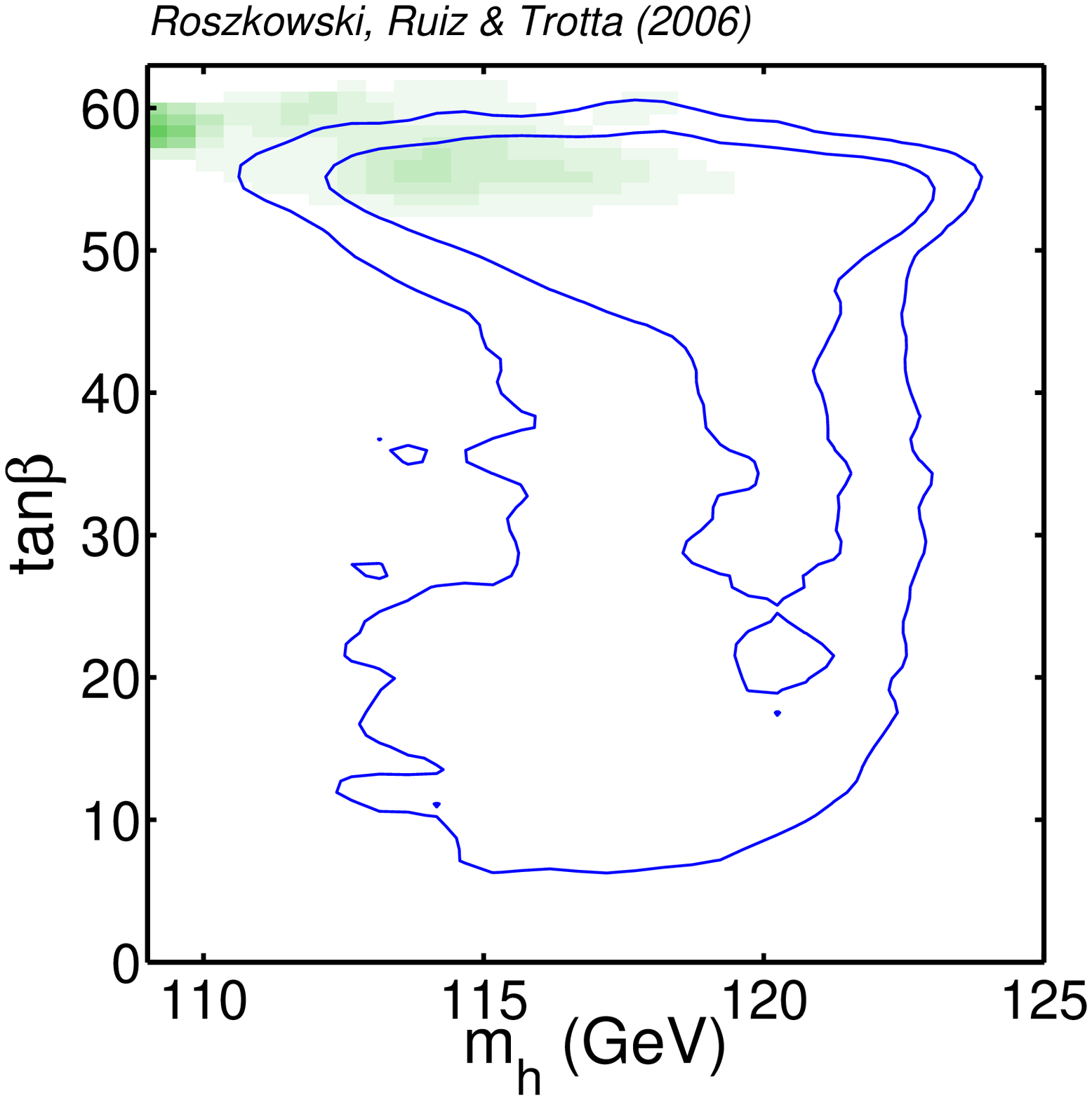}
& 	\includegraphics[height=0.3\textwidth]{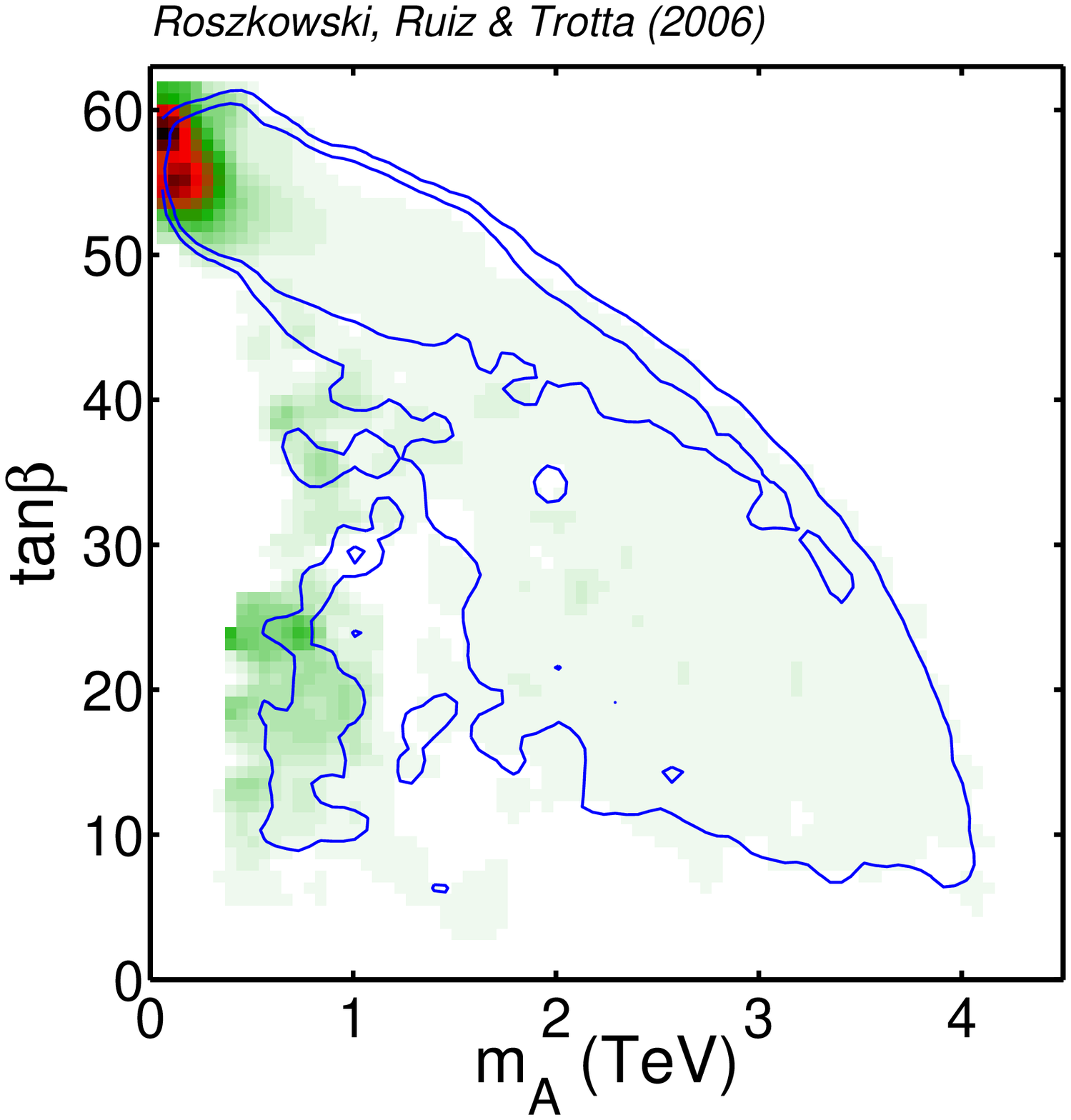}
\end{tabular}
	\includegraphics[width=0.45\textwidth]{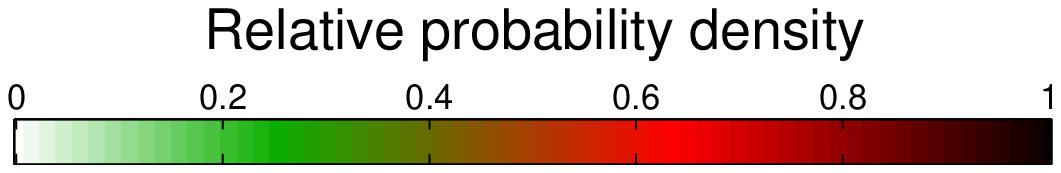}
\end{center}
\caption{The mean quality--of--fit (dashed line) in the planes of $(\mhl,\mha)$ (left
  panel), $(\mhl,\tanb)$ (middle panel) and $(\mha,\tanb)$ (right
  panel). The 68\% and 95\% posterior probability contours from
  fig.~\protect\ref{fig:rrt2-mh-mA-tanb} are also shown for
  comparison.
\label{fig:fitquality} 
}
\end{figure}

It is interesting to see a 2--dim relative probability density in the
($\mhl,\mha$) plane. This is presented in the left panel of
fig.~\ref{fig:rrt2-mh-mA-tanb}. One can see that the pseudoscalar mass
$\mha$ always remains somewhat larger than $\mhl$, but that most
of its probability density is actually concentrated at fairly low
values. (Compare also fig.~\ref{fig:heavyhiggsmass}.) For
$\mhl\lsim115\gev$ (118\gev), at 68\%~\cl\ we find $\mha\lsim0.8\tev$
(1.5\tev).  Experimental
limits on Higgs boson searches in the MSSM are often presented in the
plane spanned by the mass of the Higgs boson and $\tanb$. Our CMSSM
results for the 2--dim relative probability density are presented in
fig.~\ref{fig:rrt2-mh-mA-tanb} in the plane of $(\mhl,\tanb)$ (middle
panel) and $(\mha,\tanb)$ (right panel).  It is clear from
fig.~\ref{fig:rrt2-mh-mA-tanb} that in the CMSSM $\mha \gg \mz$ (the
decoupling regime) but we find $\mha$ predominantly in the regime of a
few hundred $\gev$, which is on the borderline of a ``mild decoupling''
regime. This will affect some relevant Higgs couplings, as we will see
shortly.

At this point we want to digress to emphasize the difference between
posterior probability (Bayesian statistics) and the mean
quality--of--fit statistics. In the absence of strong constraints from
data, the two can produce quite different distributions, which ought
to be interpreted carefully since their meaning is different. (See
ref.~\cite{rtr1} for more details and a thorough discussion.) We
illustrate this in fig.~\ref{fig:hmass-like}, for the case of Higgs
masses where we show the distribution of the mean quality--of--fit
(dashed line) in addition to the relative probability density. Next,
in fig.~\ref{fig:fitquality} we present the mean quality--of--fit in
the planes of $(\mhl,\mha)$ (left panel), $(\mhl,\tanb)$ (middle
panel) and $(\mha,\tanb)$ (right panel). This figure should be
compared with fig.~\ref{fig:rrt2-mh-mA-tanb}.  It is clear that the
largest values of the mean quality--of--fit show preference for a
smaller $\mhl$, with the peak of the distribution at $114\gev$ (below
the LEP--II bound of $114.4\gev$). The mean quality--of--fit also
favors significantly smaller $\mha\lsim 1\tev$ (as well as the other
heavy Higgs masses). We note that, according to the mean
quality--of--fit statistics, the favored region of parameter space
lies at smaller masses than the 68\% range of posterior probability,
as can be seen in fig.~\ref{fig:hmass-like}. Such a discrepancy
between the two statistical measures can only be resolved with better
data.  We also note that we have found a handful of points (about 100)
with $\zetah^2 \ll 1$ exhibiting a very good quality--of--fit at very
small values of $\mhl \sim 90\gev$. Since their statistical weight is
insignificant (compared to some $3\times10^5$ samples in our chain) we
do not display their quality--of--fit in fig.~\ref{fig:fitquality}.
\begin{figure}[!tb]
\begin{center}
\begin{tabular}{c c c}
 	\includegraphics[width=0.3\textwidth]{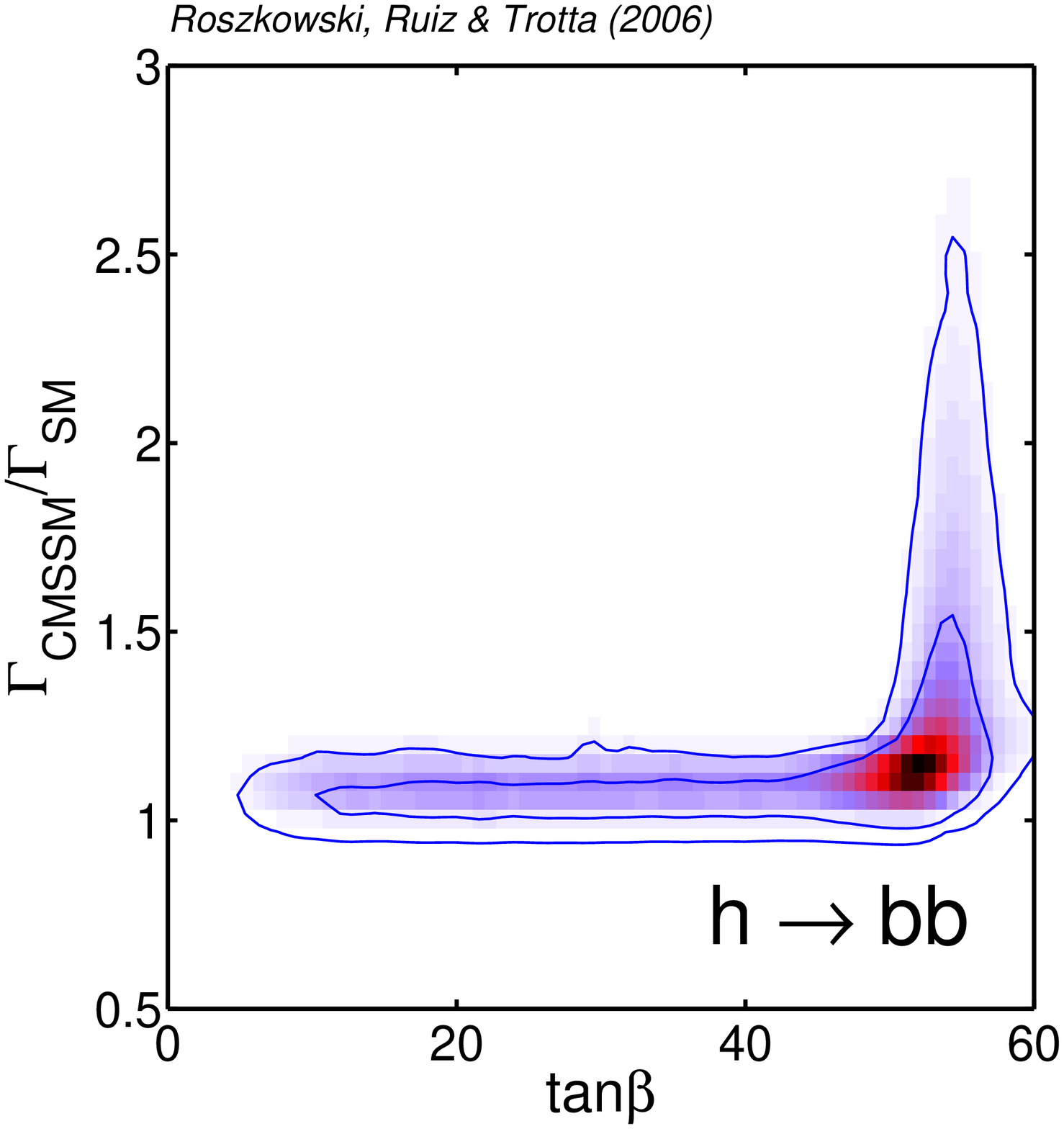}
 &	\includegraphics[width=0.3\textwidth]{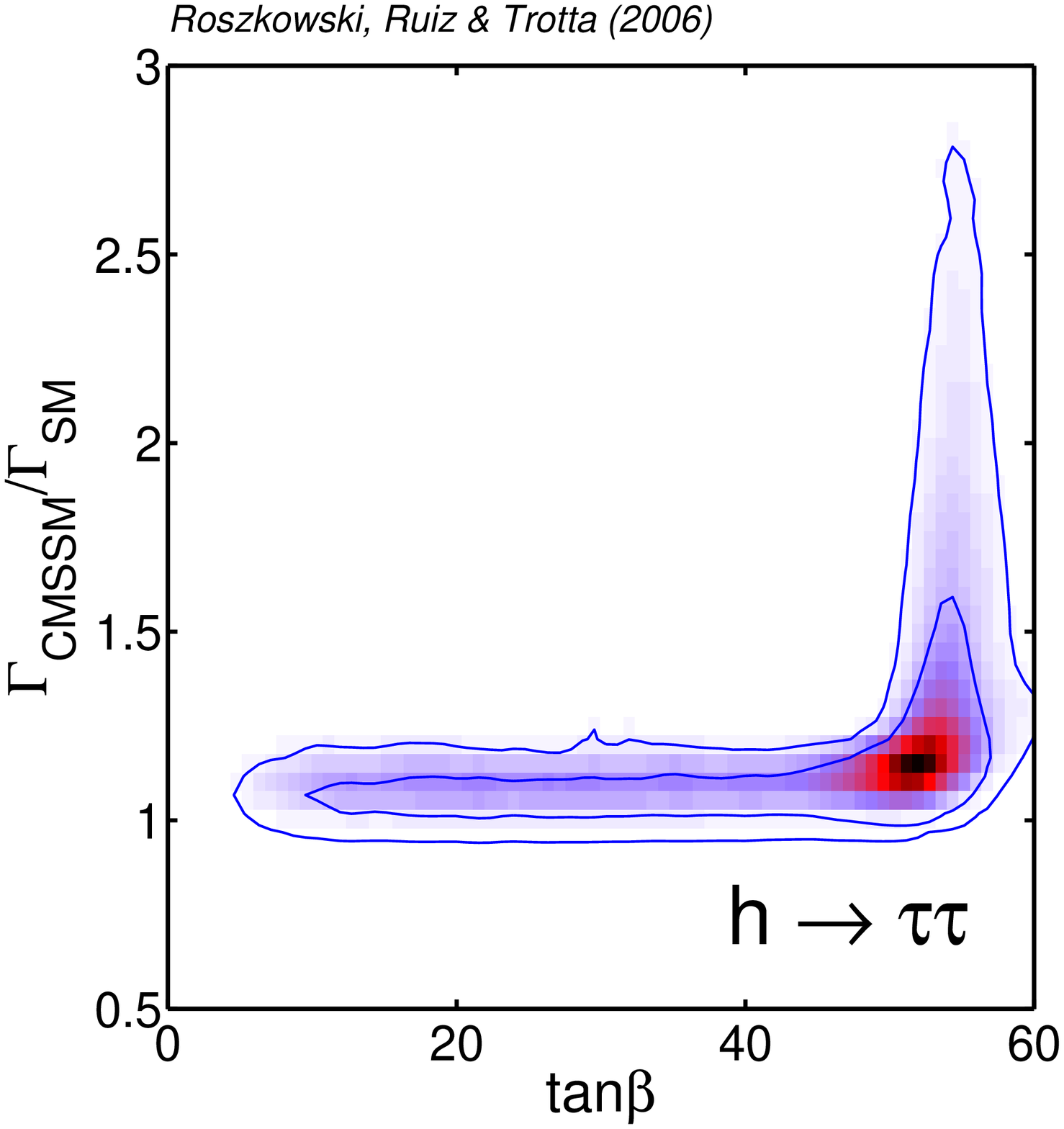}
 &	\includegraphics[width=0.3\textwidth]{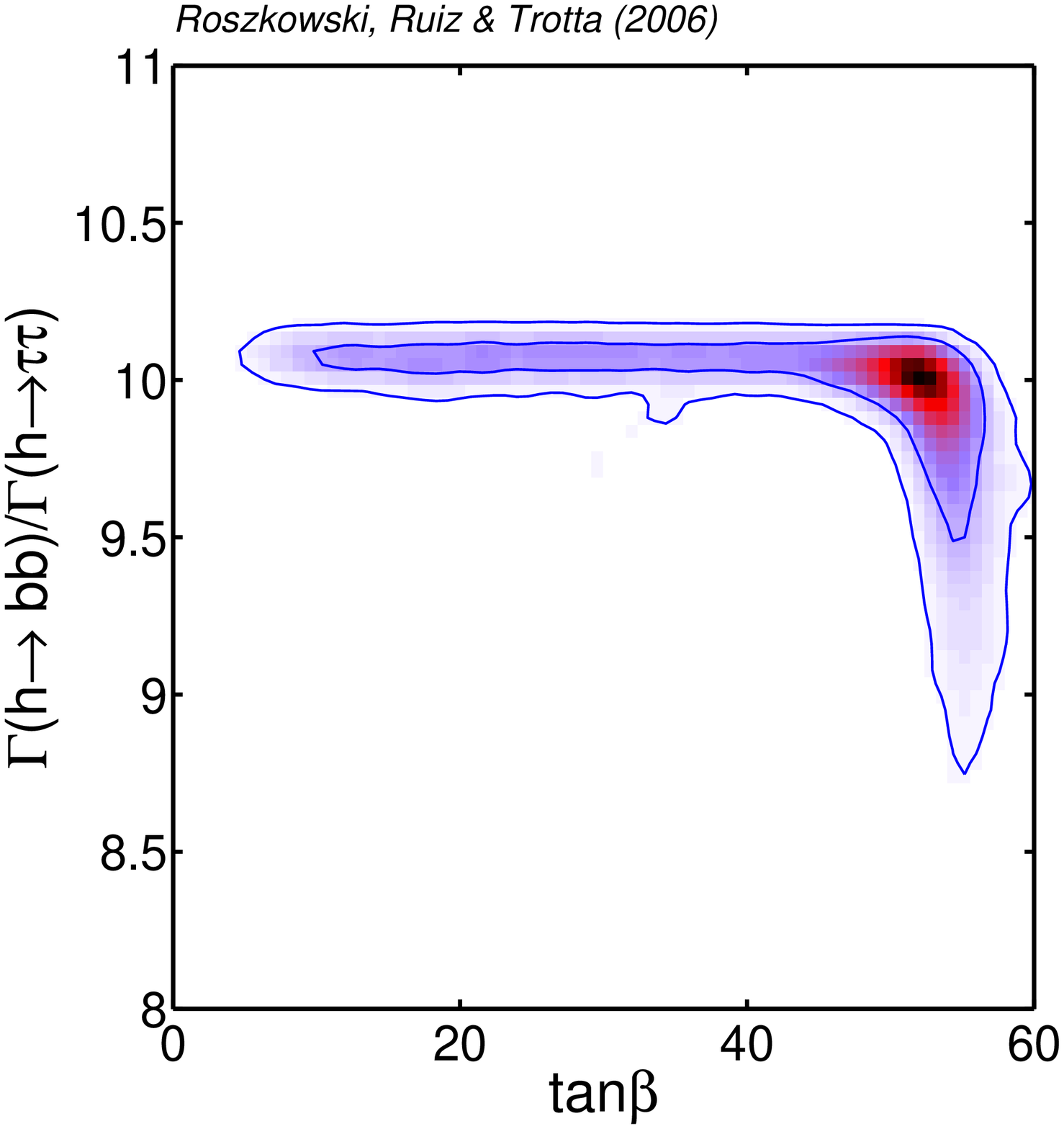}
\end{tabular}
\end{center}
\caption{The 2--dim relative probability density in the plane of
  $\tanb$ and the decay width $\Gamma(\hl\to\bbbar)_{\text{CMSSM}}$ (left panel) and
  $\Gamma(\hl\to\tauptaum)_{\text{CMSSM}}$ (middle panel) in the CMSSM, both normalized to
  the SM values. In the right panel the same quantity is plotted for
  the ratio of the two widths.  The inner (outer) solid
  contours encompass regions of the 68\% (95\%) total probability,
  respectively.
\label{fig:coups-vs-tanb} 
}
\end{figure}
\begin{figure}[!bth]
\begin{center}
\begin{tabular}{c c c}
 	\includegraphics[width=0.3\textwidth]{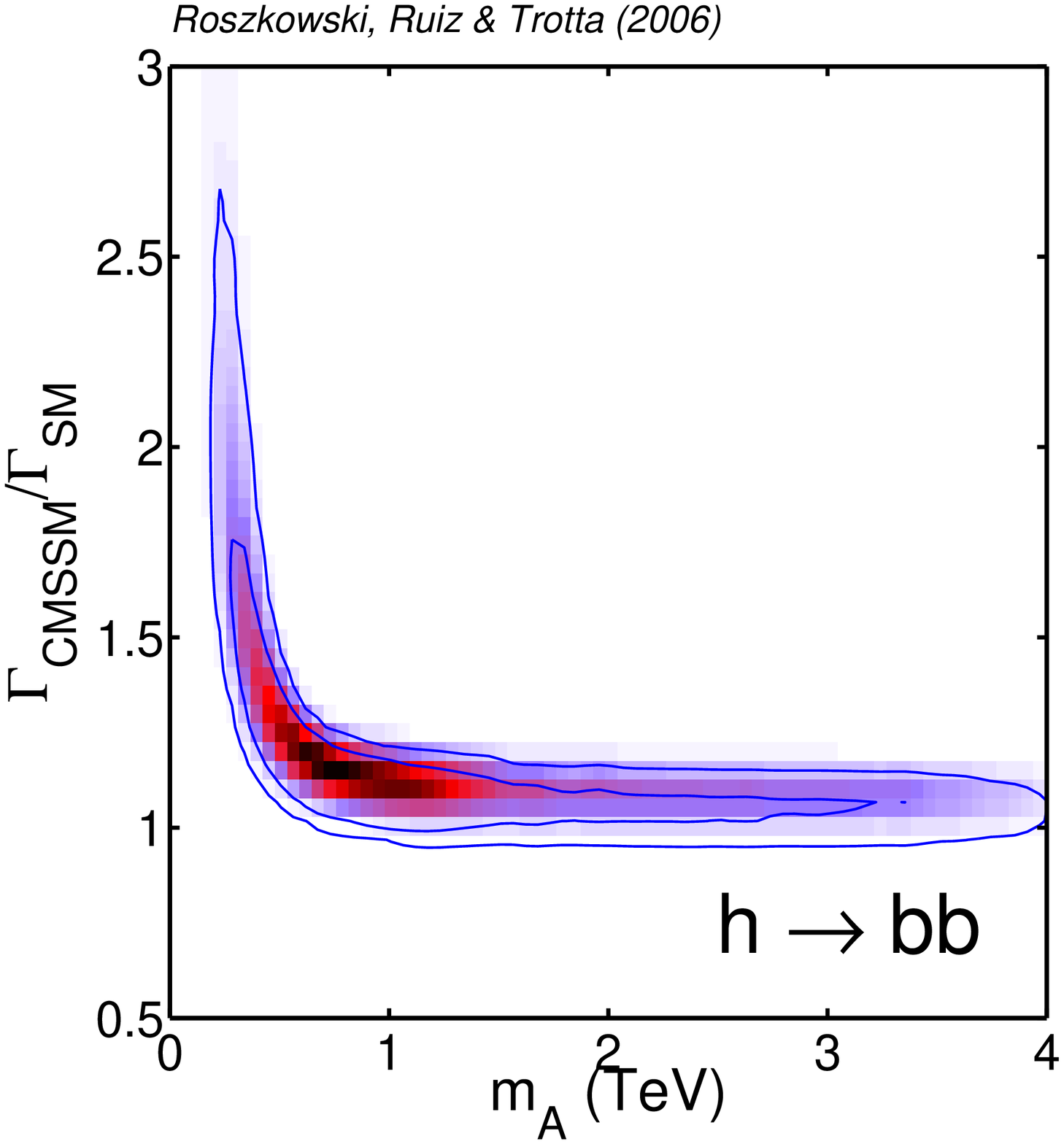}
& 	\includegraphics[width=0.3\textwidth]{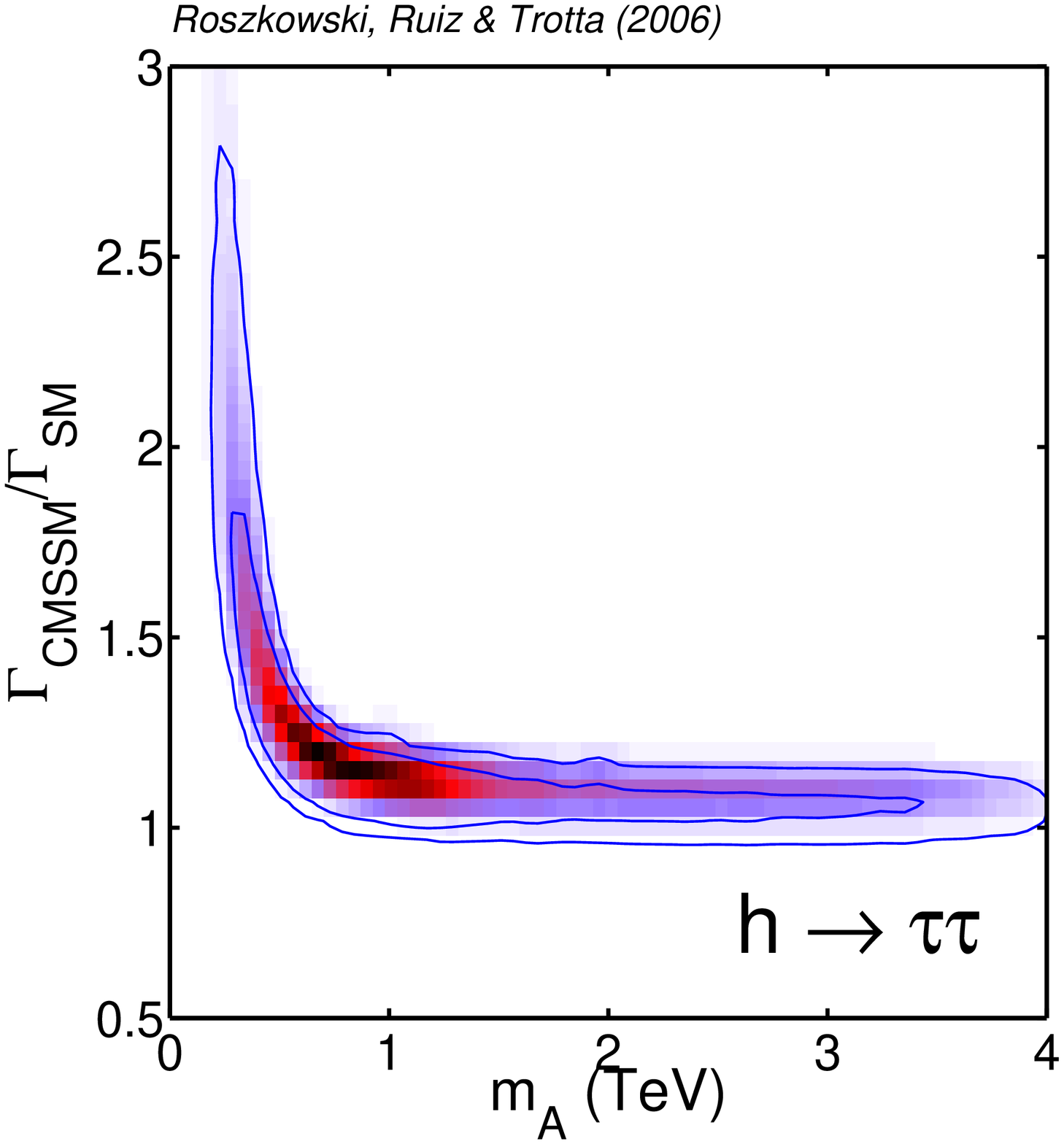}
& 	\includegraphics[width=0.3\textwidth]{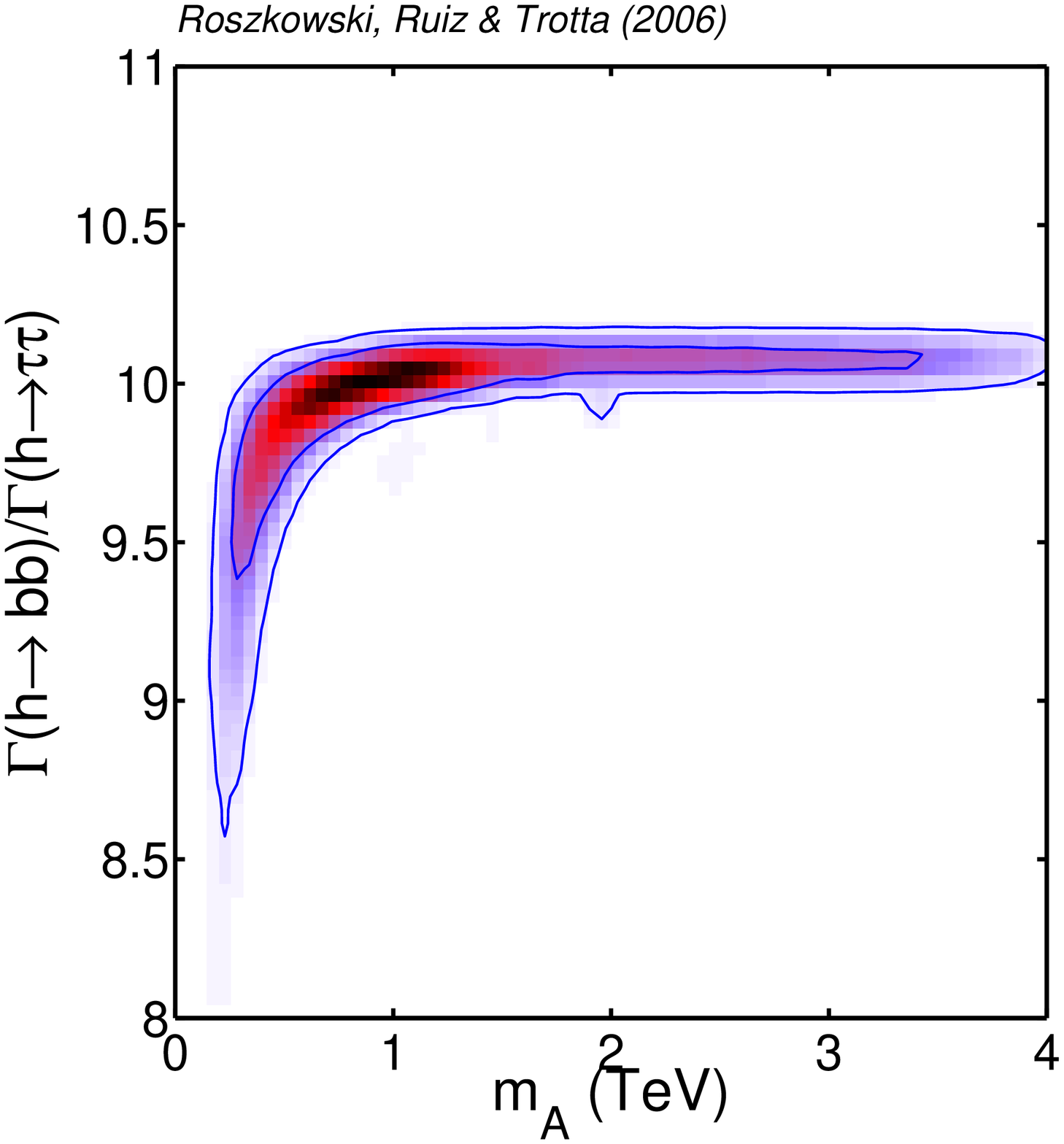}
\end{tabular}
\end{center}
\caption{The same as in fig.~\protect\ref{fig:coups-vs-tanb} but with
  $\tanb$ replaced by $\mha$.
\label{fig:coups-vs-ma} 
}
\end{figure}

Next, we discuss some relevant couplings. We compute them with the
help of FeynHiggs~v2.5.1~\cite{feynhiggs} via the corresponding decay
widths. For this purpose we translate all the Higgs and superpartner
mass spectra and other relevant parameters from the output of
SOFTSUSY. The decay widths of the neutral Higgs bosons are evaluated
with the full rotation to on--shell Higgs bosons, i.e., beyond the
effective coupling approximation.  That means that the mixing matrix
is constructed through the $Z$--factors resulting from the Higgs boson
wave function normalization~\cite{fhhhrw06}. In the case of the vector
bosons $V=Z,W$ the coupling ratio is
$g(hVV)_{\text{MSSM}}/g(hVV)_{\text{SM}}=\sin(\beta-\alphaeff)$, where
$\alphaeff$ denotes the effective (i.e., radiatively--corrected)
mixing angle in the Higgs scalar sector. The ratio is typically
strongly suppressed in the general MSSM. In contrast, in the CMSSM it
is close to 1 as we have already mentioned when discussing $\zetah^2$
in the previous section.\footnote{We do find a tiny increase, at the
level of 1.5\%, which is probably an artefact of the approximations
used in FeynHiggs for computing $\alphaeff$.} This is the case in both
the decoupling and the mild decoupling regimes.

The light Higgs couplings--squared to the third generation down--type
fermions show a more complex behavior, as shown in
figs.~\ref{fig:coups-vs-tanb} and~\ref{fig:coups-vs-ma}. In the
(C)MSSM\footnote{From now on we will denote several MSSM variables
with the subscript ``CMSSM'', in order to emphasize the fact that
their numerical values presented here are specific to the CMSSM,
rather than to the general MSSM.}  the light Higgs coupling to bottoms
$g(\hl\bbbar)_{\text{CMSSM}}$ and to taus
$g(\hl\tauptaum)_{\text{CMSSM}}$, normalized to their SM value,
$g(\hl\ffbar)_{\text{SM}}=gm_{b,\tau}/2\mW$, are given by
$-\sin\alphaeff/\cos\beta=\sin(\beta-\alphaeff)-\tanb\cos(\beta-\alphaeff)$. At
tree level both ratios are equal to 1. We note two effects
here. First, for not too large $\tanb\lsim45$ (for which in the CMSSM
there is strong preference for $\mha\gsim1\tev$) we find that the
ratio $\Gamma_{\text{CMSSM}}/\Gamma_{\text{SM}}$ becomes close to
1.01. This is probably again just a result of an approximation used in
FeynHiggs, as in the case of the $VV$ mode. 
At larger $\tanb$ the second term starts playing a bigger role. As
$\mha$ decreases to below some $0.8\tev$, corresponding to large
$\tanb\gsim50$, both coupling--squared ratios grow rather fast. The
enhancements in this region can 
be seen in the left and middle panels of figs.~\ref{fig:coups-vs-tanb}
and~\ref{fig:coups-vs-ma}.  At the $2\sigma$ level, we find that the widths
are increased relative to their SM counterparts by a factor of up to some 2.5.

The second effect on the couplings is caused by radiative corrections
from sbottom--gluino and stop--higgsino loops to the tree--level
relation between the bottom mass and its Yukawa
coupling~\cite{hrs-effect}.  At large $\tanb$ this leads to modifying
the coupling $g(\hl\bbbar)$~\cite{hrs-effect-couplings}, while the
analogous coupling to taus is not affected. (Implications of this
effect for Tevatron Higgs searches have recently been discussed in
ref.~\cite{chww05-fnal-Higgs-discovery}.)  As a result, in the CMSSM
at large $\tanb\gsim50$, in a sizable number of cases the quantity
$g^2(\hl\bbbar)_{\text{CMSSM}}$, while remaining dominant, will show a
small decrease relative to $g^2(\hl\tauptaum)_{\text{CMSSM}}$. This
feature, which is displayed in the right panels of
figs.~\ref{fig:coups-vs-tanb} and~\ref{fig:coups-vs-ma}, will give one
some chance of producing a somewhat increased number of taus in light
Higgs decays at the Tevatron, as we will see shortly.

\begin{figure}[!t]
\begin{center}
\begin{tabular}{c c c}
 	\includegraphics[width=0.3\textwidth]{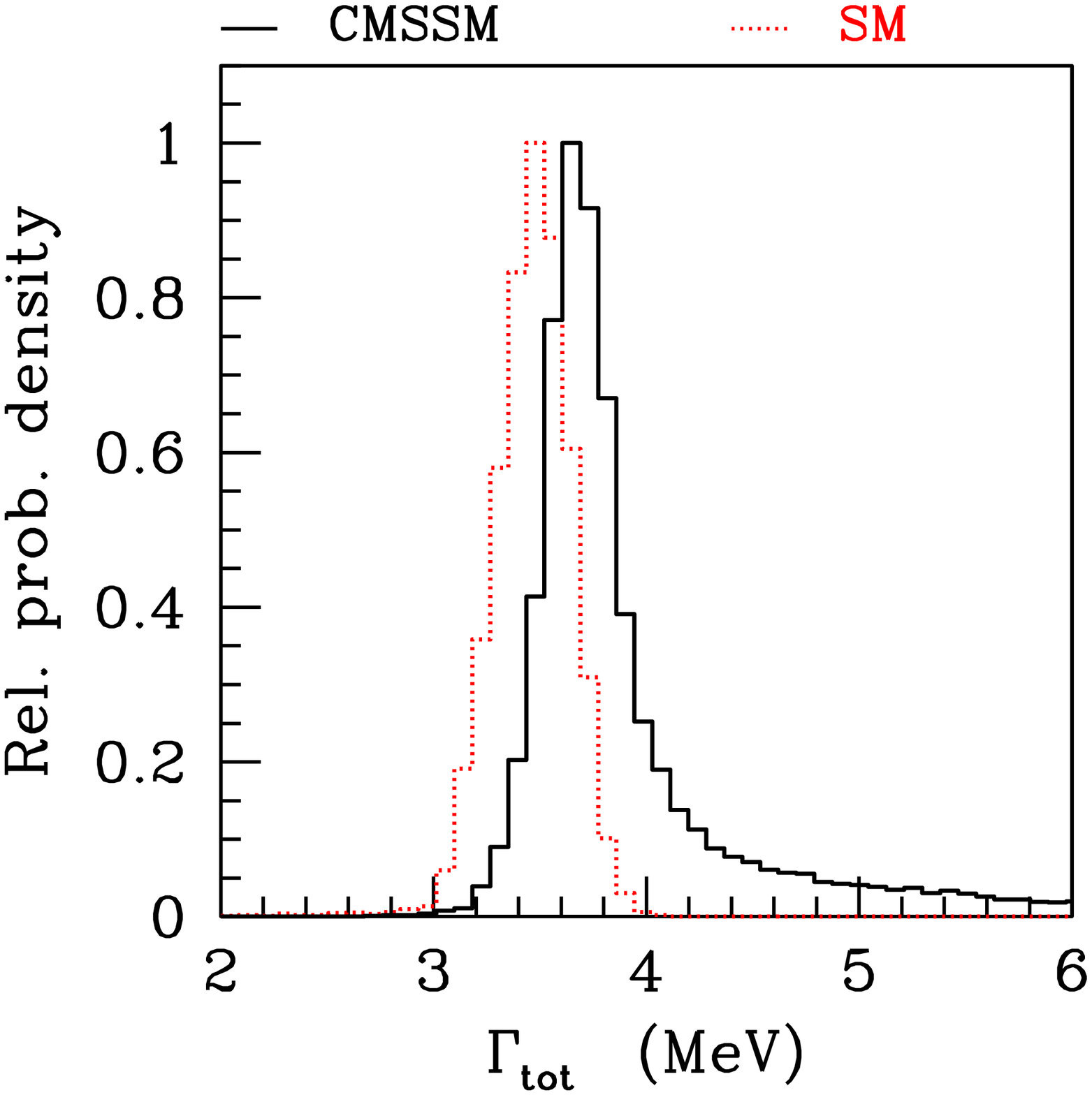}
 &	\includegraphics[width=0.3\textwidth]{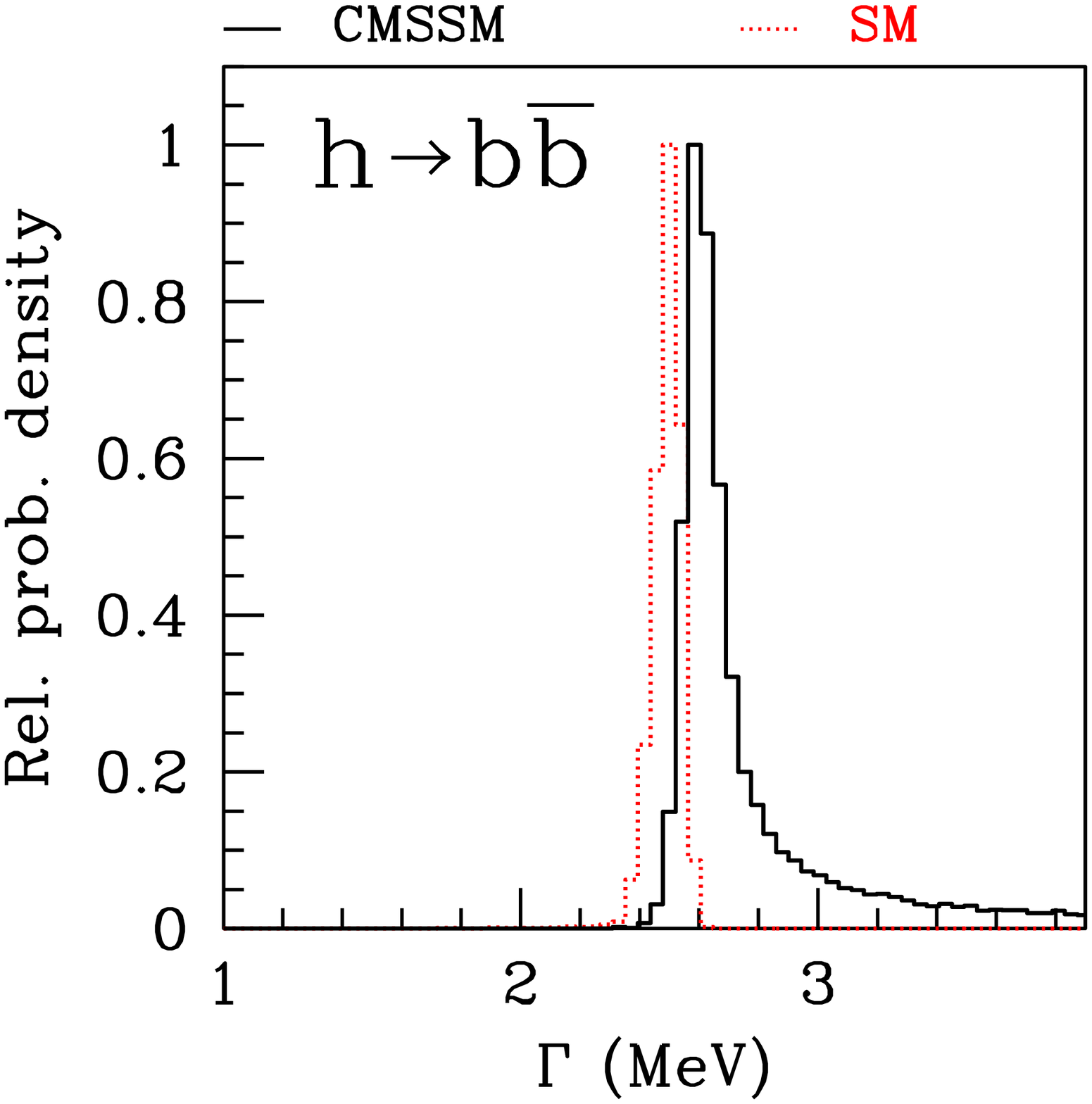}
 & 	\includegraphics[width=0.3\textwidth]{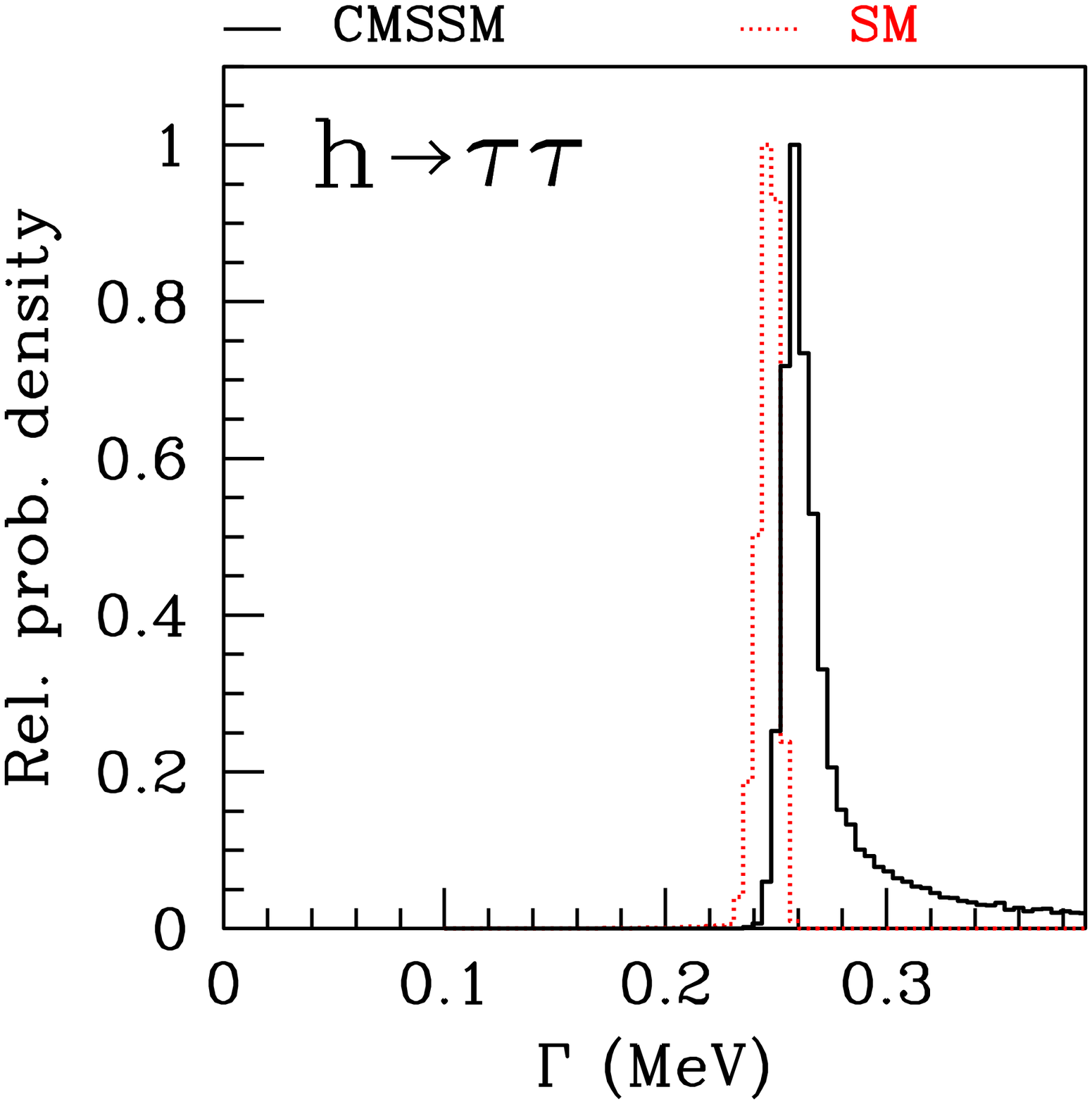}
\end{tabular}
\end{center}
\caption{The 1--dim relative probability densities for the light Higgs
  total decay width (left  panel) and partial decay widths to
  $\bbbar$ (middle panel) and $\tauptaum$ (right panel). For
  comparison, the widths of the SM Higgs boson with the same 
  mass are also shown. 
\label{fig:hlwidths} }
\end{figure}

\begin{figure}[!tbh]
\begin{center}
\begin{tabular}{c c c}
	\includegraphics[width=0.3\textwidth]{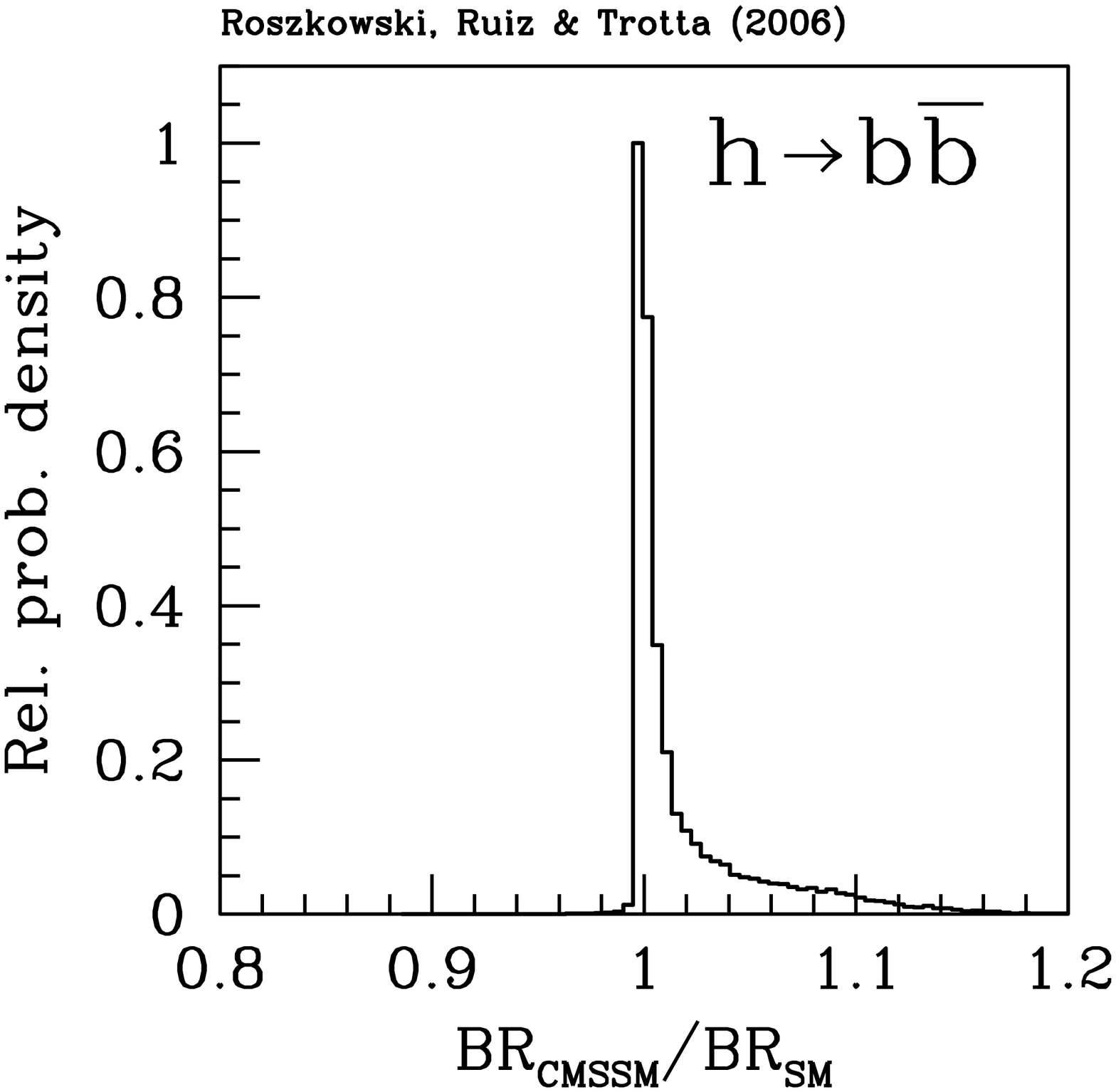}
&	\includegraphics[width=0.3\textwidth]{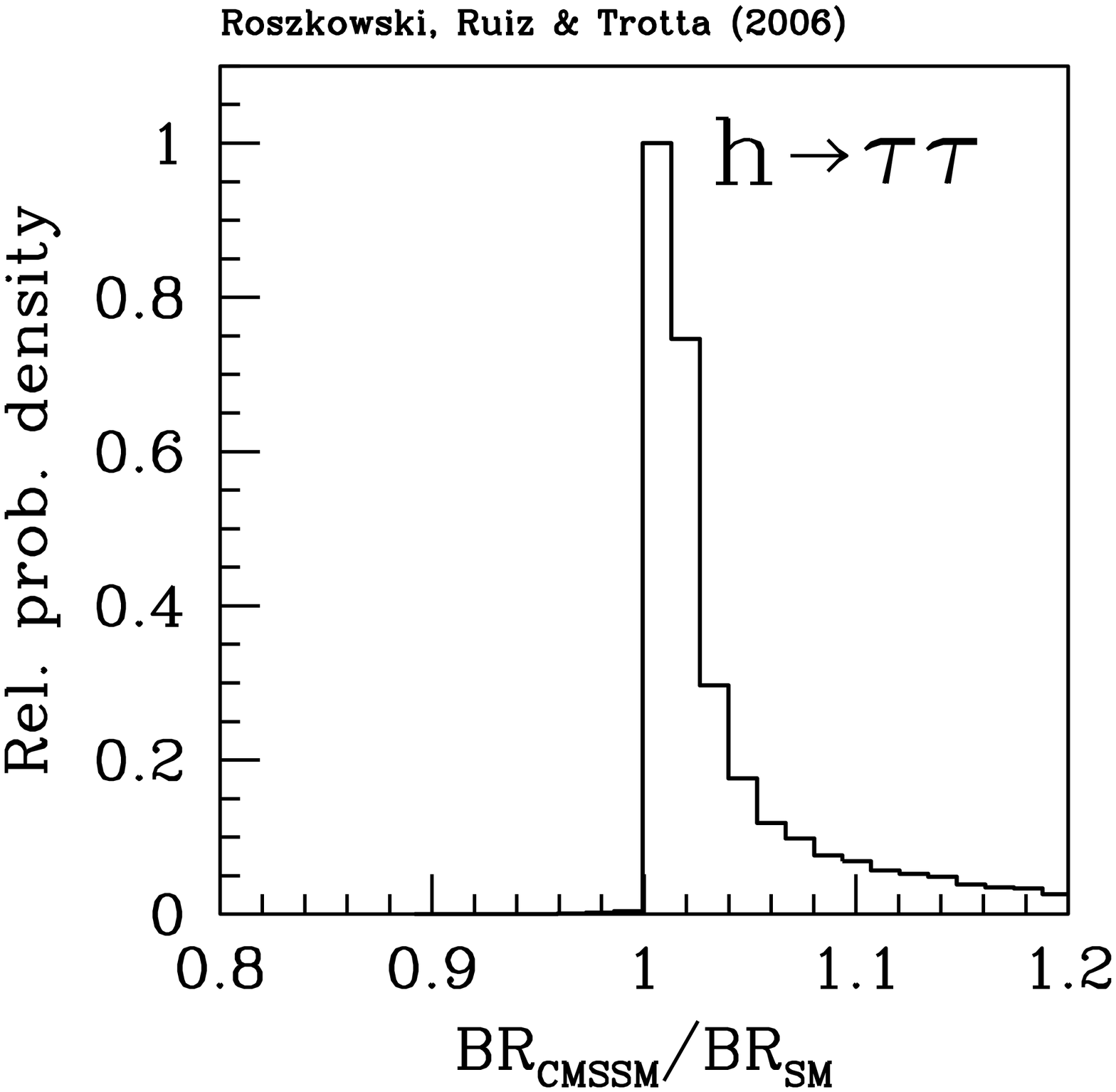}
&	\includegraphics[width=0.3\textwidth]{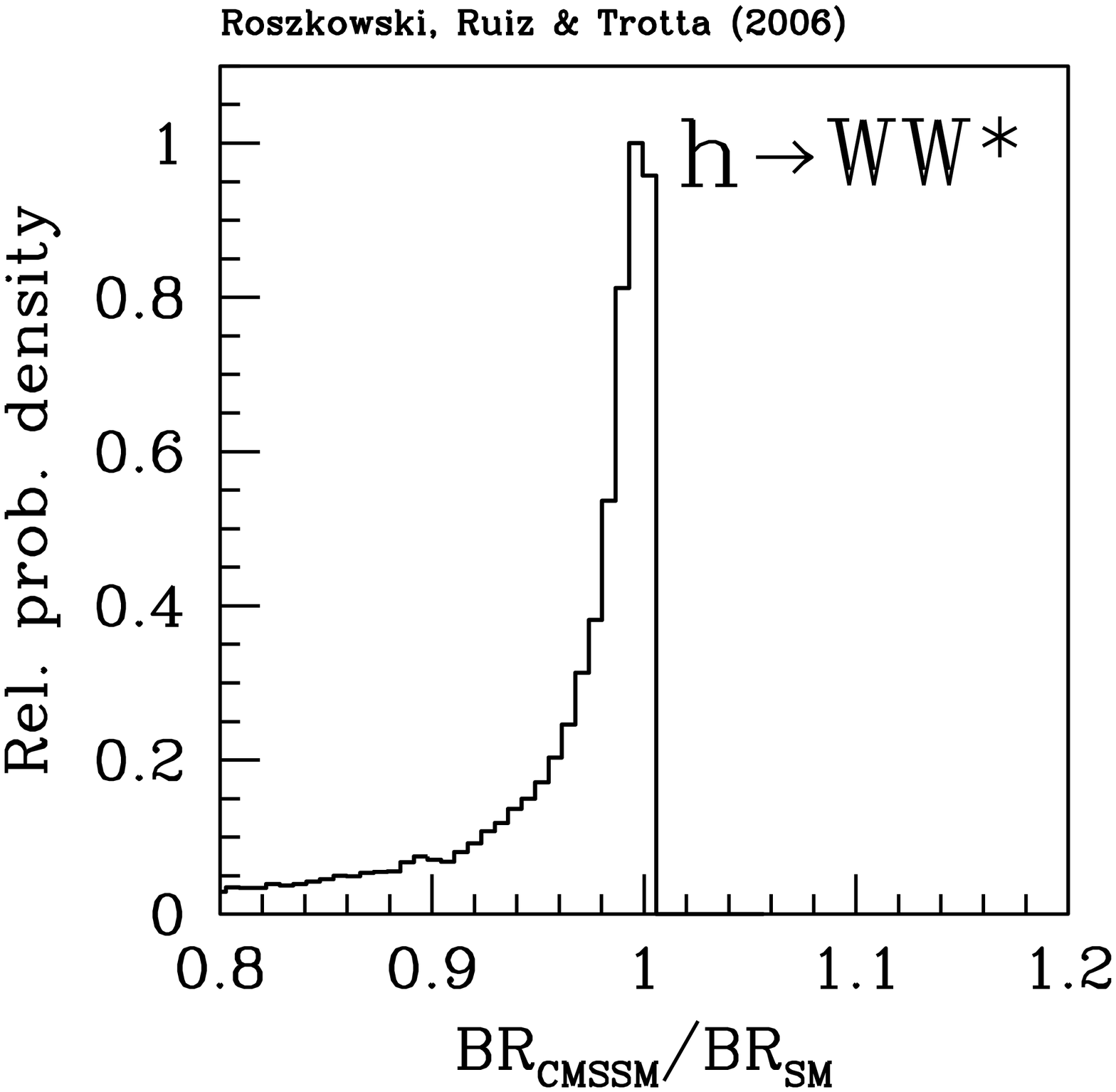}
\end{tabular}
\end{center}
\caption{The 1--dim relative probability density of the branching ratios
$BR(\hl \to b\bar{b})$ (left panel), $BR(\hl \to \tau^+\tau^-)$  (middle panel) and
$BR(\hl \to WW^\ast)$ (right panel), all normalized to their SM counterparts.
\label{fig:rrt2-hldecayratio} 
}
\end{figure}
\begin{figure}[!tbh]
\begin{center}
\begin{tabular}{c c c}
	\includegraphics[width=0.3\textwidth]{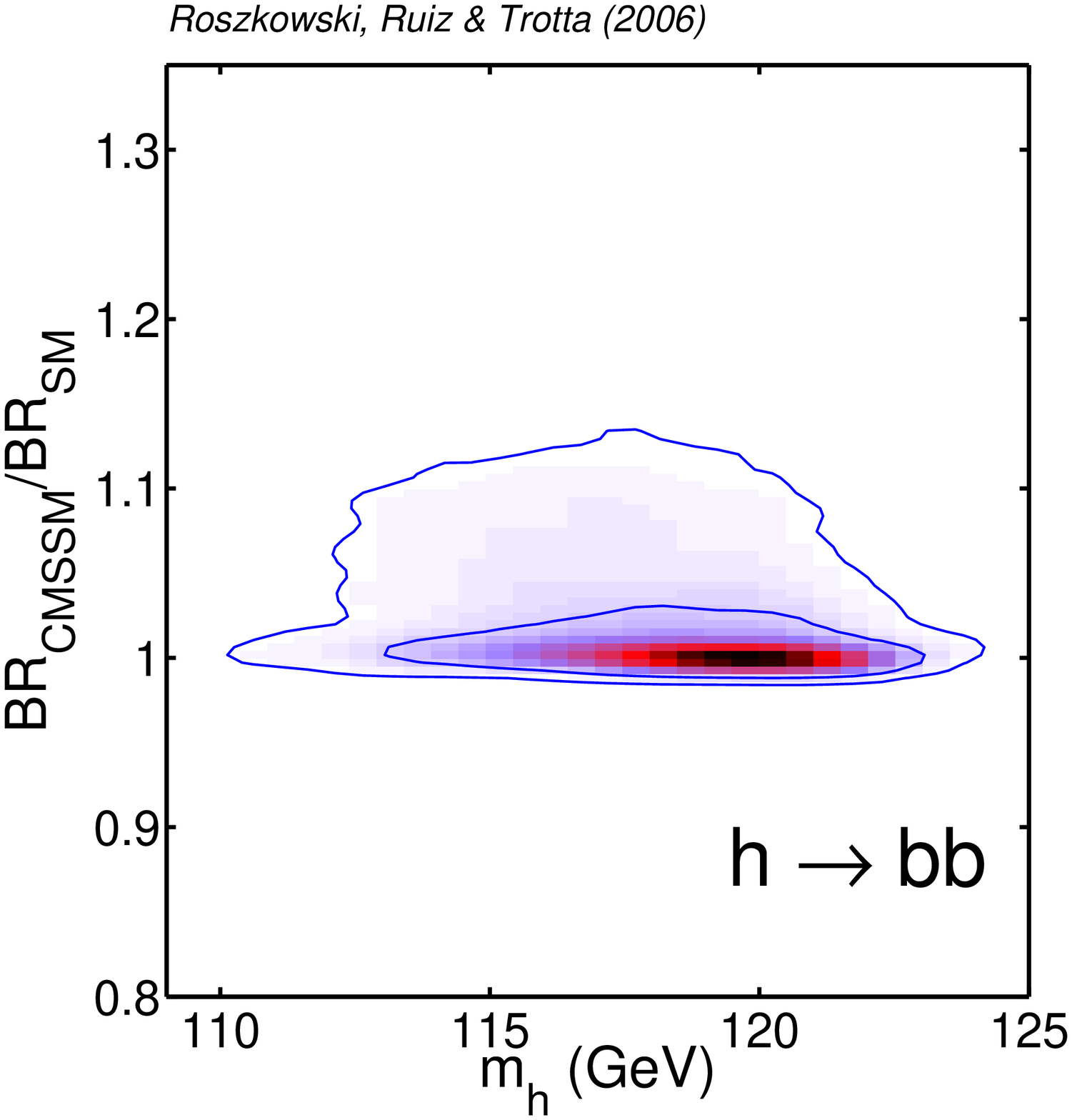}
&	\includegraphics[width=0.3\textwidth]{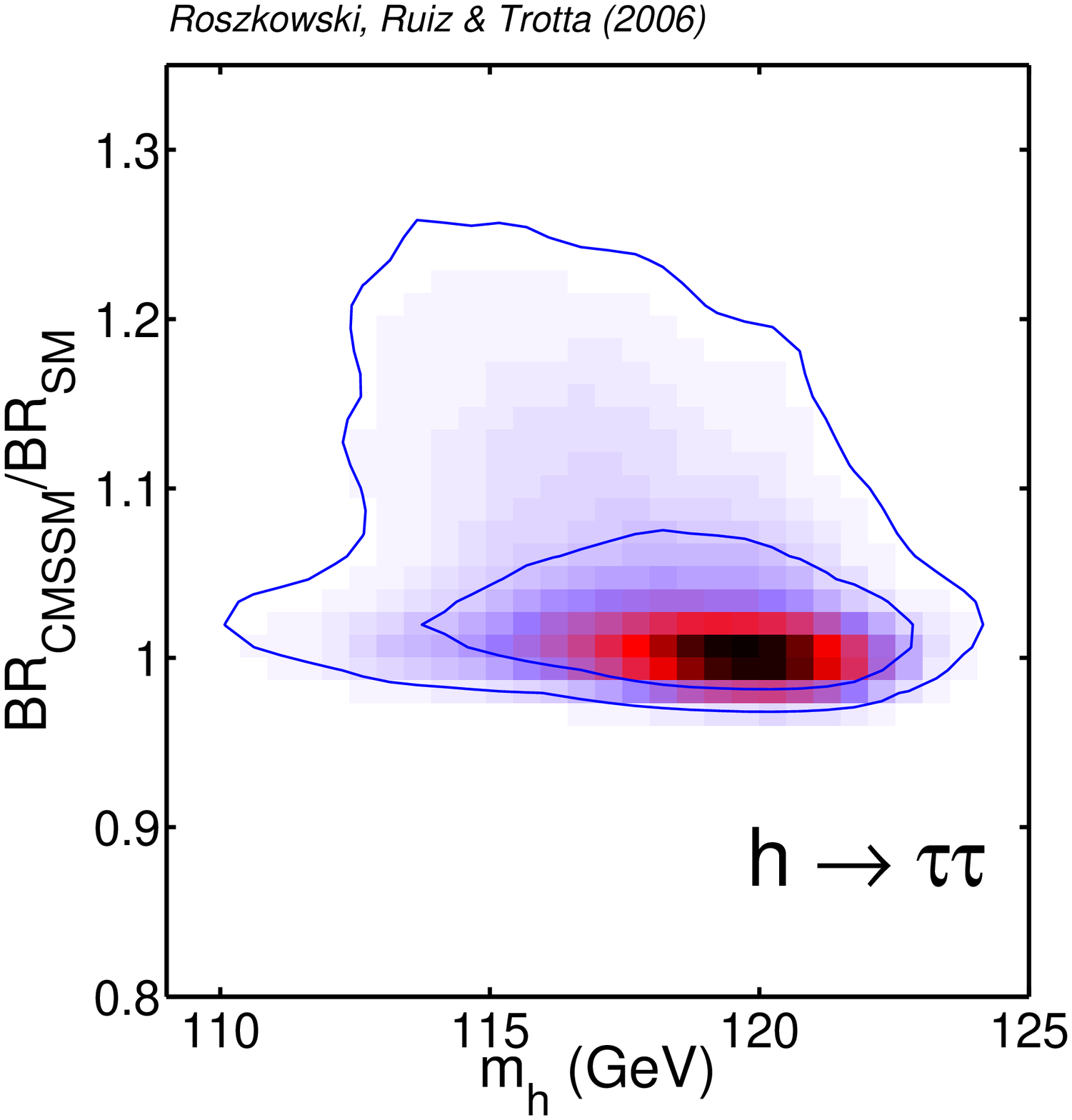}
&	\includegraphics[width=0.3\textwidth]{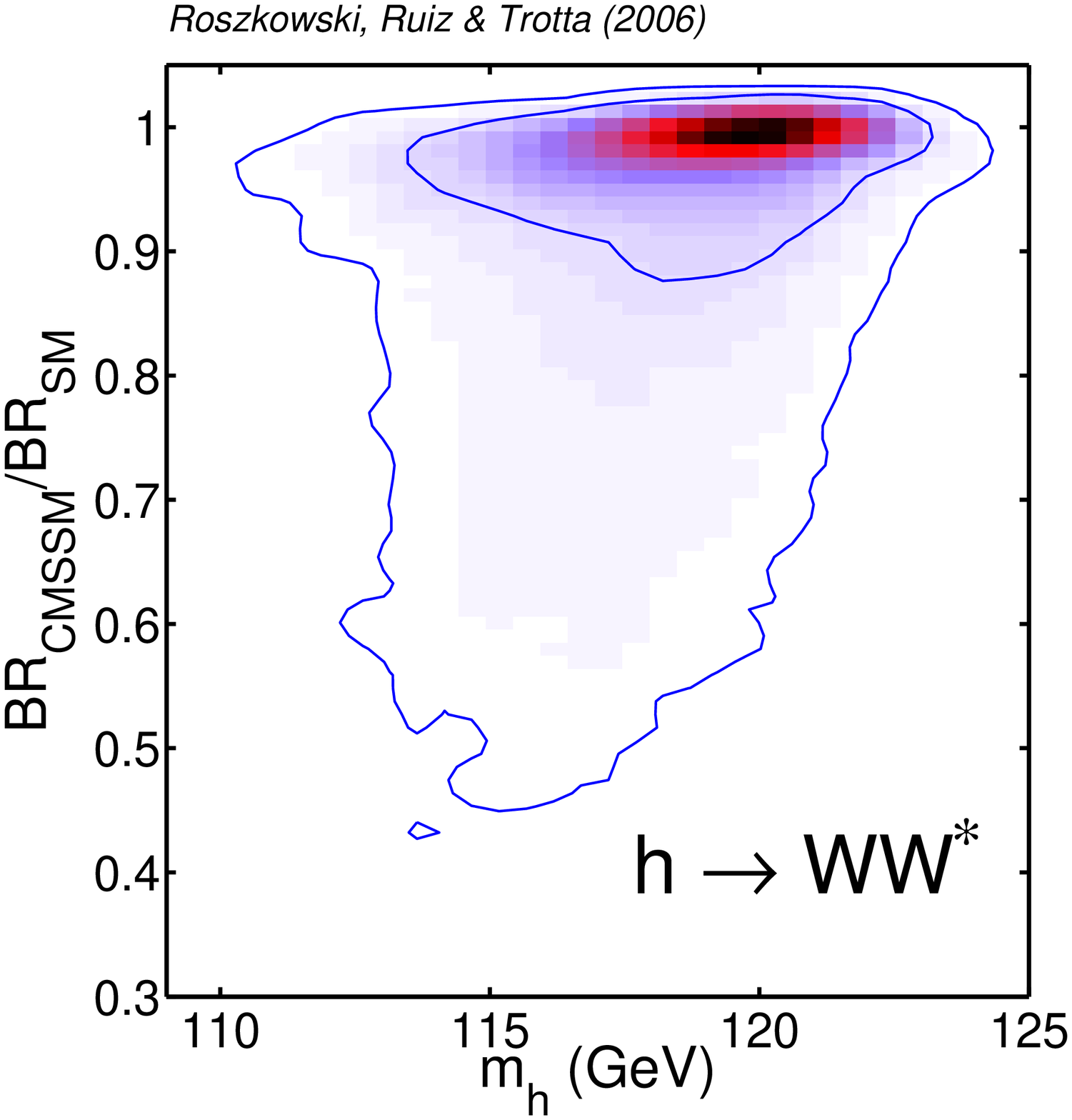}
\end{tabular}
\end{center}
\caption{The 2--dim relative probability density of the branching
 ratios $BR(\hl \to b\bar{b})$ (left panel), $BR(\hl \to
 \tau^+\tau^-)$ (middle panel) and $BR(\hl \to WW^\ast)$ (right
 panel), all normalized to their SM counterparts, as a function of $\mhl$.
\label{fig:rrt2-rd-mh} 
}
\end{figure}
\begin{figure}[!tbh]
\begin{center}
\begin{tabular}{c c c}
	\includegraphics[width=0.3\textwidth]{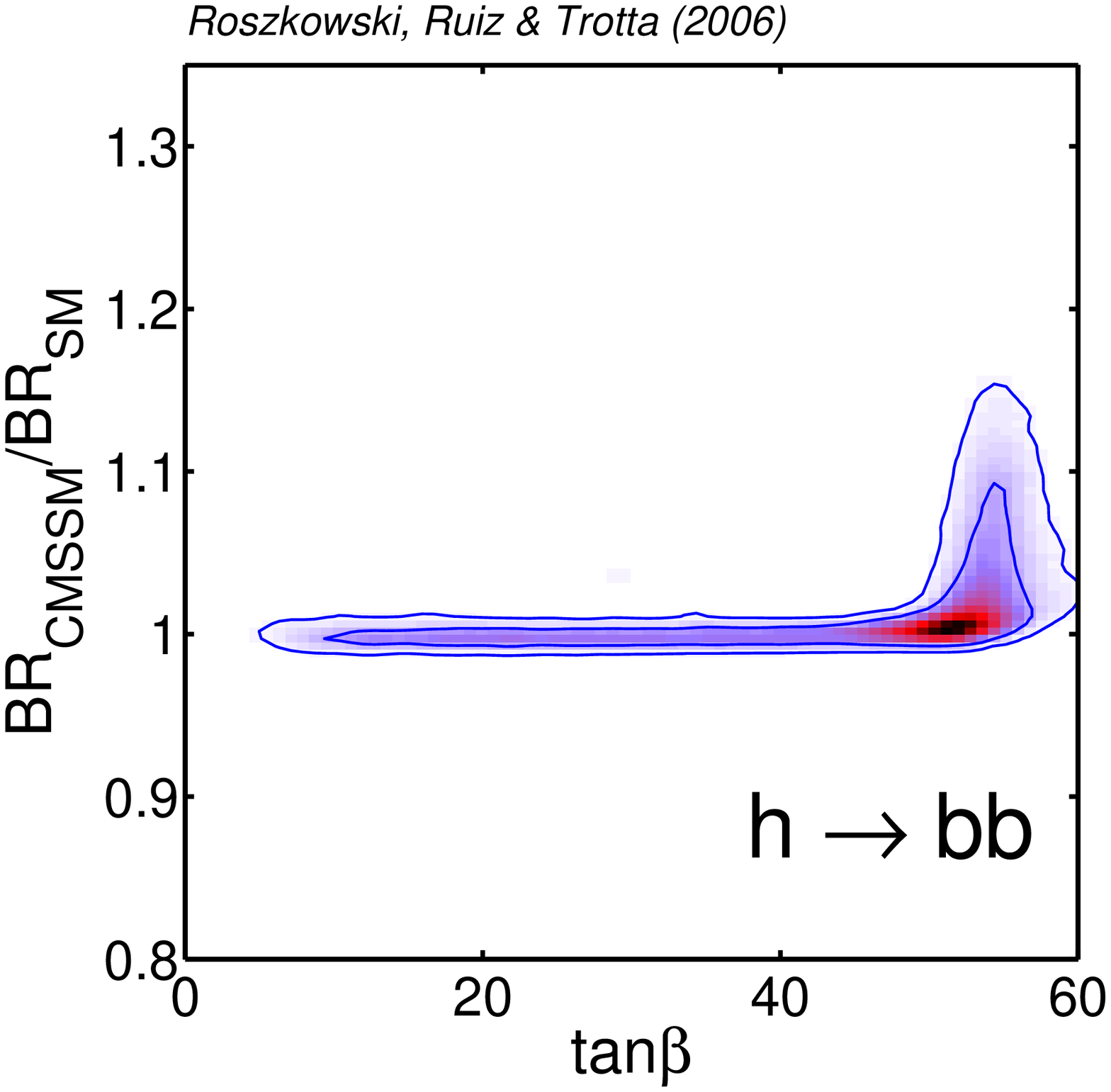}
&	\includegraphics[width=0.3\textwidth]{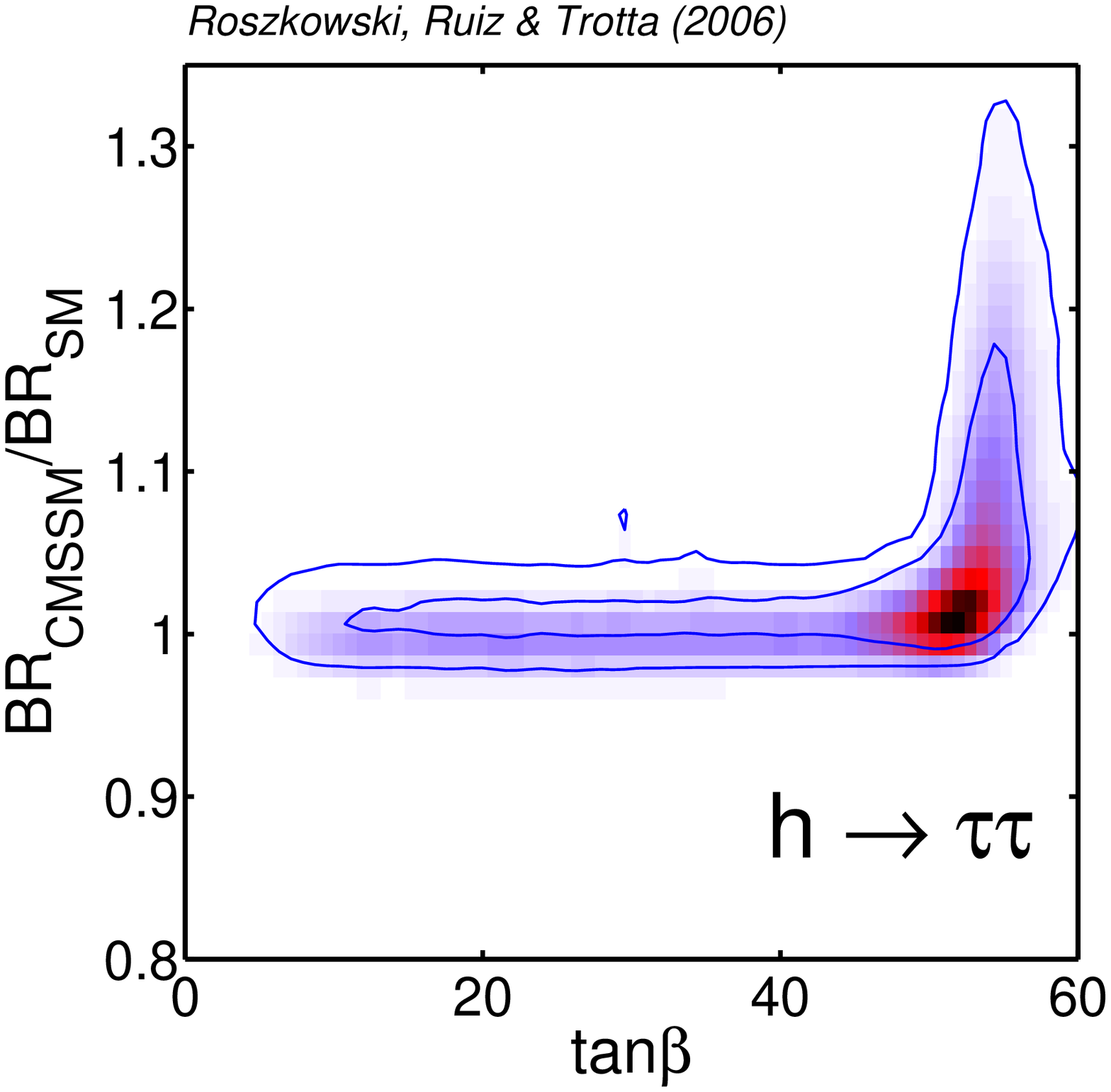}
&	\includegraphics[width=0.3\textwidth]{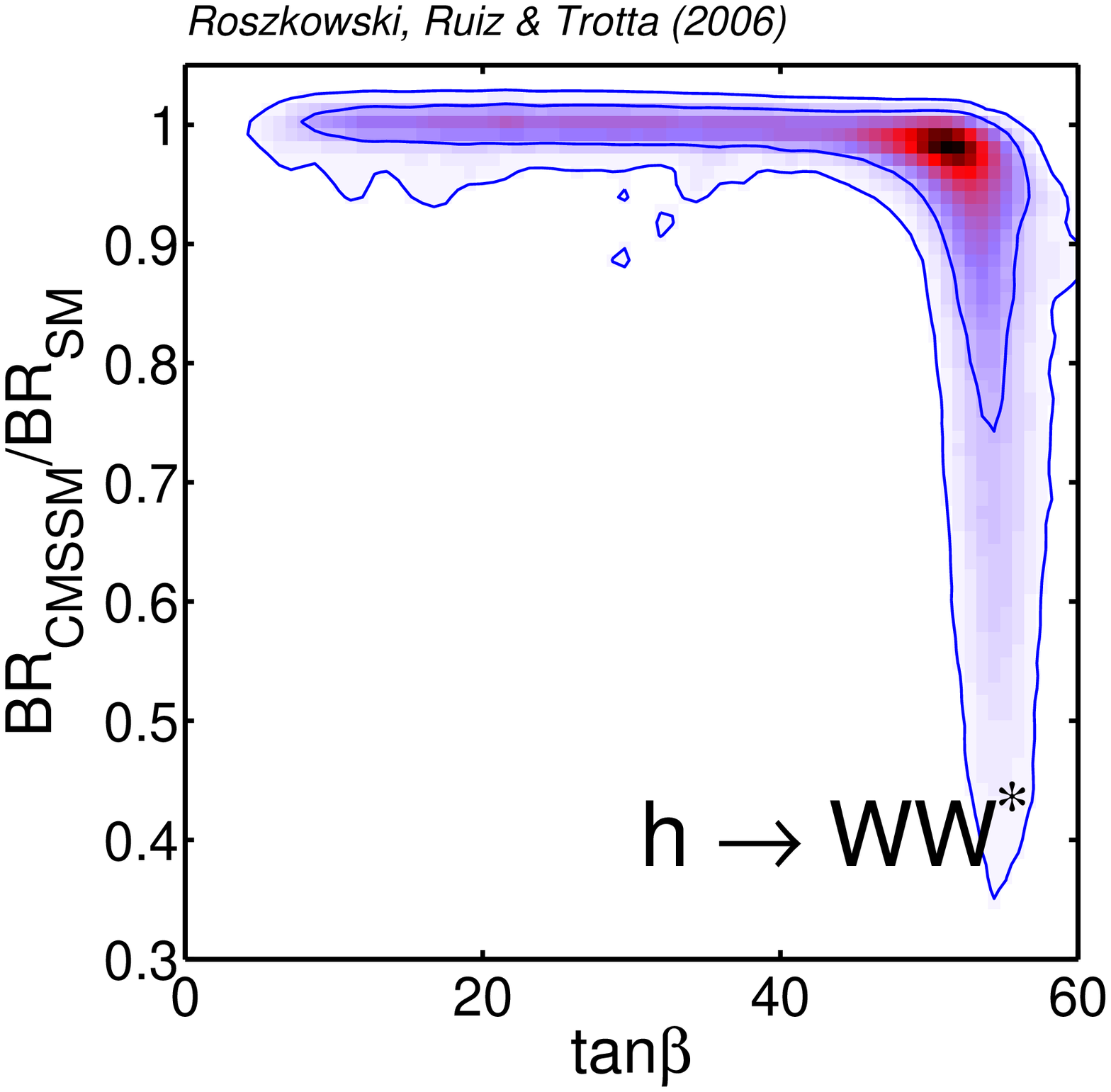}
\end{tabular}
\end{center}
\caption{The 2--dim relative probability density of the branching
 ratios $BR(\hl \to b\bar{b})$ (left panel), $BR(\hl \to
 \tau^+\tau^-)$ (middle panel) and $BR(\hl \to WW^\ast)$ (right
 panel), all normalized to their SM counterparts, as a function of $\tanb$.
\label{fig:rrt2-rd-tanb} 
}
\end{figure}

We will now examine the impact of the above properties of the light Higgs couplings
on its decays.  The varying of the couplings to bottoms and
staus is reflected in the total and partial decay widths of $\hl$ to
$\bbbar$ and $\tauptaum$ for which relative probability densities are
presented in fig.~\ref{fig:hlwidths}.  For comparison, we show the
same quantities for the SM Higgs with the same mass. Somewhat larger
widths in the $\bbbar$ and $\tauptaum$ modes are a result of both
decay channels being enhanced at large $\tanb$.  On the other hand, we
have checked that the decay width for $\hl\to W^\ast W$ (followed by
vector boson decays into light fermions) shows no deviation from the
SM.

The ensuing effect on the branching ratios is shown in
figs.~\ref{fig:rrt2-hldecayratio}--\ref{fig:rrt2-rd-tanb}.  We remind
the reader that, in the range of mass predicted in the CMSSM, the
SM--like light Higgs boson decays predominantly into $\bbbar$ pairs
($\sim90\%$), followed by $\tauptaum$ pairs ($\sim9\%$), although at
small $\tanb$ the $\hl\to WW^\ast$ branching ratio grows quickly with
$\mhl$ and at $\mhl\simeq120\gev$ it can exceed some $10\%$ (at the
expense of the above two channels).  The dominance of the $\bbbar$
mode, however, is so large that, despite some decrease of the
coupling $\hl\bbbar$ 
relative to the one of $\hl\tauptaum$ at large $\tanb$ in some parts
of the CMSSM parameter space, the branching ratio into $\bbbar$
remains basically unaffected. On the other hand it is the subdominant
modes $\tauptaum$ and $WW^\ast$ that, at large $\tanb$, can experience
either a relative increase or decrease, respectively. These effects can be seen
in fig.~\ref{fig:rrt2-hldecayratio}, where we display 1--dim relative
pdf's of the branching ratios for $\hl\to b\bar{b}$ (left panel),
$\hl\to \tau^+\tau^-$ (middle panel) and $\hl \to WW^\ast$ (right
panel), all normalized to the analogous quantities for the SM Higgs
boson. The corresponding 2--dim relative probability densities are
shown in figs.~\ref{fig:rrt2-rd-mh} and~\ref{fig:rrt2-rd-tanb}, in the
planes spanned by the above SM--normalized branching ratios and $\mhl$
and $\tanb$, respectively.  Note a small increase in the number of
produced $\tau$--leptons (which is rather small to start with), which
may help in Higgs searches in that important decay channel.

In conclusion, in the CMSSM with flat prior ranges as given at the
 beginning of sec.~\ref{sec:constraints} (most importantly
 $\mzero<4\tev$) and with observables as in tables~\ref{tab:meas}
 and~\ref{tab:measderived},
the mass of the light Higgs lies
predominantly in the range shown in~\eqref{eq:lighthiggsmassrange} and
fig.~\ref{fig:hlmass}, while the other Higgs bosons are typically
somewhat, but not necessarily much, heavier. The light Higgs coupling
to $VV$ remains basically SM--like, while the ones to $\bbbar$ and
$\tauptaum$ do show some increase relative to the SM ones at large
$\tanb$. This has some effect on on light Higgs decays and, in the
case of the bottoms, also on its production, as we shall see in the
next section.

\section{Light Higgs production and decay }\label{sec:higgsattev}

We will now assess the discovery prospects of the light CMSSM Higgs
boson at the Tevatron. To this end we will consider the following
production and decay processes:

\begin{itemize}
\item
{\bf vector boson bremsstrahlung}: $V^{\ast} \to V \hl$ (where $V =
  W, Z$), followed by $\hl \to b \bar{b}$, $\hl \to \tau^+\tau^-$ or, in
  the case of $W\hl$, also $\hl \to W W^{\ast}$;

\item
{\bf gluon--gluon fusion}: $gg \to \hl$, followed
by either $\hl \to WW^{\ast}$ or $\hl \to \tau^+\tau^-$;

\item
{\bf associated bottom production}: $\hl b (b)$, with a $b$
quark tagged on a hard spectrum ($p_T>15\gev$ and $\eta<2.5$),
followed by either $h \to b \bar{b}$ or $\hl \to \tau^+\tau^-$;

\item
{\bf inclusive production}: $p\bar{p} \to \hl $, followed by $h \to
\tau^+\tau^-$.

\end{itemize}

We have also considered $VV$ fusion and $t
\bar{t}\hl$ Higgs production processes, but in the CMSSM these are
subdominant. 

Some additional comments about the processes that we consider are in
order. The vector boson bremsstrahlung process is determined by the
effective coupling $g(\hl VV)\sim\sin(\beta-\alphaeff)$. On the other
hand, the other three processes are to a large extent determined by
the behavior of the effective coupling $g(\hl\bbbar)$ which, as we
have seen above, for $\tanb\gsim50$ and $\mha\lsim1\tev$, can markedly
deviate from the SM value.

In the gluon--gluon fusion process $gg \to \hl$ we include diagrams
with top or bottom quark (and their superpartner) lines in the
loop. In the associated bottom production process $\hl b (b)$
we compute the cross section of $bg\to b\hl$. An alternative, and
effectively equivalent, way would be to consider the process $gg\to \hl
\bbbar$ with the momentum of one of the bottoms integrated
out~\cite{hwg03}. The inclusive process $p\bar{p} \to \hl$ can
likewise be computed in two ways. At NLO in the four--flavor
scheme one can add the (dominant) process $gg\to \hl \bbbar$ and the
(subdominant) one $\qqbar\to \hl \bbbar$, and integrate out the momenta
of the bottoms. Alternatively, one obtains very similar results by
computing the process $\bbbar\to \hl$ at NNLO in the five--flavor
scheme~\cite{hwg03}. Here we follow the latter approach.  Relative
strengths of the above processes depend on several parameters,
especially on $\tanb$ (when large) but typically, in the MSSM in the
regime of light Higgs mass and large $\tanb$, the gluon--gluon fusion
process is dominant and associated bottom production is a factor of a
few smaller but otherwise comparable. For a comprehensive review of
Higgs properties and collider search prospects, see, for instance, ref.~\cite{djouadi05}.

Since, as we have seen above, the light Higgs boson is SM--like, it
will be convenient to normalize our results to the corresponding
processes involving the SM Higgs boson with the same mass. We expect
most ratios to be close to 1 and in fact it is some possible
departures from the SM case that we will attempt to identify.

We compute the light Higgs production cross sections in the CMSSM,
normalized to their SM counterparts,
$\sigma_{\text{CMSSM}}/\sigma_{\text{SM}}$, with the help of
FeynHiggs~v2.5.1~\cite{feynhiggs}. The package implements the Higgs
production cross sections at the Tevatron and the LHC, both evaluated in the
effective coupling approximation using the SM cross sections provided
in ref.~\cite{maltoni}.  The calculation of the branching ratios is
based on ref.~\cite{hhw00}. The code also includes SUSY corrections to
Higgs couplings to the bottom quarks, which can be substantial at large
$\tanb$, as discussed earlier.


\begin{figure}[!tb]
\begin{center}
\begin{tabular}{c c}
 	\includegraphics[width=0.3\textwidth]{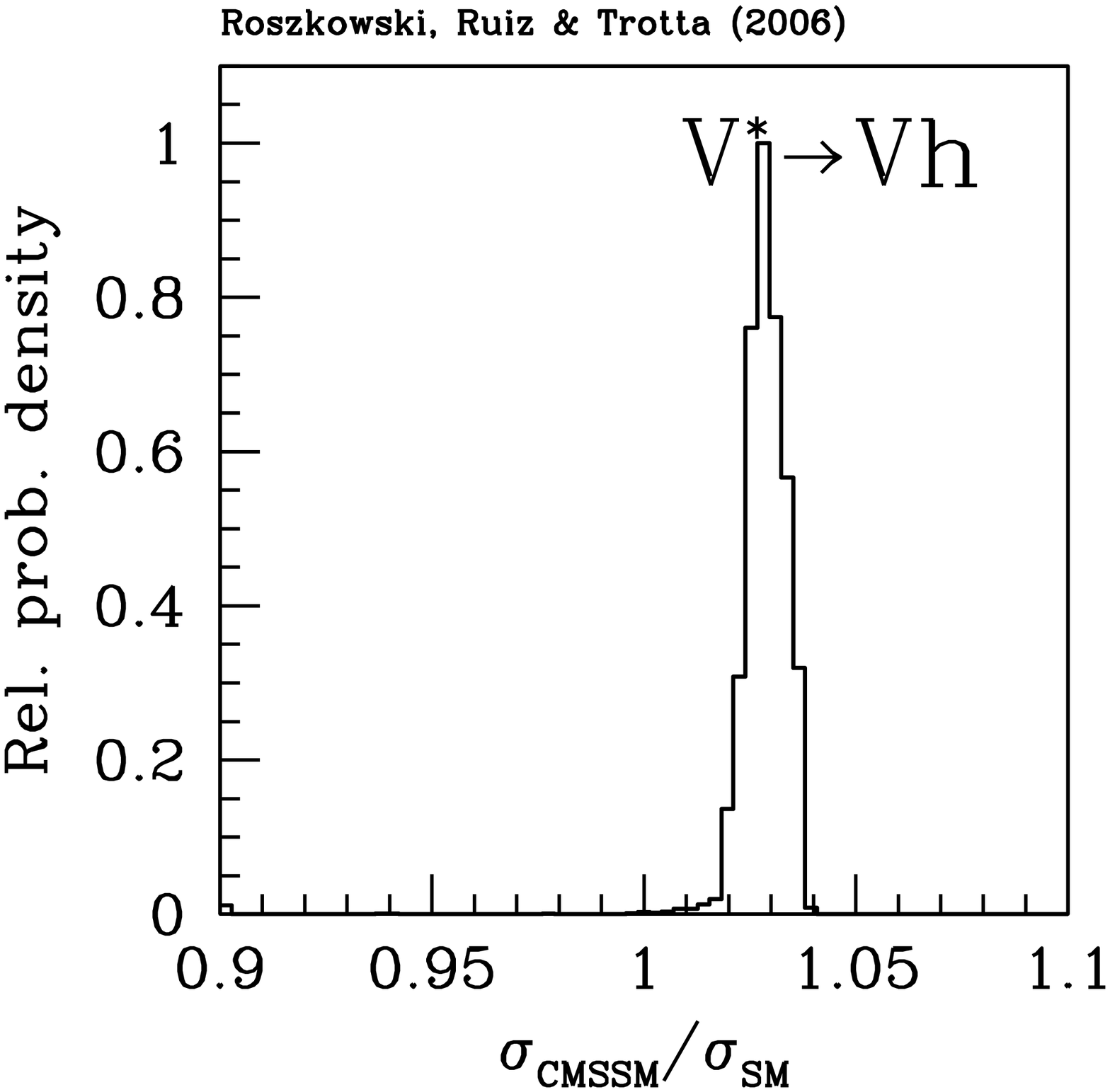}
 &	\includegraphics[width=0.3\textwidth]{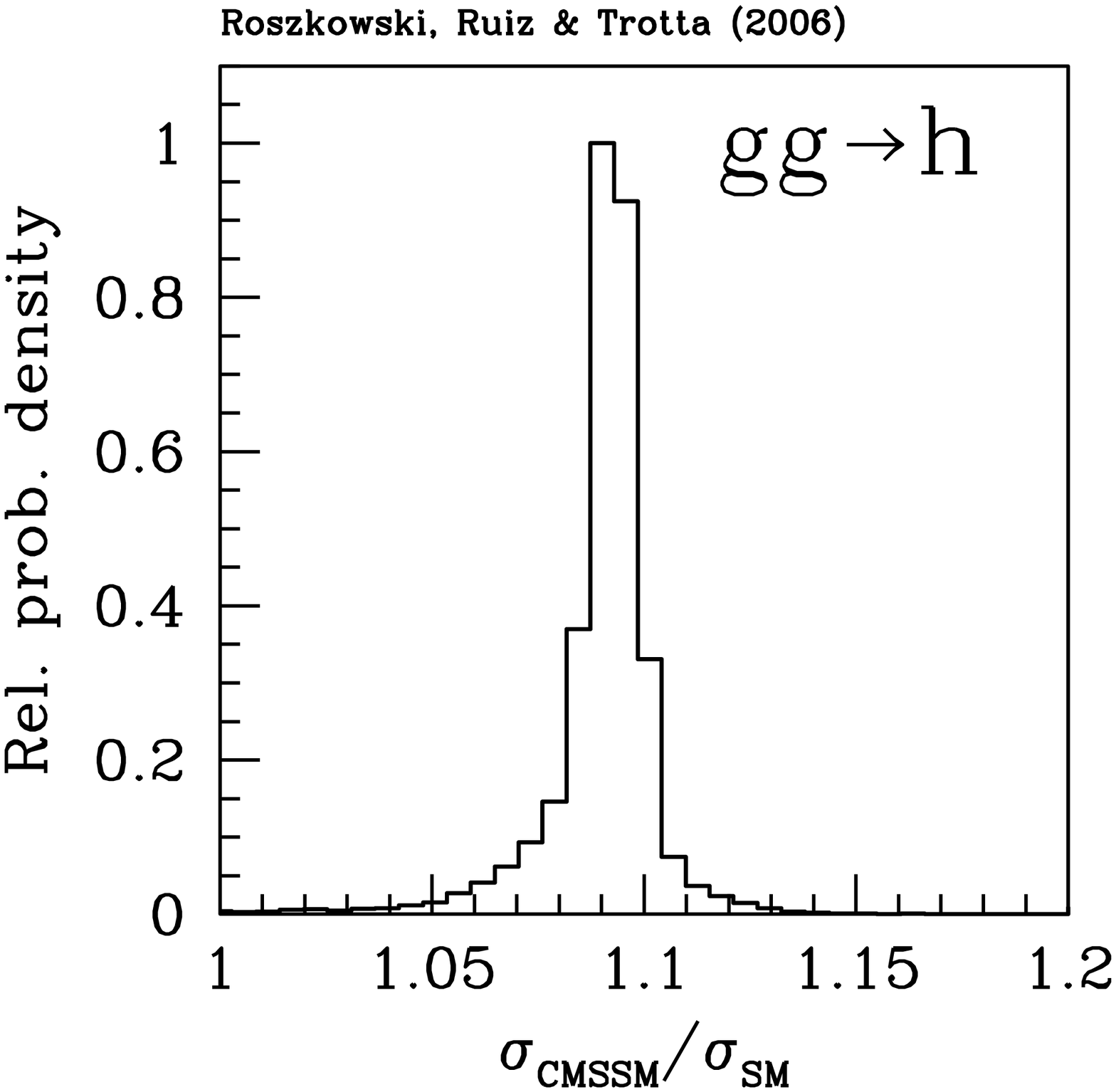}\\
 	\includegraphics[width=0.3\textwidth]{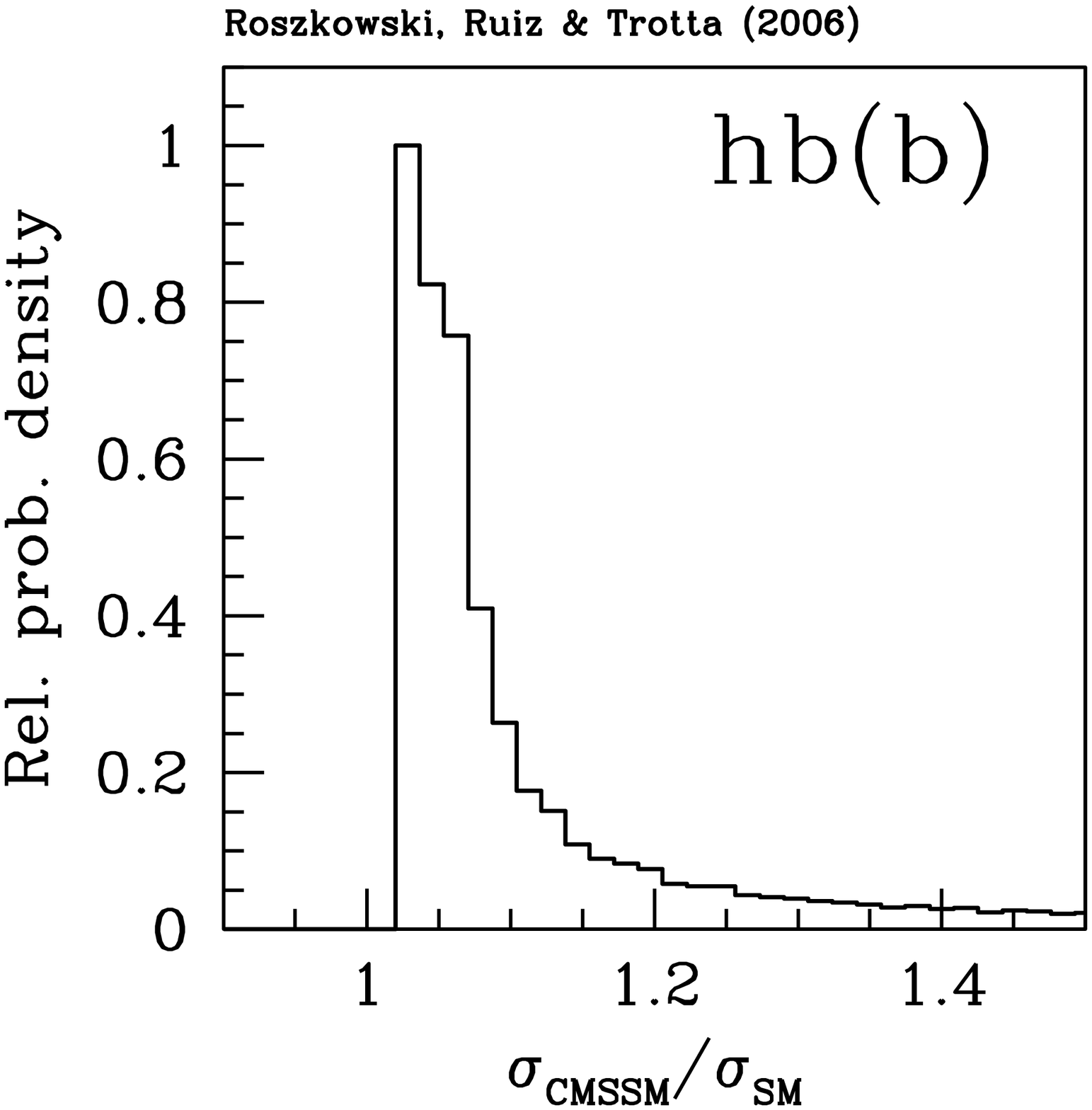}
 &	\includegraphics[width=0.3\textwidth]{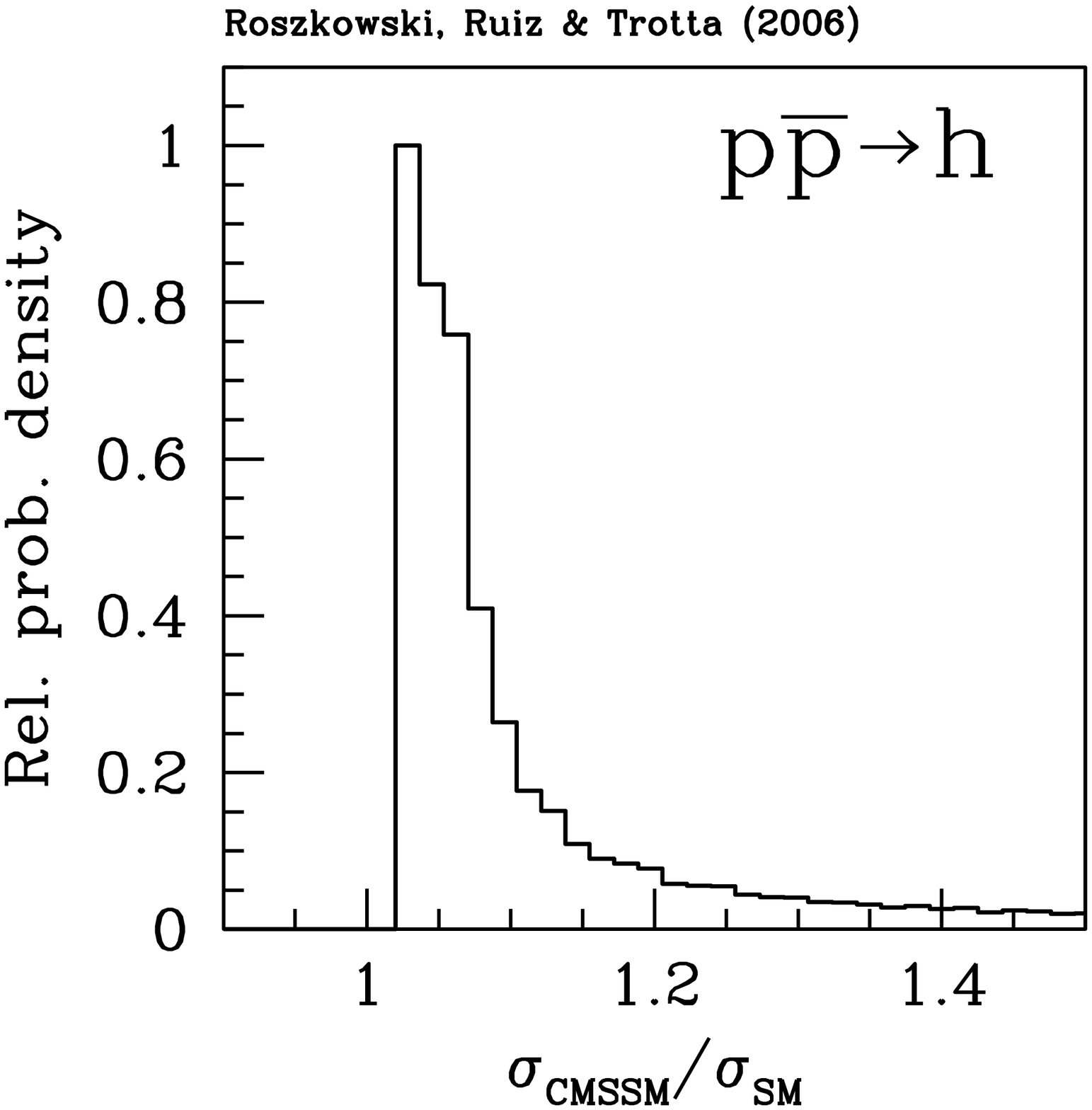}
\end{tabular}
\end{center}
\caption{The 1--dim relative probability density for light Higgs
  production cross sections at the Tevatron (normalized to the SM
  case) for $V^{\ast} \to V\hl$, where $V=Z,W$ (upper left panel), $gg
  \to \hl$ (upper right panel), $\hl b (b)$ (lower left panel) and
  $p\bar{p} \to \hl$ (lower right panel).
\label{fig:rrt2-hlsigma}
}
\end{figure}
\begin{figure}[!htb]
\begin{center}
\begin{tabular}{c c}
	\includegraphics[width=0.3\textwidth]{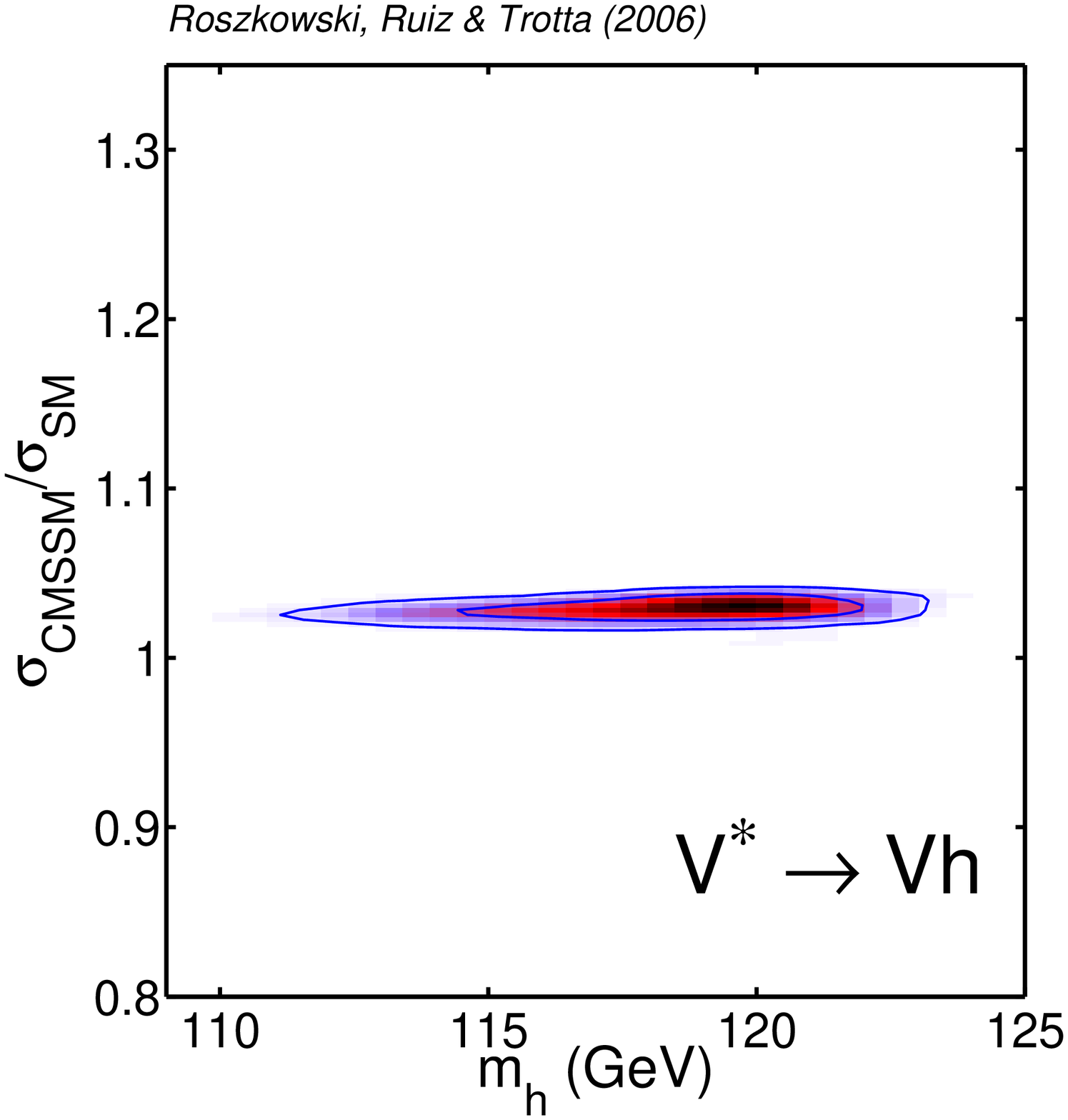}
 &	\includegraphics[width=0.3\textwidth]{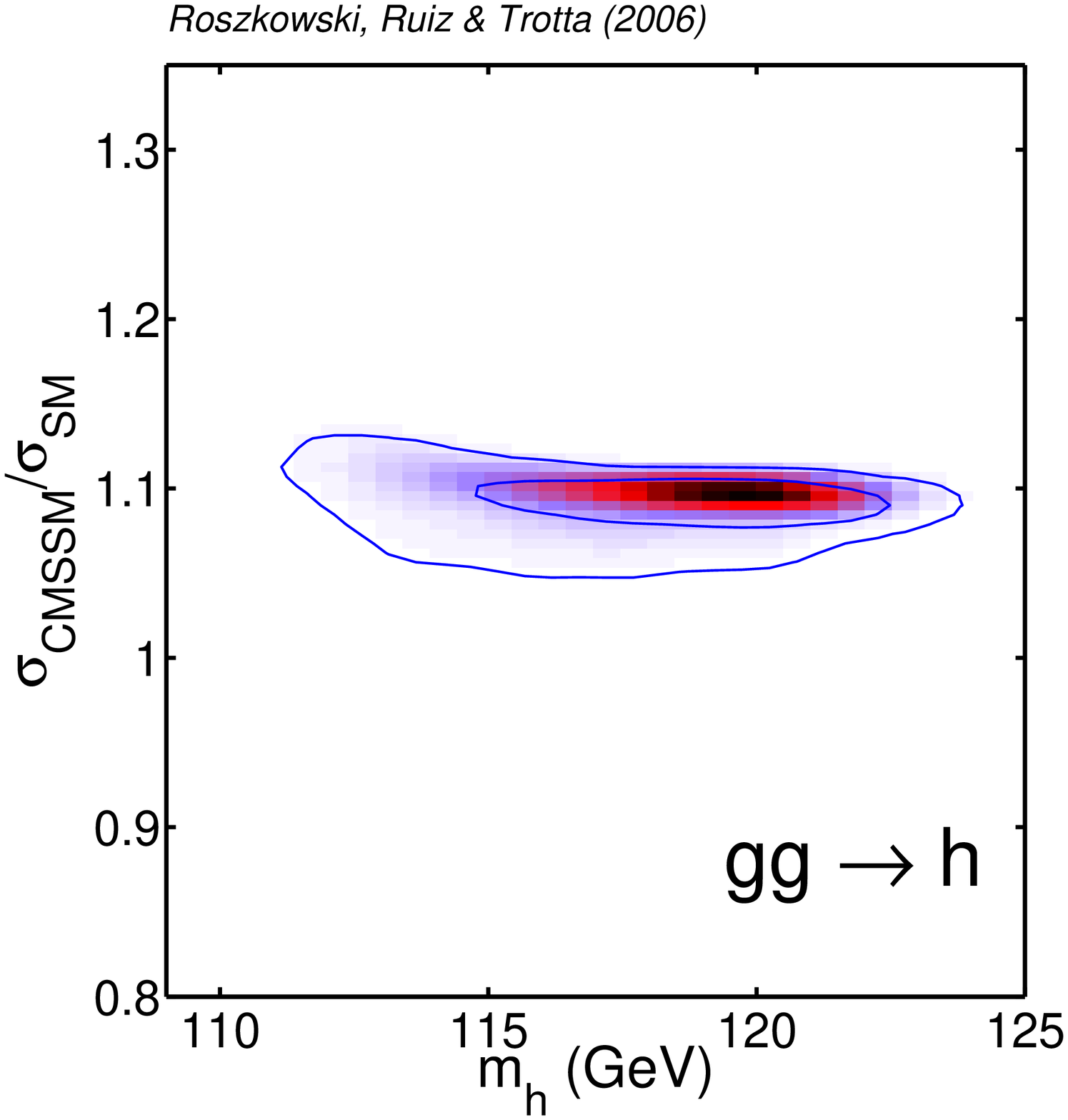}\\
	\includegraphics[width=0.3\textwidth]{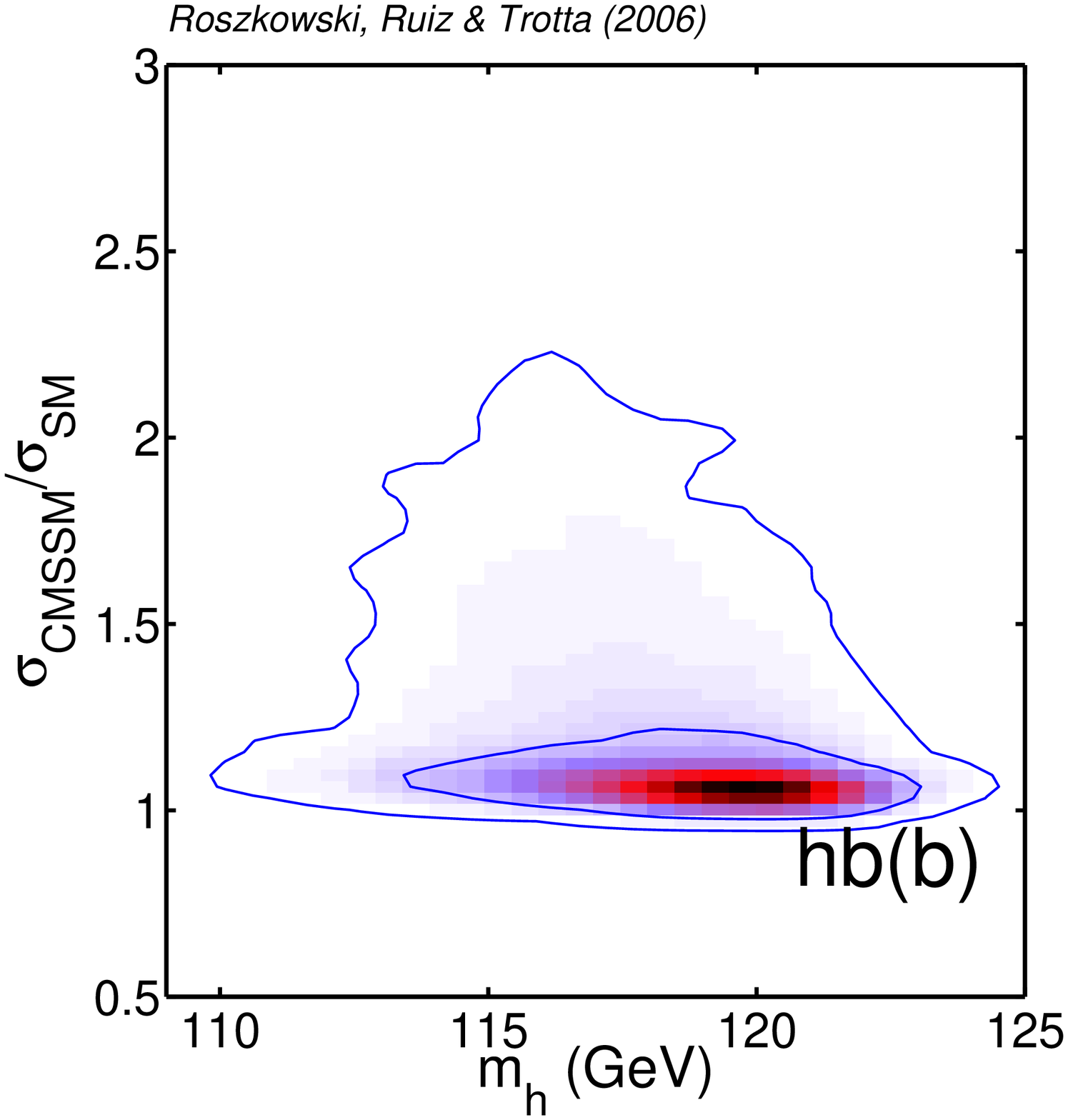}
 &	\includegraphics[width=0.3\textwidth]{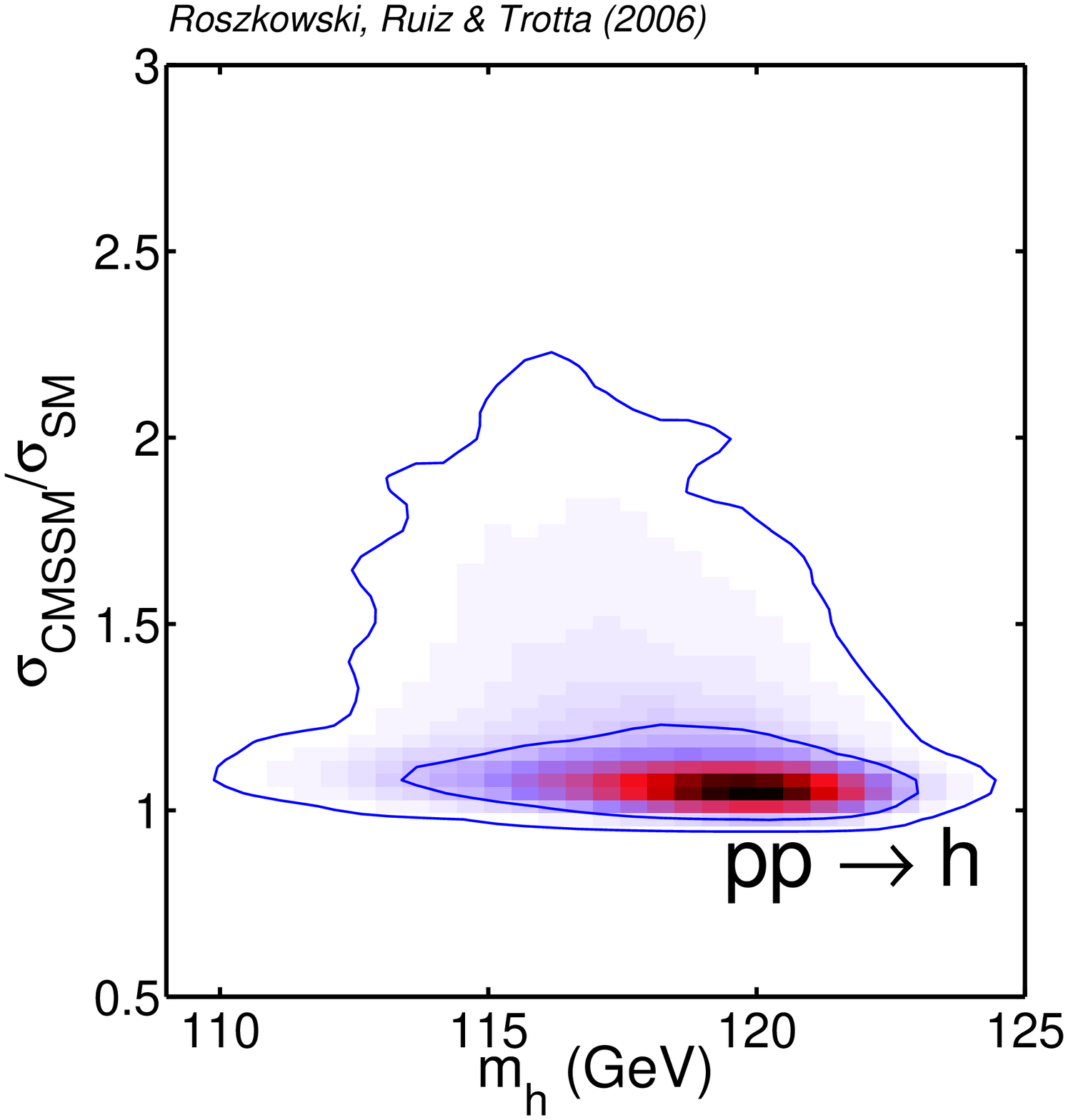}
\end{tabular}
\end{center}
\caption{The 2--dim relative probability density of light Higgs
  production cross sections at the Tevatron (normalized to the SM
  case) as a function of its mass $\mhl$.
\label{fig:rrt2-hlsigmavsmhl}
}
\end{figure}
\begin{figure}[!bt]
\begin{center}
\begin{tabular}{c c}
	\includegraphics[width=0.3\textwidth]{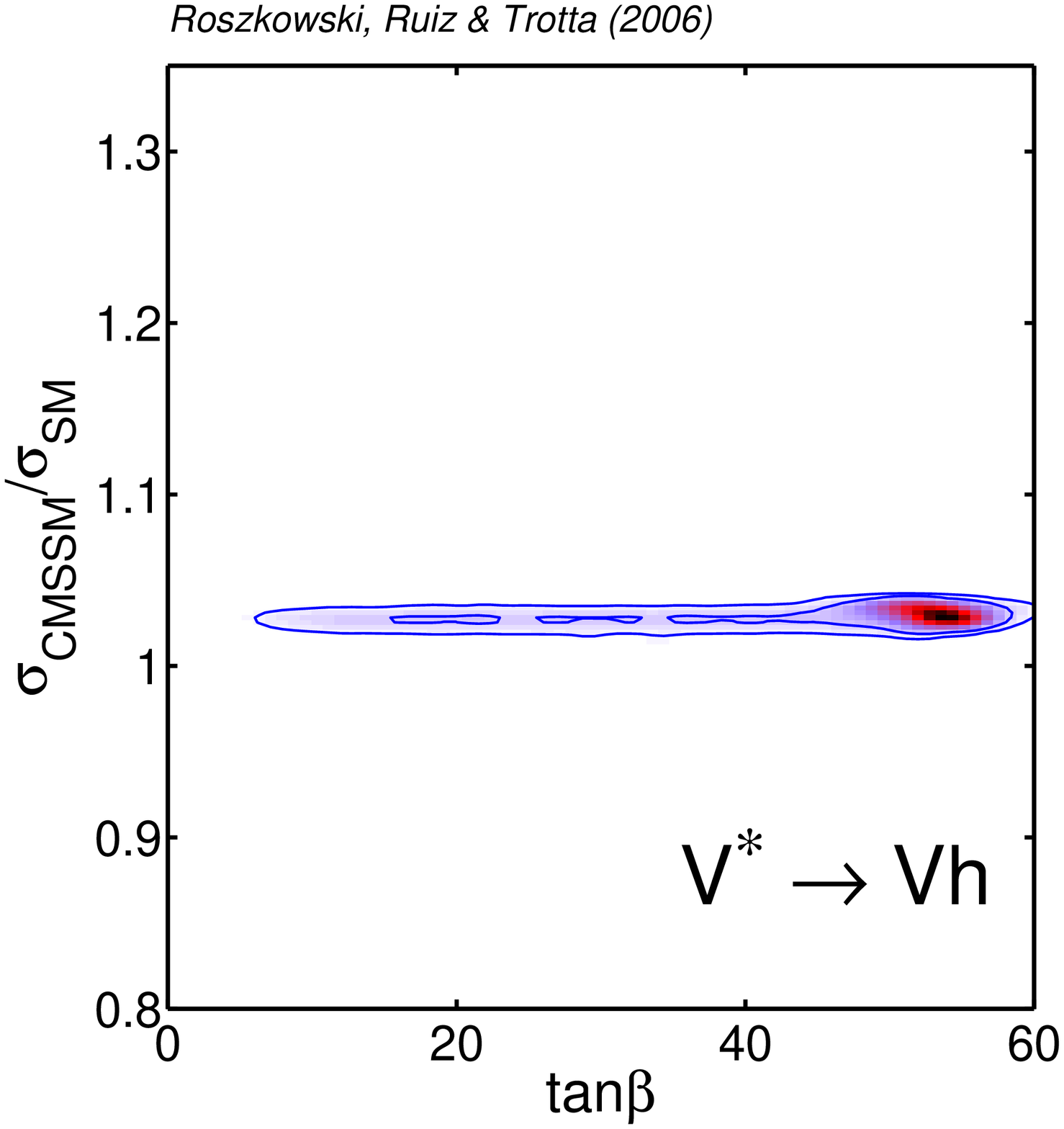}
 &	\includegraphics[width=0.3\textwidth]{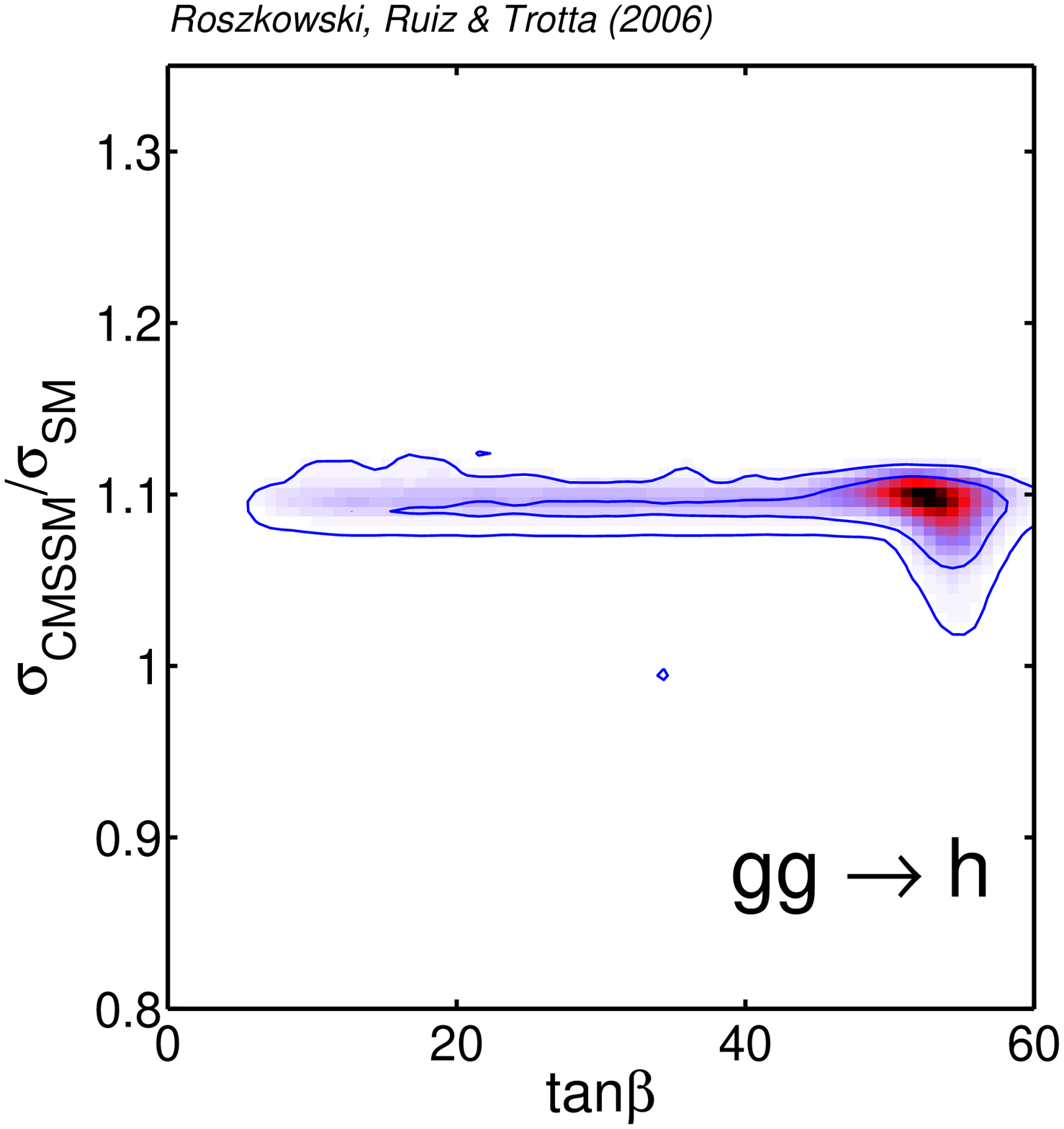}\\
	\includegraphics[width=0.3\textwidth]{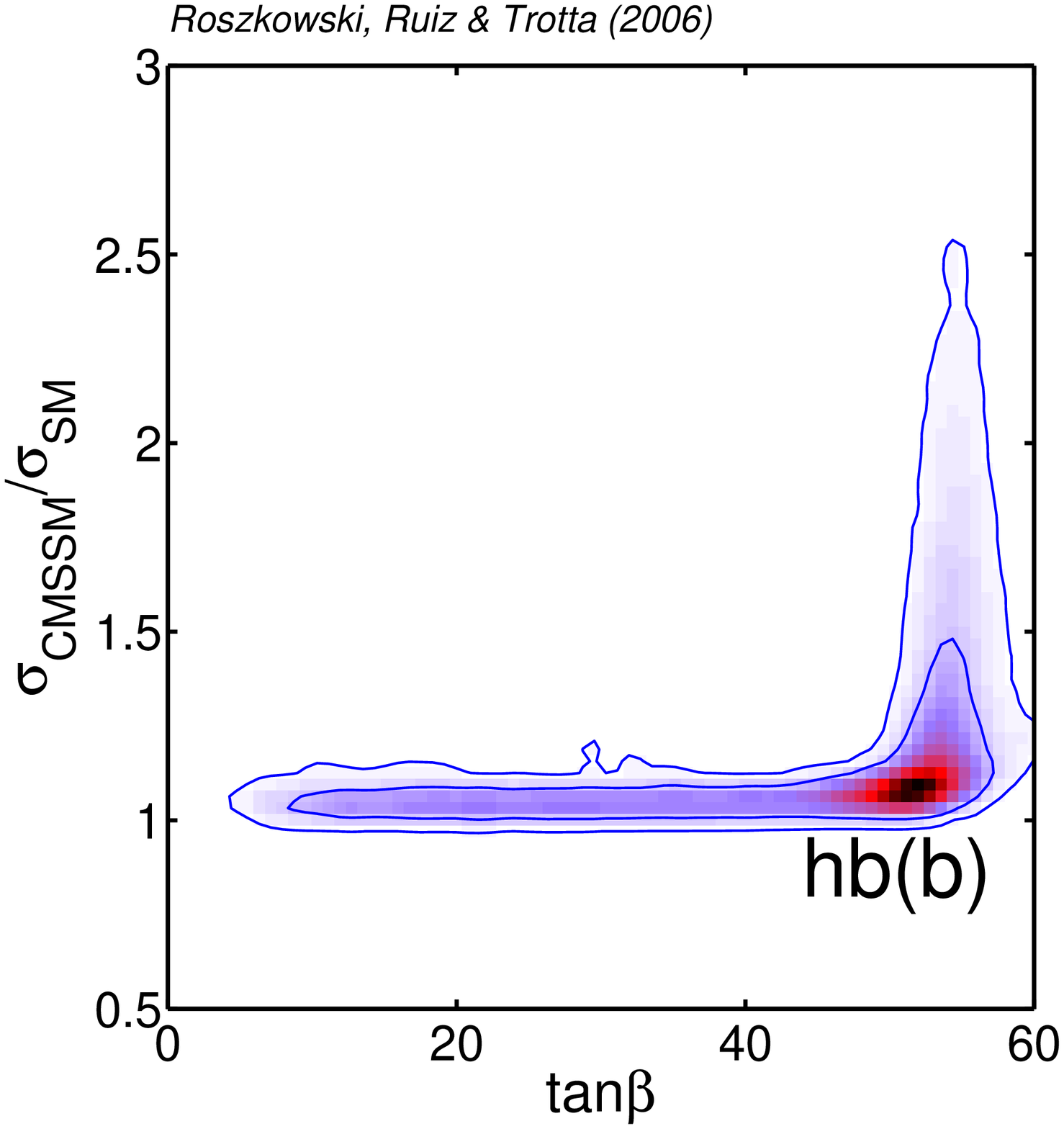}
 &	\includegraphics[width=0.3\textwidth]{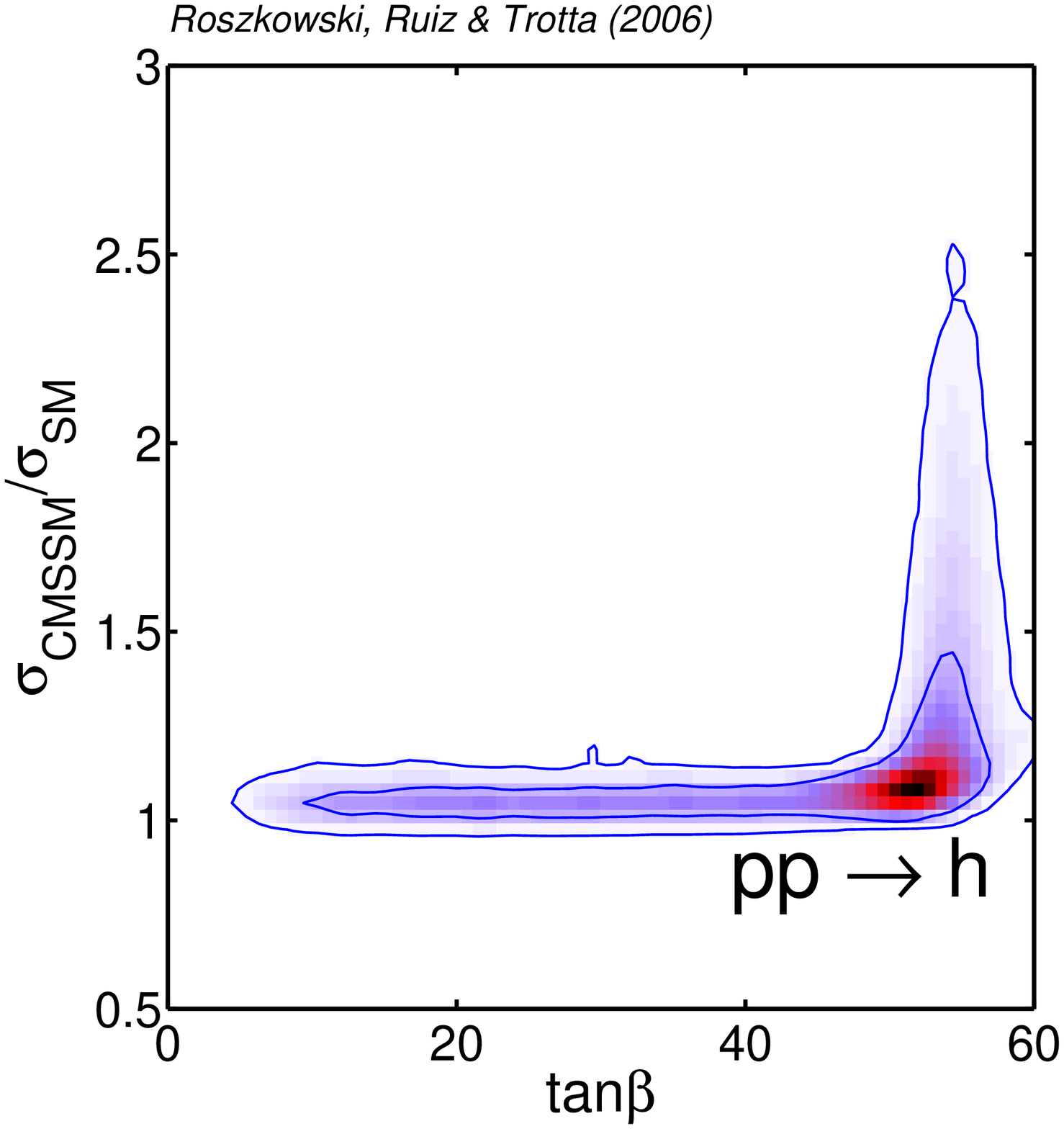}
\end{tabular}
\end{center}
\caption{The 2--dim relative probability density for light Higgs
  production cross sections at the Tevatron (normalized to the SM
  case) as a function of $\tanb$.
\label{fig:rrt2-hlsigmavstanb}
}
\end{figure}
\begin{figure}[!tb]
\begin{center}
\begin{tabular}{c c}
	\includegraphics[width=0.3\textwidth]{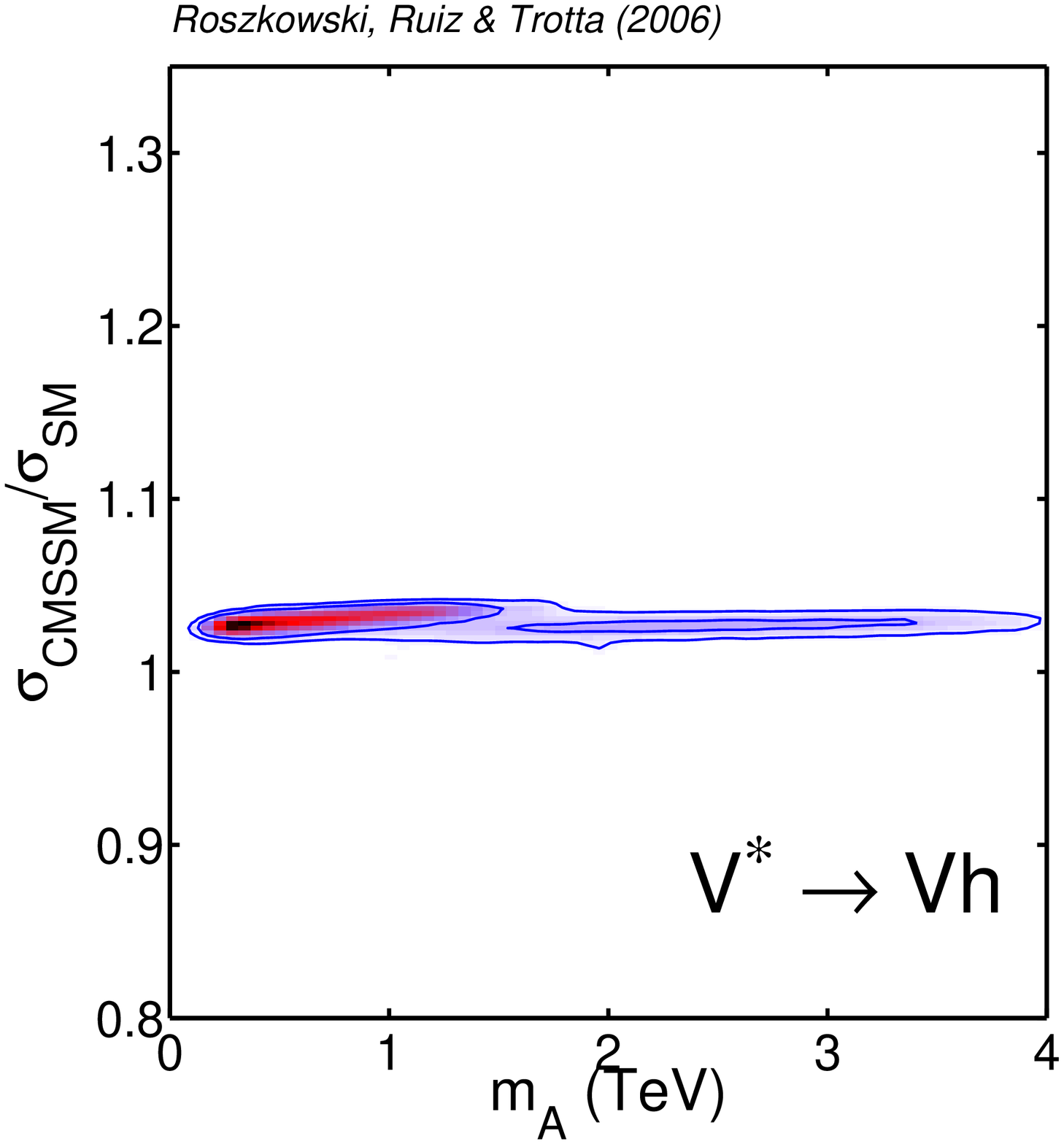}
 &	\includegraphics[width=0.3\textwidth]{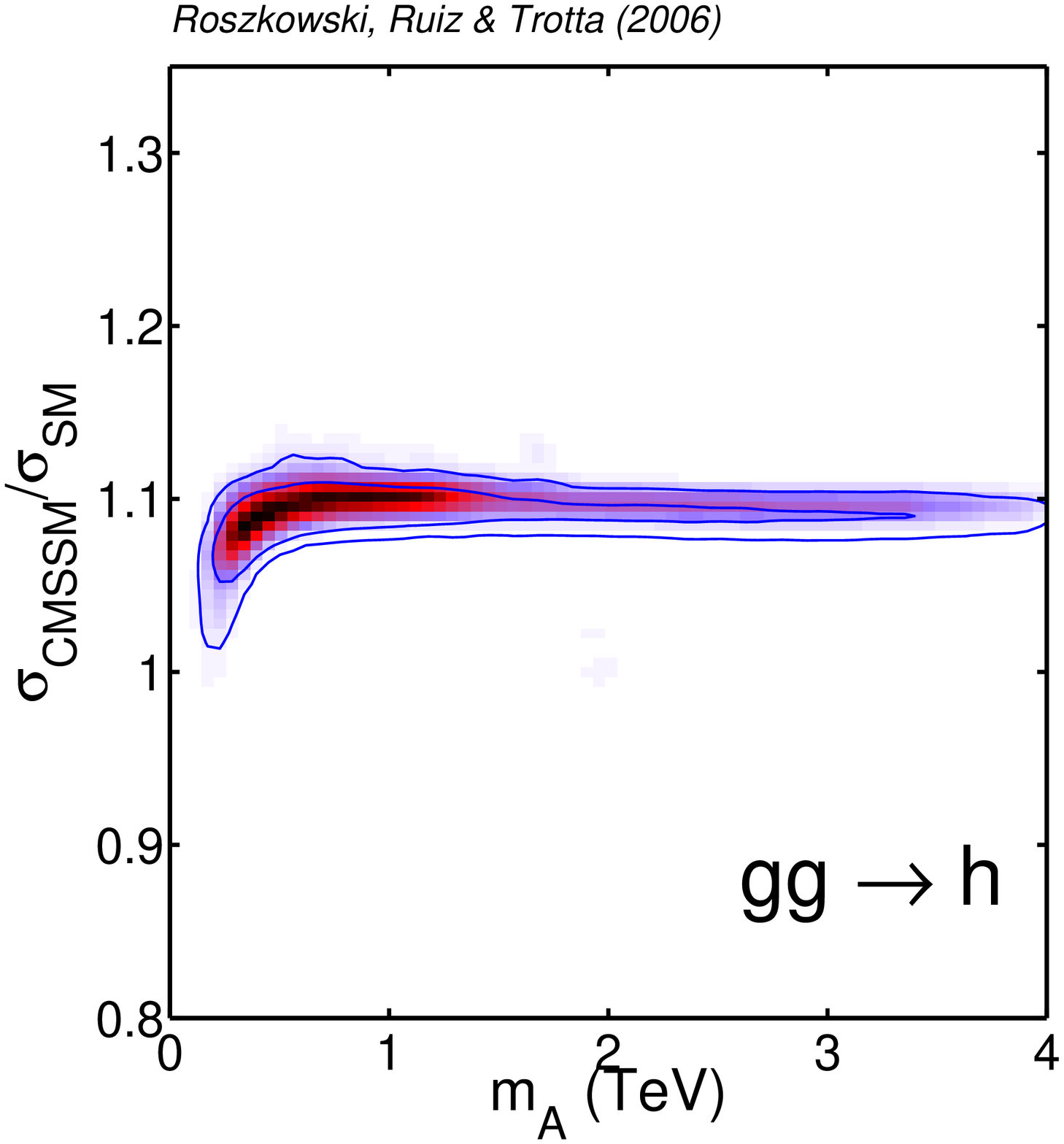}\\
	\includegraphics[width=0.3\textwidth]{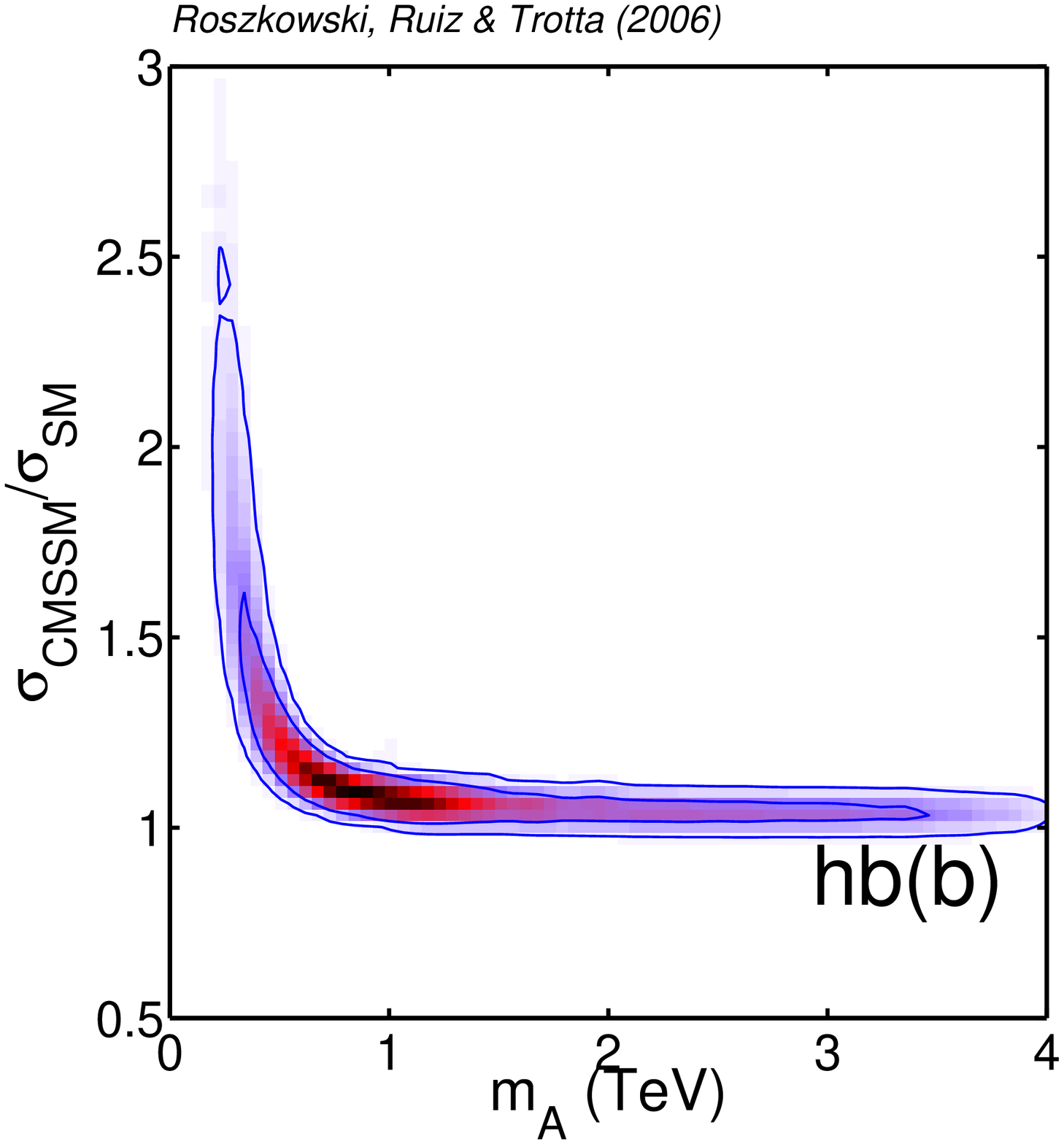}
 &	\includegraphics[width=0.3\textwidth]{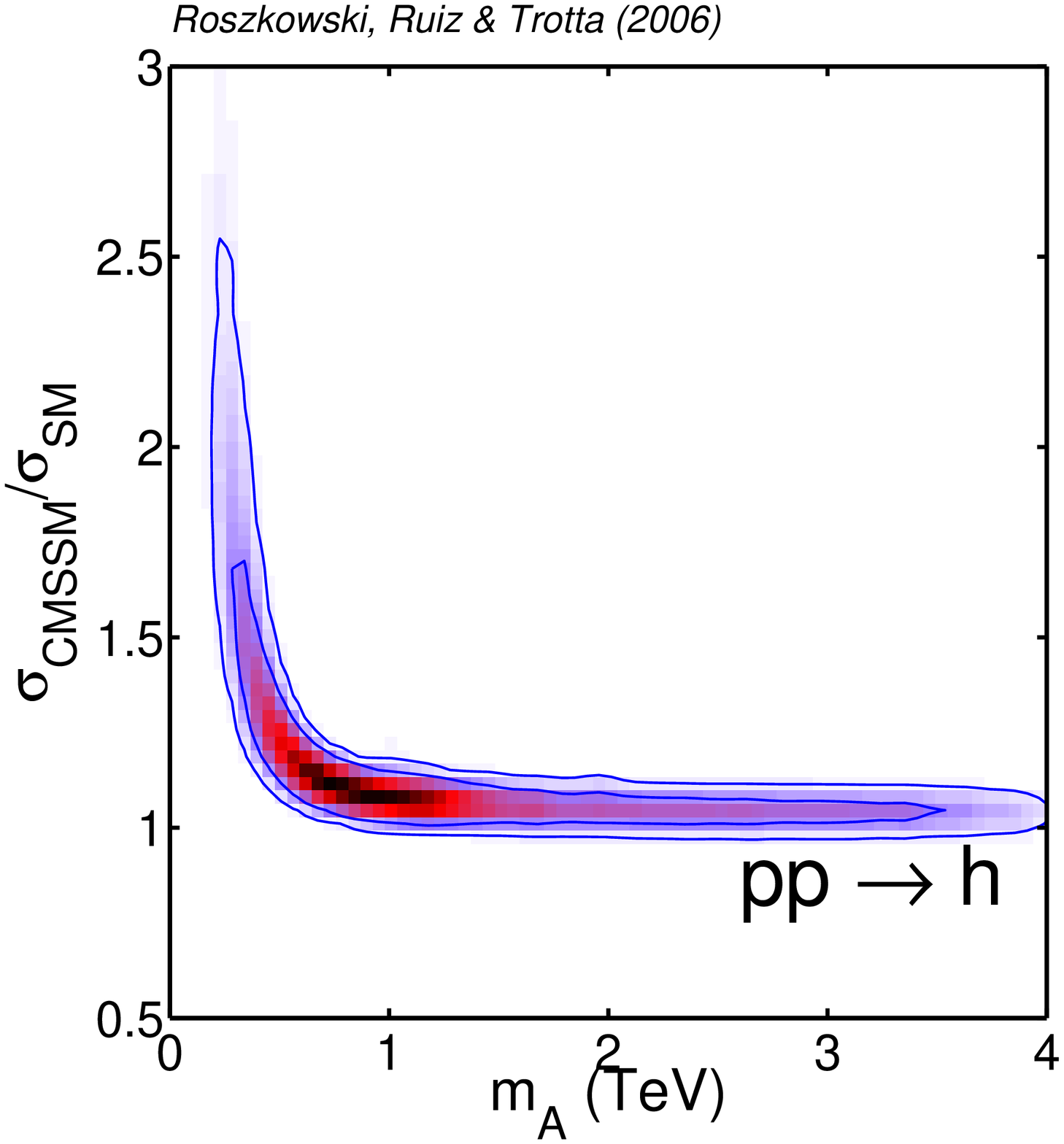}
\end{tabular}
\end{center}
\caption{The 2--dim relative probability density for light Higgs
  production cross sections at the Tevatron (normalized to the SM
  case) as a function of $\mha$.
\label{fig:rrt2-hlsigmavsmha}
}
\end{figure}
\begin{figure}[!tb]
\begin{center}
\begin{tabular}{c c}
 	\includegraphics[width=0.3\textwidth]{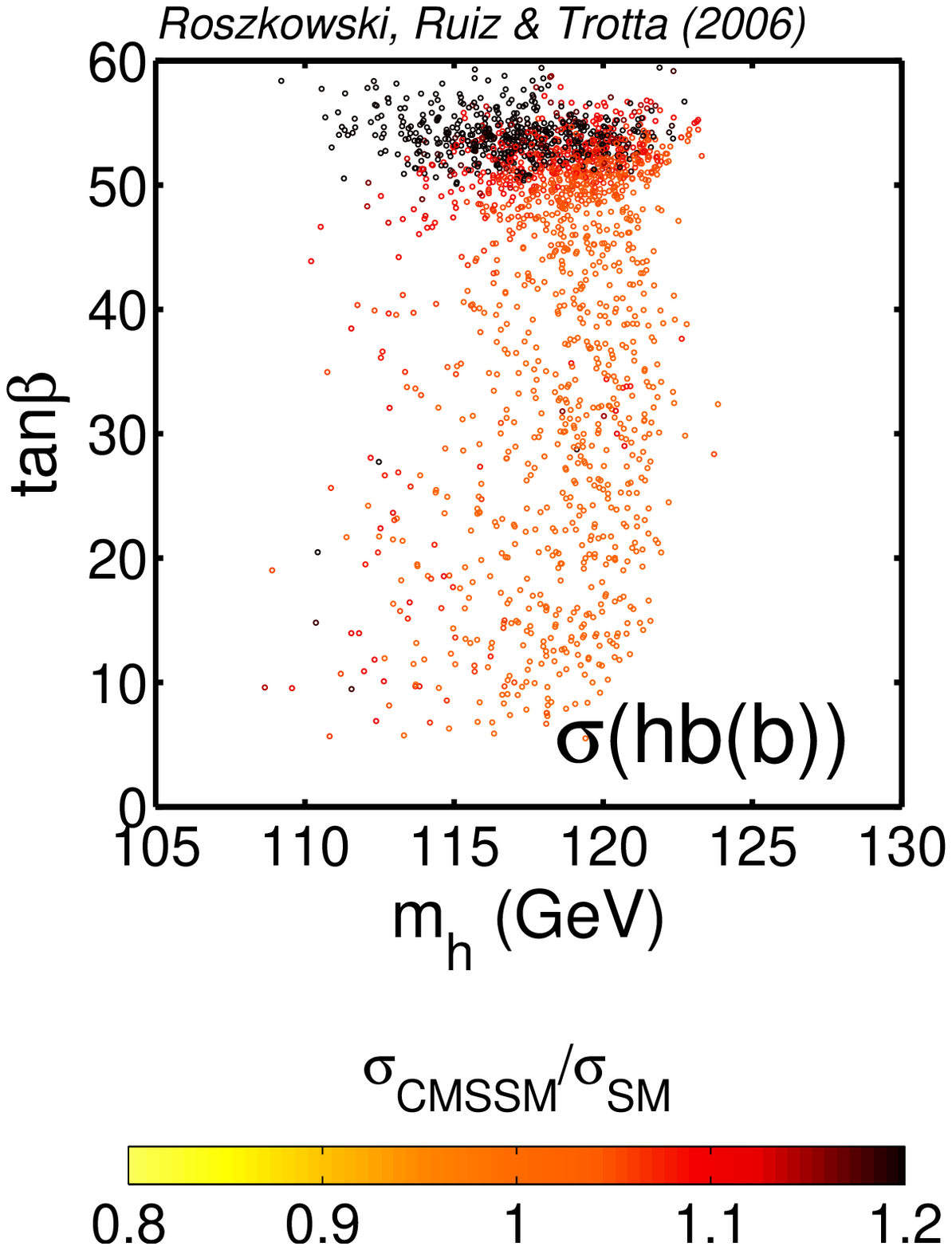}
 &	\includegraphics[width=0.3\textwidth]{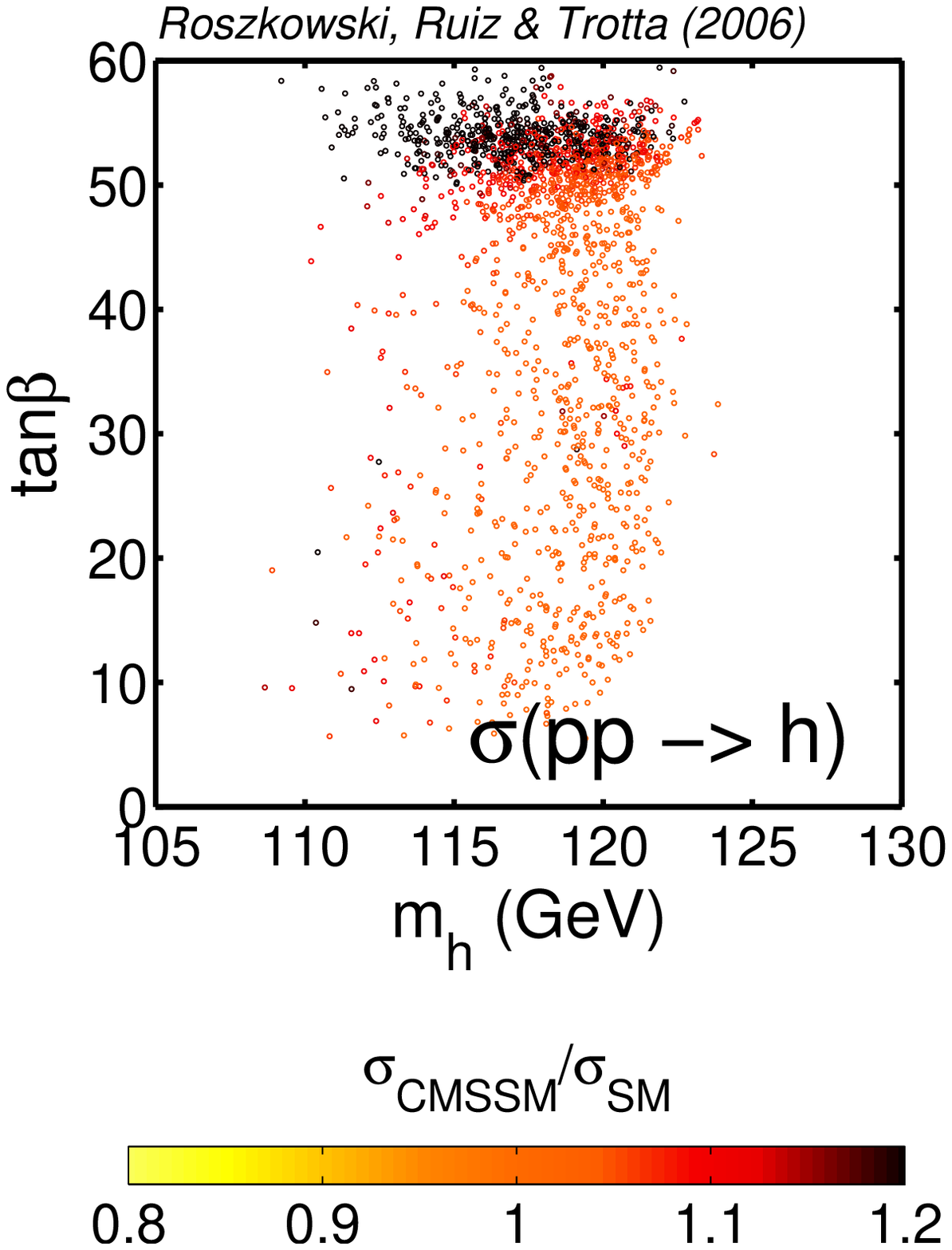}
\end{tabular}
\end{center}
\caption{Values in the $(\mhl,\tanb)$ plane of light Higgs production
  cross sections at the Tevatron (normalized to the SM case) for the
  processes $\hl b(b)$ and $\ppbar\to \hl$. For the processes
  $V^{\ast} \to V \hl$, where $V = Z, W$, and $gg\to \hl$ the values
  are almost independent of $\tanb$; compare
  fig.~{\protect\ref{fig:rrt2-hlsigmavstanb}}.
\label{fig:rrt2-hlsigma-tanbvsmhl}
}
\end{figure}
\begin{figure}[!tb]
\begin{center}
\begin{tabular}{c c}
 	\includegraphics[width=0.3\textwidth]{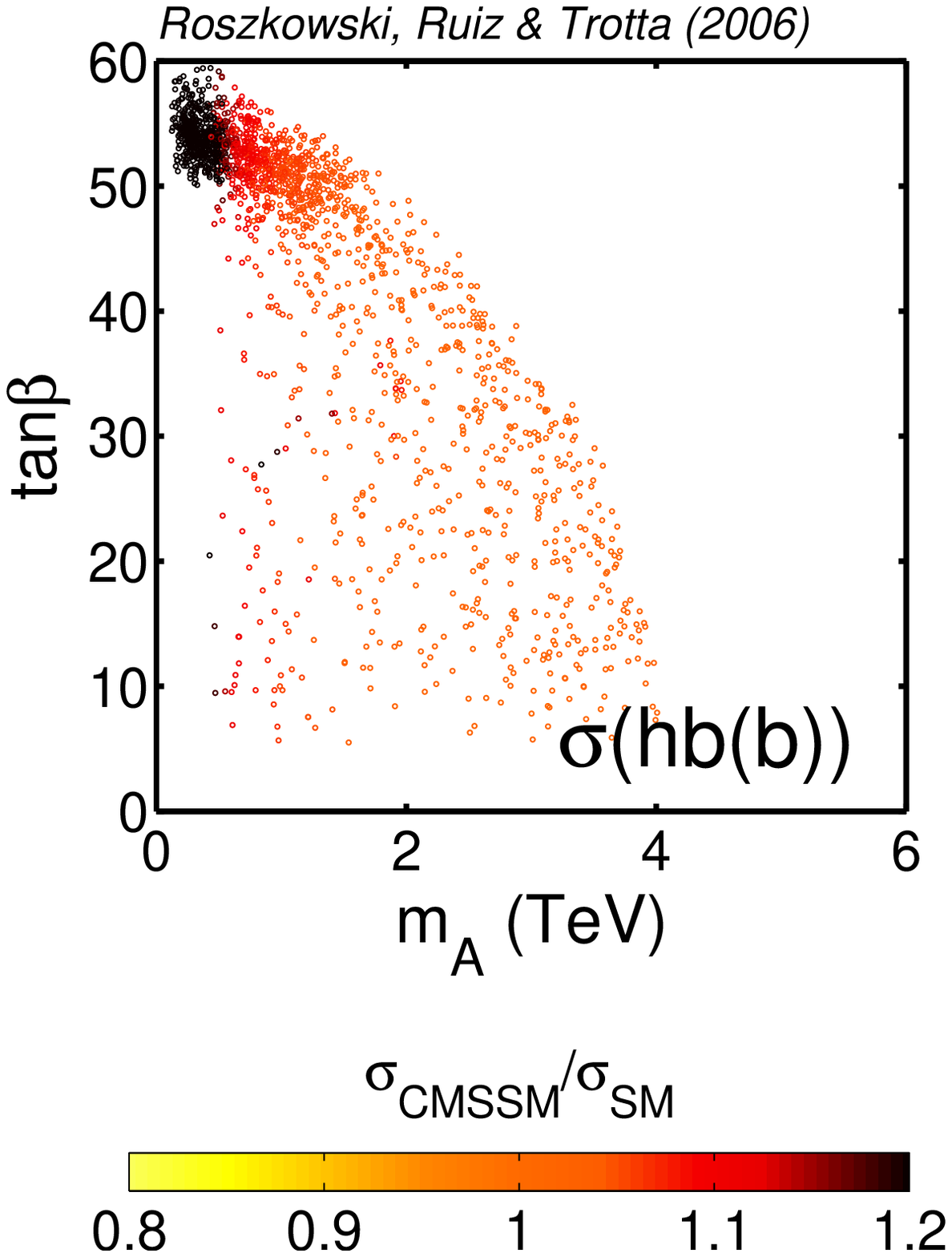}
 &	\includegraphics[width=0.3\textwidth]{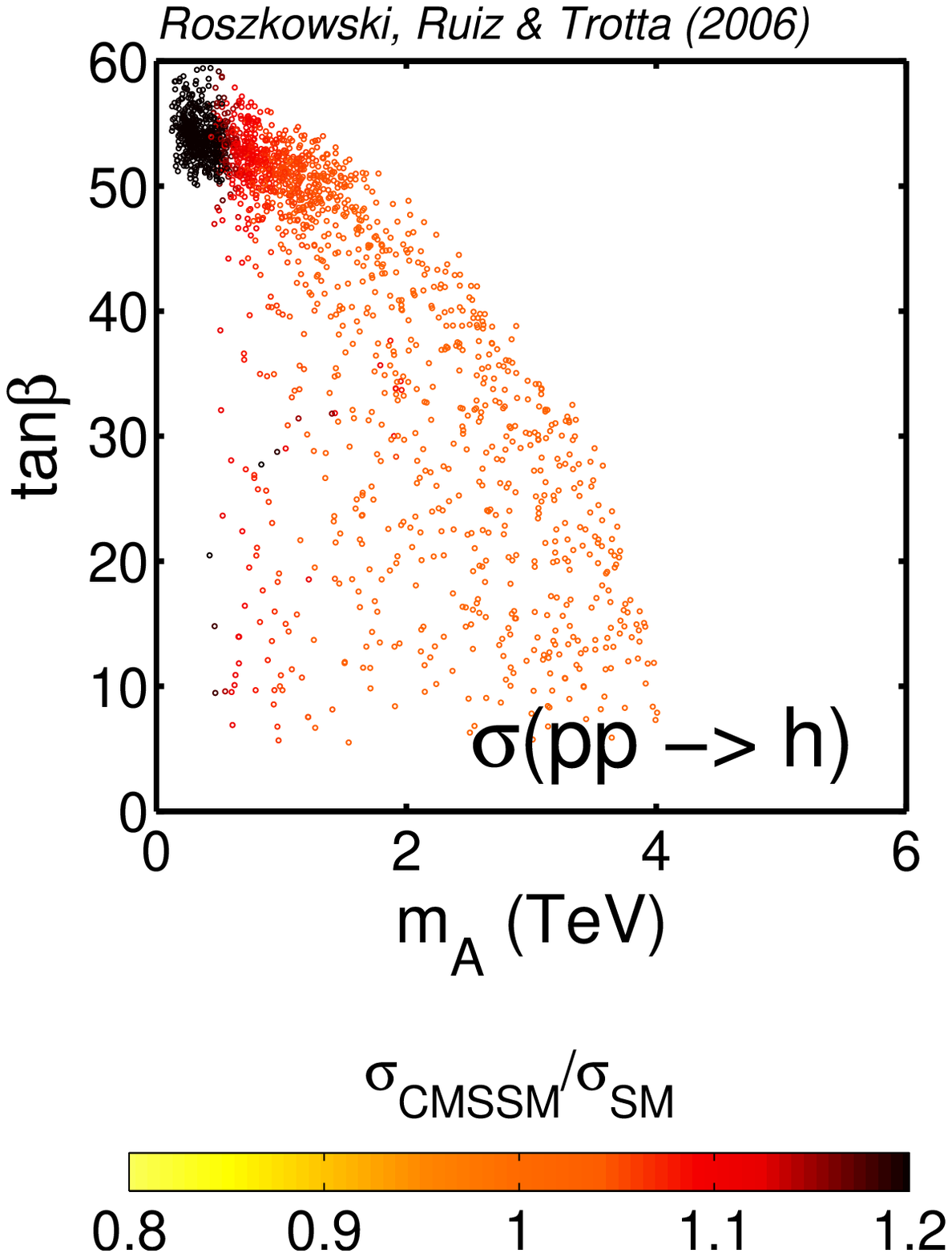}
\end{tabular}
\end{center}
\caption{Values in the $(\mha,\tanb)$ plane of light Higgs production
cross sections at the Tevatron (normalized to the SM case) for the
processes $\hl b(b)$ and $\ppbar\to \hl$. For the processes $V^{\ast}
\to V \hl$, where $V = Z, W$, and $gg\to \hl$ the values are almost
independent of $\tanb$ and $\mha$; compare
figs.~{\protect\ref{fig:rrt2-hlsigmavstanb}}
and~{\protect\ref{fig:rrt2-hlsigmavsmha}}.
\label{fig:rrt2-hlsigma-tanbvsmha}
}
\end{figure}

As above, we will follow the procedure developed in~\cite{rtr1} in
presenting our results in terms of relative posterior pdf, here simply
called relative probability density, for various variables.  First, in
fig.~\ref{fig:rrt2-hlsigma} we display relative probability densities
for $\sigma_{\text{CMSSM}}/\sigma_{\text{SM}}$ for the processes of
primary interest at the Tevatron. All the ratios are close to 1 but
only in the case of $V^\ast\to V\hl$ ($V=Z,W$) is a pdf very strongly
peaked very close to 1. The pdf for the gluon--gluon fusion SM--normalized
cross section is peaked around 1.1, with rather little variation. This is a
reflection of the behavior of the (radiatively corrected) coupling
$g(\hl\bbbar)_{\text{CMSSM}}$; compare the
left panels of figs.~\ref{fig:coups-vs-tanb}
and~\ref{fig:coups-vs-ma}. For the remaining two processes we observe
some more variation, and increase relative to the SM, than in $gg\to\hl$. Actually,
because of the way these processes are computed (as described above),
their pdf's will in most cases be very similar. We nevertheless
present them both for completeness.

In order to display the behavior of the production cross sections in
more detail, we present in fig.~\ref{fig:rrt2-hlsigmavsmhl} a 2--dim
relative probability density of the SM--normalized cross sections
as a function of $\mhl$, in fig.~\ref{fig:rrt2-hlsigmavstanb} of $\tanb$ and
in fig.~\ref{fig:rrt2-hlsigmavsmha} of the pseudoscalar Higgs mass
$\mha$. As regards the gluon--gluon fusion process, the bottom quark
exchange contribution to the cross section is subdominant relative to
the top quark one (by a factor of a few). This explains why there is
much less variation in the corresponding pdf than for $\hl b (b)$ and
$p\bar{p} \to \hl $. On the other hand, the enhancement of the
coupling $g(\hl \bbbar)_{\text{CMSSM}}$ at
$\tanb\gsim50$ and $\mha\lsim1\tev$, cause a slight increase of a few
per cent in the cross section relative to the SM. Otherwise,
unsurprisingly, the pdf's for the three processes mirror the behavior
of the $g(\hl\bbbar)_{\text{CMSSM}}$ coupling and we include them for
completeness.  To finish our discussion of light Higgs production, we present in
fig.~\ref{fig:rrt2-hlsigma-tanbvsmhl} the ranges in the
CMSSM of the SM--normalized cross sections for the above two processes
in the plane of $\tanb$ and $\mhl$, while in
fig.~\ref{fig:rrt2-hlsigma-tanbvsmha} the same is shown in the plane
of $\tanb$ and $\mha$.


\begin{figure}[!tb]
\begin{center}
\begin{tabular}{c c c}
 	\includegraphics[width=0.3\textwidth]{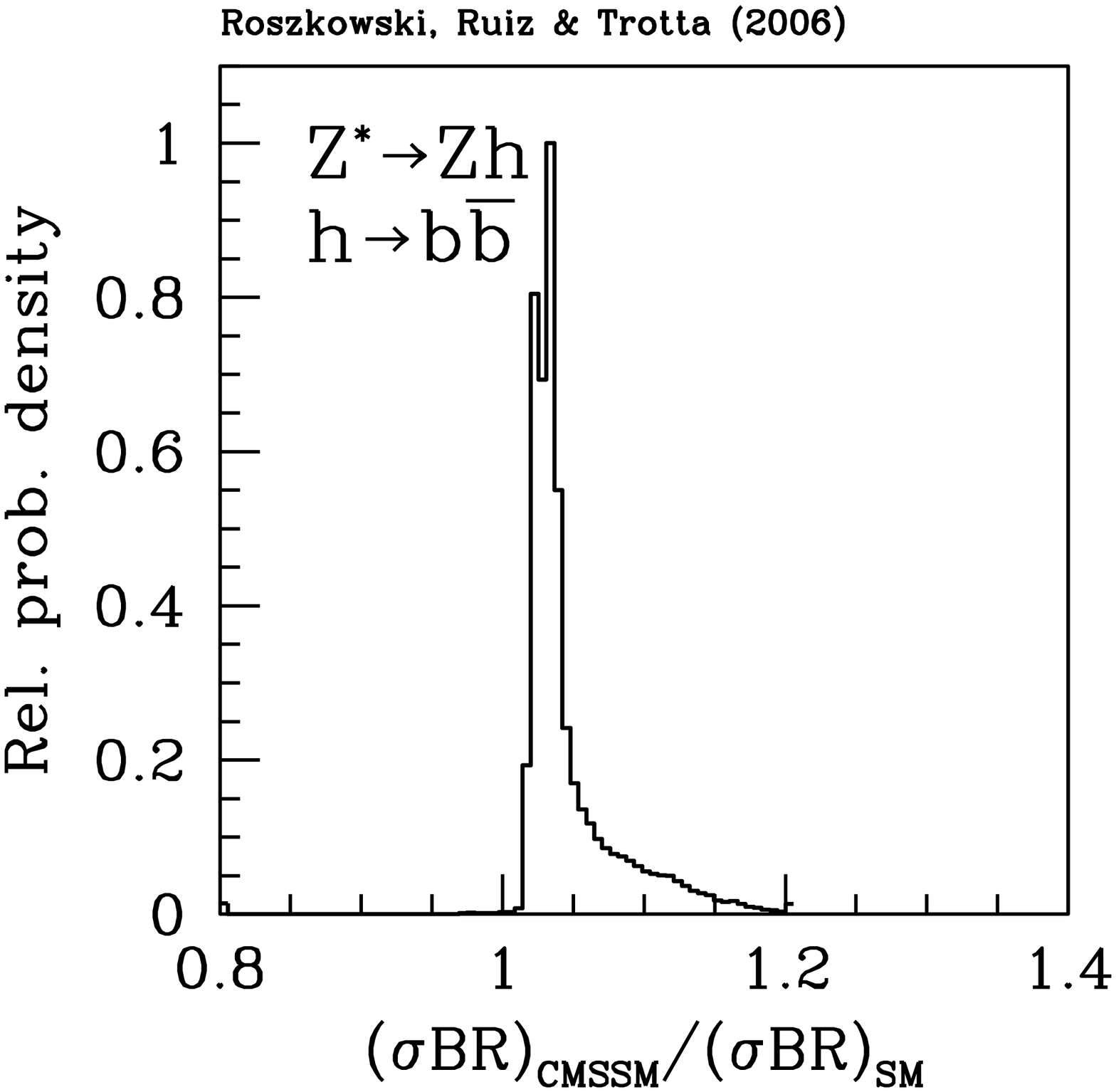}
 &	\includegraphics[width=0.3\textwidth]{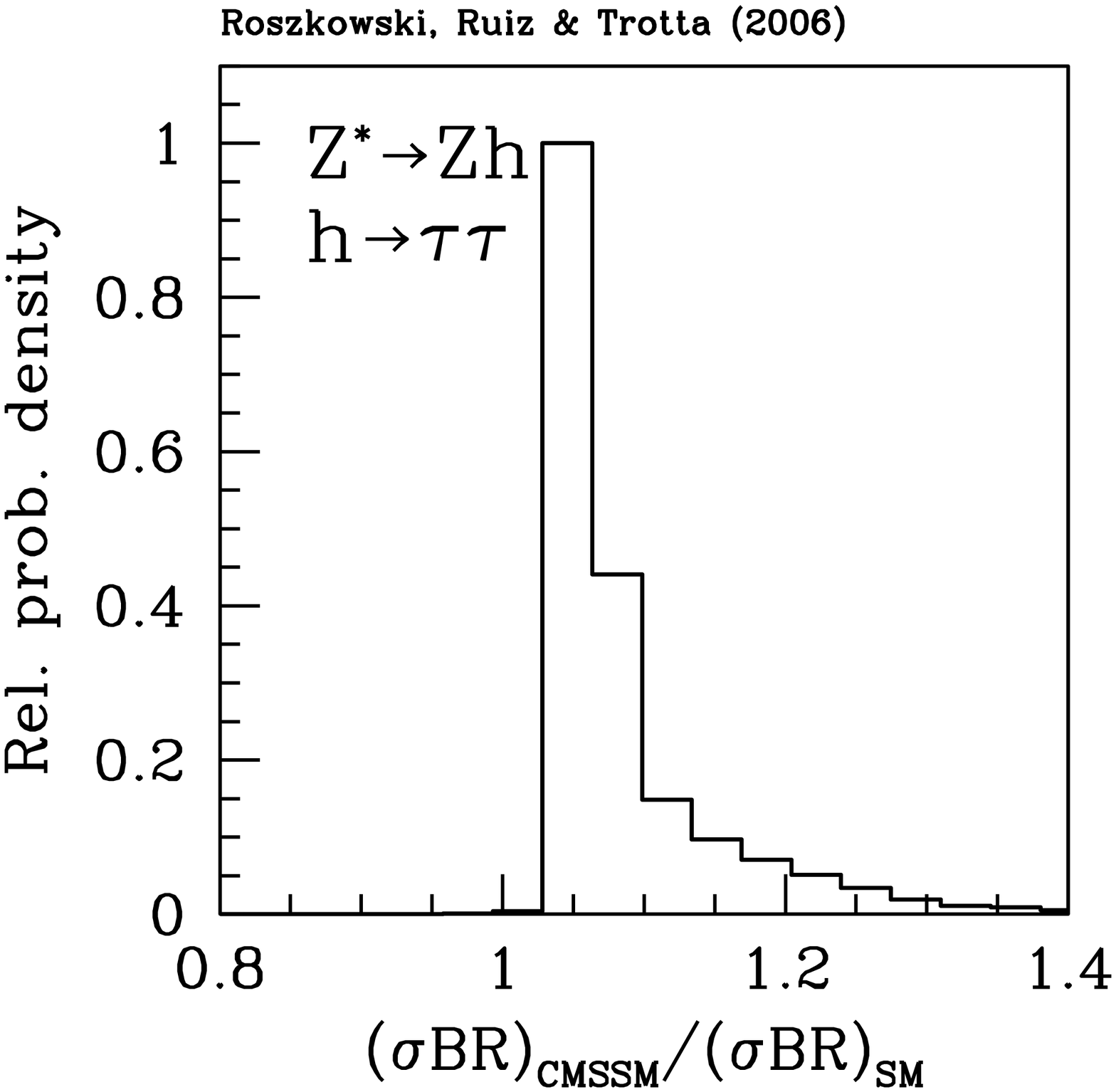}
 &	\includegraphics[width=0.3\textwidth]{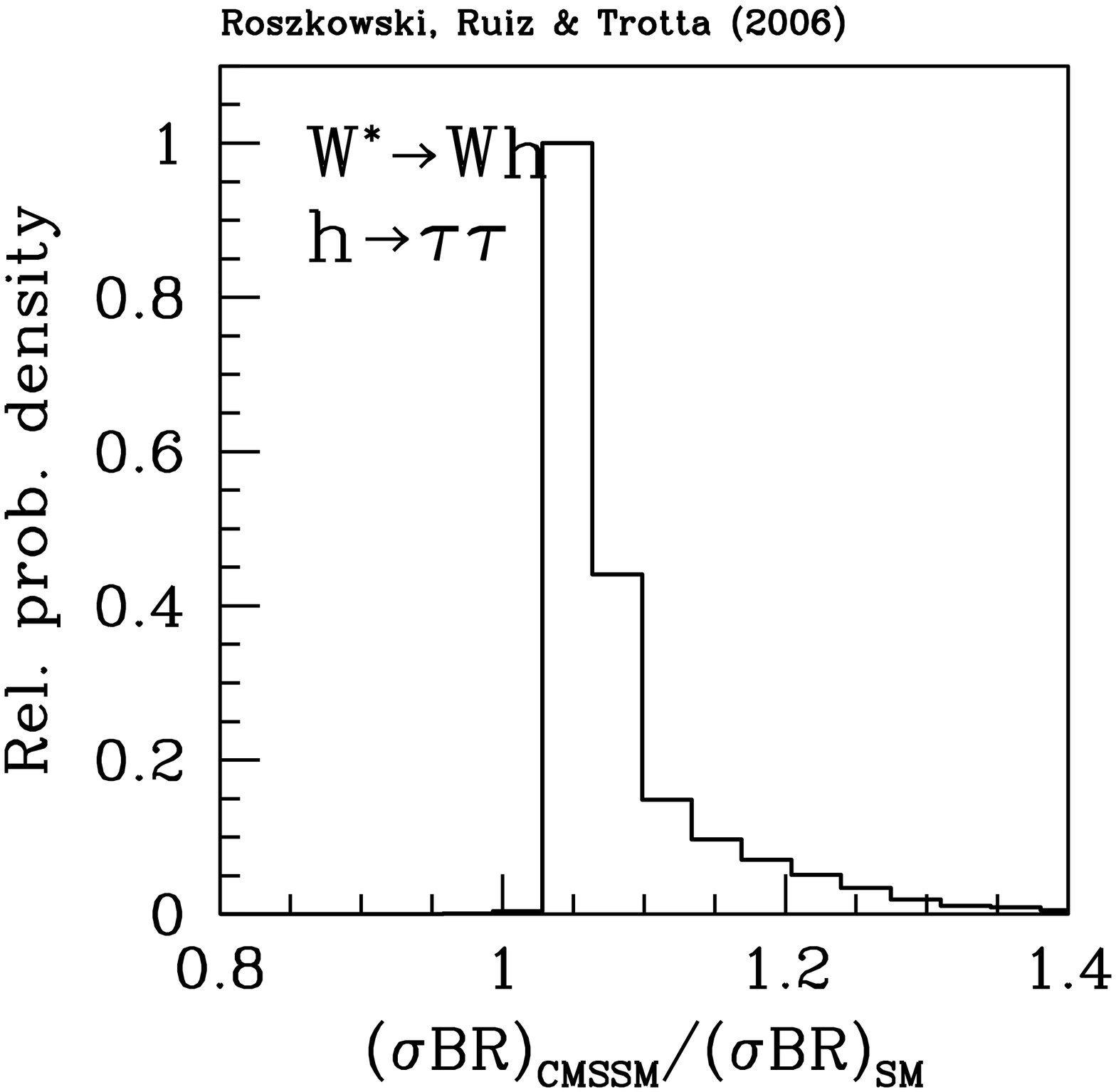}\\
	\includegraphics[width=0.3\textwidth]{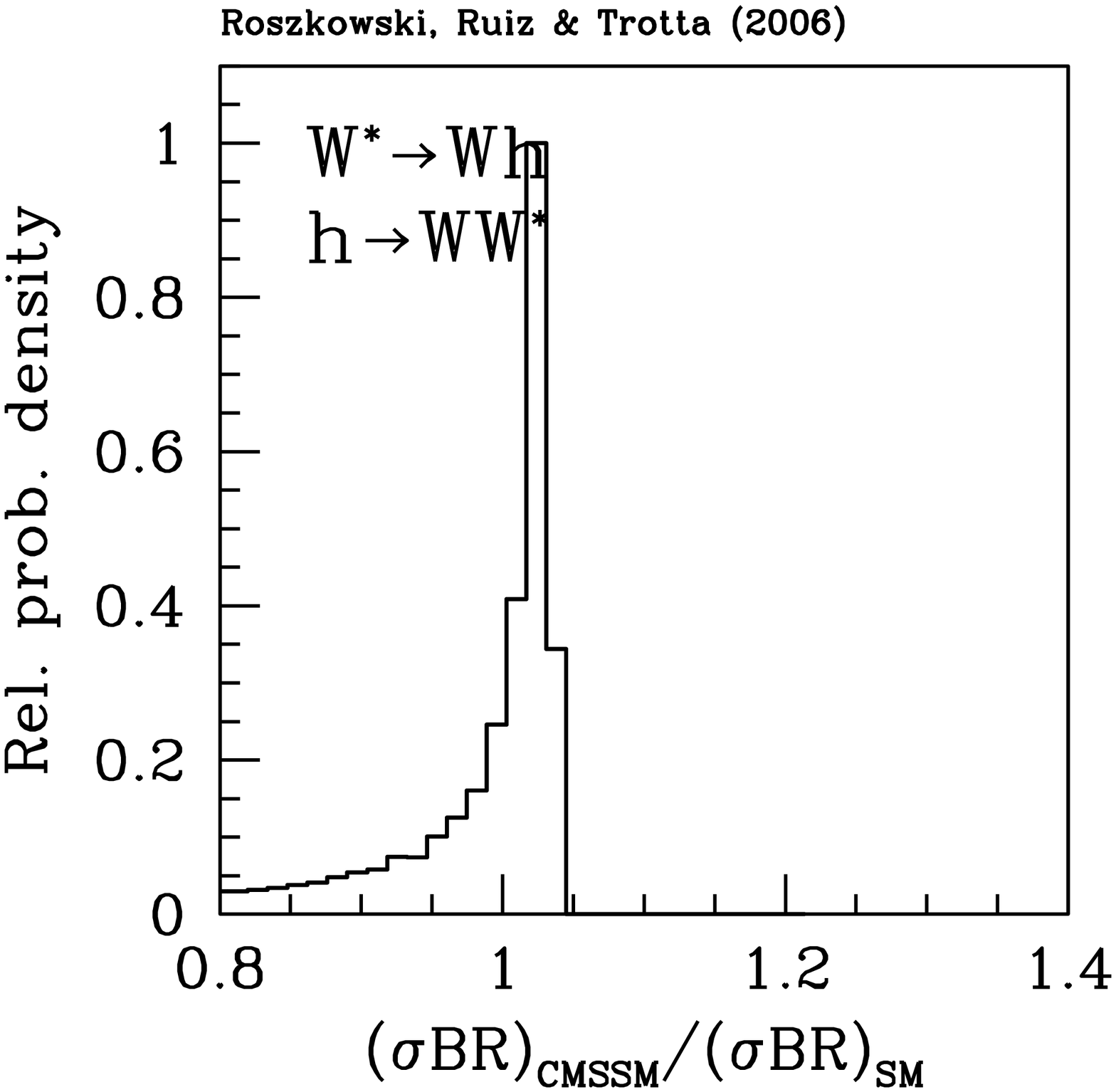}
 &	\includegraphics[width=0.3\textwidth]{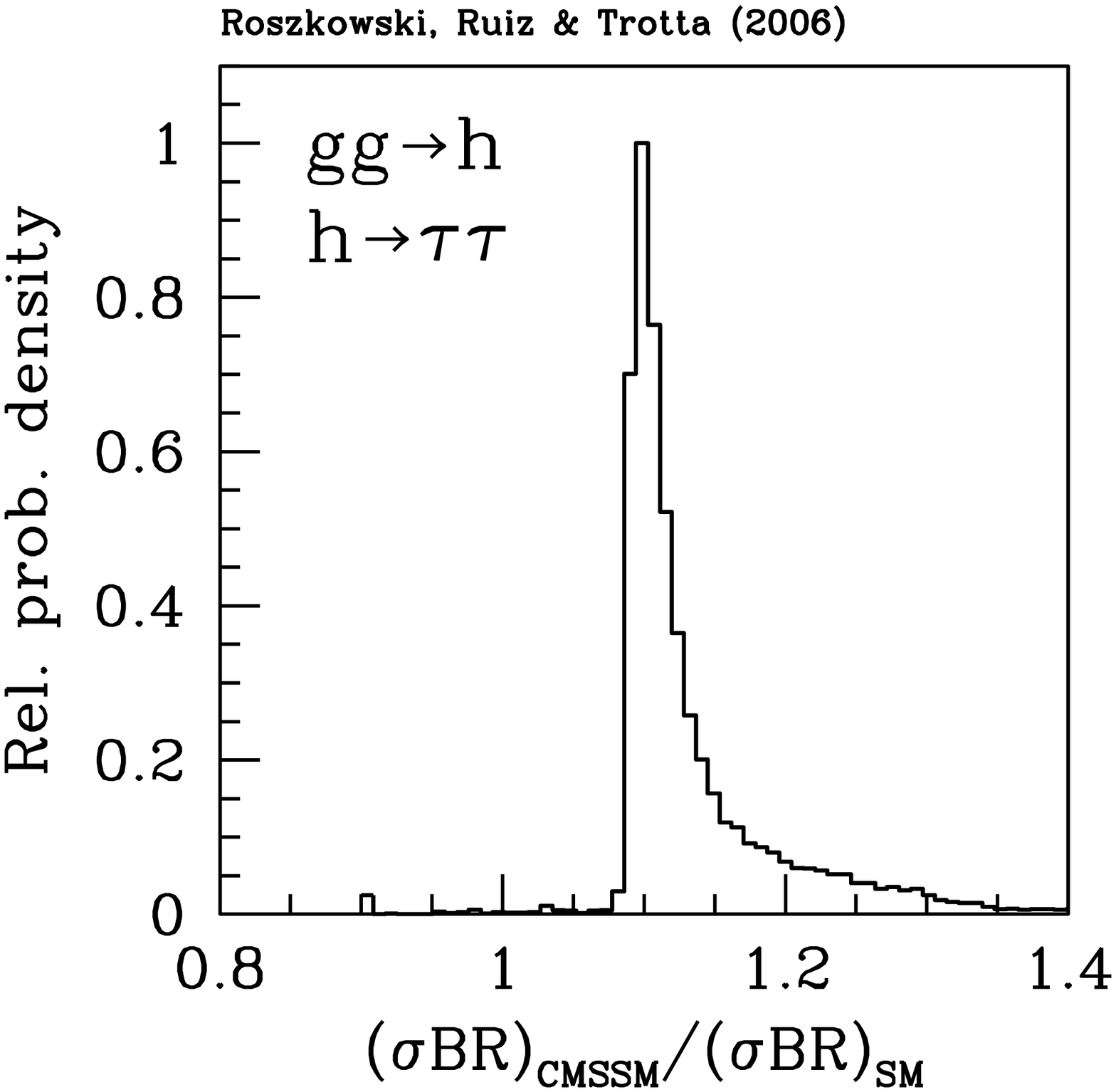}
 & 	\includegraphics[width=0.3\textwidth]{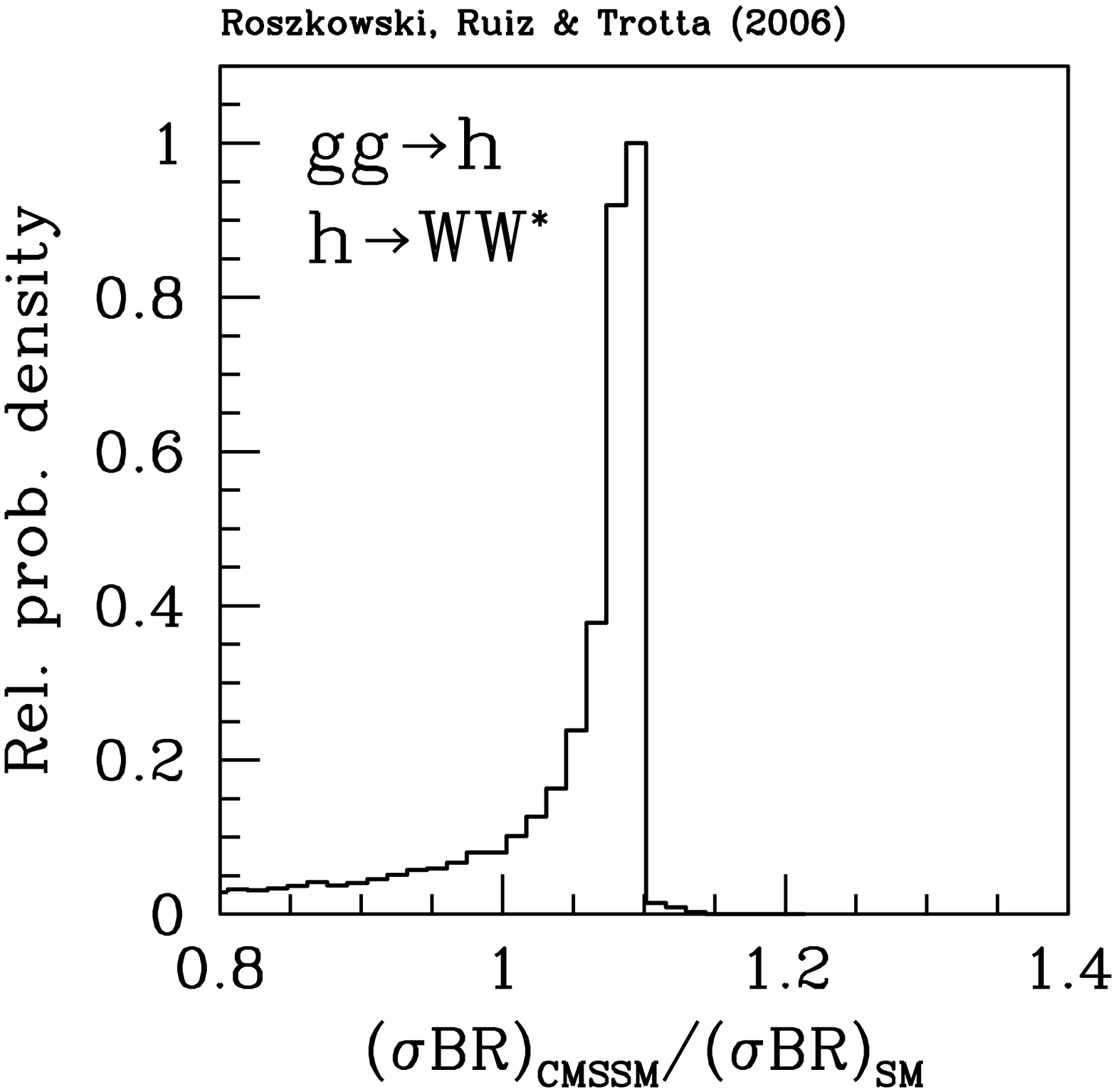}\\
	\includegraphics[width=0.3\textwidth]{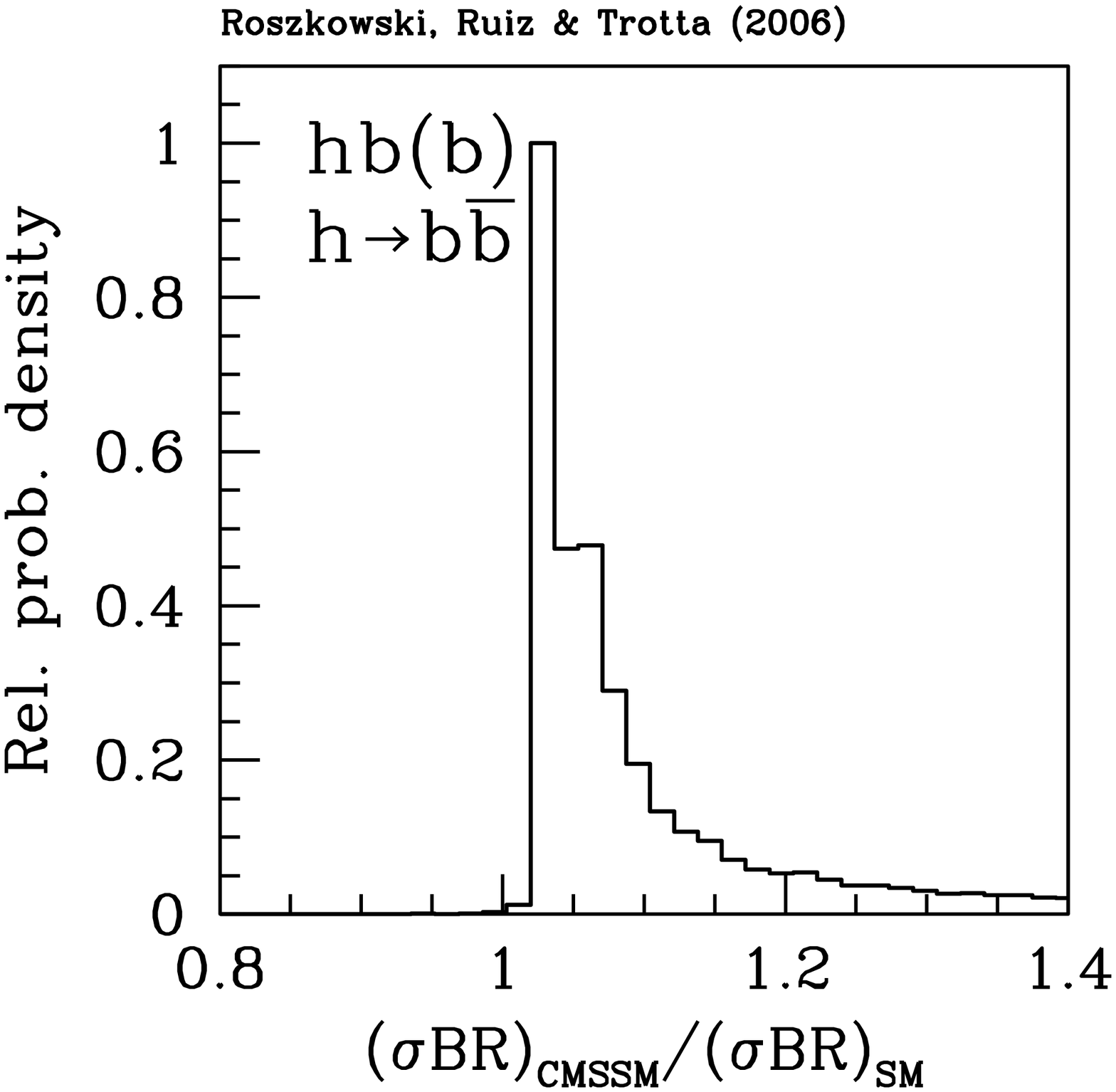}
 & 	\includegraphics[width=0.3\textwidth]{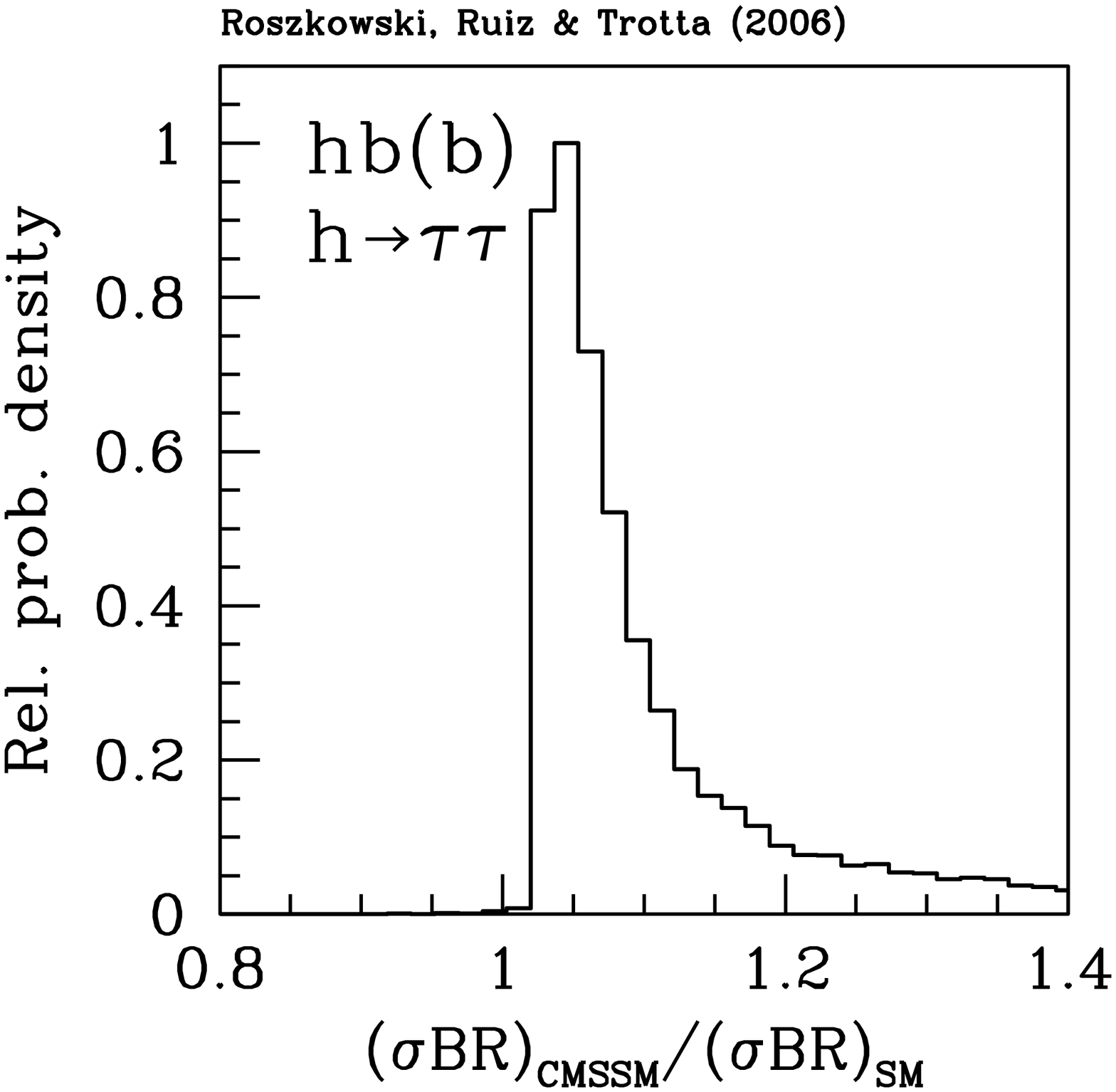}
 &	\includegraphics[width=0.3\textwidth]{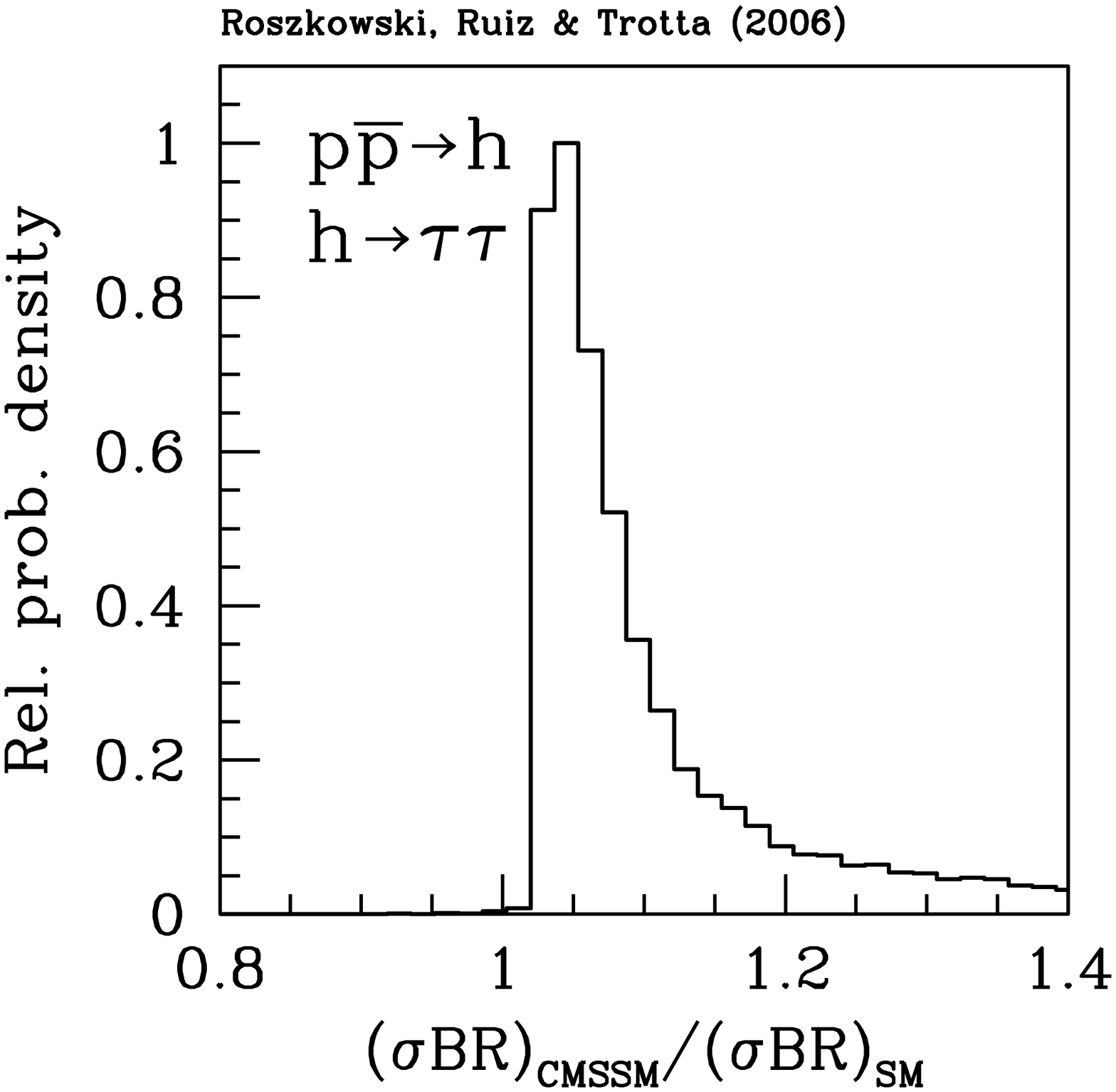}
\end{tabular}
\end{center}
\caption{The 1--dim relative probability density for light Higgs
  production cross sections times decay branching ratios at the
  Tevatron (normalized to the SM case).
\label{fig:rrt2-hlsigmadecay}
}
\end{figure}
\begin{figure}[!tb]
\begin{center}
\begin{tabular}{c c c}
	\includegraphics[width=0.3\textwidth]{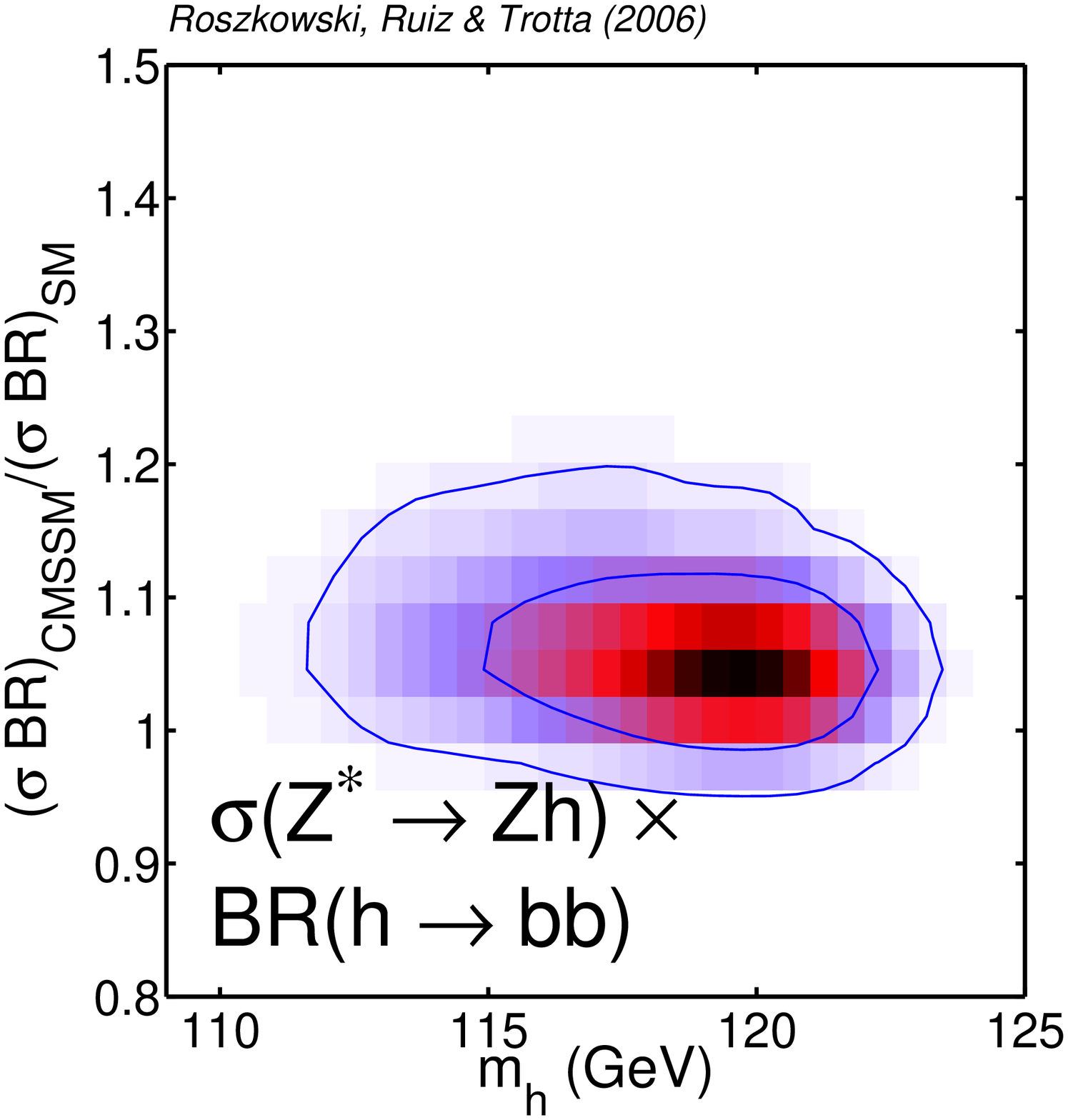}
 &	\includegraphics[width=0.3\textwidth]{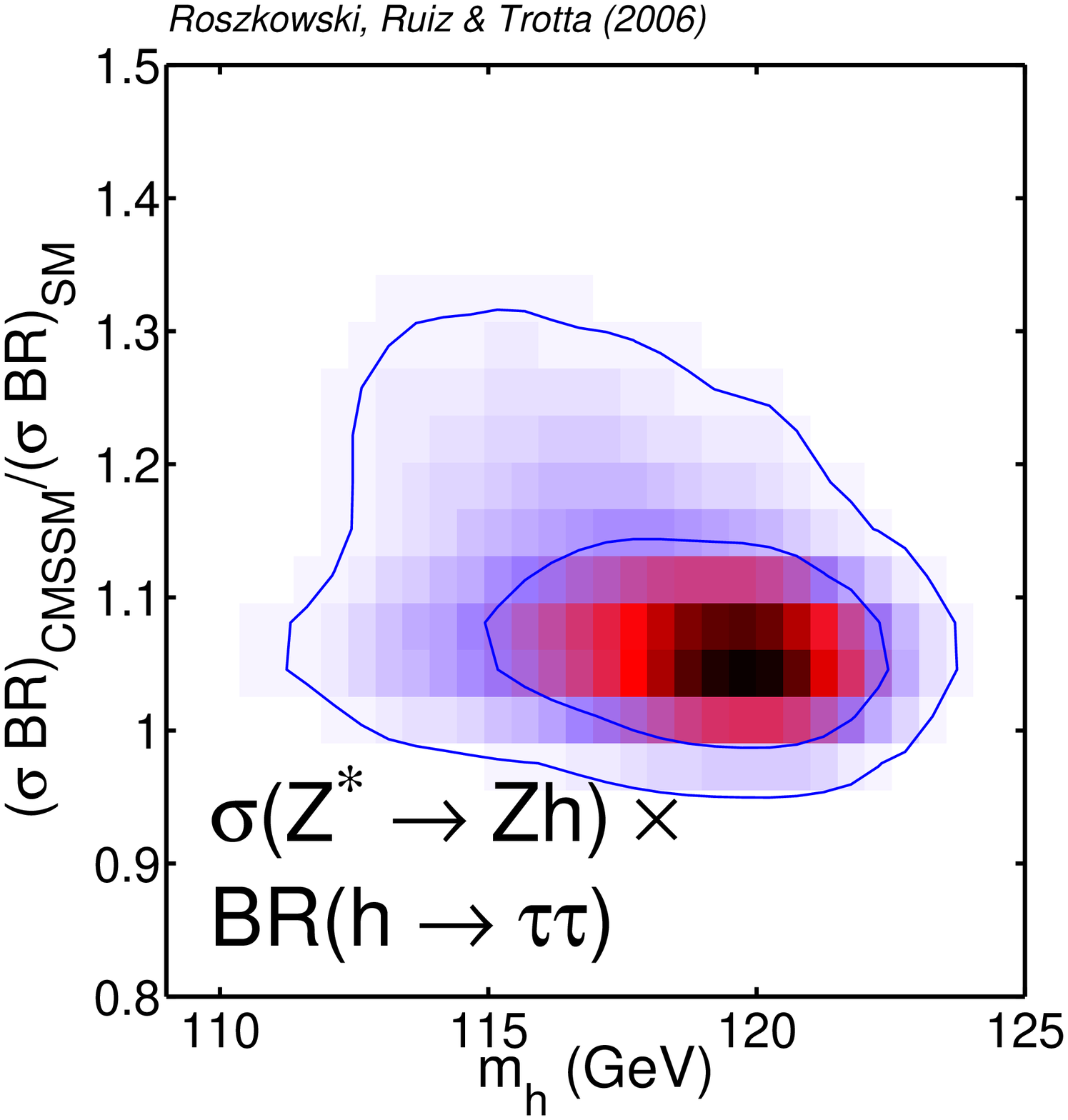}
 &	\includegraphics[width=0.3\textwidth]{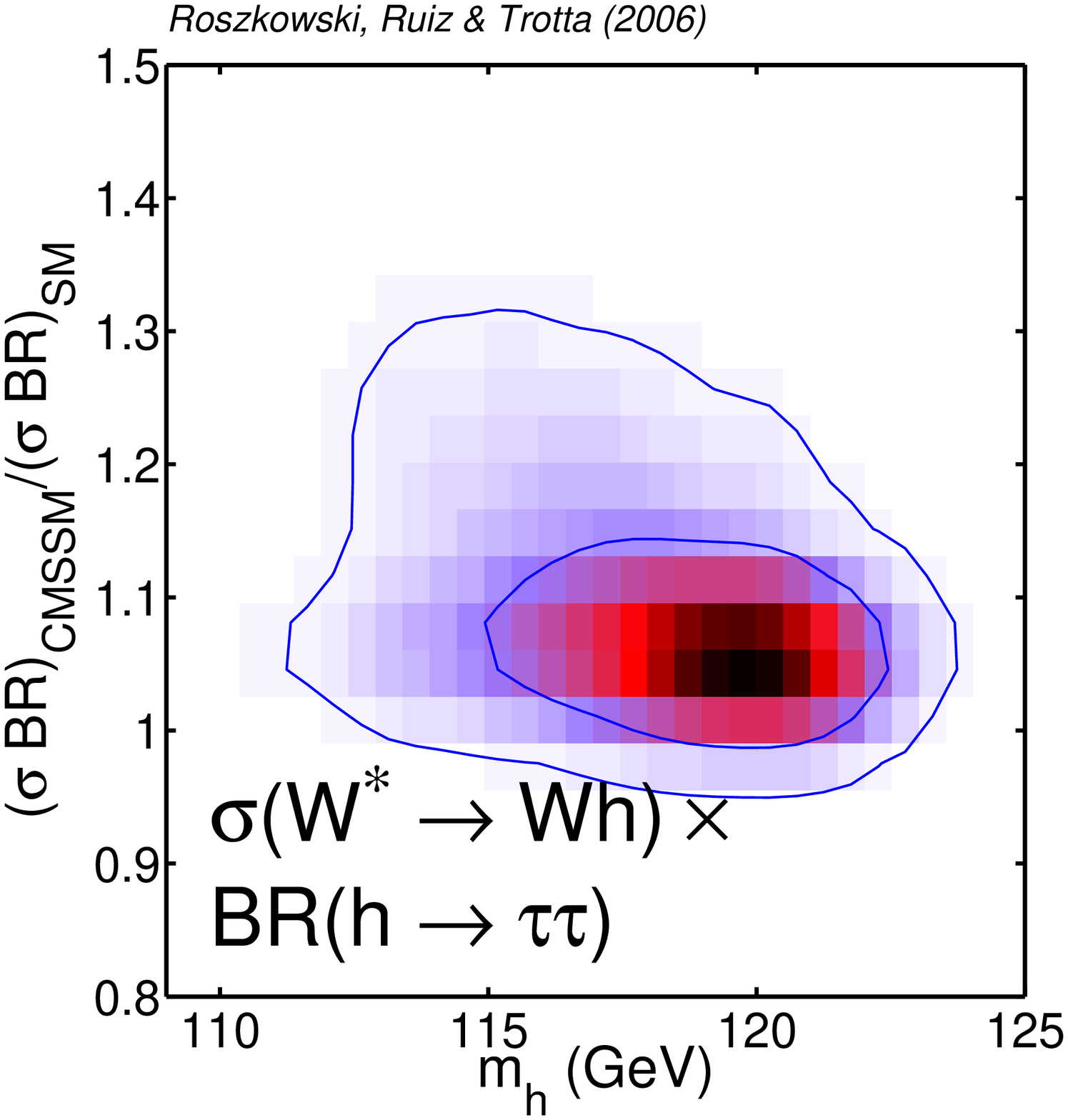}\\
	\includegraphics[width=0.3\textwidth]{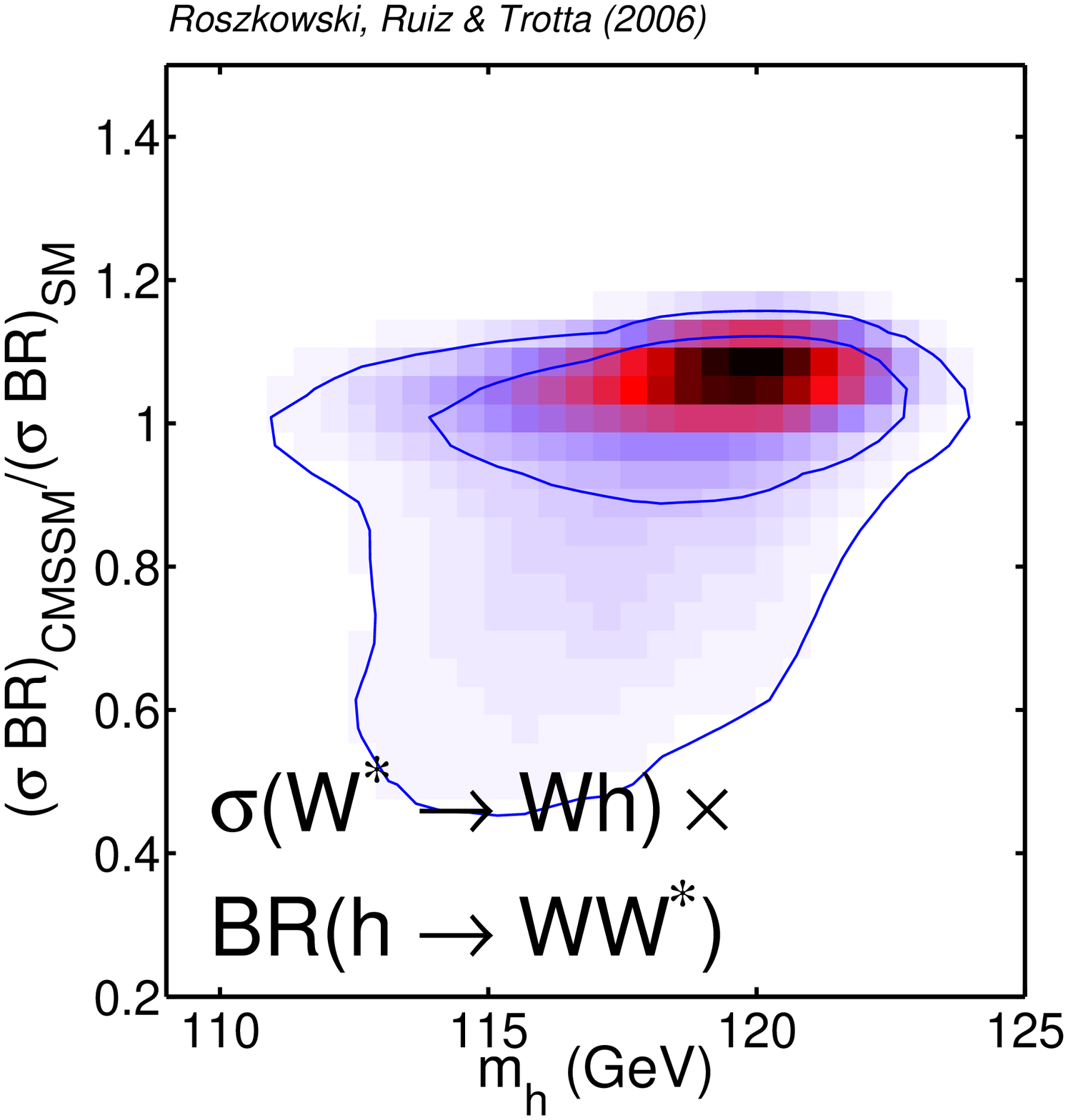}
 &	\includegraphics[width=0.3\textwidth]{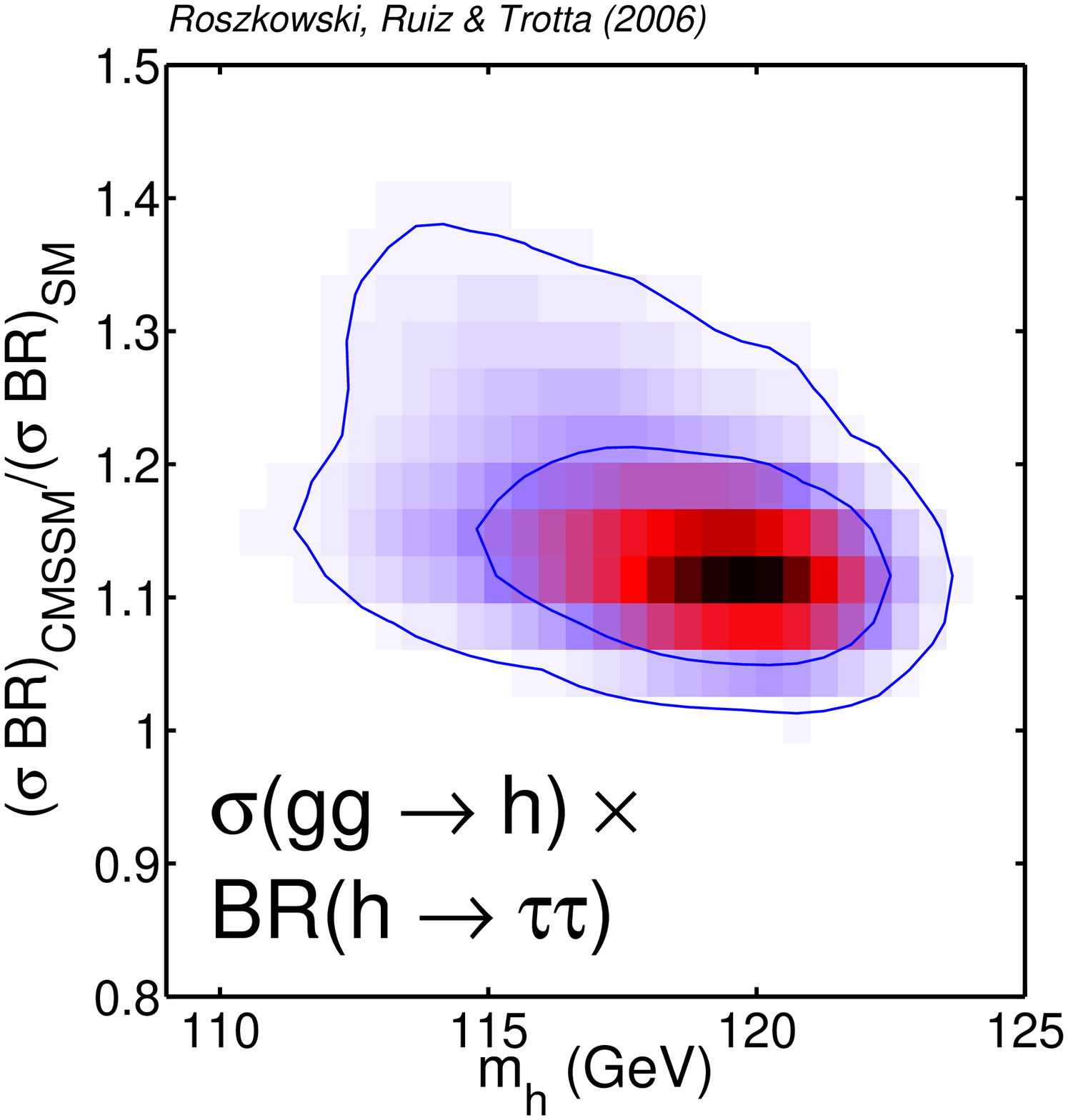}
 &	\includegraphics[width=0.3\textwidth]{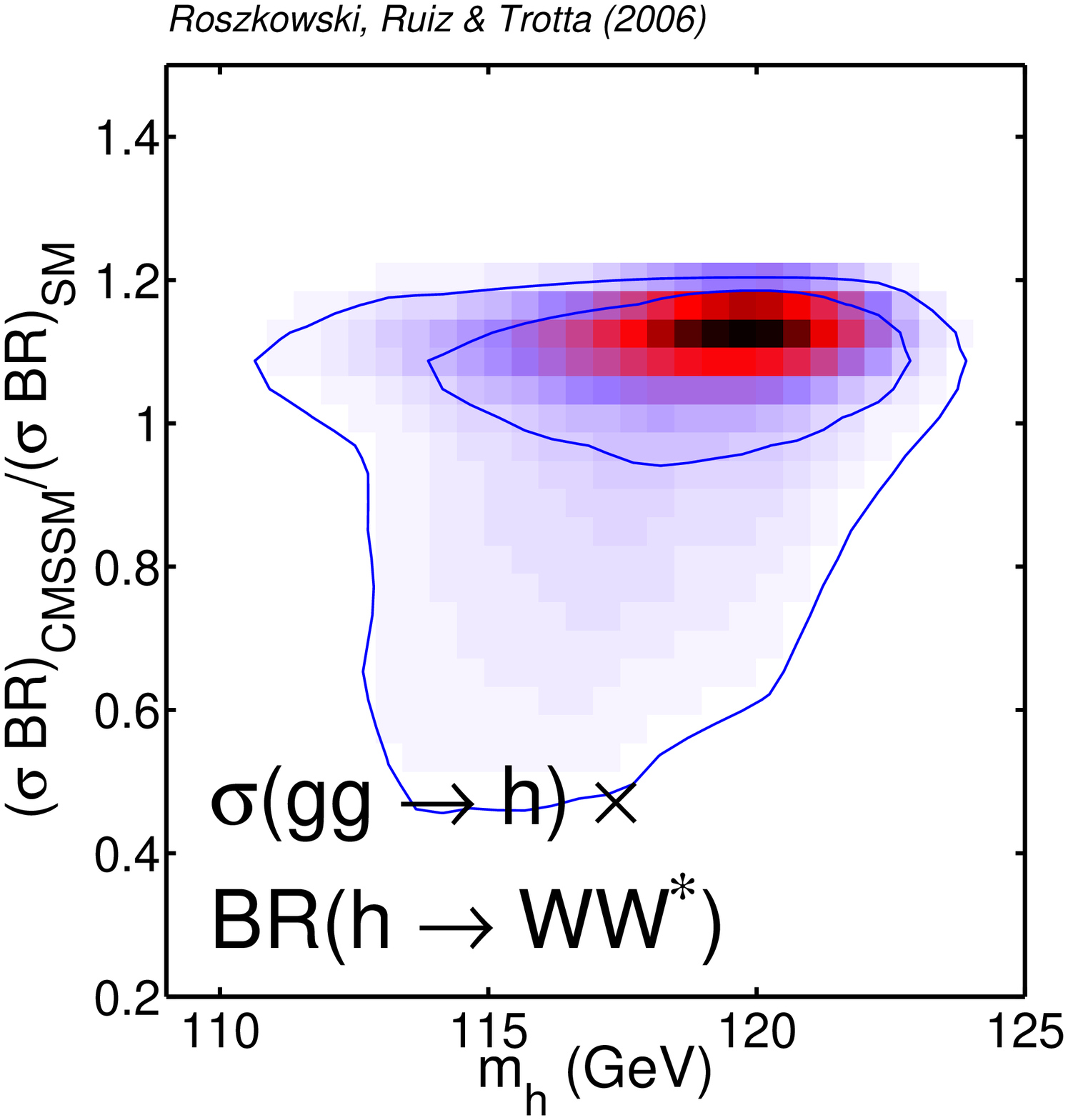}\\
	\includegraphics[width=0.3\textwidth]{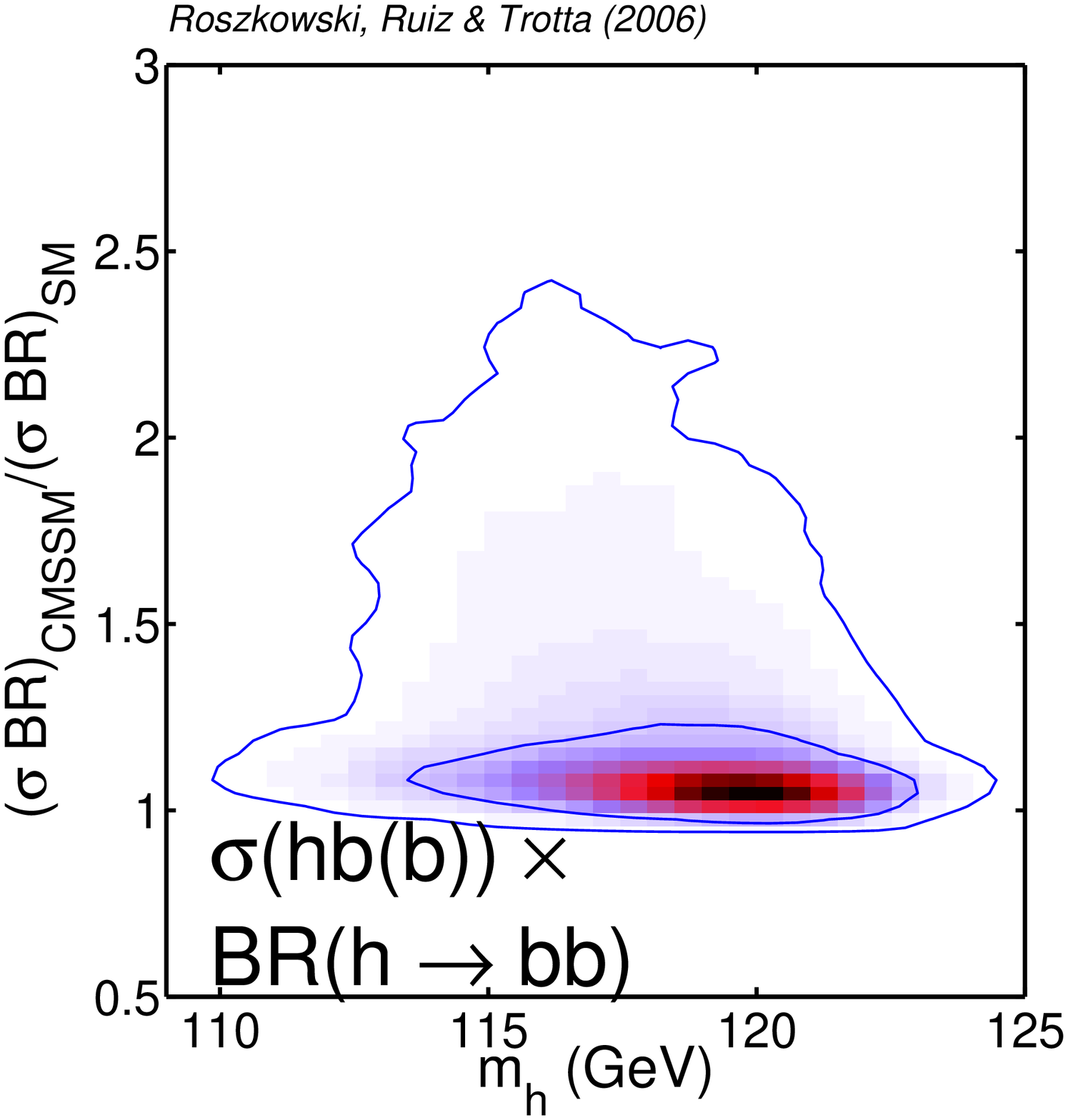}
 &	\includegraphics[width=0.3\textwidth]{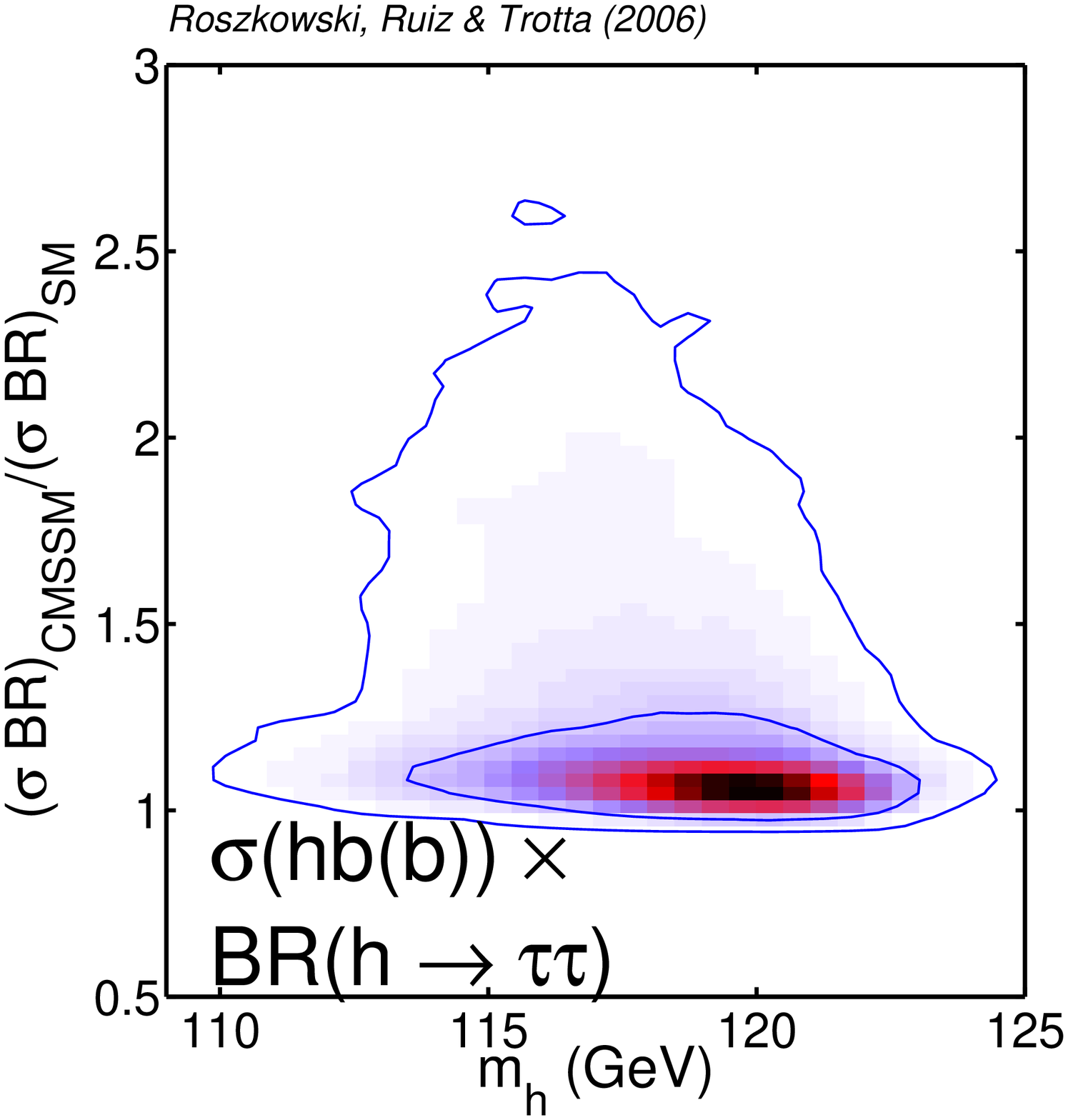}
 &	\includegraphics[width=0.3\textwidth]{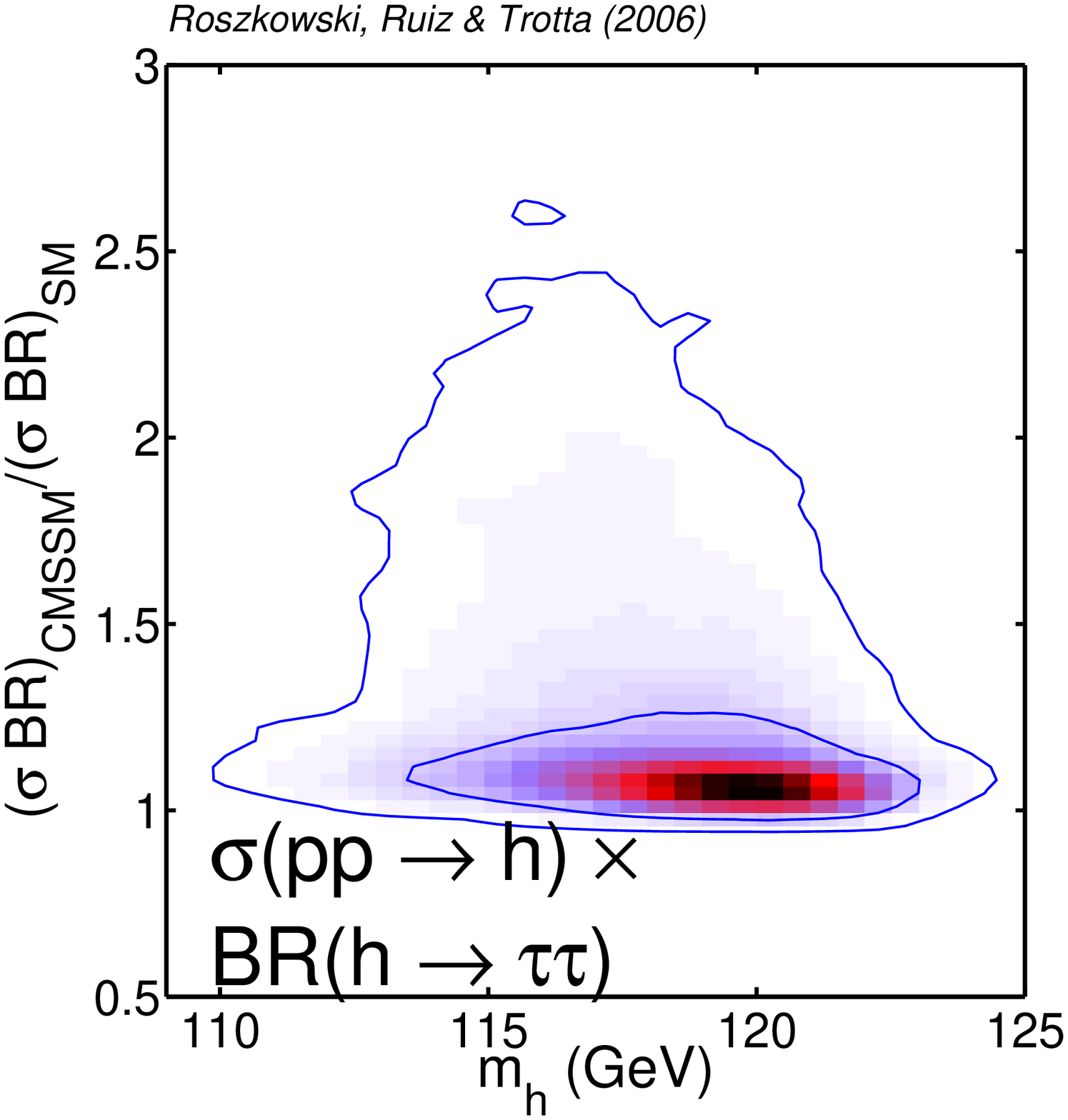}
\end{tabular}
\end{center}
\caption{The 2--dim relative probability density of light Higgs
  production cross sections times decay branching ratios at the
  Tevatron (normalized to the SM case) as a function of its mass $\mhl$.
\label{fig:rrt2-hlsigmabrvsmhl}
}
\end{figure}
\begin{figure}[!tb]
\begin{center}
\begin{tabular}{c c c}
	\includegraphics[width=0.3\textwidth]{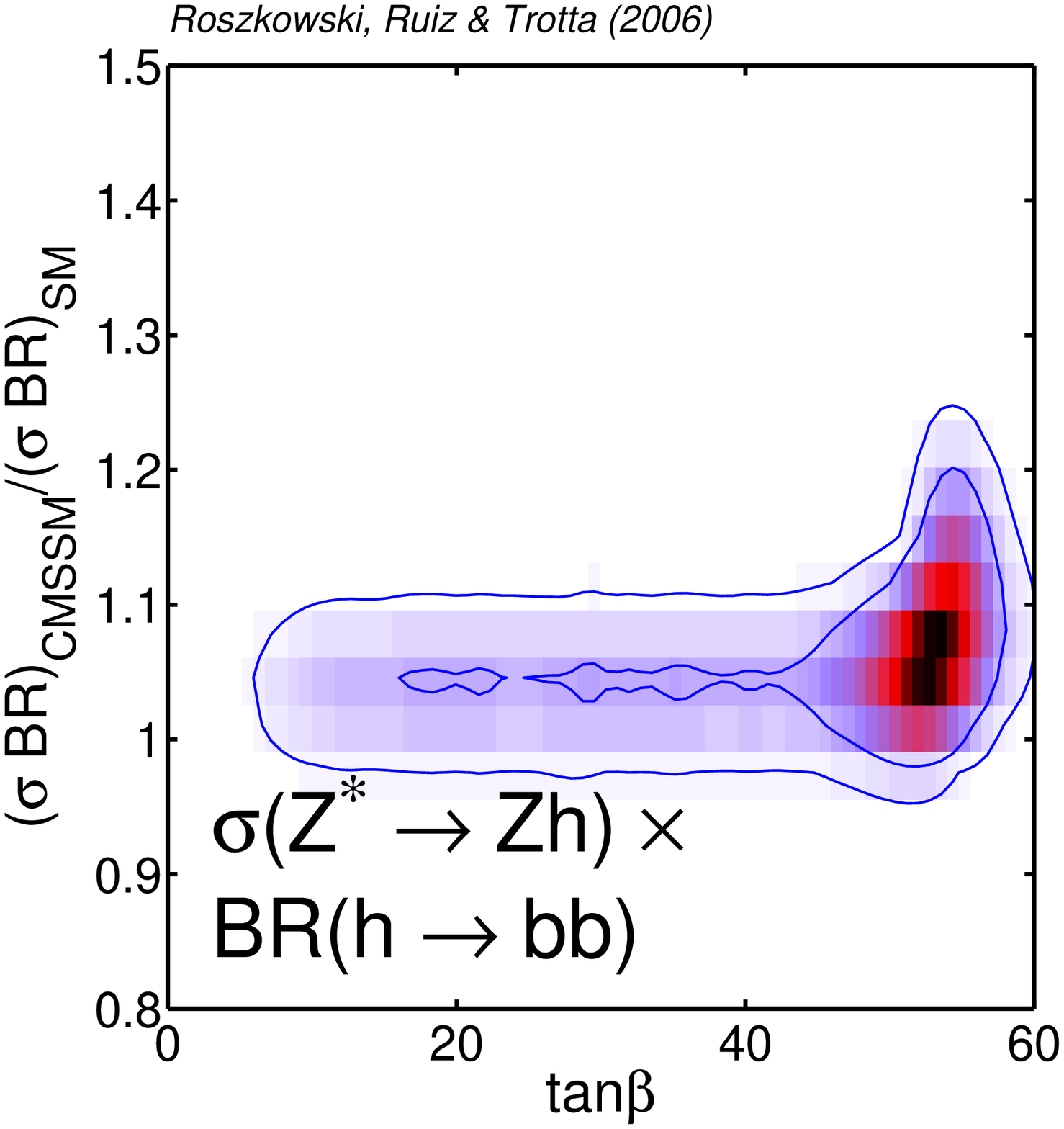}
 &	\includegraphics[width=0.3\textwidth]{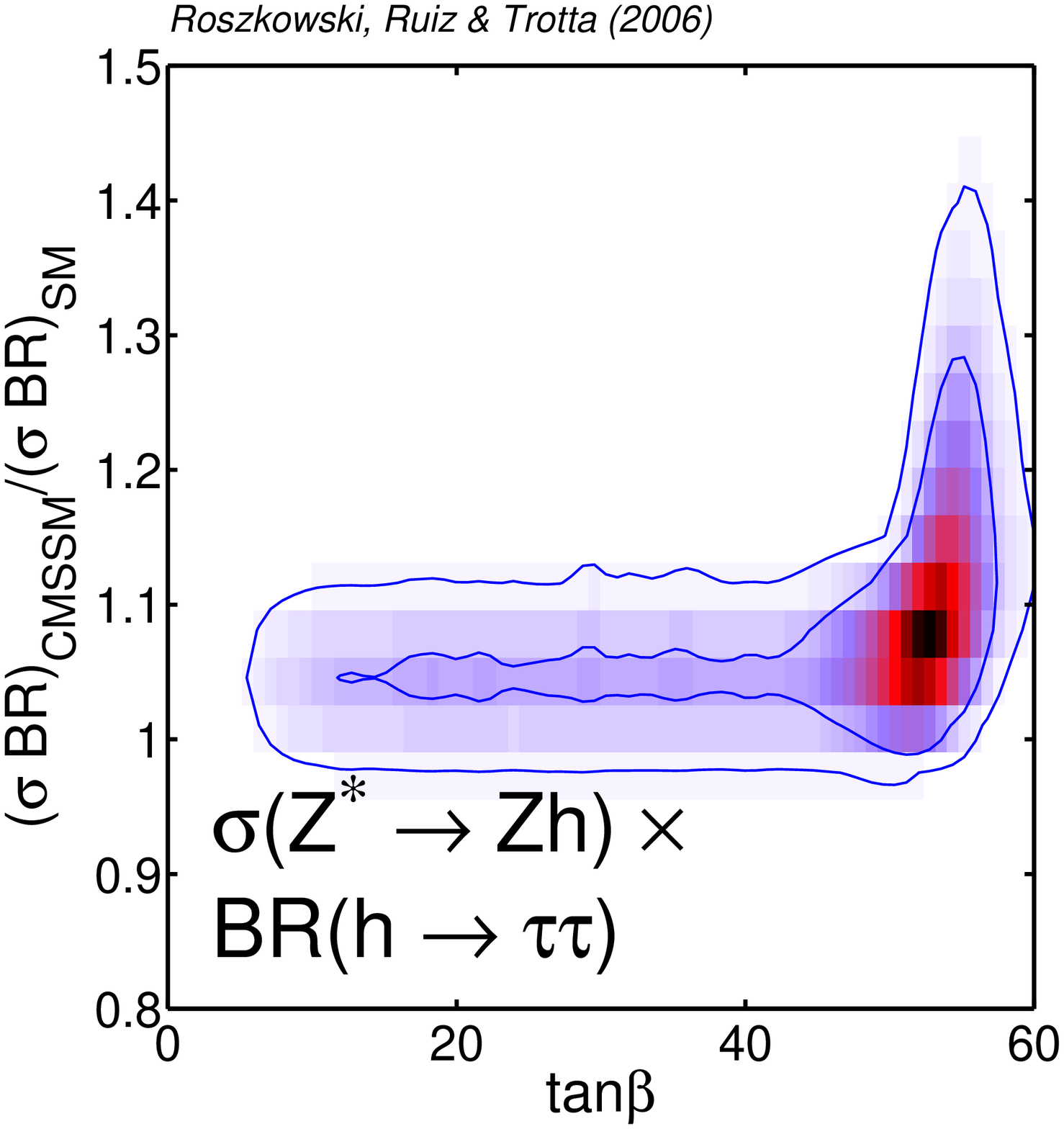}
 &	\includegraphics[width=0.3\textwidth]{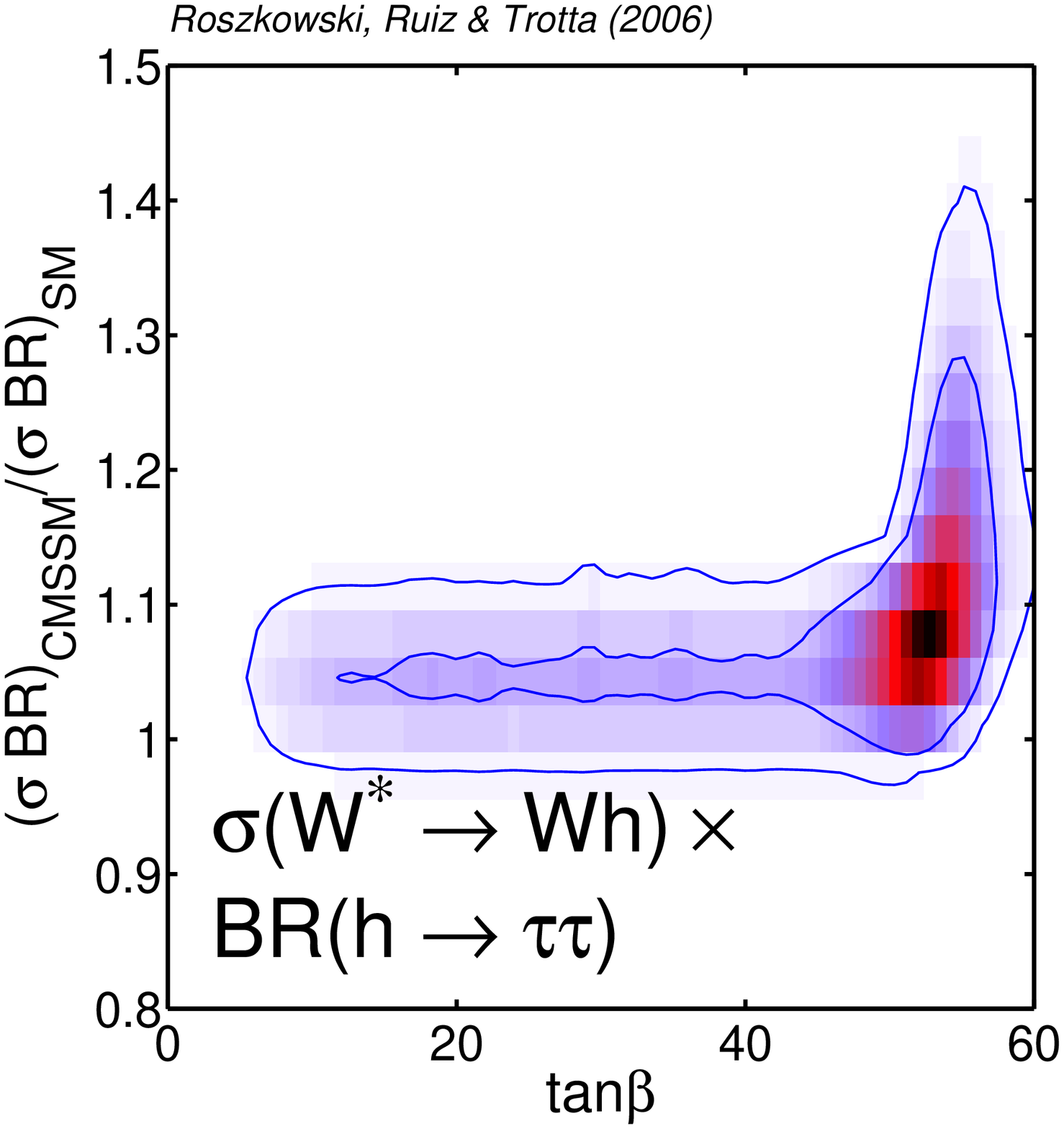}\\
	\includegraphics[width=0.3\textwidth]{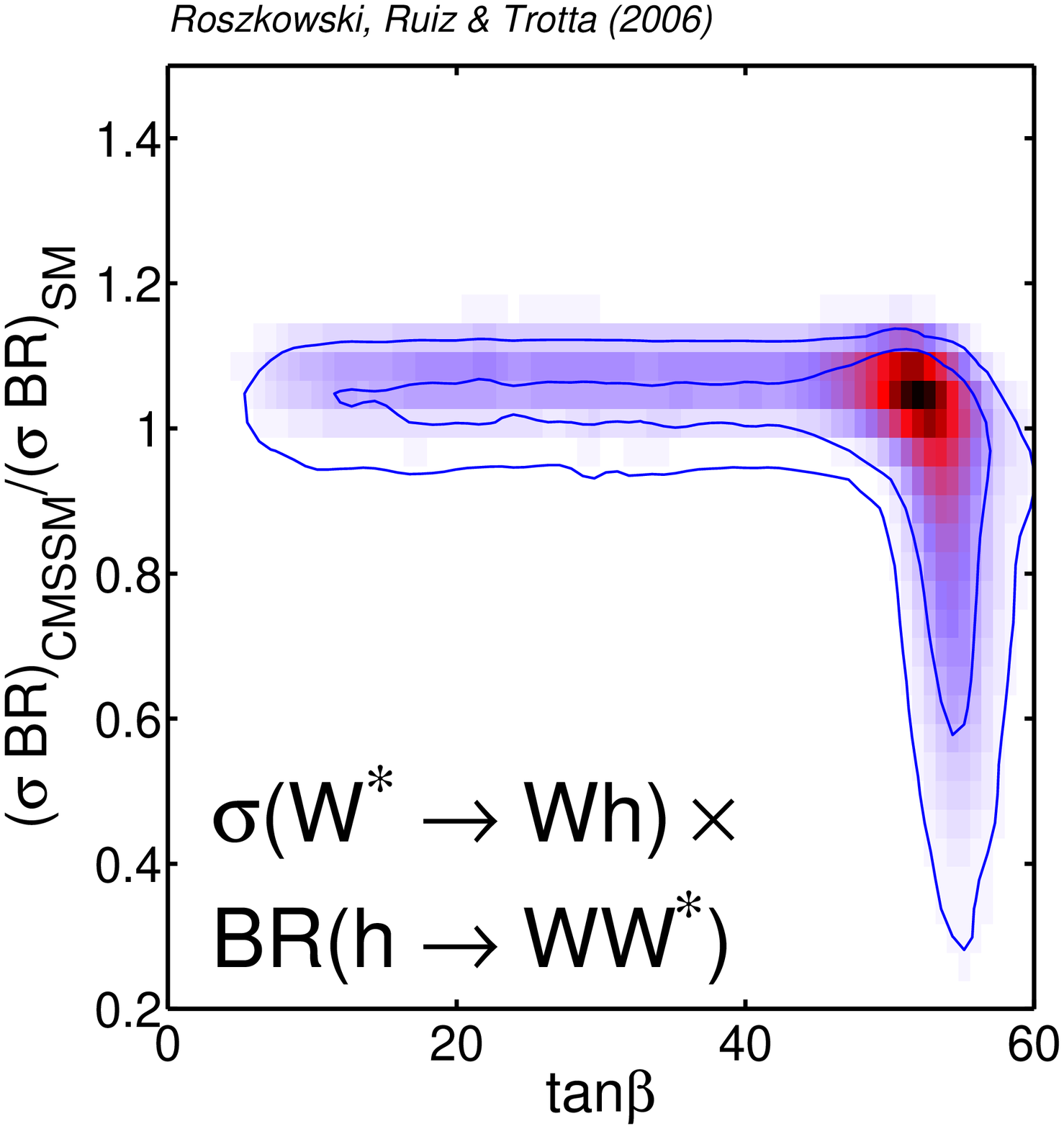}
 &	\includegraphics[width=0.3\textwidth]{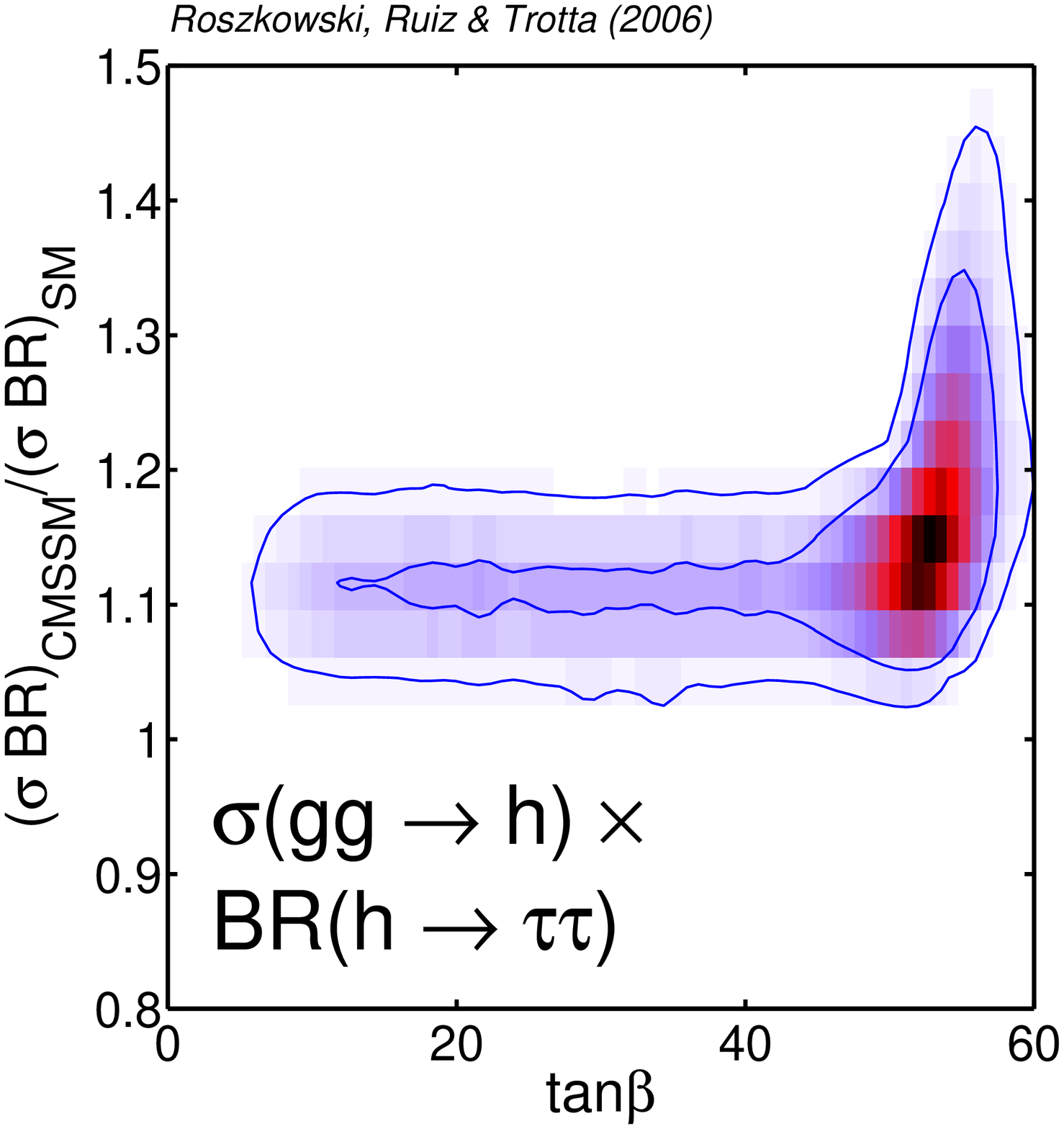}
 &	\includegraphics[width=0.3\textwidth]{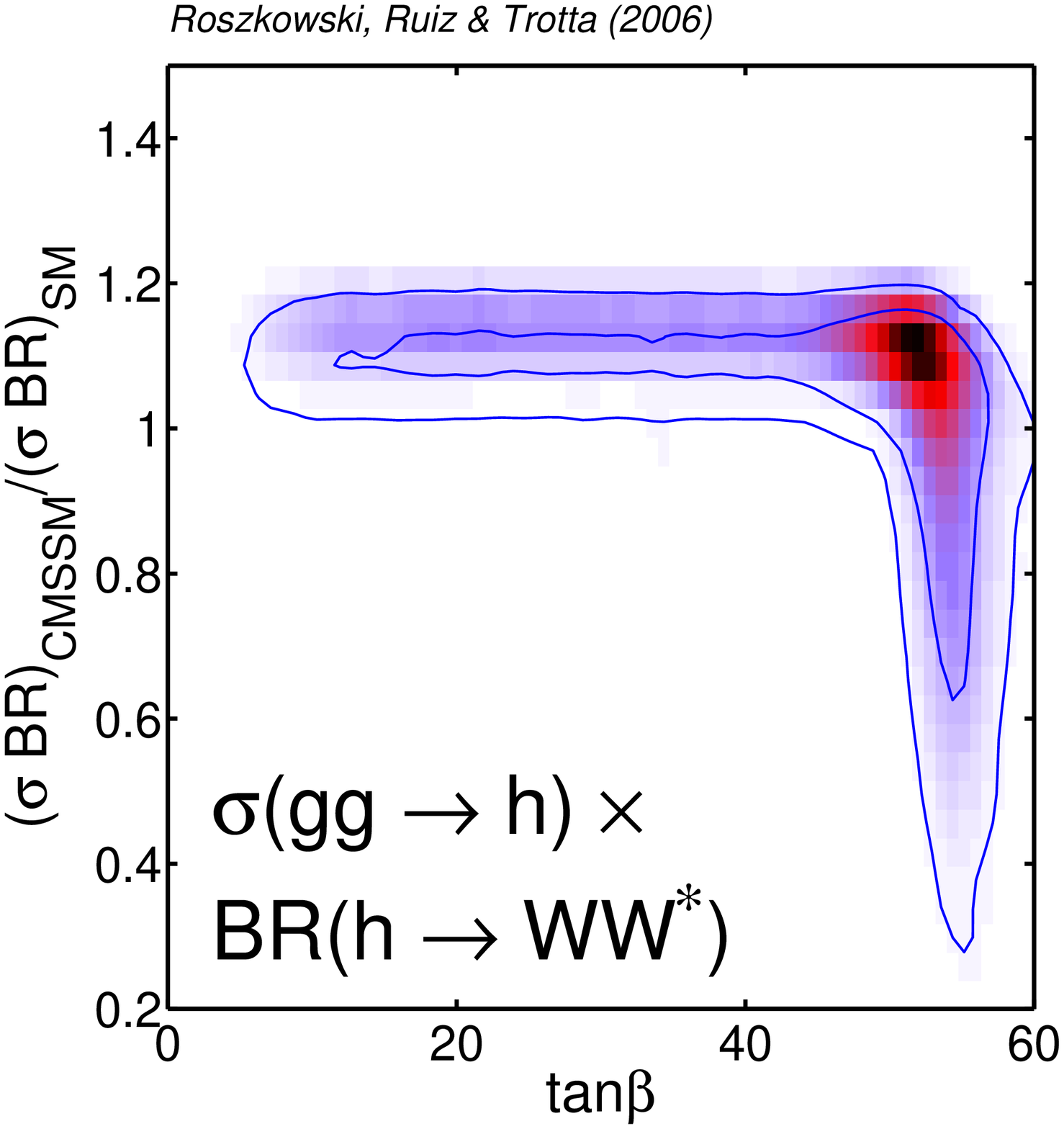}\\
	\includegraphics[width=0.3\textwidth]{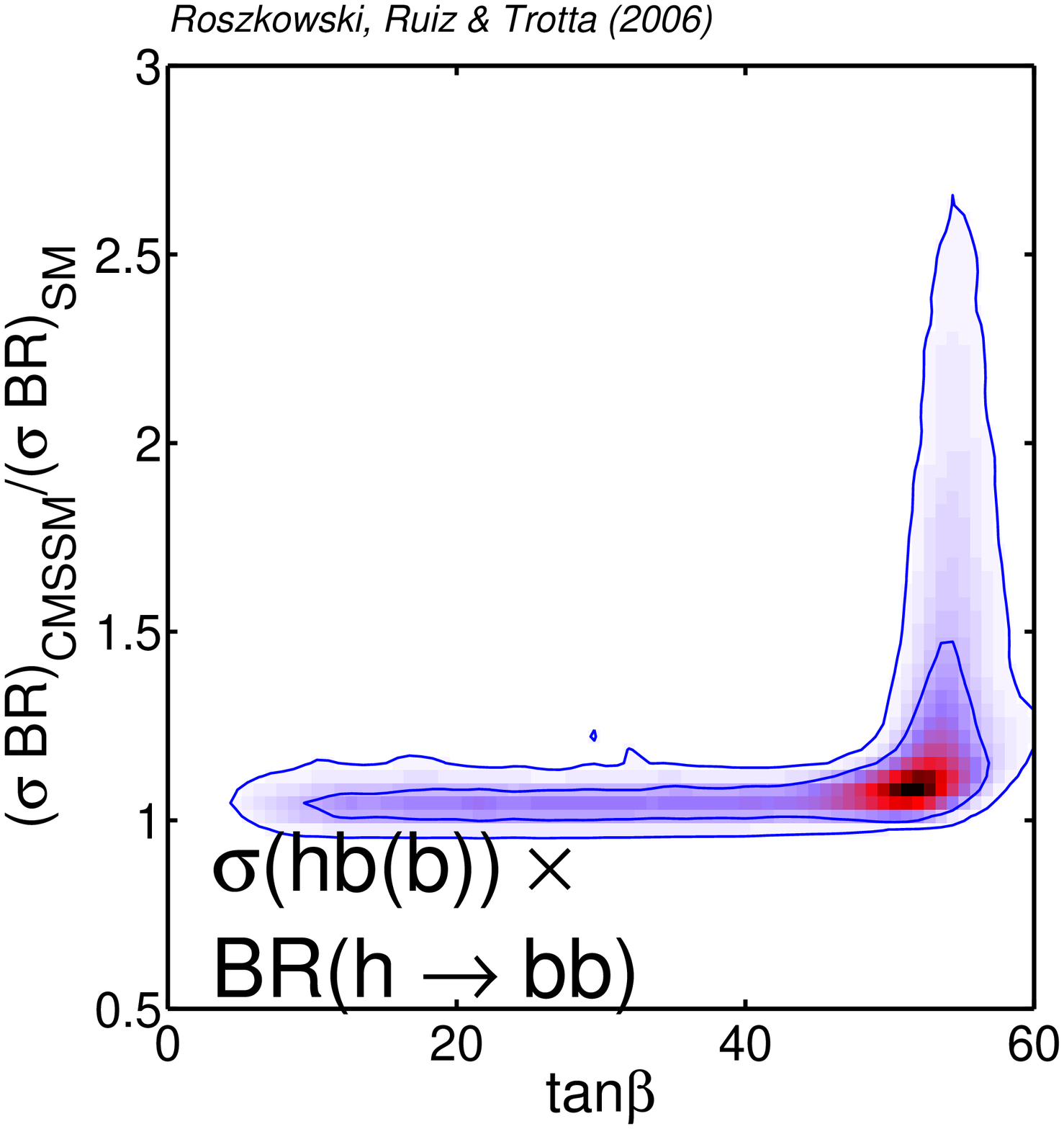}
 &	\includegraphics[width=0.3\textwidth]{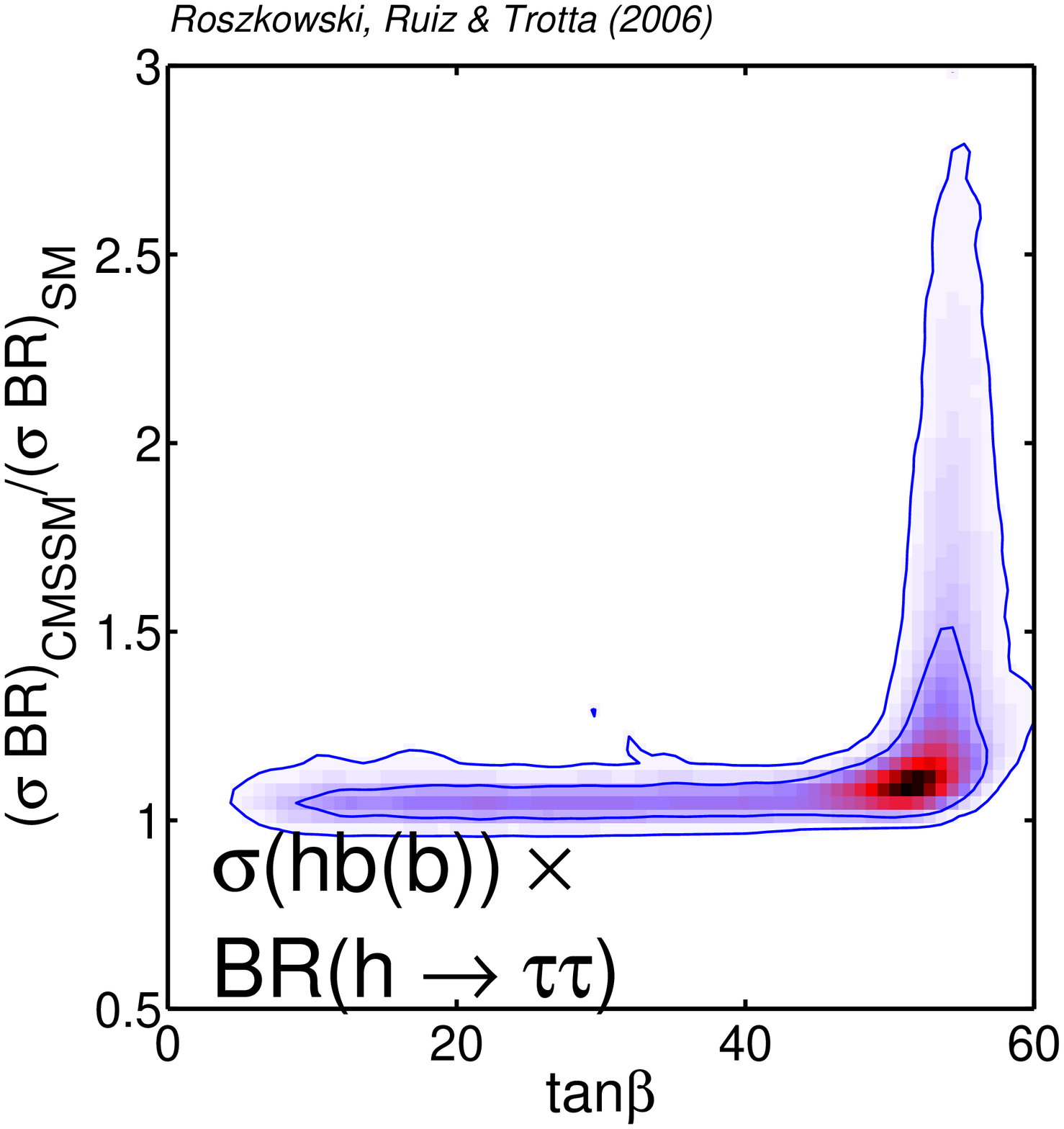}
 &	\includegraphics[width=0.3\textwidth]{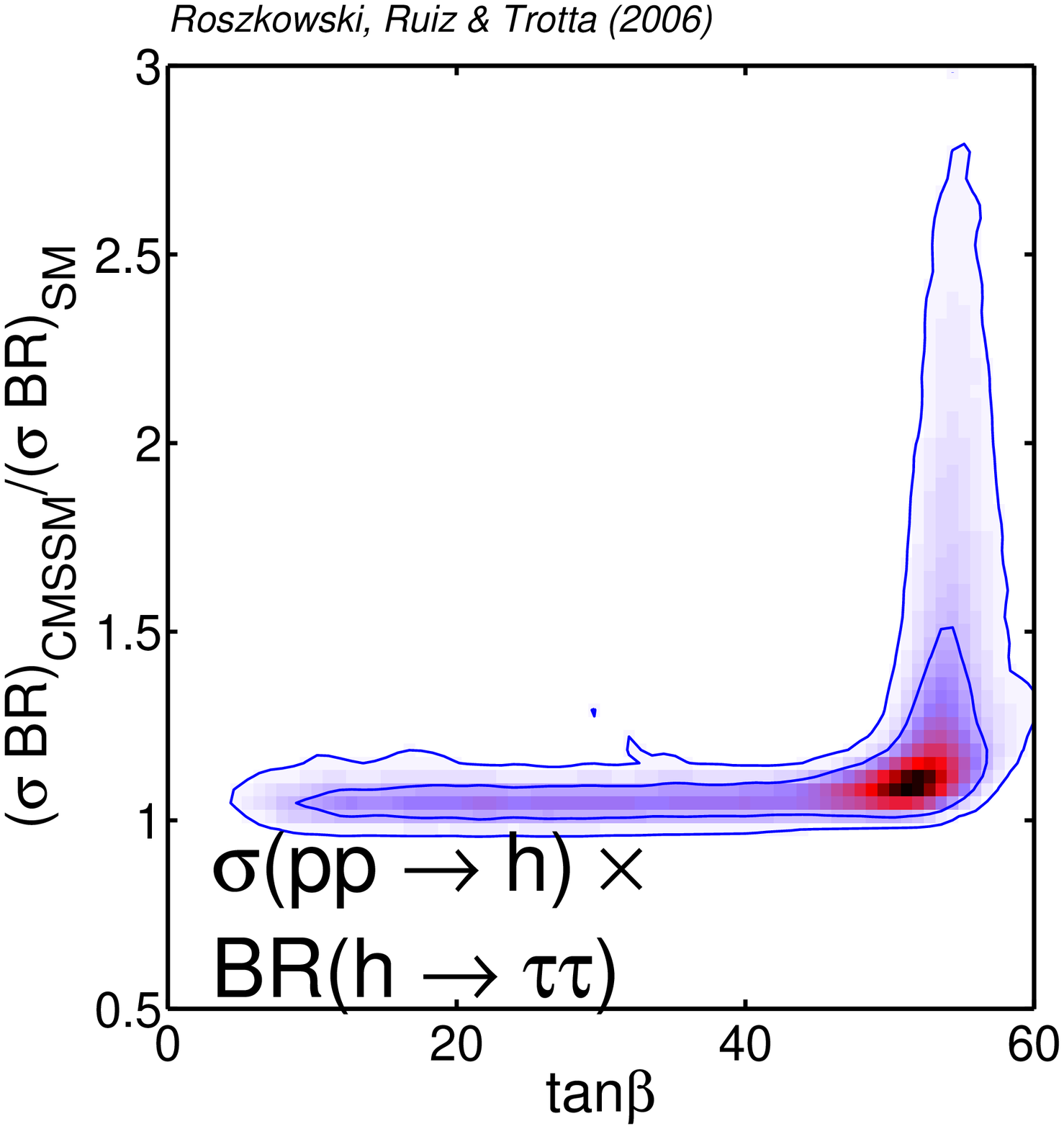}
\end{tabular}
\end{center}
\caption{The 2--dim relative probability density for light Higgs
  production cross sections times decay branching ratios at the
  Tevatron (normalized to the SM case) as a function of $\tanb$.
\label{fig:rrt2-hlsigmabrvstanb}
}
\end{figure}
\begin{figure}[!tb]
\begin{center}
\begin{tabular}{c c c}
	\includegraphics[width=0.3\textwidth]{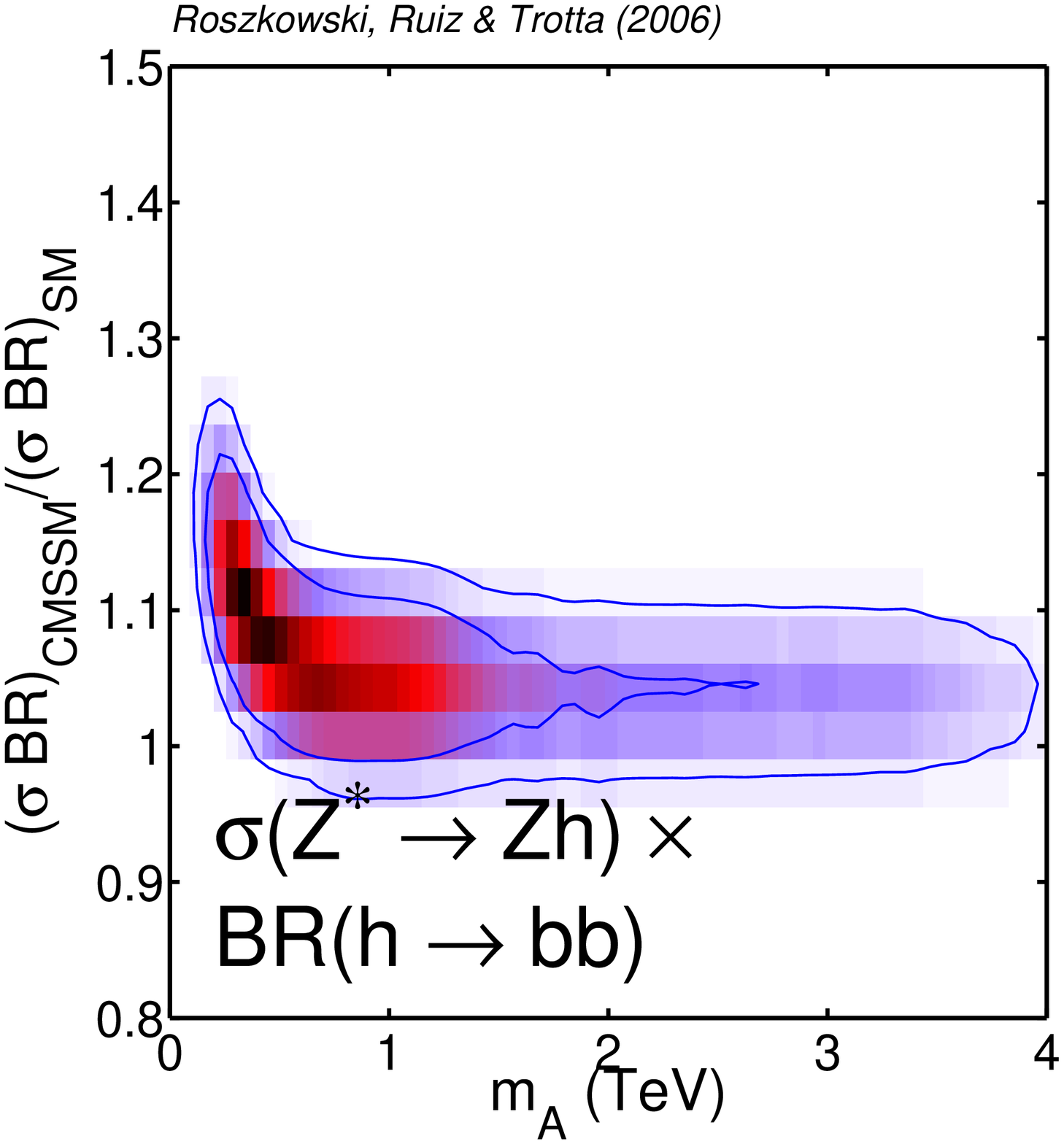}
 &	\includegraphics[width=0.3\textwidth]{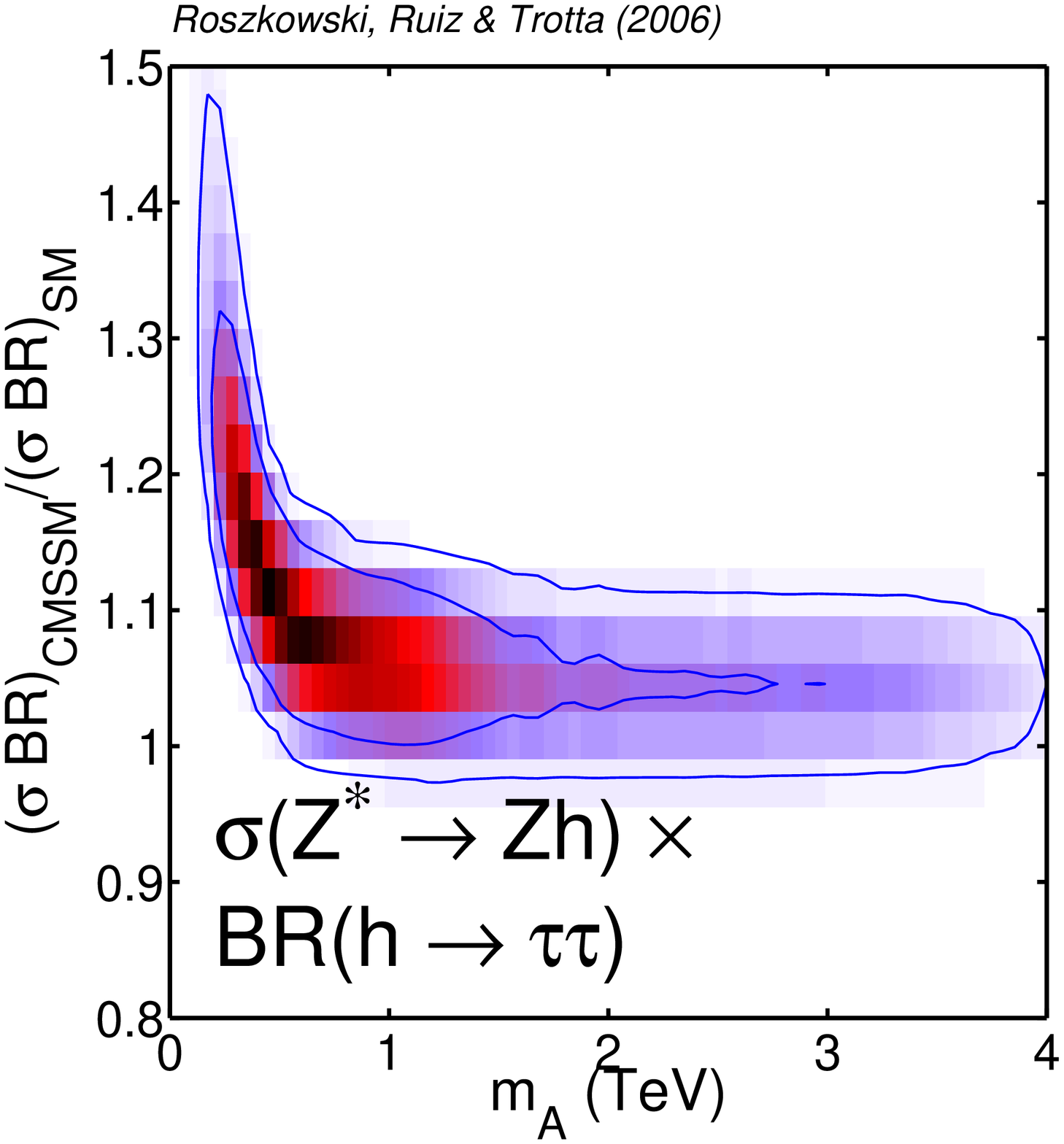}
 &	\includegraphics[width=0.3\textwidth]{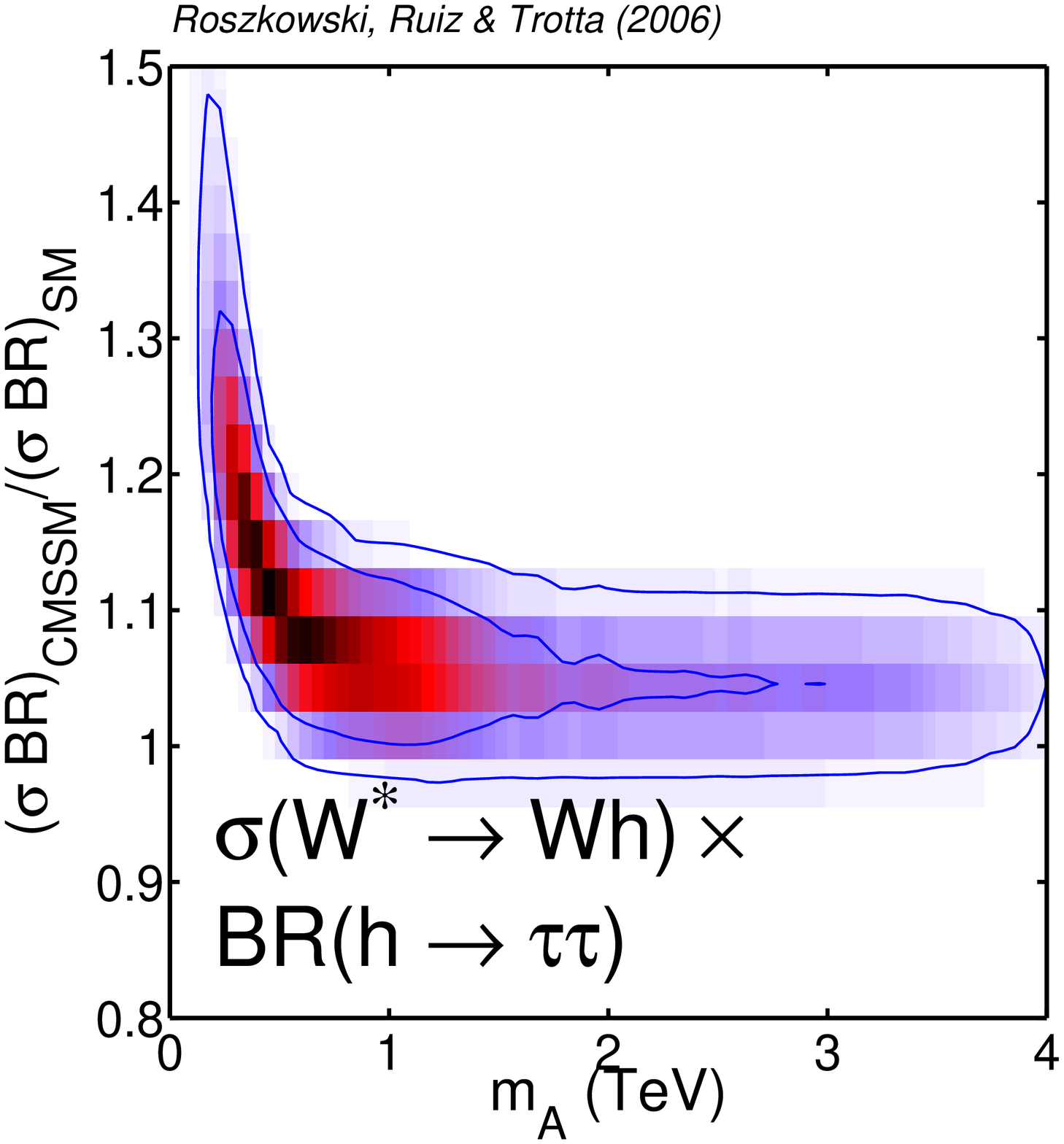}\\
	\includegraphics[width=0.3\textwidth]{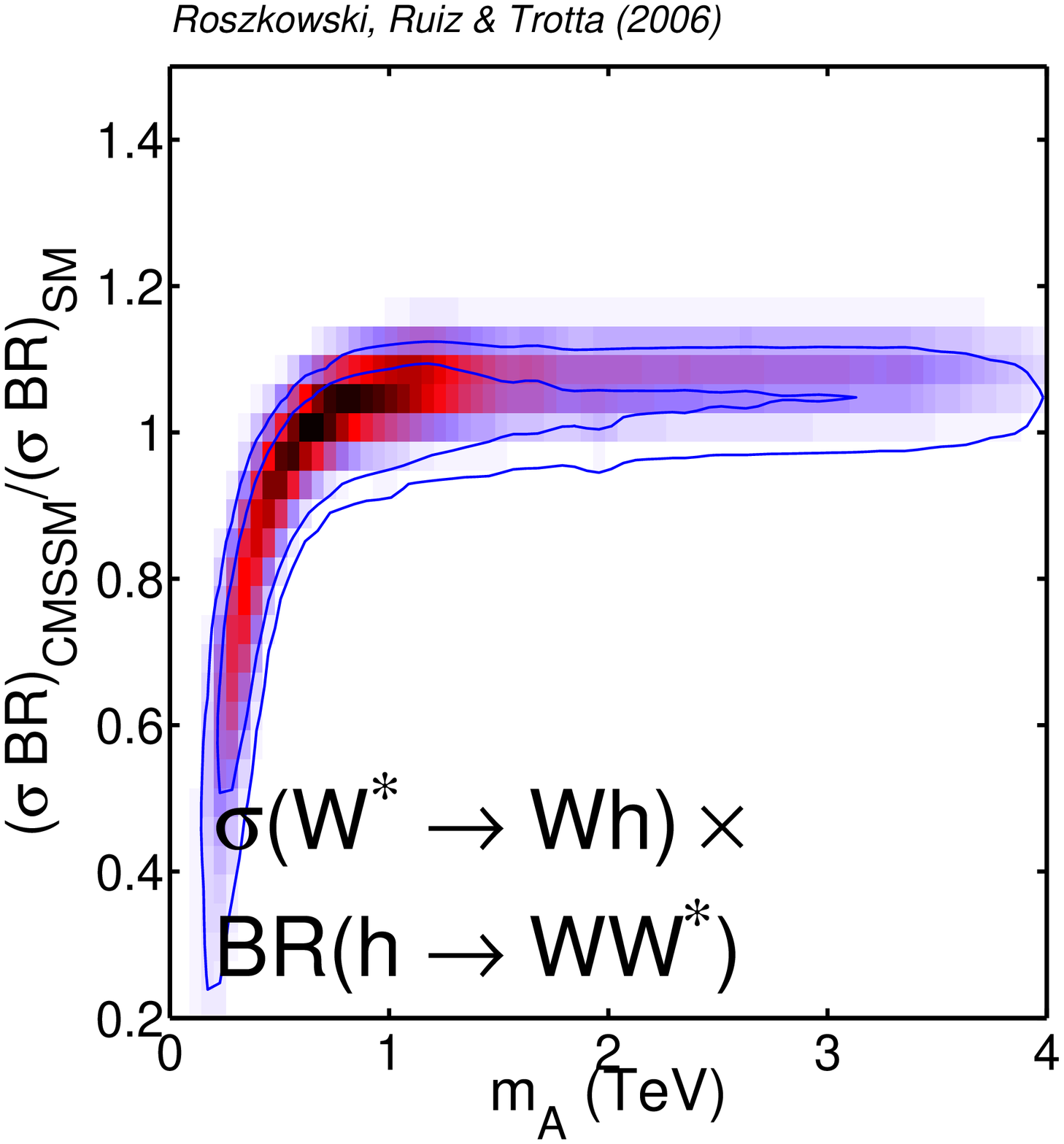}
 &	\includegraphics[width=0.3\textwidth]{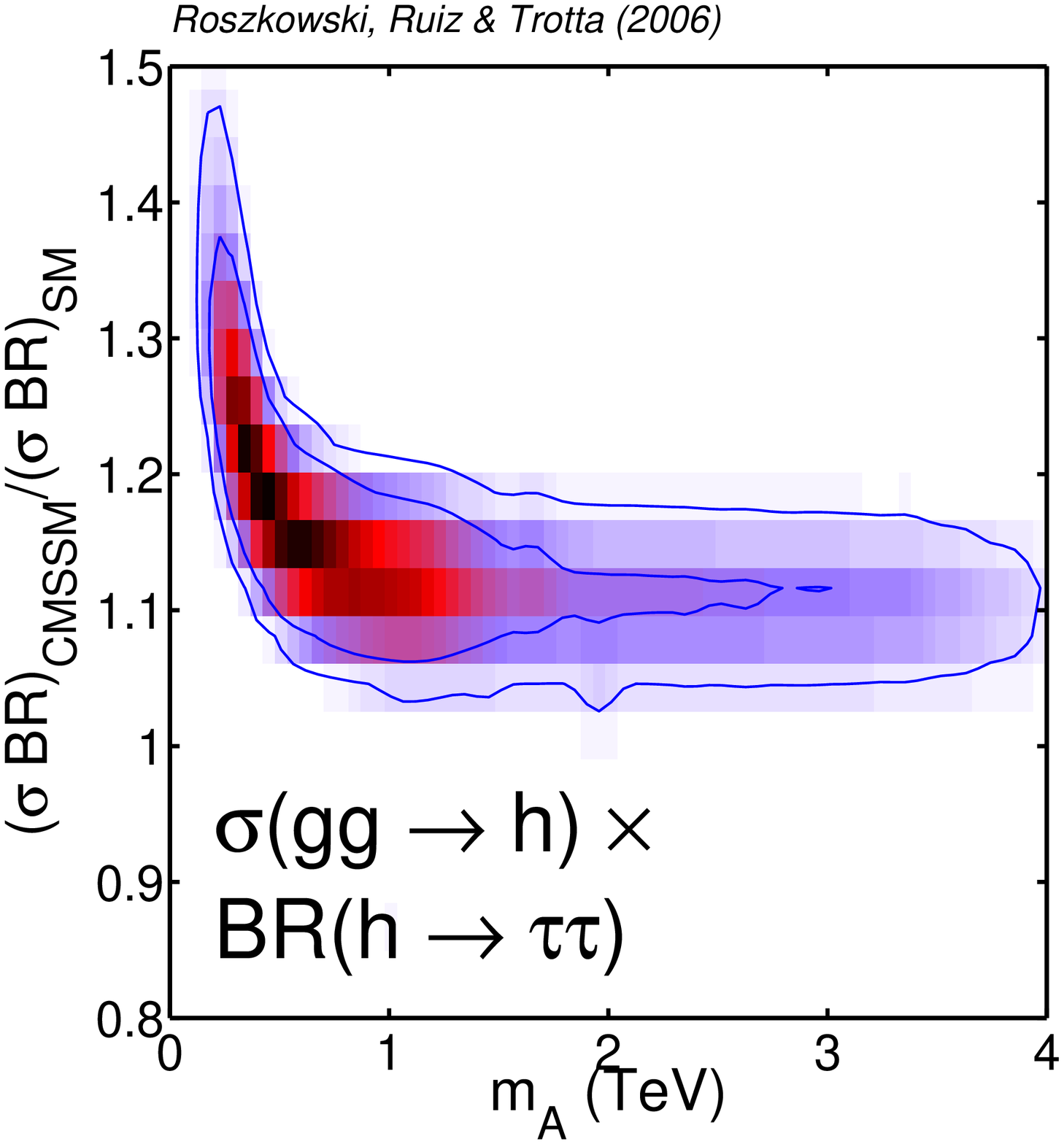}
 &	\includegraphics[width=0.3\textwidth]{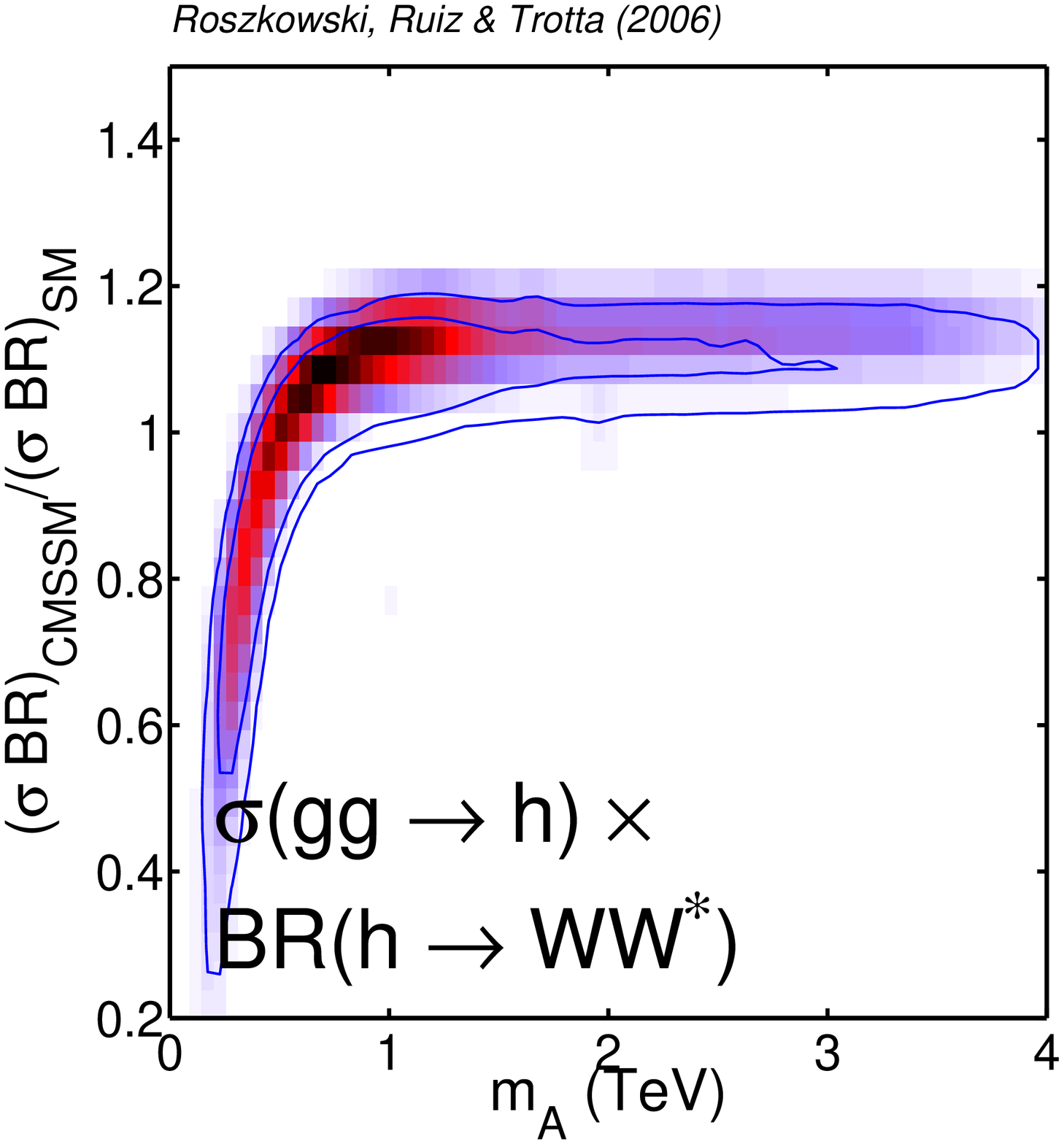}\\
	\includegraphics[width=0.3\textwidth]{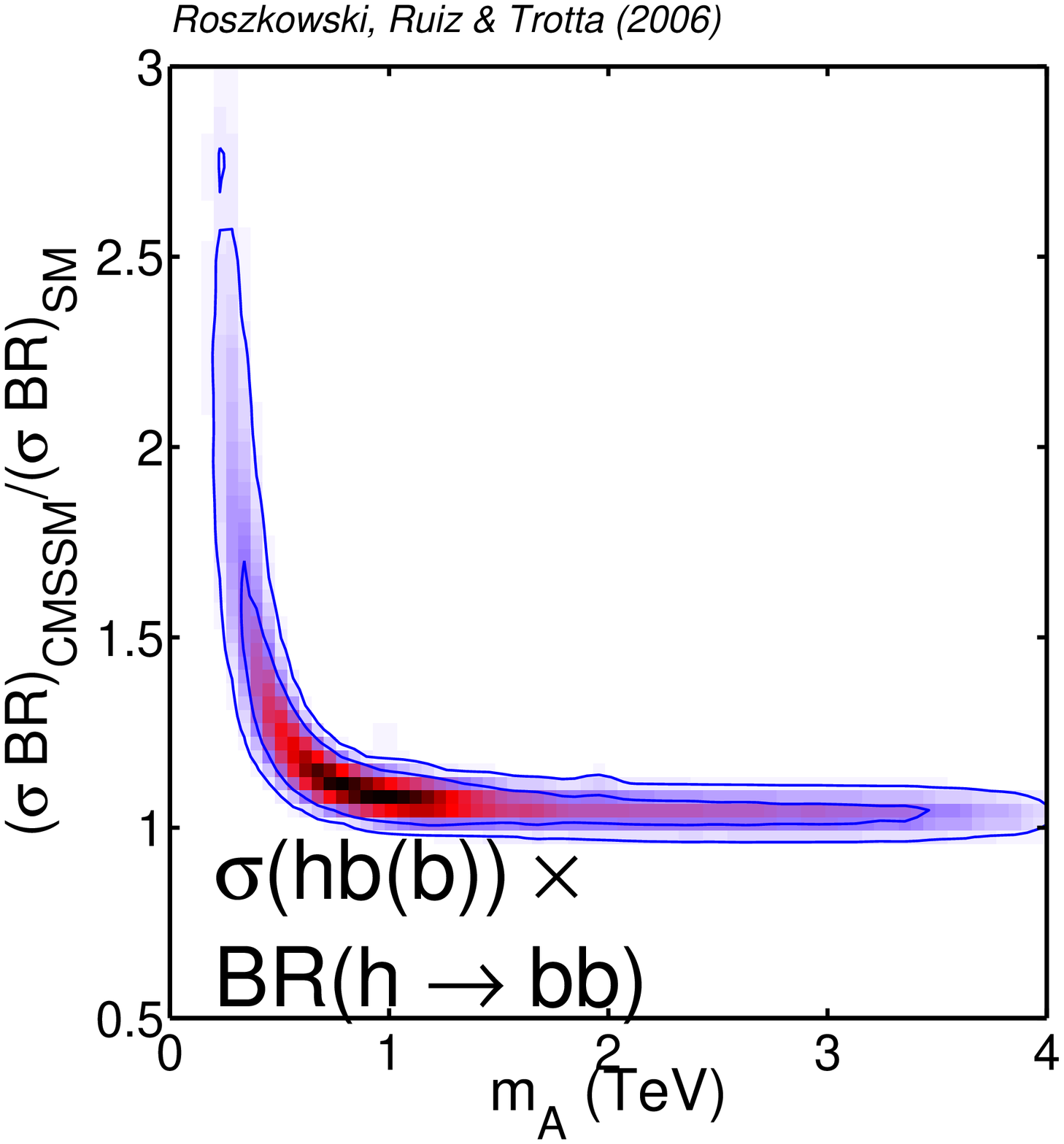}
 &	\includegraphics[width=0.3\textwidth]{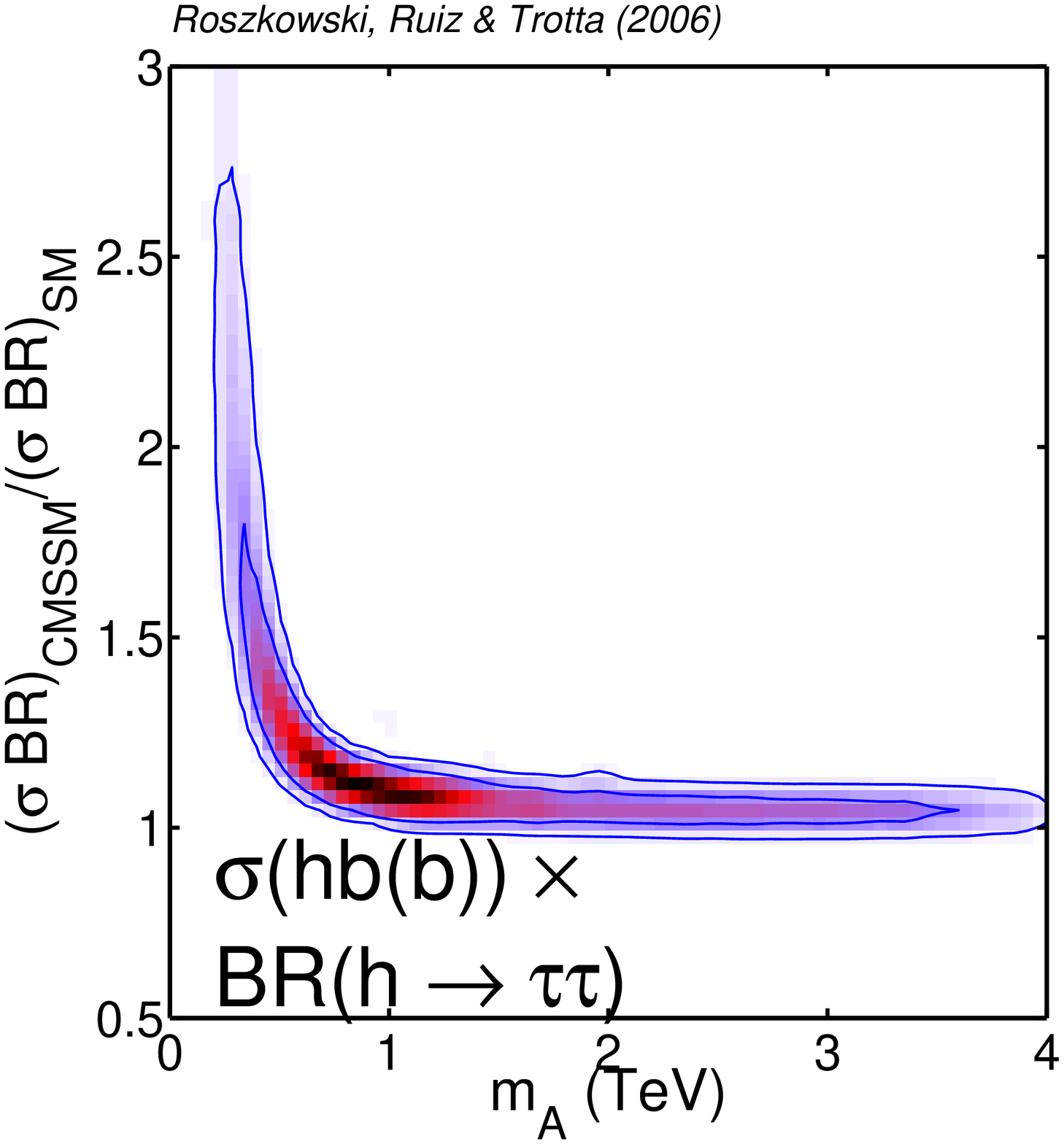}
 &	\includegraphics[width=0.3\textwidth]{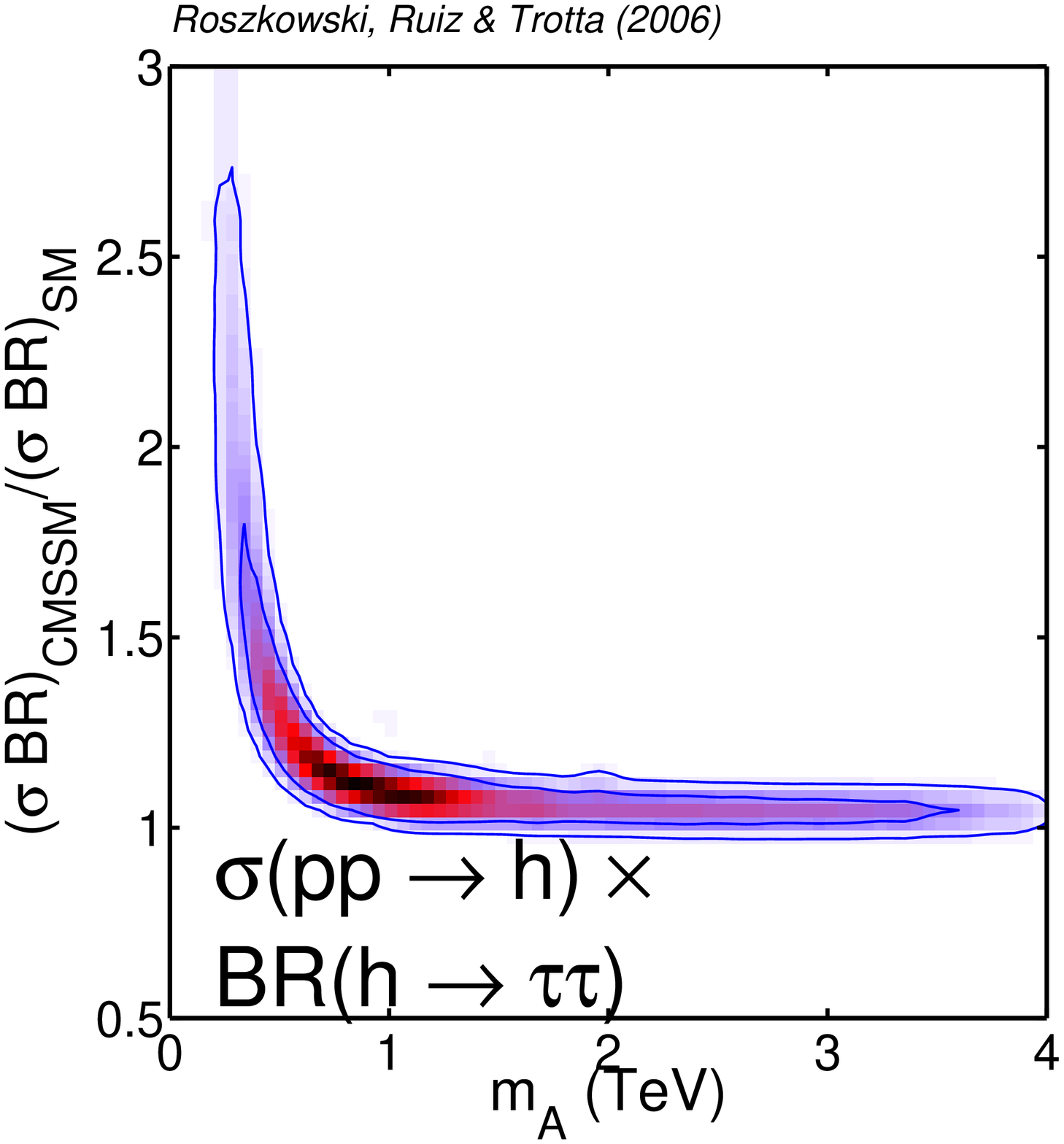}
\end{tabular}
\end{center}
\caption{The 2--dim relative probability density for light Higgs
  production cross sections times decay branching ratios at the
  Tevatron (normalized to the SM case) as a function of $\mha$.
\label{fig:rrt2-hlsigmabrvsmha}
}
\end{figure}
\begin{figure}[!tb]
\begin{center}
\begin{tabular}{c c c}
	\includegraphics[width=0.3\textwidth]{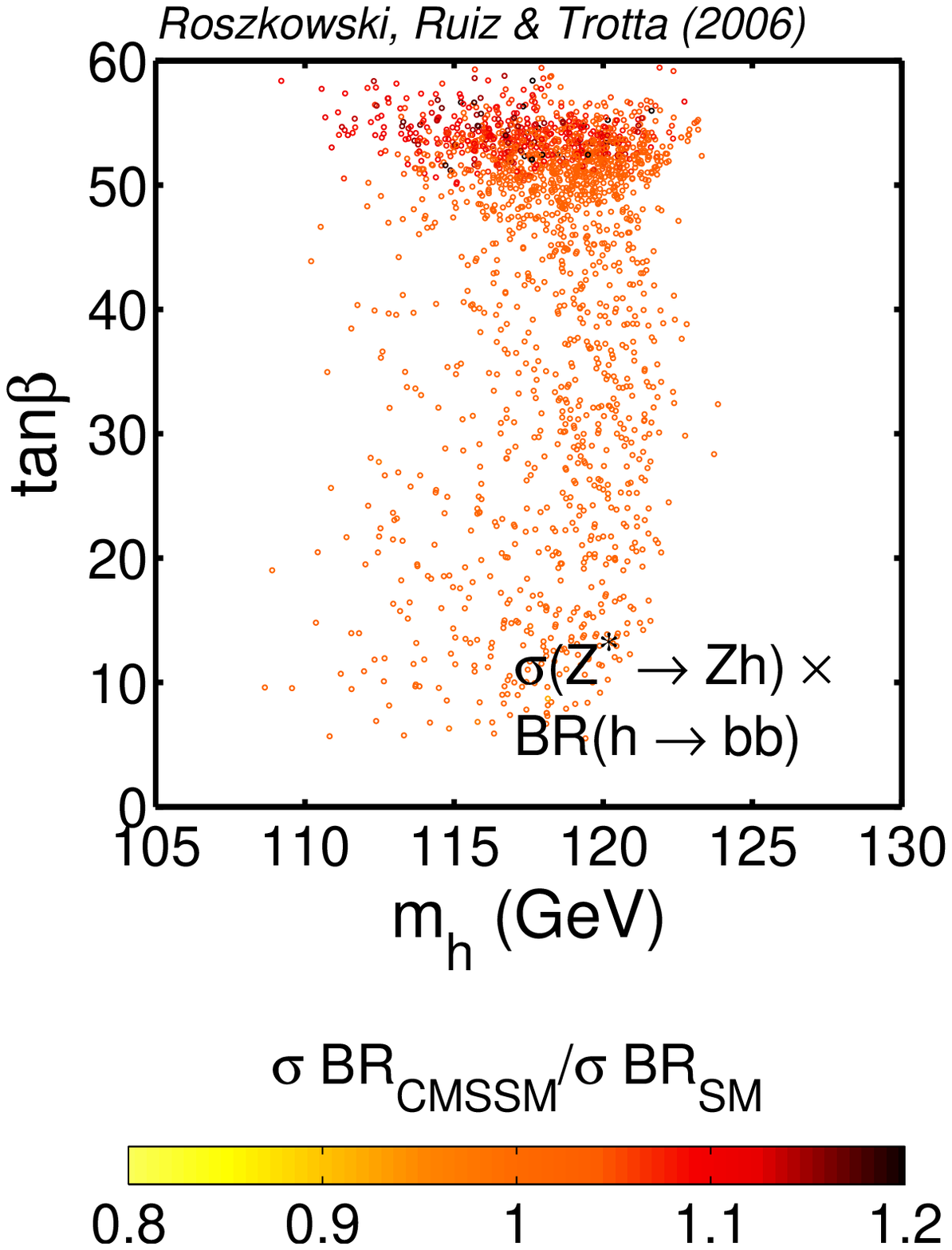}
&	\includegraphics[width=0.3\textwidth]{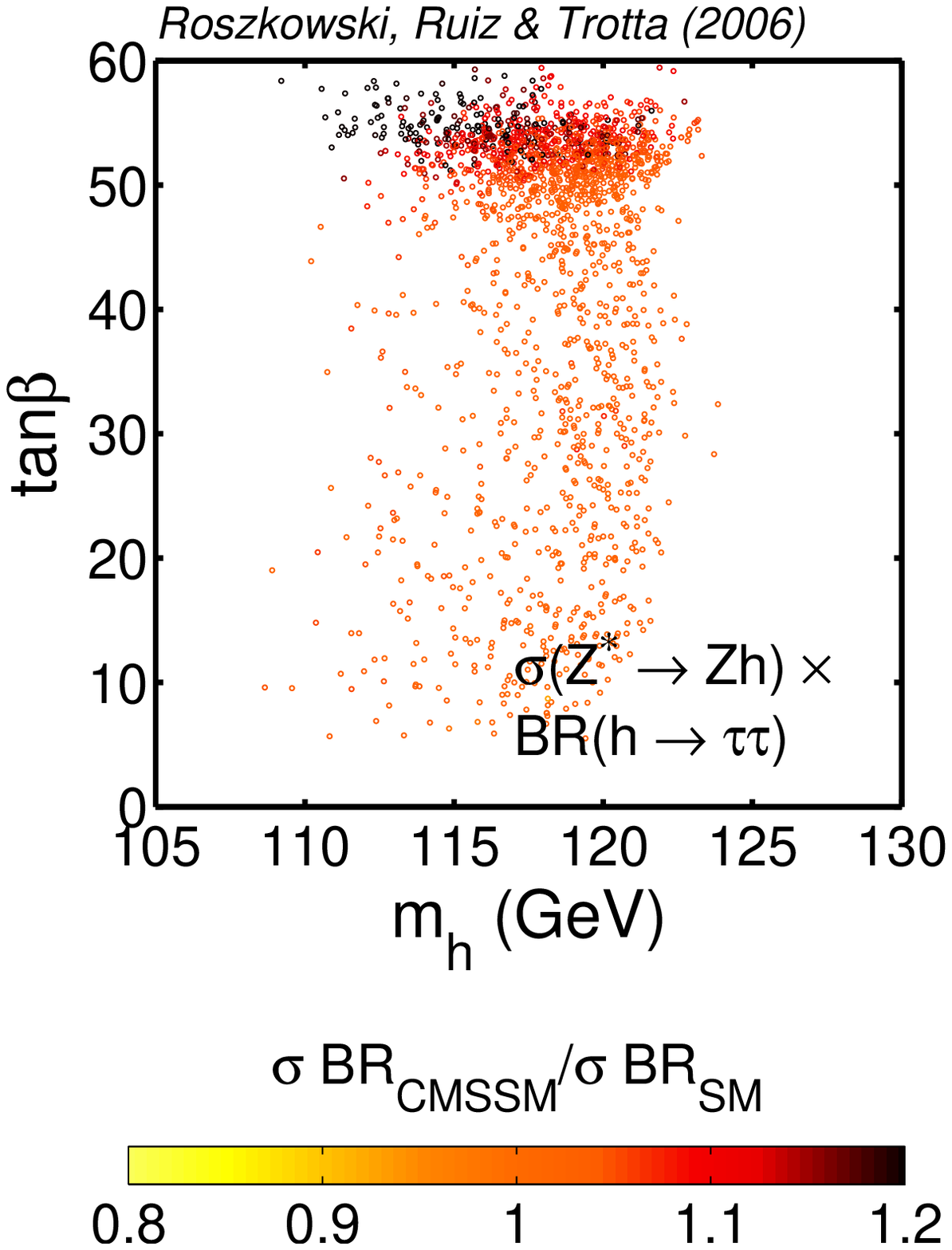}
&	\includegraphics[width=0.3\textwidth]{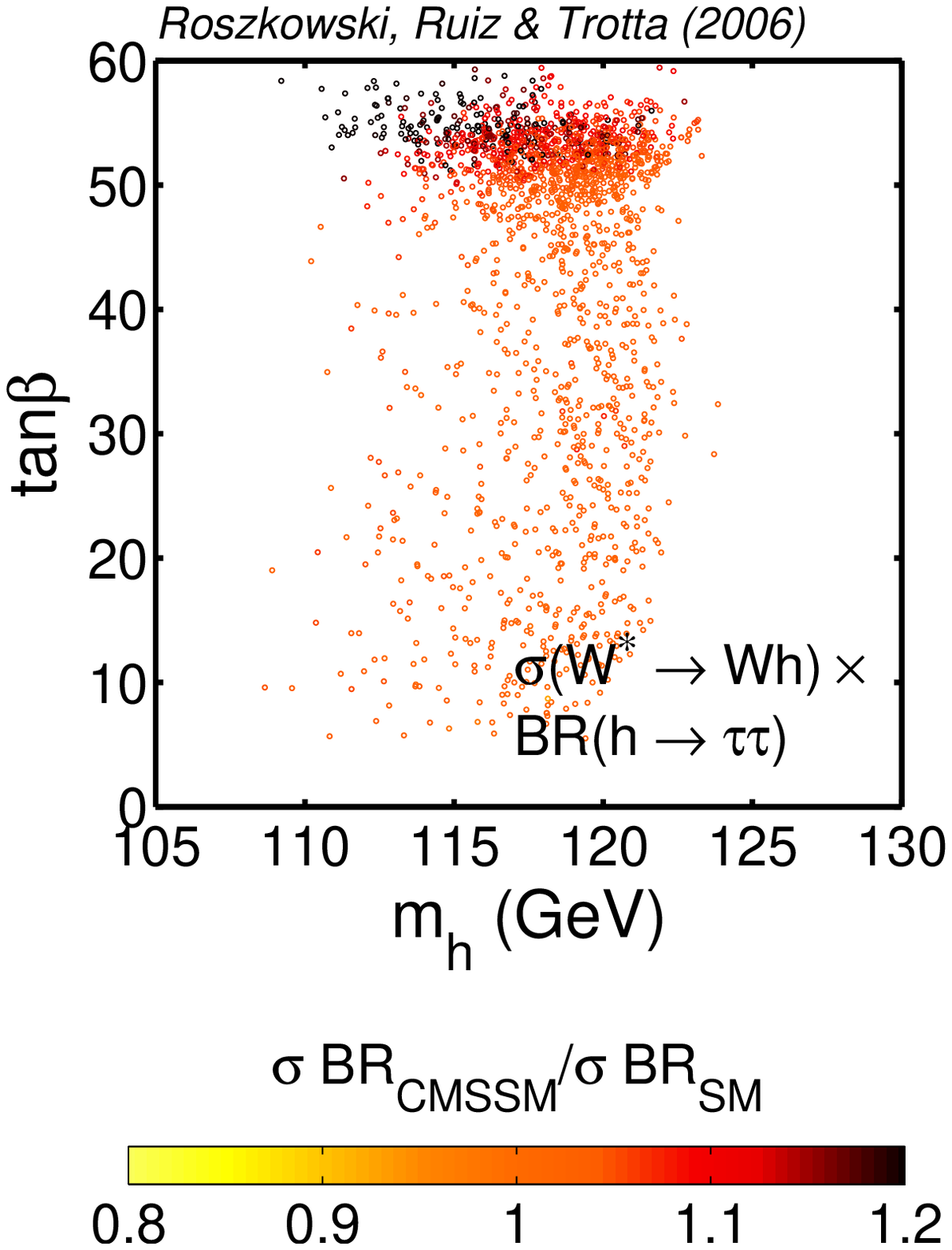}\\
	\includegraphics[width=0.3\textwidth]{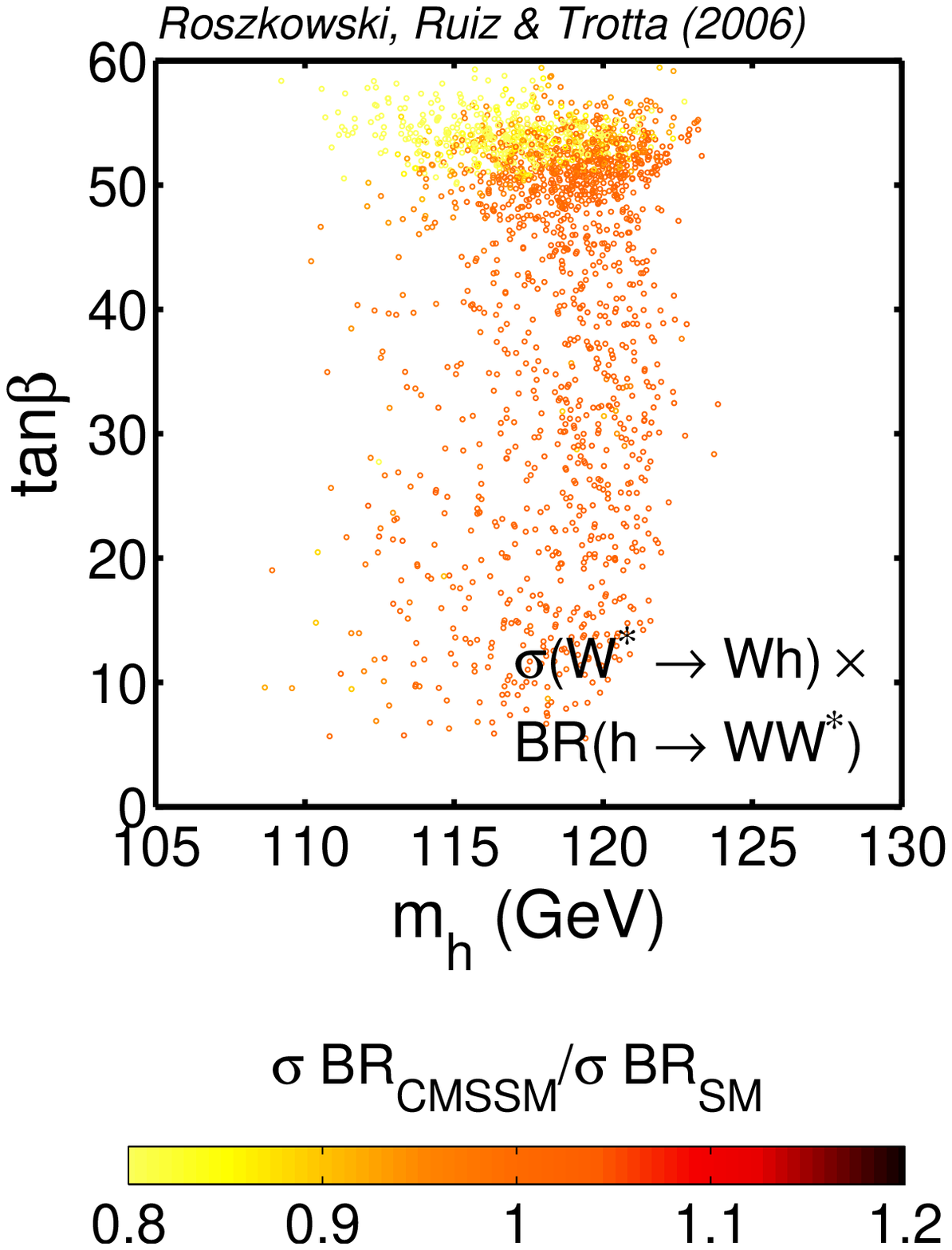}
&	\includegraphics[width=0.3\textwidth]{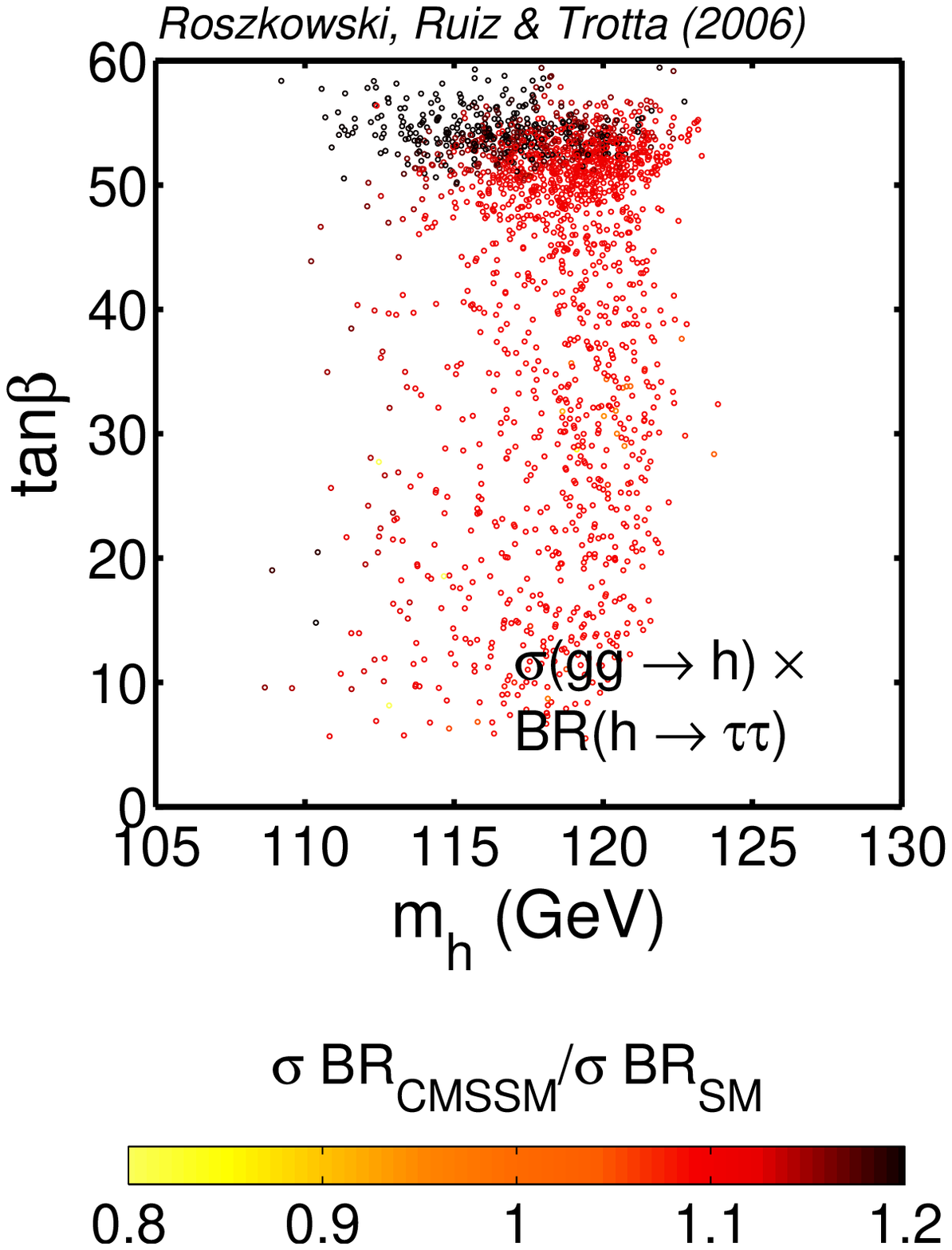}
&	\includegraphics[width=0.3\textwidth]{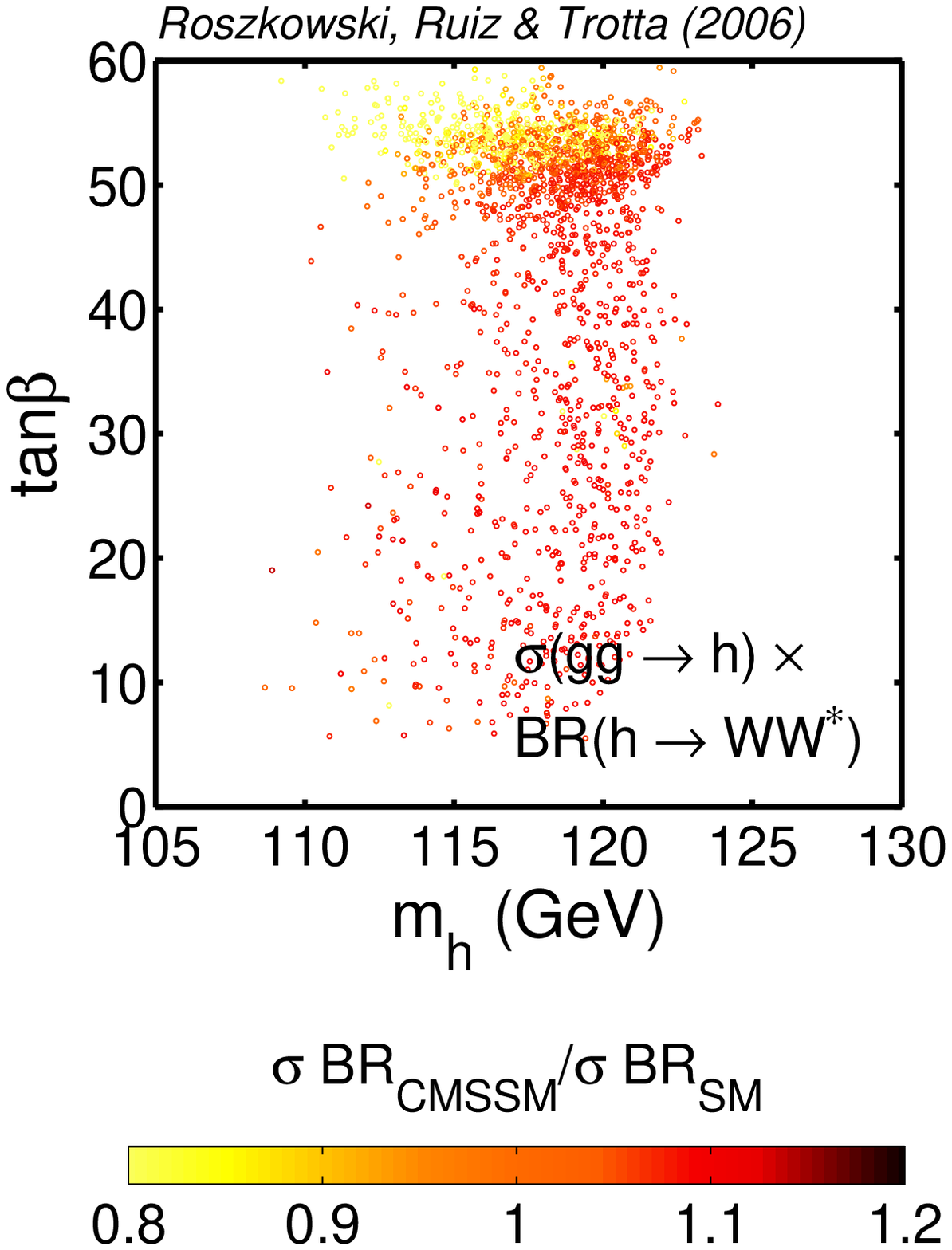}\\
	\includegraphics[width=0.3\textwidth]{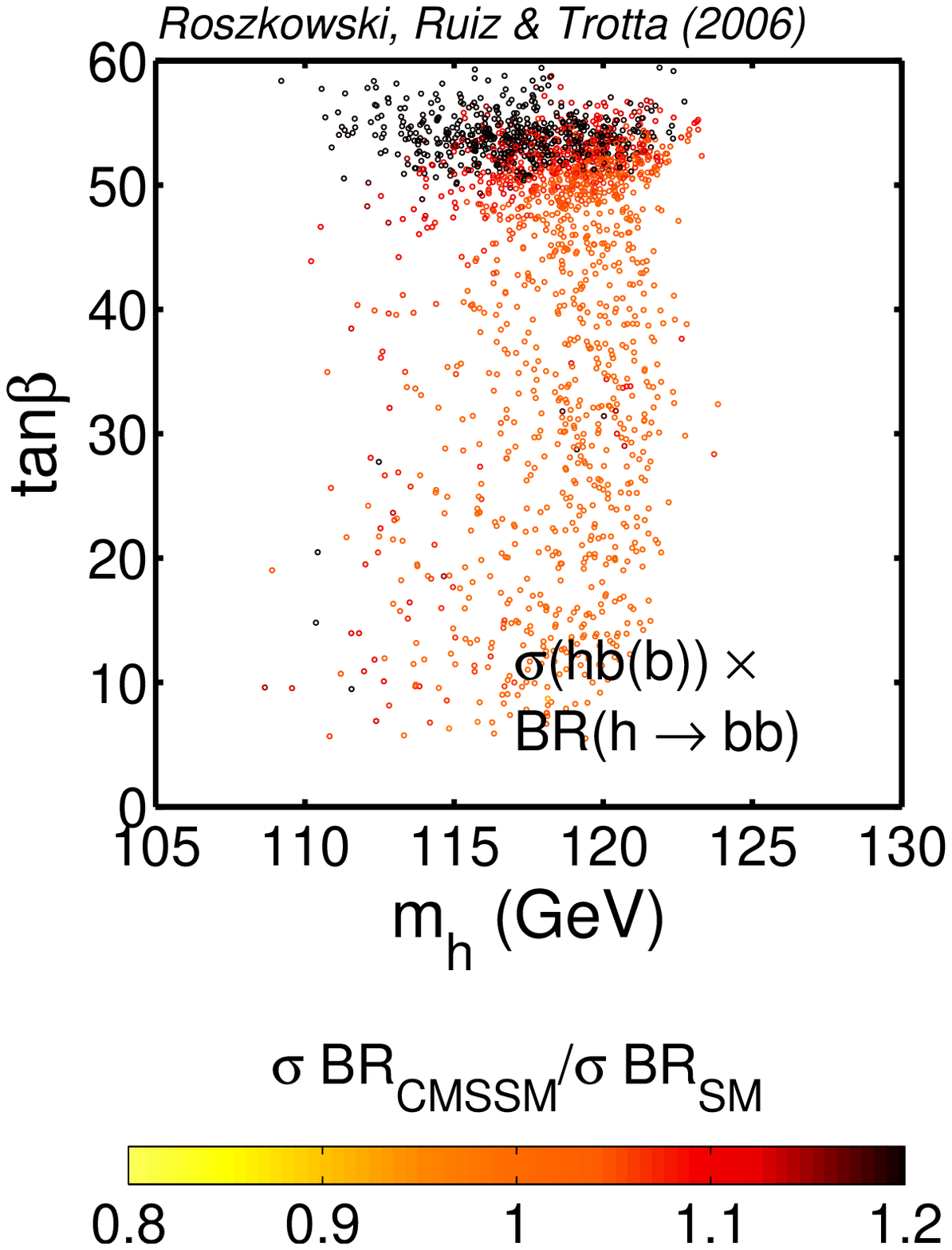}
&	\includegraphics[width=0.3\textwidth]{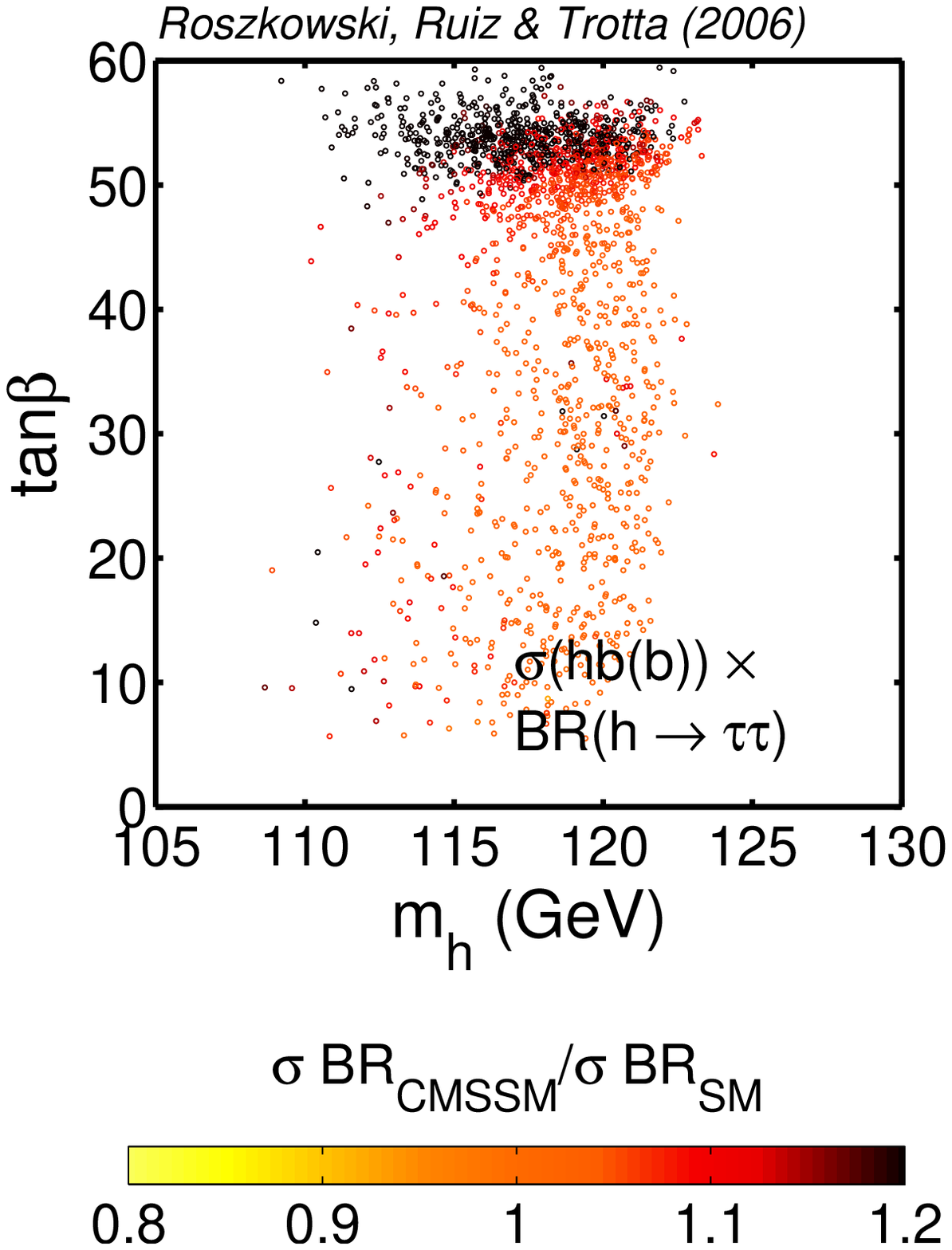}
&	\includegraphics[width=0.3\textwidth]{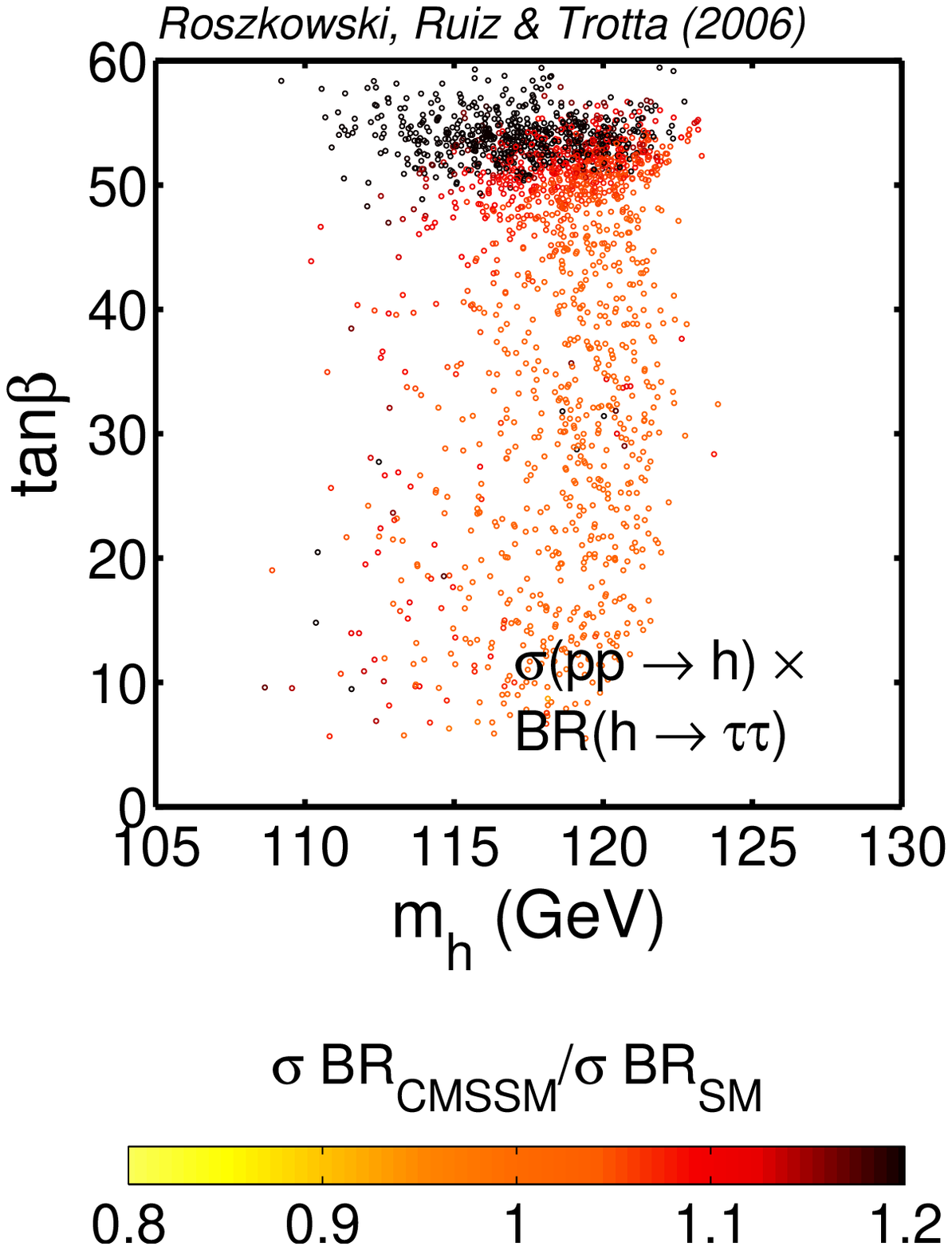}
\end{tabular}
\end{center}
\caption{Values in the $(\mhl,\tanb)$
plane of production cross section times the branching ratio,
normalized to their SM values, for the main light CMSSM Higgs
production and decay channels at the Tevatron in the $(\mhl,\tanb)$
plane.
\label{rrt2:3d-mhlsigmabr-mhlvstanb}
}
\end{figure}
\begin{figure}[!tb]
\begin{center}
\begin{tabular}{c c c}
	\includegraphics[width=0.3\textwidth]{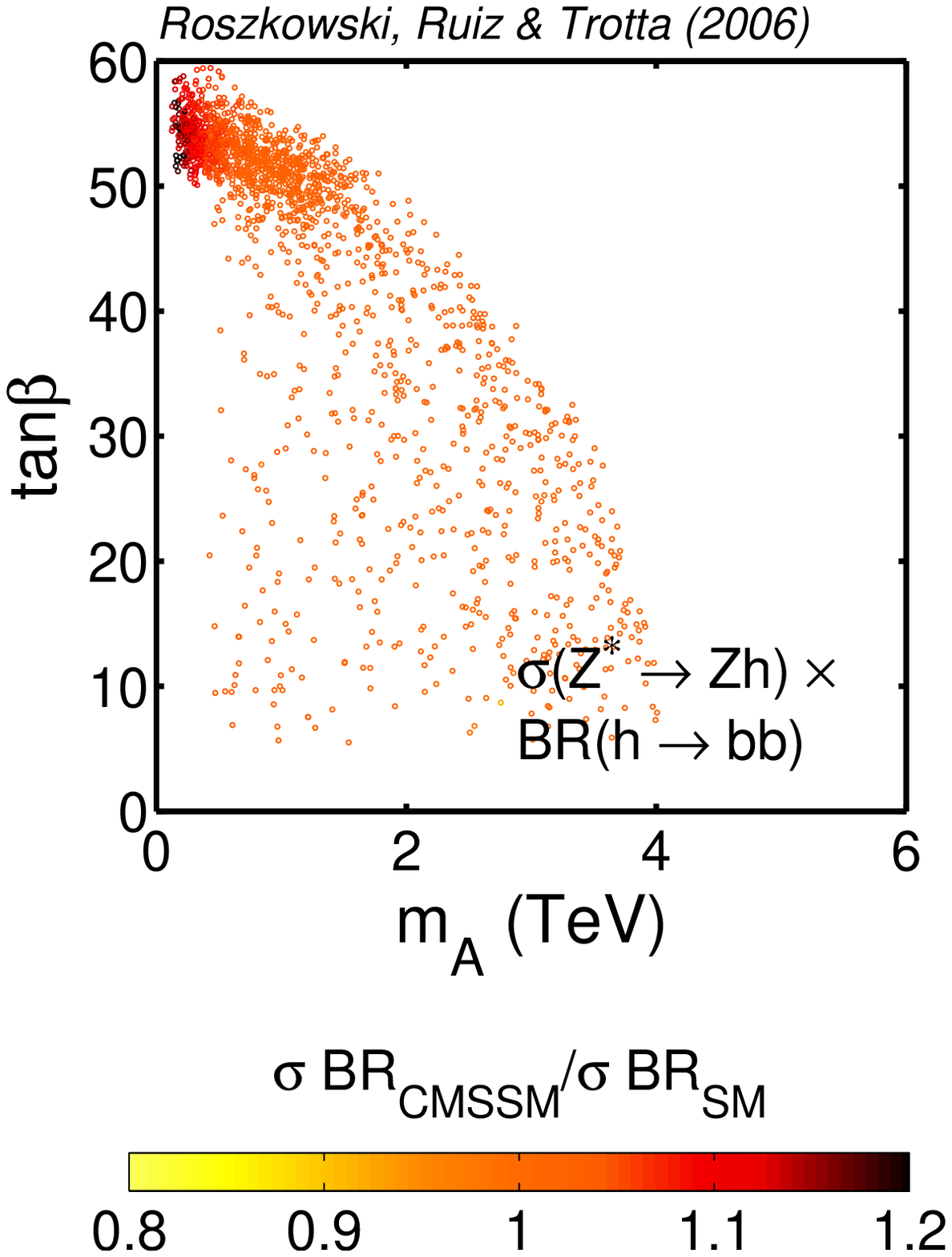}
&	\includegraphics[width=0.3\textwidth]{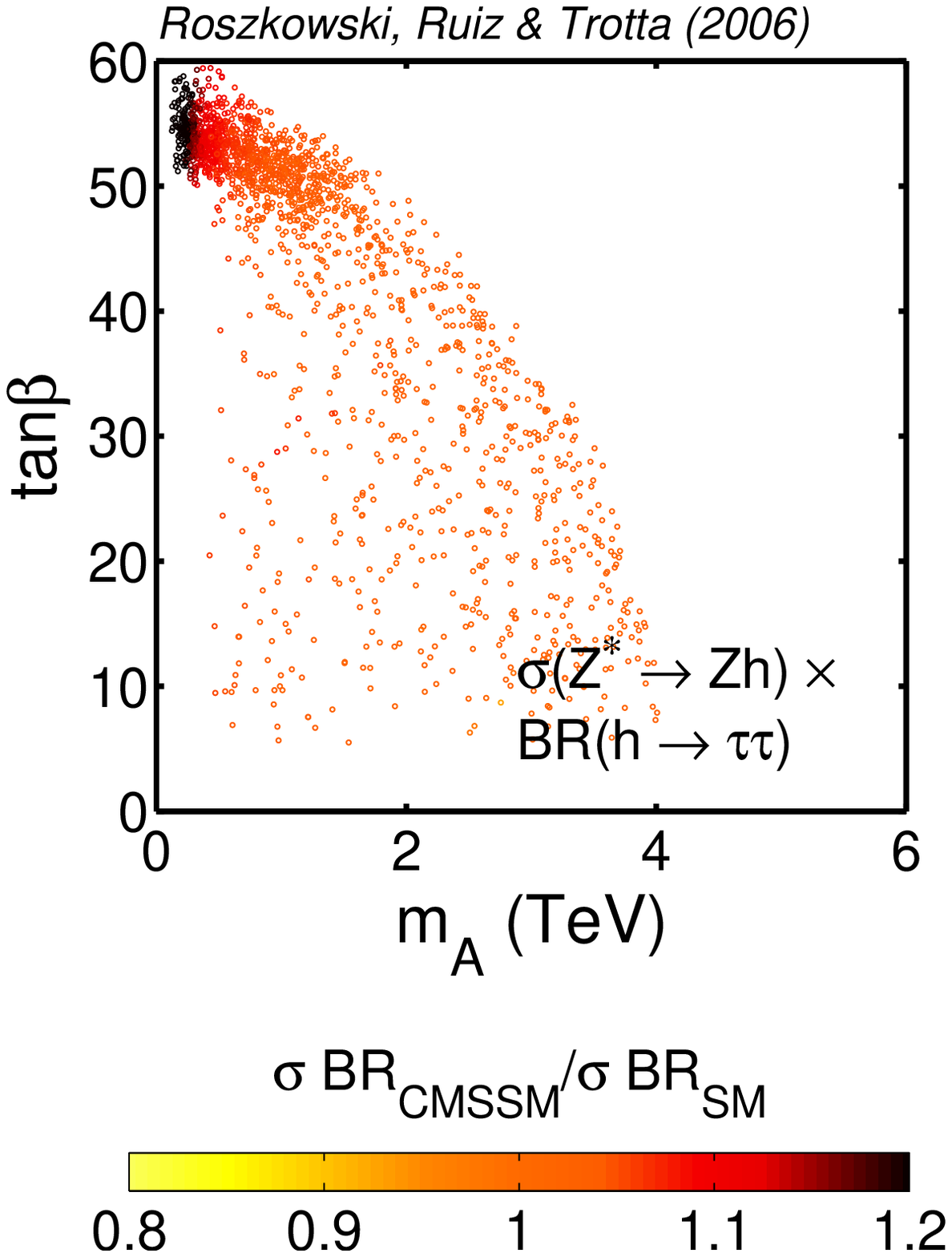}
&	\includegraphics[width=0.3\textwidth]{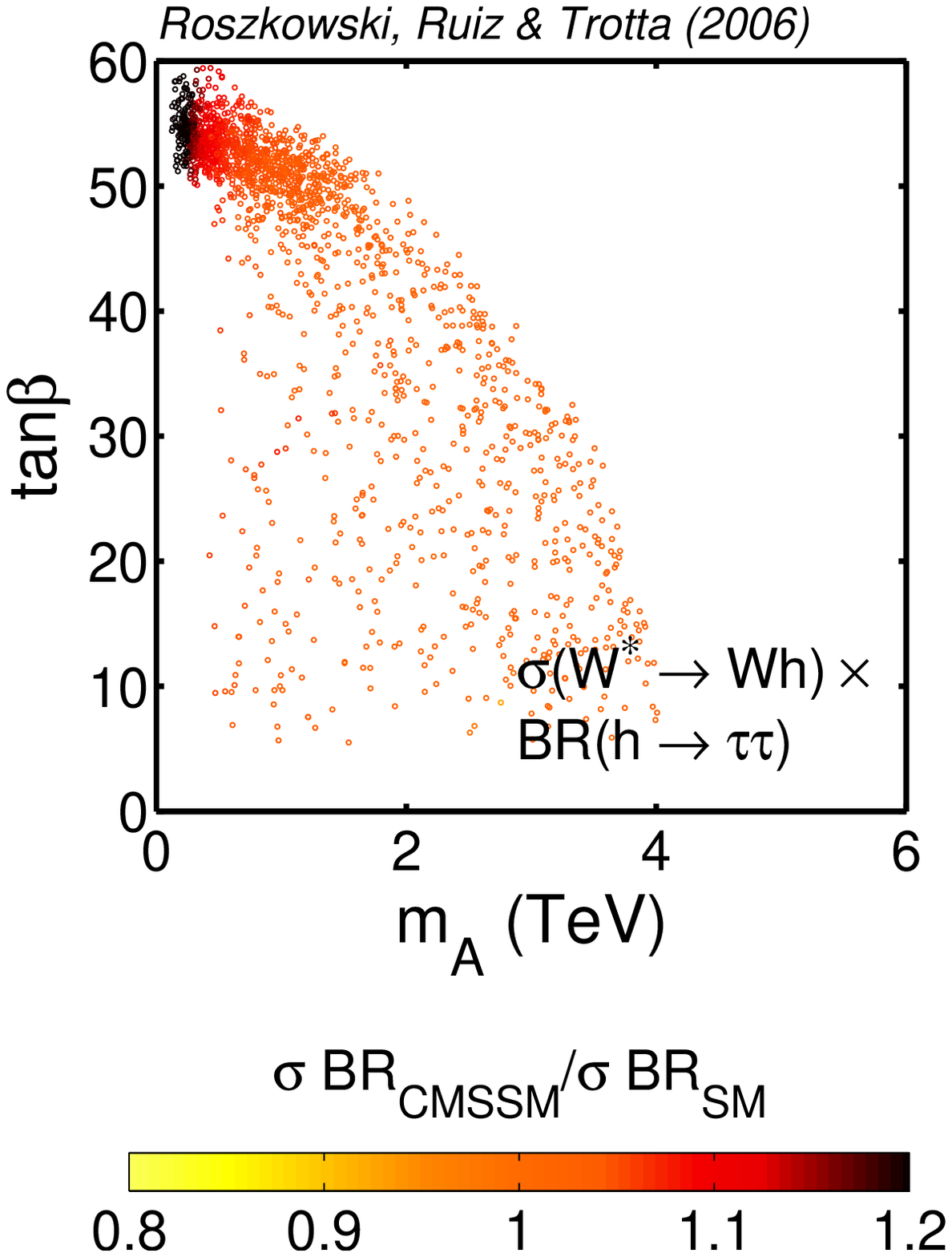}\\
	\includegraphics[width=0.3\textwidth]{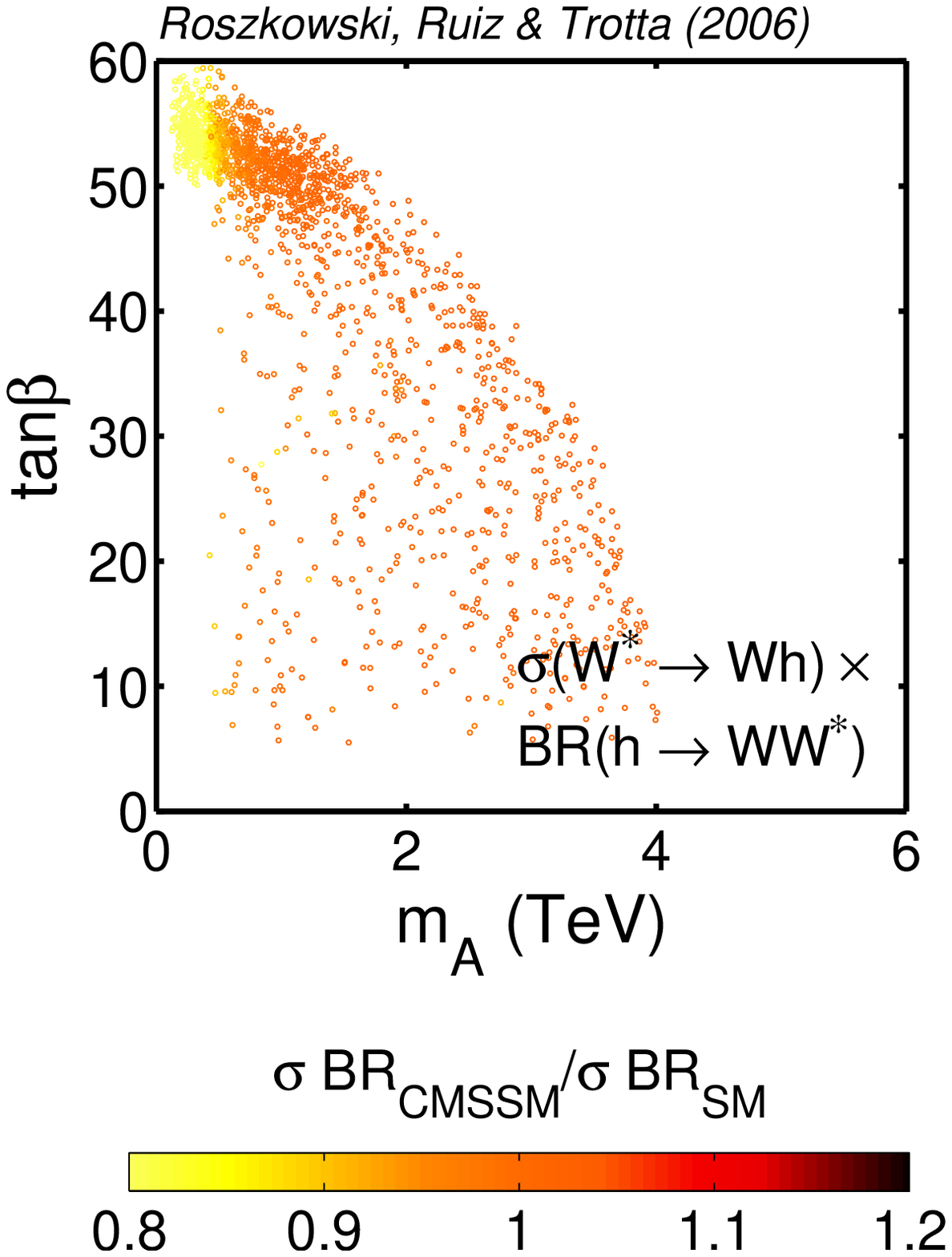}
&	\includegraphics[width=0.3\textwidth]{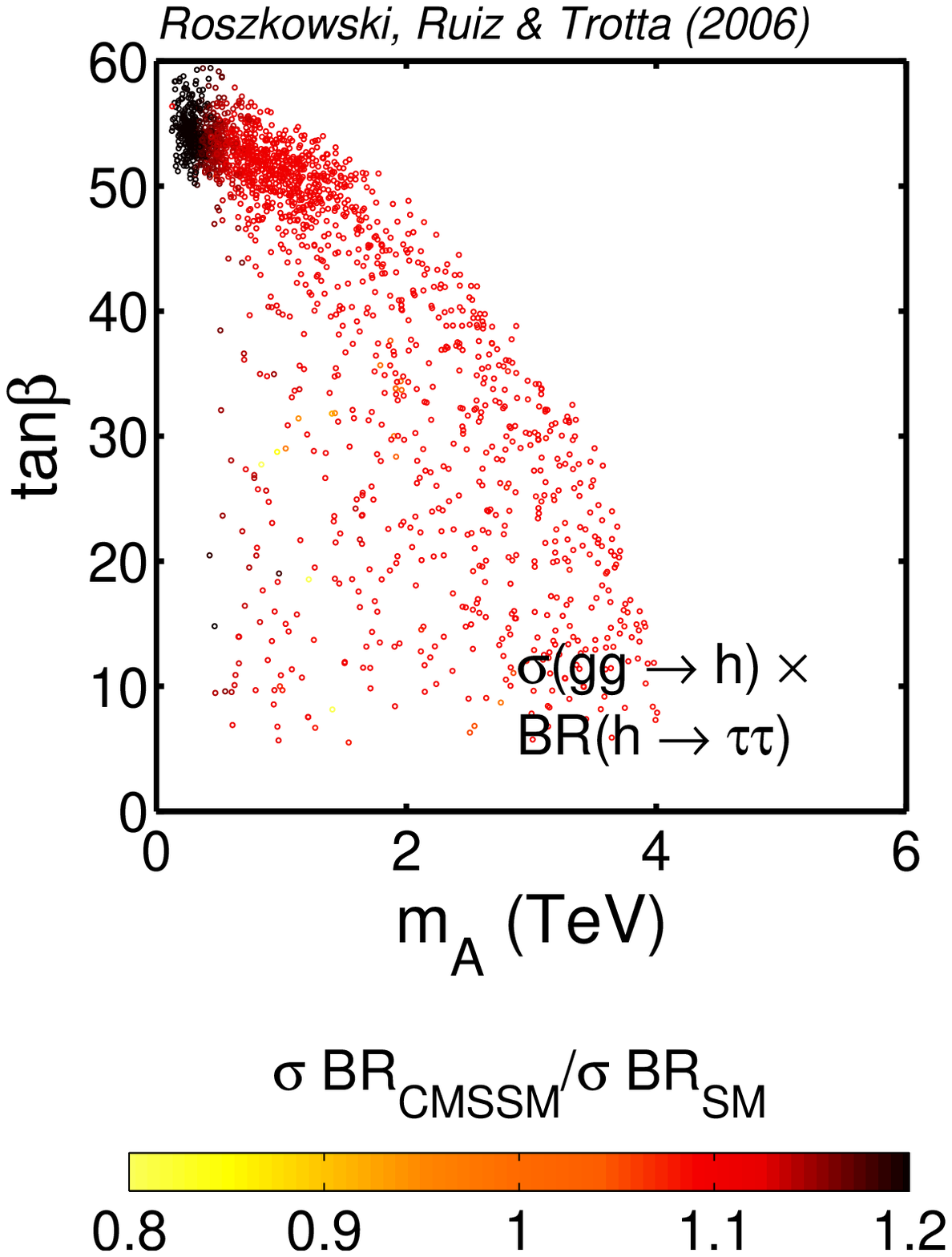}
&	\includegraphics[width=0.3\textwidth]{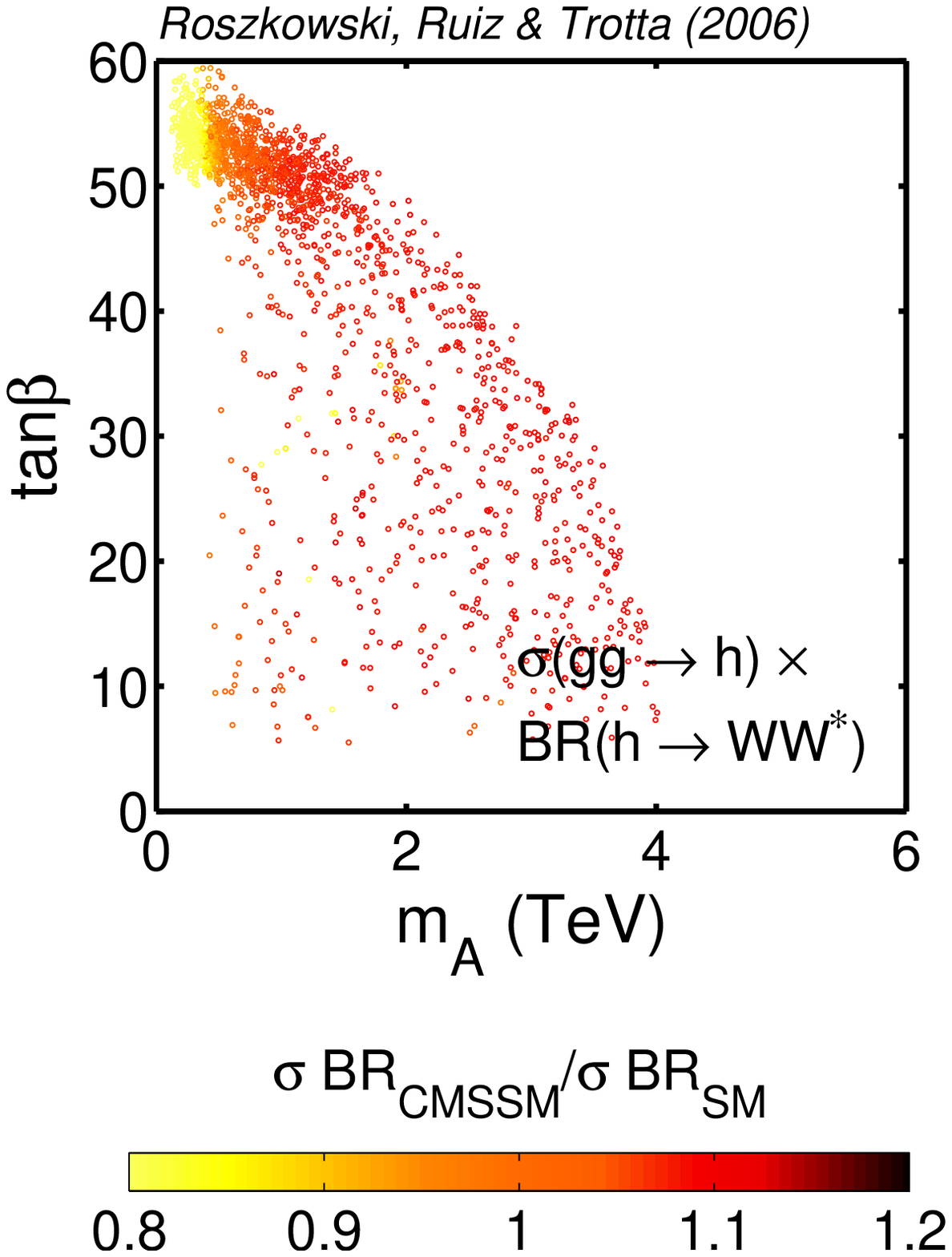}\\
	\includegraphics[width=0.3\textwidth]{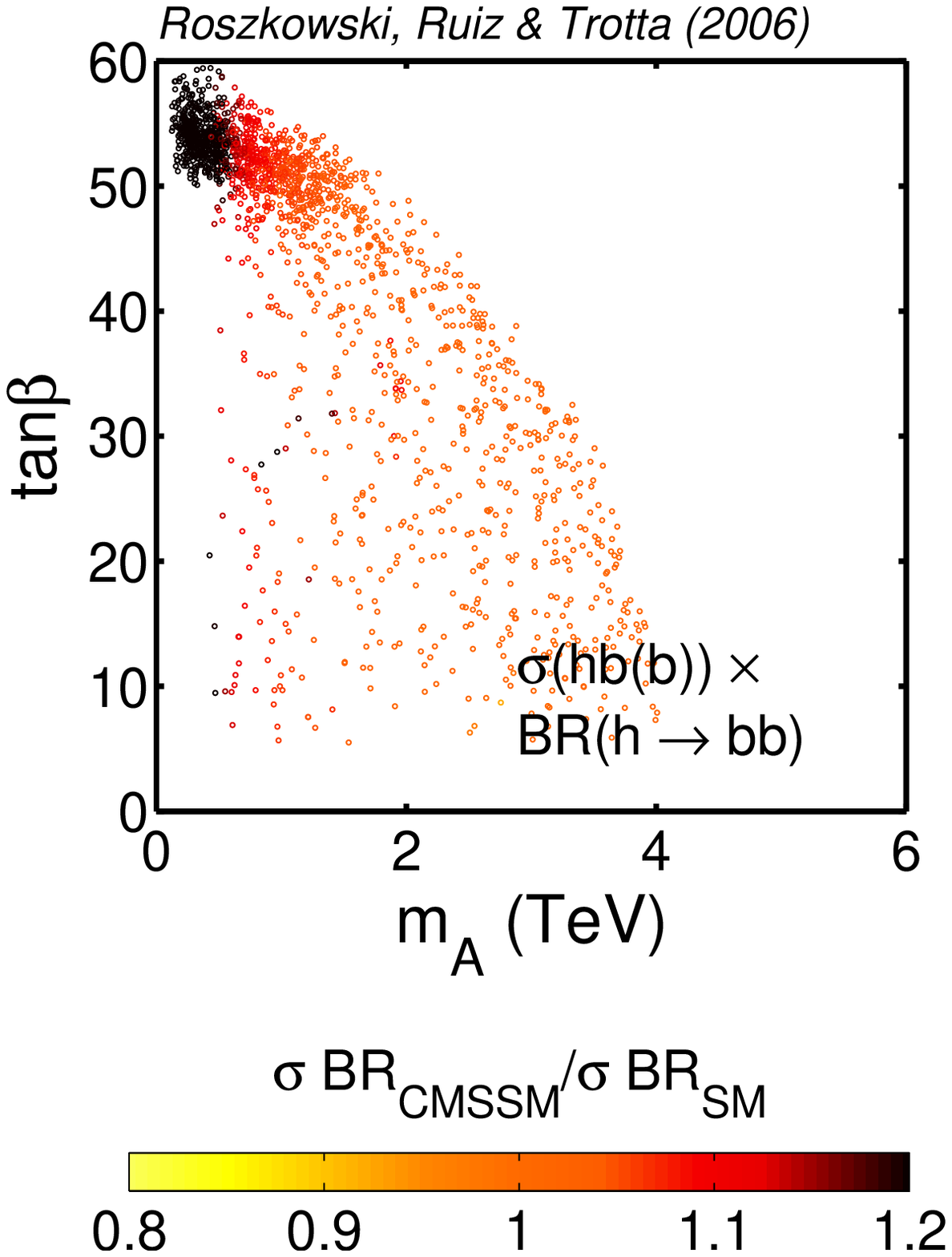}
&	\includegraphics[width=0.3\textwidth]{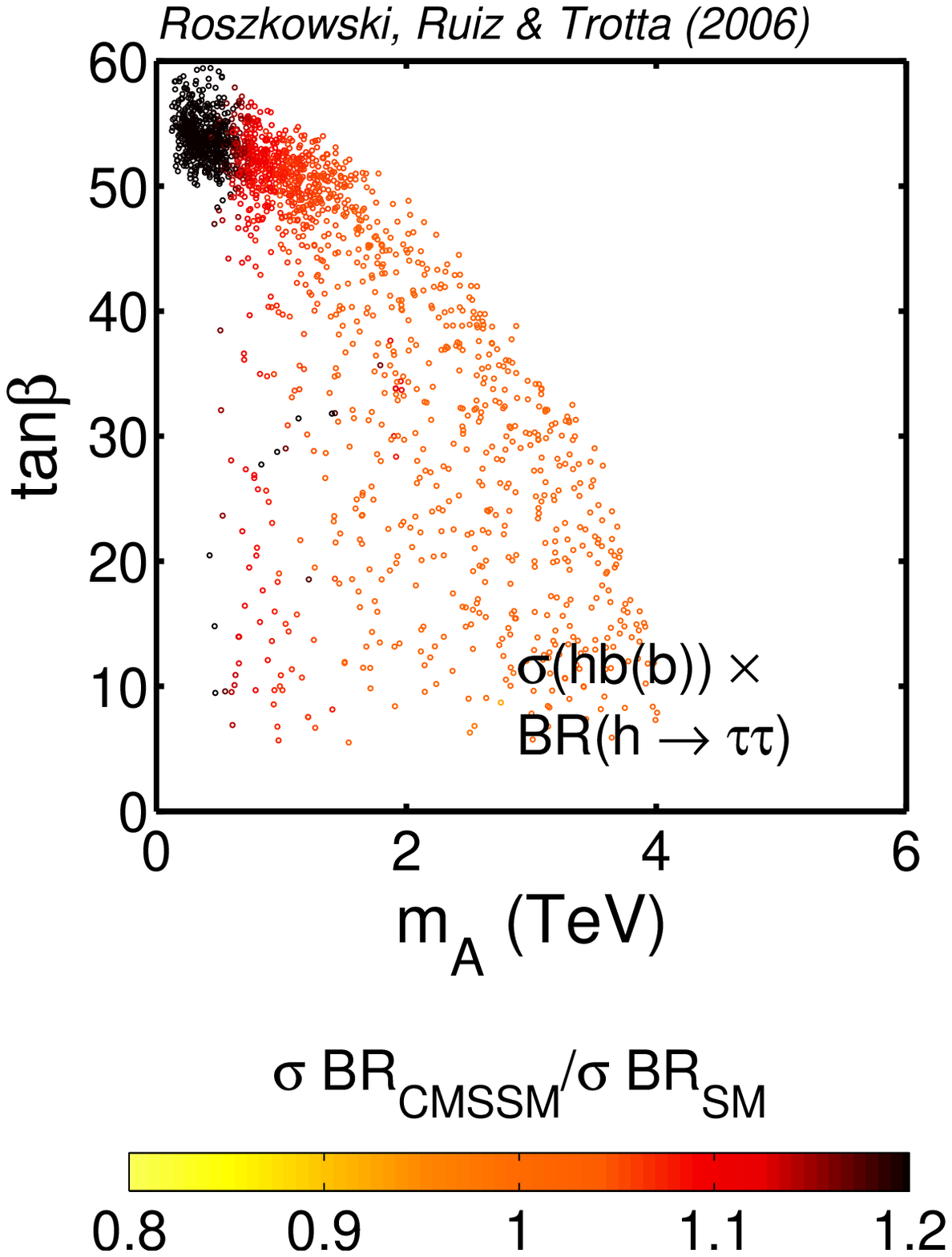}
&	\includegraphics[width=0.3\textwidth]{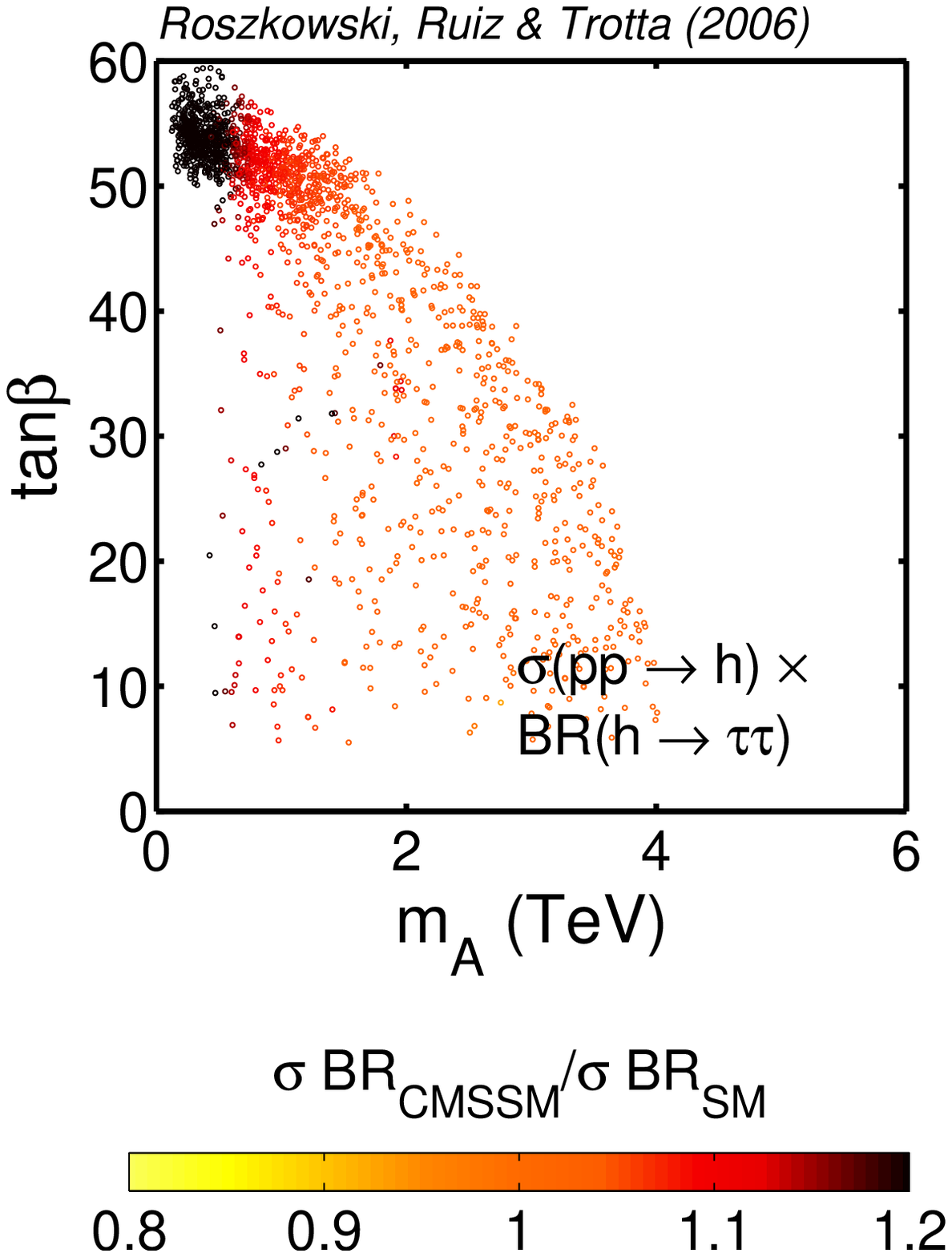}
\end{tabular}
\end{center}
\caption{Values in the $(\mha,\tanb)$ plane of production cross
section times the branching ratio, normalized to their SM values, for
the main light CMSSM Higgs production and decay channels at the
Tevatron.
\label{rrt2:3d-mhlsigmabr-mhavstanb}
}
\end{figure}

We now combine the above results for the light Higgs production cross
sections and decay branching ratios at the Tevatron. In
fig.~\ref{fig:rrt2-hlsigmadecay} we show 1--dim relative probability
densities for SM--normalized light Higgs production cross section
times decay branching ratio $(\sigma\times BR)_{\text{CMSSM}}
/(\sigma\times BR)_{\text{SM}}$, while the ratio's dependence on
$\mhl$ is displayed in fig.~\ref{fig:rrt2-hlsigmabrvsmhl}, on $\tanb$
in fig.~\ref{fig:rrt2-hlsigmabrvstanb}, and on $\mha$ in
fig.~\ref{fig:rrt2-hlsigmabrvsmha}.  In
fig.~\ref{rrt2:3d-mhlsigmabr-mhlvstanb} we show a distribution of
values of the above product in the plane spanned by $\mhl$ and
$\tanb$, while in fig.~\ref{rrt2:3d-mhlsigmabr-mhavstanb} the same
quantities are shown in the usual plane of $\mha$ and $\tanb$.  As
before, all parameters other than the ones shown in each figure have
been marginalized over.

The emerging picture is rather clear. As expected, in the CMSSM, for all
the considered processes, we generally find very similar light Higgs
search prospects as for the SM Higgs boson with the same mass. We note,
however, some differences, which may help optimize search strategies. To
start with, in the vector boson bremsstrahlung process $V^\ast\to
V\hl$ ($V=Z,W$), $\hl\to\bbbar, \tauptaum$ modes are almost
indistinguishable from the SM Higgs case with the same mass. This is
caused by the fact that in the CMSSM the coupling $g(\hl
VV)_{\text{CMSSM}}$ is very close to its SM value.  We note, however,
that we do find a slight enhancement of the $\bbbar$ and $\tauptaum$ final states,
which may be of some help in these important search channels. The same
is of course true for all the other processes considered here. On the
other hand, we have found in the CMSSM parameter space some deviations
in the $g(\hl\bbbar)_{\text{CMSSM}}$ coupling from the SM value, which
may significantly change production cross section of all the modes
except $V^\ast\to V\hl$. On the other hand, the $\hl\to WW^\ast$ mode
is typically reduced by up to some 5\% within the 68\% posterior
probability region. Within the 95\% posterior
probability  region, it can actually be significantly reduced, especially in the
region of $\tanb\gsim50$ and $\mha\lsim1\tev$.

We emphasize that the results presented in
figs.~\ref{fig:rrt2-hlsigmadecay}--\ref{rrt2:3d-mhlsigmabr-mhavstanb}
have been derived in the framework of the CMSSM. This should be kept
in mind when comparing them with existing experimental Higgs search
limits. Many of them have been set for specific (e.g., maximal and no
mixing) scenarios and/or choices of parameters in the general MSSM,
which are basically never realized in the CMSSM. Conversely, it would be helpful
to add to experimental results limits applicable to the CMSSM.

\section{Summary and conclusions}\label{sec:summary}

In this work we have performed a global scan of the CMSSM parameter
space by employing a powerful MCMC technique. We have then analyzed
our results for light Higgs properties and discovery prospects at the
Tevatron, mostly in terms of Bayesian statistics, although we have
demonstrated that an alternative mean quality--of--fit analysis can lead
to rather different results. In particular, while the former favors
the light Higgs mass range above the final LEP--II 95\%~\cl, the
latter points more towards values below it for a large part.

The couplings of the light Higgs of the CMSSM to vector bosons and
bottoms and taus are basically very similar to those of the SM Higgs boson
with the same mass. Small enhancements, at the level of a few per cent,
have been found in most of the CMSSM parameter space, although at
large $\tanb\gsim50$ and $\mha\lsim1\tev$ (the  preferred region),
the differences can be substantial. Our intention was to provide
experimentalists involved in Higgs searches at
the Tevatron detailed information about detection prospects of the CMSSM light Higgs
boson. Despite the fact that the pdf's for associate bottom production and
inclusive Higgs production modes are basically indistinguishable, we
have displayed them separately for the sake of completeness and
convenience.

At the Tevatron, the sensitivity to a SM--like Higgs boson in the mass range
of up to some $122\gev$ (compare~\eqref{eq:lighthiggsmassrange}) seems
excellent. According to ref.~\cite{tev-higgs-sensitivity-study01+03},
with about $2\fb^{-1}$ of integrated luminosity per experiment
(expected by the end of 2006), a 95\%~\cl\ exclusion limit can be set
for the whole 95\% posterior probability light Higgs mass range given ***
in~\eqref{eq:lighthiggsmassrange} ****. 
We stress again that this conclusion
depends on our assumed ranges of flat priors, especially on
$\mzero<4\tev$, as discussed in the text. (Extending the range of
$\mzero$ up to $8\tev$, and accordingly $\mhl$ up to $\lsim 125.6
\gev$ (95\%~\cl),  would require instead about $2.5\fb^{-1}$ of
integrated luminosity per experiment, which is again well within
Tevatron's reach.) 
While keeping this in mind, we still find it remarkable
that negative Higgs searches at the Tevatron should allow one to make
definitive conclusions about the ranges of CMSSM parameters, in
particular $\mzero$, which extend well beyond the reach of even the LHC in direct
searches for superpartners.

Should a signal (hopefully) start being seen in this mass
range, a $3\sigma$ evidence ($5\sigma$ discovery) can be claimed with
about $4\fb^{-1}$ ($12\fb^{-1}$) per experiment.  On the other hand,
with about $8\fb^{-1}$ of integrated luminosity ultimately expected
per experiment, a $5\sigma$ discovery will be possible, should the
light Higgs mass be around $115\gev$. While such low mass is just
below the 68\% region of posterior probability according to Bayesian
statistics, it is actually favored by an alternative mean
quality--of--fit analysis. In other words, according to this measure,
light Higgs search appears more promising than in the more
conservative Bayesian probability scenario.  A conclusive search for
the CMSSM light Higgs boson at the Tevatron seems therefore fully
feasible.


{\bf Acknowledgements} \\ L.R. thanks the Theoretical Astrophysics
Group at Fermilab and the CERN Theory Division for hospitality and
support during his visits when some of the work was done. L.R. is
grateful to P.~Slavich, A.~Sopczak and D.~Tovey for helpful comments,
and to A.~Anastassov, G.~Bernardi, T.~Junk, J.~Nachtman, S.M.~Wang and
T.~Wright for their clarifications regarding Higgs searches at the
Tevatron.  R.RdA. thanks  S.~Heinemeyer and T.~Hahn for their
help with FeynHiggs and A.L.~Read for providing LEP--II Higgs mass
bounds data. We thank M.~Misiak for providing us with his code and
helpful information. R.T. thanks L.~Lyons and G.~Nicholls for useful discussions.
 R.RdA. is supported by the program ``Juan de la Cierva''
of the Ministerio de Educaci\'{o}n y Ciencia of Spain. R.T.\ is
supported by the Royal Astronomical Society through the Norman Lockyer
Fellowship. R.T. thanks the Galileo Galilei Institute for Theoretical
Physics for hospitality during the completion of this work and the
INFN for partial support.  We acknowledge support from ENTApP, part of
ILIAS, contract number RII3-CT-2004-506222. L.R. acknowledges support
from the EC 6th Framework Programme MRTN-CT-2004-503369.  The use of
the Glamdring cluster of Oxford University and the HEP cluster of
Sheffield University are gratefully acknowledged. Parts of our
numerical code are based on the publicly available package
\texttt{cosmomc}.\footnote{Available from \texttt{cosmologist.info}.}


\end{document}